\documentclass[preprint,10pt]{elsarticle}

\usepackage[letterpaper,left=1.5cm,right=1.5cm,top=3cm,bottom=3cm]{geometry}

\usepackage{lineno}
\modulolinenumbers[5]

\usepackage{hyperref} 
\hypersetup{ 
  colorlinks   = true, 
  urlcolor     = blue, 
  linkcolor    = blue, 
  citecolor    = blue, 
} 
\usepackage[fleqn]{amsmath}
\usepackage{amsfonts}
\usepackage{amssymb}
\usepackage{amsthm}

\usepackage{pifont}
\usepackage{natbib}
\usepackage{geometry}
\usepackage{graphicx}
\usepackage{txfonts}

\usepackage{setspace}
\usepackage[utf8]{inputenc}
\usepackage{caption}
\usepackage{subcaption}
\graphicspath{ {images/} }
\usepackage[usenames, dvipsnames]{color}
\usepackage{color,soul}
\setstcolor{red}
\usepackage[dvipsnames]{xcolor}

\journal{Journal of Computational Physics}

\biboptions{authoryear} 

\begin{document}

\begin{frontmatter}

\title{Dynamics of a vortex column of supercritical fluid across the pseudo-boiling line}

\author{Jordi Poblador-Ibanez\fnref{myfootnote1}\corref{mycorrespondingauthor}}
\ead{J.PobladorIbanez@tudelft.nl}
\address{Delft University of Technology, Delft, 2628 CD, The Netherlands}
\author{Fazle Hussain\fnref{myfootnote4}}
\address{Texas Tech University, Lubbock, TX 79409, United States}
\fntext[myfootnote1]{Postdoctoral Researcher, Department of Maritime and Transport Technology.}
\fntext[myfootnote4]{Professor, Department of Mechanical Engineering, Texas Tech University.}


\cortext[mycorrespondingauthor]{Corresponding author}


\begin{abstract}
The evolution of an axisymmetric vortex column in a weakly compressible supercritical fluid is analysed. A thermal layer is imposed to radially stratify the fluid and uncover effects of the large fluid property variations across the pseudo-boiling line. A multi-dimensional flow solver based on a low-Mach approximation is employed. Using supercritical carbon dioxide as the fluid, we examine axisymmetric configurations at low Reynolds number with the vortex core hotter or colder than the surrounding fluid and for different thermodynamic pressures close to the critical pressure. Vorticity evolution depends strongly on the core temperature and ambient pressure, differing substantially from the classical Oseen solution during the thermal mixing process under highly varying fluid properties. Viscous effects dominate the vorticity evolution. Beyond diffusion, three additional viscous mechanisms are identified, which become significant across the pseudo-boiling line: (1) a vorticity stretching term, (2) an alignment of vorticity and viscosity/density gradients, and (3) a vorticity source due to the interaction between the fluid swirl and the viscosity and density gradients. The first two mechanisms alter existing vorticity, while the latter injects new vorticity. In fact, the third mechanism can generate reverse vorticity, locally increasing circulation and substantially modifying the temporal evolution of the vortex.
\end{abstract}


\end{frontmatter}

\setlength\abovedisplayshortskip{0pt}
\setlength\belowdisplayshortskip{-5pt}
\setlength\abovedisplayskip{-5pt}


\section{Introduction} 
\label{sec:introduction}

Supercritical fluids (SCF) feature notably in the green energy transition because of their unique thermophysical and flow properties. Relevant applications include thermal management using supercritical carbon dioxide (sCO\(_2\)) for the cooling of microchips \citep{2021_JSF_Saeed} or as a working fluid in heat pumps \citep{2022_ER_Song}, sCO\(_2\) power cycles \citep{2020_SNAS_Marchionni}, and supercritical water nuclear reactors \citep{2022_PNE_Wu}. Moreover, SCF are also found in space propulsion with high operating pressures in combustion chambers of rocket engines \citep{2021_PECS_Jofre}. In each of such applications, turbulence is key to the efficiency of the heat transfer or fuel-oxidizer mixing processes. Consequently, a thorough understanding of the fundamental mechanisms governing turbulent mixing in SCF is essential. \par

Turbulence is rotational, and vorticity dynamics are central to the energy transfer across various length and time scales. Vorticity stretching is crucial for generating extreme events at small scales in turbulent flows; thus, \citet{2020_PRF_Buaria} quantified the relationship between vortex stretching and enstrophy (vorticity squared) production, whereas \citet{2020_PRF_Johnson,2021_JFM_Johnson} demonstrated that strain self-amplification may contribute more to the energy transfer across scales than vorticity stretching. More recently, \citet{2022_ARFM_Yao} reviewed the relationship between vortex reconnection in viscous fluids and the turbulence cascade, showing that numerous small-scale structures are successively generated in such events, leading to a local increase in enstrophy. The long-standing related question of finite-time singularity of the Navier-Stokes equations continues to haunt turbulence researchers \citep{2019_JFM_Moffatt,2020_JFM_Moffatt,2020b_JFM_Yao,2023_FCM_Hou} who have often taken recourse to vortex stretching to attempt to answer this puzzle. A new mechanism of vorticity generation, the pushover, was discovered by \citet{2002_JFM_Schoppa} to be the reason for destabilising stable turbulent profiles causing transient growth of the near-wall streaks. Vortex stretching and tilting have been discovered to be responsible for the enigmatic phenomena of negative production, i.e., the sustained transfer of kinetic energy from turbulent to the mean flow in sizeable flow domains \citep{1980_JFM_Zaman,2024_JFM_Hussain,2025_JFM_Garcia}. \par

\begin{figure}
\centering
\begin{subfigure}{0.5\textwidth}
  \centering
  \includegraphics[width=0.9\linewidth]{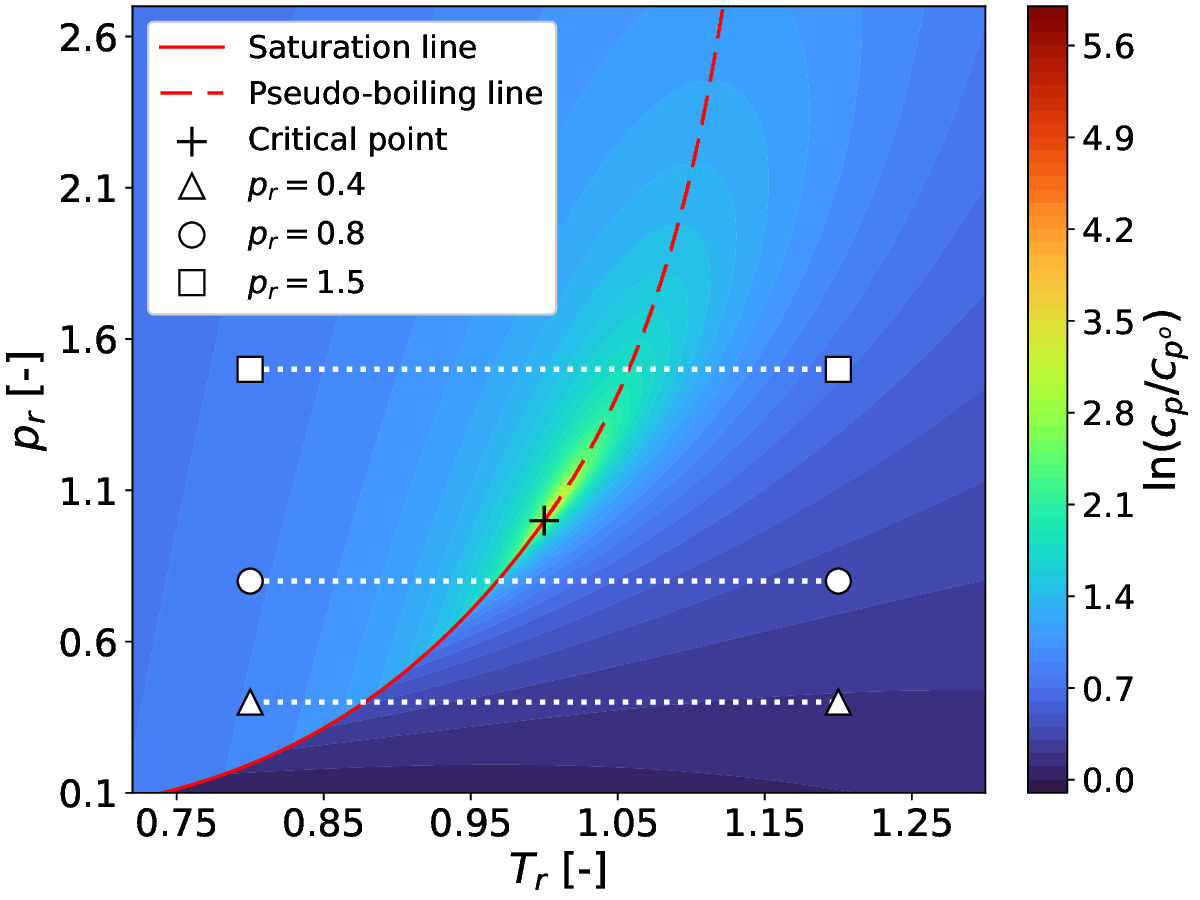}
  \caption{} 
  \label{subfig:cp_CO2}
\end{subfigure}%
\begin{subfigure}{0.5\textwidth}
  \centering
  \includegraphics[width=0.9\linewidth]{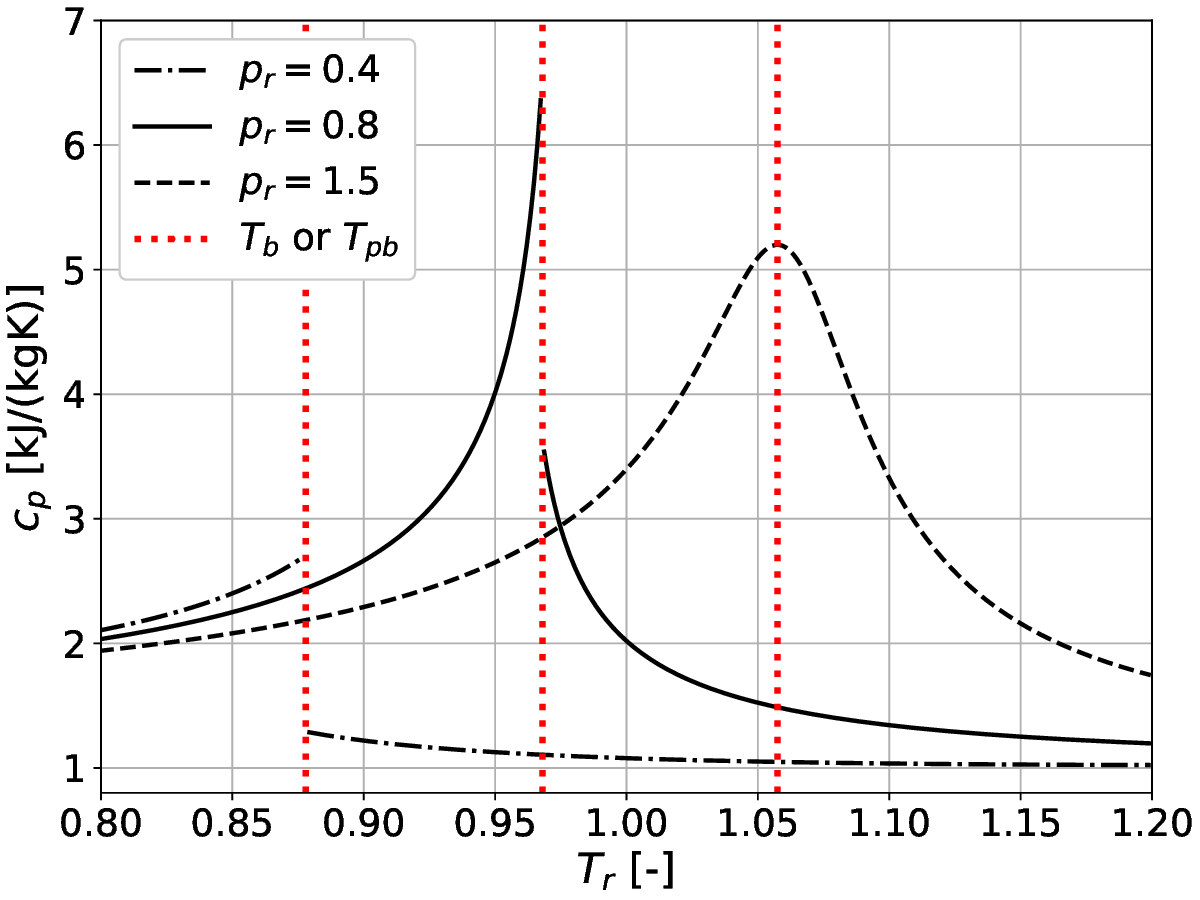}
  \caption{} 
  \label{subfig:cp_CO2_pressures}
\end{subfigure}%
\caption{Map of \(c_p\) for CO\(_2\) as a function of \(p_r\) and \(T_r\). The real-fluid model described in section \ref{sec:setup} is used to obtain \(c_p\). (a) contour plot of \(\ln(c_p/c_{p^o})\) showing the saturation line, pseudo-boiling line and critical point. \(c_{p^o}=846\) J/(kgK); and (b) \(c_p\) distributions at different \(p_r\). The (pseudo-)boiling temperatures \(T_b\) and \(T_{pb}\) are shown.}
\label{fig:Fig1}
\end{figure}

However, these works focus on incompressible flows; a lack of fundamental understanding persists in flows with highly varying properties, such as SCF undergoing thermal mixing, where compressibility effects are dominant even at low Mach numbers and the fluid can show features of phase transition, i.e., pseudo-boiling, across the Widom line or pseudo-boiling line \citep{2015_JSF_Banuti,2017_PRE_Banuti}. This is visualised in figure \ref{fig:Fig1} where the specific heat, \(c_p\), of CO\(_2\) is shown as a function of reduced pressure (\(p_r=p/p_c\)) and reduced temperature (\(T_r=T/T_c\)). \(p_c\) and \(T_c\) are the critical pressure and critical temperature, respectively. Figure \ref{subfig:cp_CO2} depicts \(\ln(c_p/c_{p^o})\) due to the extreme \(c_p\) values near the critical point (\(p_r=1\), \(T_r=1\)), where \(c_{p^o}\) is a reference specific heat for gaseous CO\(_2\) given by 846 J/(kgK) at 298 K and 1 atm. Along the saturation line for subcritical conditions (\(p_r<1\), \(T_r<1\)), discontinuities in fluid properties occur as two distinct phases (vapour and liquid) co-exist, leading to the classical boiling phenomena with a distinct fluid interface. Closer to the critical point, the properties of each phase resemble each other more closely and, for supercritical states (\(p_r>1\), \(T_r>1\)), a continuous single-phase transition from liquid-like to gas-like properties occurs. As such, this transition is referred to as pseudo-boiling and is identified by the pseudo-boiling line, defined here as the temperature where the specific heat is maximised at a given pressure \(p_r>1\) \citep{2015_JSF_Banuti}. This is shown in figure \ref{subfig:cp_CO2_pressures} where the \(c_p\) distribution as a function of \(T_r\) is plotted for \(p_r=0.4\), \(0.8\) and \(1.5\). At sufficiently high pressures, i.e., \(p_r>3\), pseudo-boiling effects are mitigated and eventually vanish as much smoother variations of fluid properties occur. \par 

These challenging thermodynamic conditions render the experimental extraction of detailed flow information difficult. In the context of cryogenic liquid jet injection into a hotter gaseous environment, one observes that the distinction between liquid and gas is lost at supercritical conditions \citep{1998_JPP_Mayer,2006_CST_Oschwald,2012_IJAE_Chehroudi}. Surface tension vanishes and turbulent mixing resembles that of single-phase flows but in a region of strong density gradients. Experimental techniques relying on optical access, e.g., high-speed imaging or shadowgraphy, struggle to capture such turbulent mixing regions due to the increased optical distortions. Similarly, the same behaviour is observed when injecting liquid hydrocarbon fuels into air, including now the effects of vapour-liquid equilibrium and species mixing between two very different fluids \citep{2016_Fuel_Falgout}. Given these limitations, researchers have relied on numerical campaigns to analyse the details of the structure of turbulent flows in SCF and the critical effects of the large variations of fluid properties across the pseudo-boiling line in a broad range of configurations covering turbulent jets \citep{2018_PoF_Lapenna,2019_PoF_Lagarza}, wall-bounded flows \citep{2018_PoF_Azih,2018_PRF_Ma,2025_JFM_Wan,2026_PoF_Barea}, boundary layers \citep{2019_JFM_Kawai} and homogeneous isotropic turbulence \citep{2026_PRF_Martin}. In particular, it has been shown that localised baroclinic torques become important generators of flow rotation and instability under supercritical conditions \citep{2018_PoF_Azih,2023_PoF_Bernades,2023_PRF_Barea,2024_JFM_Poblador}, directly affecting the turbulent flow. \par 

Taking a more fundamental approach, this study focuses on a canonical axisymmetric vortex column to obtain insights into the vorticity dynamics of SCF. Since vortex columns emerge from the roll-up of vortex sheets or during the separation of boundary layers, scenarios where a temperature difference exists across the sheet are considered. As a result, such vortex columns may exhibit a core temperature different from that of the surrounding fluid. For example, Fig. 2 in \citet{2026_PoF_Barea} clearly shows localised pockets of colder or hotter fluid in a differentially heated turbulent channel flow of sCO2, presumably driven by vortical structures. Although these vortices are subject to additional mechanisms in three-dimensional (3D) flows, e.g., alignment with the flow direction resulting in an axial velocity component, studying the dynamics of an idealized two-dimensional (2D) vortex column provides relevant insights into the physical mechanisms determining the evolution of vorticity in SCF. In particular, it allows us to isolate the effects of compressible vortex stretching or viscous stresses from baroclinicity, which may quickly destabilise vortices in SCF and conceal other vorticity production mechanisms. For example, \citet{2025_IJHMT_Wan} show that the effects of viscosity gradients in the vorticity budget in a differentially-heated turbulent channel supercritical flow is an order of magnitude lower than the baroclinic torque. However, when accounting for viscous diffusion, the vorticity budget of viscous terms can be larger than baroclinicity \citep{2023_PRF_Barea,2025_PoF_ElMansy}. Yet, viscosity fluctuations remain important \citep{2019_JFM_Kawai}, meaning viscosity gradients can strongly affect the vortical structure of SCF. \par

For this purpose, an axisymmetric weakly-compressible unbounded flow at supercritical conditions based on a 2D Oseen-like vortex column, i.e., without an axial velocity, and radially stratified via a temperature distribution is studied using the multi-dimensional flow solver described in \citet{2022_PoF_Poblador}. As such, the evolution of all flow quantities is expected to occur only in the radial direction, i.e., the baroclinic torque vanishes. Axisymmetric solutions of this kind of vortex columns in the limit of low-Mach number, relatively high Reynolds number \(Re \sim \mathcal{O}(10^4)\), and under the perfect gas assumption have already been studied, e.g., \citet{1991_JFM_Colonius} and \citet{2000_JFM_Ellenrieder}. However, these works omit the large variations in fluid properties across the pseudo-boiling line characteristic of SCF \citep{2022_JHT_Bernades}. \par 

The study of such supercritical flow with a multi-dimensional solver requires meticulous attention. In multi-dimensional solvers, the computational domain's shape, size, and boundary conditions (BC) can affect the evolution of axisymmetric flows. These issues are well documented in the literature \citep{2004_JFM_Pradeep,2014_JCP_Dong}. Typically, sufficiently large domains are imposed to mimic unbounded flow conditions, limiting non-physical instabilities. Here, the highly varying fluid properties may exacerbate this numerical problem, breaking flow symmetry more easily. Particularly, sharp density gradients occur and the baroclinic torque can become notably destabilising even at low \(Re\) if density and pressure gradients are not exactly aligned. Note that our work aims at producing the strictly axisymmetric solution of the vortex numerically. Instability analyses brought about by stratification in 3D domains with pre-imposed perturbations are addressed in, e.g., \citet{1996_PoF_Mahesh} or \citet{2005_JFM_Sipp}. Directly applied to SCF and the effects of large gradients of fluid properties across the pseudo-boiling line, \citet{2024_JFM_Bugeat} study the emergence of instabilities in stratified Couette flows and \citet{2019_JFM_Ren} discuss the instability mechanisms in boundary layers. \par

This paper is structured as follows. Section \ref{sec:setup} describes the governing equations, the thermodynamic model, the numerical method, and the computational setup. Next, section \ref{subsec:initialisation_sCO2} introduces the thermodynamic states and the corresponding properties of the working fluid, i.e., CO\(_2\); section \ref{subsec:solution_collapse} verifies the axisymmetric behaviour of the numerical solution; section \ref{subsec:cold_vs_hot} compares and quantifies various vortex-related quantities for different thermodynamic pressure and temperature distributions across the vortex column; section \ref{subsec:other_variables} describes the behaviour of the SCF as it undergoes pseudo-boiling and pseudo-condensation; and section \ref{subsec:vorticity_analysis} details the mechanisms of vorticity transport, generation and diffusion under such fluid behaviour with emphasis on viscous mechanisms and the strengthening of pseudo-boiling effects as the fluid approaches its critical point. Lastly, section \ref{sec:conclusions} summarises the main findings. \par

\section{Methodology and Computational Setup}
\label{sec:setup}

The governing equations for a weakly compressible SCF are the mass (\ref{eqn:cont}), momentum (\ref{eqn:mom}), and energy (\ref{eqn:energy}) equations

\begin{equation}
\label{eqn:cont}
\frac{\partial \rho}{\partial t} + \boldsymbol{\nabla} \boldsymbol{\cdot} (\rho\boldsymbol{u})=0
\end{equation}
\begin{equation}
\label{eqn:mom}
\frac{\partial }{\partial t}(\rho\boldsymbol{u}) + \boldsymbol{\nabla} \boldsymbol{\cdot} (\rho\boldsymbol{uu}) = -\boldsymbol{\nabla} p + \boldsymbol{\nabla} \boldsymbol{\cdot}\boldsymbol{\tau}
\end{equation}
\begin{equation}
\label{eqn:energy}
\frac{\partial }{\partial t}(\rho h) + \boldsymbol{\nabla} \boldsymbol{\cdot} (\rho h\boldsymbol{u}) = \boldsymbol{\nabla} \boldsymbol{\cdot} \bigg(\frac{k}{c_p}\boldsymbol{\nabla} h \bigg)
\end{equation}

\noindent
where \(\rho\), \(\boldsymbol{u}(=u,v)\), \(p\), \(h\), \(k\) and \(c_p\) are density, velocity, pressure, enthalpy, thermal conductivity, and isobaric specific heat, respectively. The viscous stress tensor is \(\boldsymbol{\tau}=\mu [\boldsymbol{\nabla}\boldsymbol{u}+\boldsymbol{\nabla}\boldsymbol{u}^\text{T}-\frac{2}{3}(\boldsymbol{\nabla}\boldsymbol{\cdot}\boldsymbol{u})\boldsymbol{I}]\), with \(\mu\) being the dynamic viscosity. Note that bulk viscosity, \(\zeta\), is neglected in (\ref{eqn:mom}), i.e., Stokes hypothesis \(\zeta=\lambda+\frac{2}{3}\mu\approx0\) where \(\lambda\) is the second coefficient of viscosity. Moreover, the viscous dissipation and pressure terms are neglected in (\ref{eqn:energy}) due to the low velocities involved in this work.  \par

The flow solver described in \citet{2022_PoF_Poblador} is used without its multi-component and multiphase features. It contains the necessary framework to model SCF and can be readily used for more complex vortex studies. A real-fluid model (RFM) based on a volume-translated Soave-Redlich-Kwong equation of state \citep{2006_IECR_Lin} is integrated to obtain the fluid properties under high-pressure conditions. \(h\) and \(c_p\) are evaluated with departure functions from the ideal state \citep{2001_Poling}, while \(\mu\) and \(k\) are obtained from the generalized correlation given by \citet{1988_IECR_Chung}. The solver considers a weakly compressible formulation where pressure and density are uncoupled under the assumption that variations in the dynamic pressure driving the flow are negligible compared to the high ambient pressure in SCF. Thus, density becomes a function of the thermodynamic ambient pressure \(p_0\) and temperature \(T\). A conservative formulation of momentum (\ref{eqn:mom}) and energy (\ref{eqn:energy}) is implemented numerically in which both equations are spatially discretized with a finite-volume method using second-order central differences and integrated in time with a second-order Adams-Bashforth scheme. The pressure-velocity coupling is addressed with the predictor-projection method using the split pressure gradient technique to handle the variable density in the pressure Poisson equation (PPE) and use a fast direct solver \citep{2014_JCP_Dodd,2018_CMA_Costa}. \par

The computational domain consists of a 2D box of size \([-L/2,L/2]\times[-L/2,L/2]\) m with the Oseen-like vortex centred at \(\boldsymbol{x_0}=(0,0)\) m. The vortex is initialised with a compact support radial vorticity profile given by (\ref{eqn:initialvorticity}), which differs from a Gaussian distribution by spatially localizing the vorticity \citep{1988_CTR_Melander}. As a result, physically consistent dynamics are better captured, e.g., the reconnection of two antiparallel vortices only occurs once the compact vortex cores touch each other \citep{2020_JFM_Yao,2022_ARFM_Yao}. In an effort to maintain consistency with the vortex parameters detailed in \citet{2020_JFM_Yao}, the vortex initial core radius is \(r_c=2/3\) m and \(g(r^+)=\exp[-(K/r^+)\exp(1/(r^+-1))]\), where \(r^+=r/r_c\), \(r=||\boldsymbol{x}-\boldsymbol{x_0}||\), and \(K=0.5\exp(2)\log(2)\). Note that in the compact vorticity distribution, the vortex core radius has been defined as the distance from the vortex centre that contains all vorticity.

\begin{equation}
\label{eqn:initialvorticity}
\omega_z(r)=
    \begin{cases}
        \omega_0[1-g(r^+)] & \text{if $r\leq r_c$}\\
        0 & \text{if $r>r_c$}\\
    \end{cases}
\end{equation}

The velocity field is initialised by solving (\ref{eqn:initialvelocity}) with the same direct solver used for the PPE to avoid the costly integration of Biot-Savart equation \citep{1973_CaF_Wu}, which is especially time-consuming in 3D. Using (\ref{eqn:initialvelocity}), the initialisation assumes \(\boldsymbol{\nabla\cdot u}=0\). This might affect the symmetry breaking of the vortex at high \(Re\) due to sensitivities to grid resolution in the vicinity of sharp gradients of \(\boldsymbol{\omega}\) when solving (\ref{eqn:initialvelocity}) (see \ref{apn:A}). Under stable conditions, \(\boldsymbol{u}\) quickly relaxes to the solution of the weakly compressible governing equations despite the inherent incompressibility assumption of the initial condition. \par 

Lastly, the temperature field is initialised following a hyperbolic tangent profile centred around \(r_c\), given by (\ref{eqn:initialtemperature}), to impose a radial gradient; thus, resulting in an axisymmetric configuration where the vortex column is radially stratified due to the sharp density variations across the pseudo-boiling line. Two possible configurations are considered whether the vortex core is hotter or colder than the surrounding fluid, ensuring that the pseudo-boiling line is crossed in each case. Periodic BC are used in (\ref{eqn:initialvelocity}) to initialise \(\boldsymbol{u}\) following \citet{2004_JFM_Pradeep}, but homogeneous Neumann BC are used for \(p\), \(\boldsymbol{u}\) and \(h\) during the simulation to approximate outflow BC for the compressible flow.

\begin{equation}
\label{eqn:initialvelocity}
\boldsymbol{\nabla^2 u}=-\boldsymbol{\nabla \times \omega}
\end{equation}
\begin{equation}
\label{eqn:initialtemperature}
T(r) = 
\begin{cases}
    T_\text{min}+\frac{1}{2}[T_\text{max}-T_\text{min}][\tanh(10[r-r_c])+1] \quad \text{if cold core} \\
    T_\text{max}-\frac{1}{2}[T_\text{max}-T_\text{min}][\tanh(10[r-r_c])+1] \quad \text{if hot core}
\end{cases}
\end{equation}

\section{Results and Discussion}
\label{sec:results}

\subsection{Axisymmetric Initialisation for Supercritical Carbon Dioxide}
\label{subsec:initialisation_sCO2}

\begin{figure}
\centering
\begin{subfigure}{0.25\textwidth}
  \centering
  \includegraphics[width=1.05\linewidth]{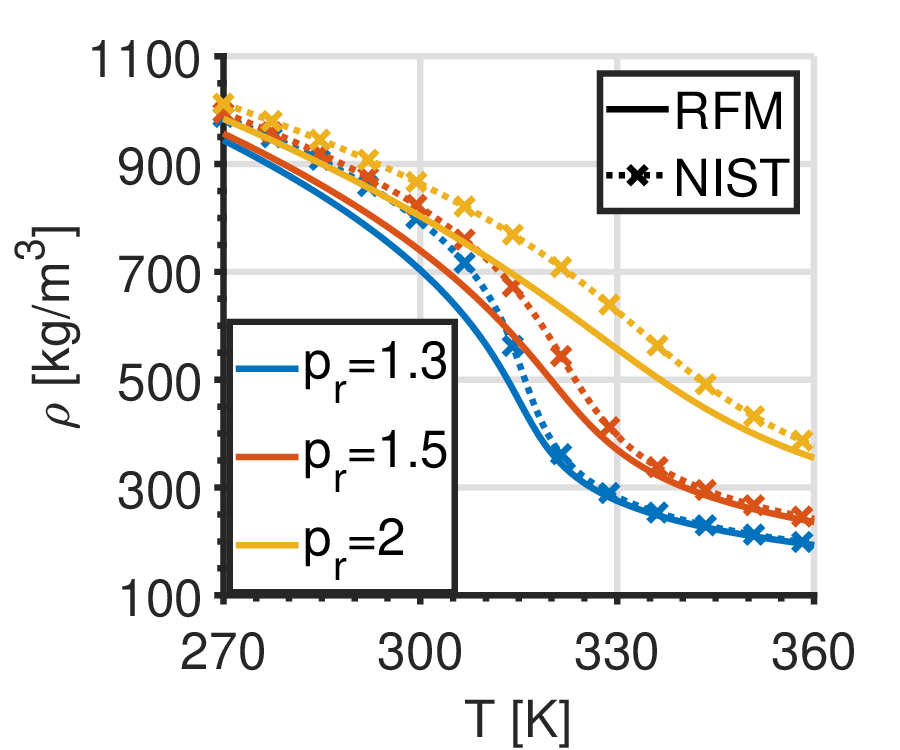}
  \caption{} 
  \label{subfig:validation_den}
\end{subfigure}%
\begin{subfigure}{0.25\textwidth}
  \centering
  \includegraphics[width=1.05\linewidth]{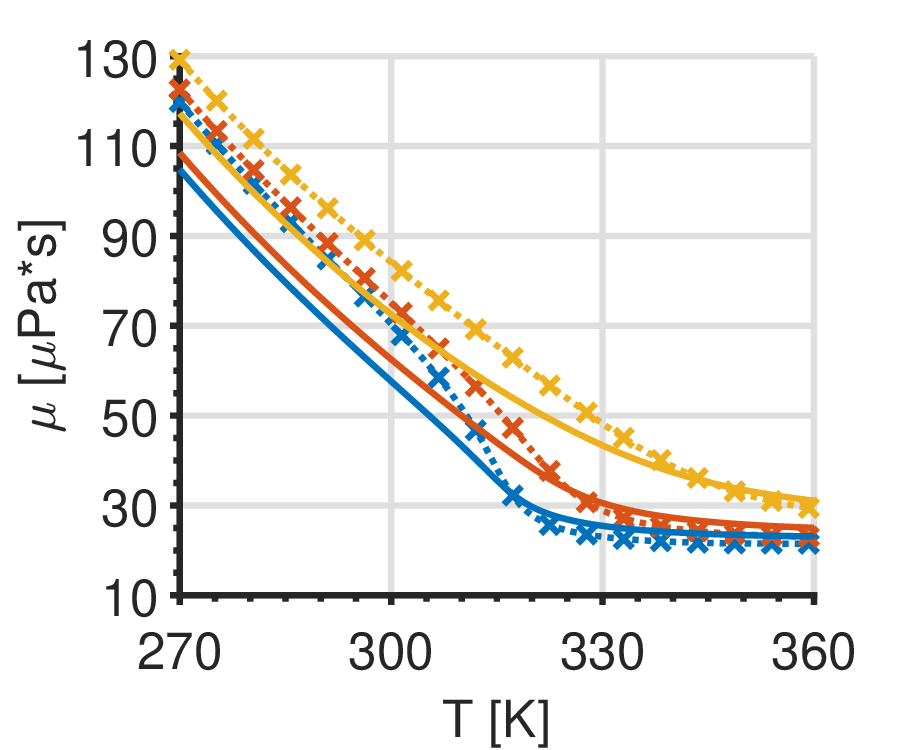}
  \caption{} 
  \label{subfig:validation_vis}
\end{subfigure}%
\begin{subfigure}{0.25\textwidth}
  \centering
  \includegraphics[width=1.05\linewidth]{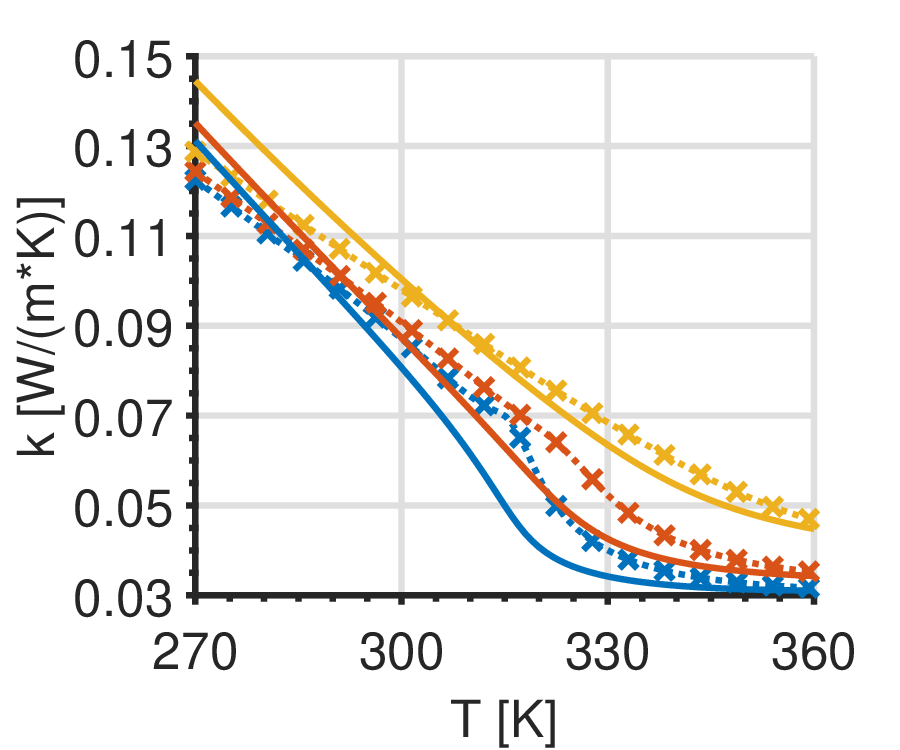}
  \caption{} 
  \label{subfig:validation_cond}
\end{subfigure}%
\begin{subfigure}{0.25\textwidth}
  \centering
  \includegraphics[width=1.05\linewidth]{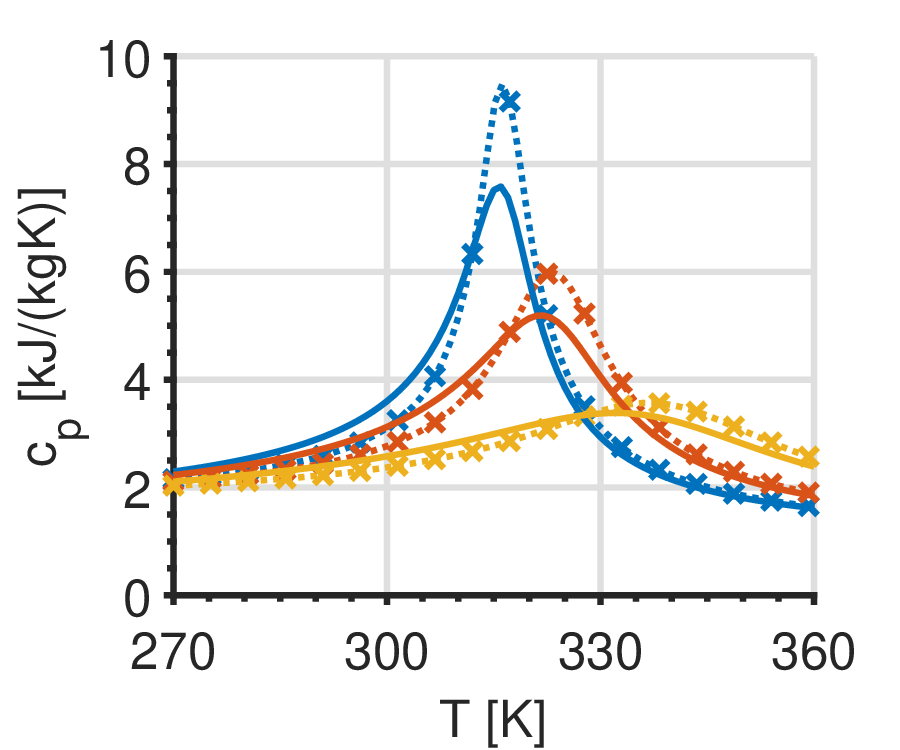}
  \caption{} 
  \label{subfig:validation_cp}
\end{subfigure}%
\caption{Validation of the RFM against NIST data for sCO\(_2\) at different reduced pressures \(p_r\) across the pseudo-boiling line between \(T_\text{min}\) and \(T_\text{max}\). (a) \(\rho\); (b) \(\mu\); (c) \(k\); and (d) \(c_p\).}
\label{fig:Fig2}
\end{figure}

Supercritical carbon dioxide (sCO\(_2\)) is considered as the working fluid. The critical pressure and temperature for CO\(_2\) are \(p_c = 73.78\) bar and \(T_c= 304.13\) K. The RFM under thermodynamic conditions relevant to sCO\(_2\) across the pseudo-boiling line \citep{2024_Energy_Draskic} is validated in figure \(\ref{fig:Fig2}\) against reference data from the National Insitute of Standards and Technology (NIST) for \(\rho\), \(\mu\), \(k\), and \(c_p\) \citep{NIST_Webbook}. Different reduced pressures, i.e., \(p_r=p/p_c\), are considered with temperature ranging between \(T_\text{min}=273.72\) K and \(T_\text{max}=349.75\) K (or a reduced temperature \(T_r=T/T_c\) between 0.9 and 1.15). The RFM captures well the varying fluid properties, but underestimates density and viscosity at the lower temperatures corresponding to the liquid-like state. This uncertainty is accepted in this work. Moreover, the specific heat ratio \(\gamma=c_p/c_v\) behaves similarly to \(c_p\) since \(c_v\) remains approximately constant for the analysed conditions. \par 

\begin{figure}
\centering
\includegraphics[width=0.4\linewidth]{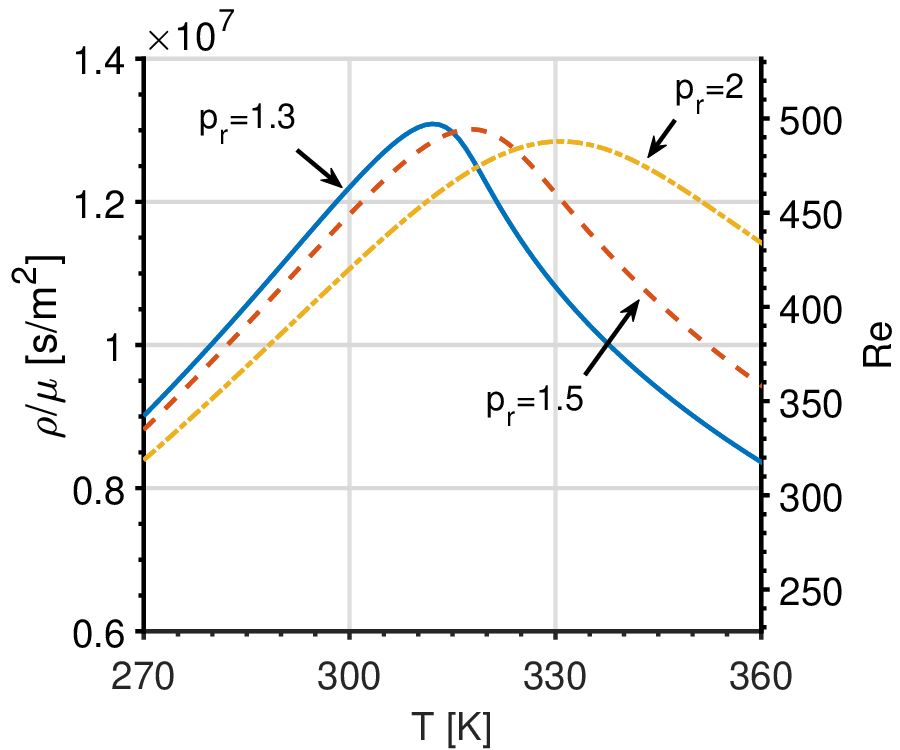}
\caption{Evolution of \(\rho/\mu\) and \(Re\) for sCO\(_2\) at different reduced pressures \(p_r\) across the pseudo-boiling line between \(T_\text{min}\) and \(T_\text{max}\), with \(Re_\text{max}=500\).}
\label{fig:Fig3}
\end{figure}

The Reynolds number is defined as \(Re\equiv(\rho\Gamma_0)/\mu\), where the vortex circulation is \(\Gamma_0=\iint_S \boldsymbol{\omega \cdot} d\boldsymbol{S}\). Since \(\rho/\mu\) varies with temperature, we choose a \(\Gamma_0\) for a given \(Re_\text{max}\) at \(p_r=1.3\). Figure \ref{fig:Fig3} shows \(\rho/\mu\) for different \(p_r\) and the variation of \(Re\) with \(T\) for \(Re_\text{max}=500\). \((\rho/\mu)_\text{max}=1.31657\times 10^{7}\) s/m\textsuperscript{2}, or a minimum in kinematic viscosity, occurs around the pseudo-boiling line and does not vary much for the analysed conditions. Further, \((\rho/\mu)_\text{max}\) shifts to higher temperatures with increasing pressure, following the pseudo-boiling line. Table \ref{tab:initial_vorticity} shows the values of \(\Gamma_0\) and \(\omega_0\) used to initialise the vortex at different \(Re_\text{max}\). \par

Note that the vortex parameters are very small given the size of the vortex in the computations, i.e., \(r_c\), and the properties of sCO\(_2\). Regardless, one could easily scale down the problem maintaining the same \(Re\). For example, vortices defined in the micron scale near the pseudo-boiling line in microconfined turbulence in supercritical differentially-heated channel flows \citep{2023_PoF_Bernades} may be approximated by local vorticity magnitudes based on the following. For typical friction Reynolds numbers between 180 and 5000 and channel thicknesses of \(\mathcal{O}(100\) \(\mu\text{m})\), the friction velocity \(u_\tau\) is of \(\mathcal{O}(0.1-1\) m/s\()\) using \((\rho/\mu)_\text{max}\) for sCO\(_2\). Assuming the vorticity at the micron range scales as \(\omega_0\sim u^+/l\) where \(u^+\) is the velocity fluctuation in the flow of \(\mathcal{O}(1-10\) \(u_\tau)\) and \(l\) is a vortex diameter of \(\mathcal{O}(1-10\) \(\mu\text{m})\), \(\omega_0\) falls within \(\mathcal{O}(10^4-10^7\) s\(^{-1})\). When calculating \(\Gamma_0\) based on a vortex size of \(l\), one obtains \(\Gamma_0\) of \(\mathcal{O}(10^{-8}-10^{-3}\) m\(^2\)/s\()\), or \(Re\) of \(\mathcal{O}(1-10^4)\). This broad range of \(Re\) implies that some vortices rapidly dissipate under viscous diffusion while others are more easily affected by the turbulent flow. More importantly, our target conditions in table \ref{tab:initial_vorticity} fall within the expected range of \(Re\) values in relevant problem setups. In particular, we focus on the axisymmetric evolution of the unperturbed vortex column at low Reynolds numbers (or locally laminar). Additionally, the assumption that the dynamic pressure balancing centrifugal force in the vortex is negligible compared to the thermodynamic pressure holds true when considering scaling. Given the dimensionality of our problem, \(p(r)\ll p_0\) since azimuthal velocities \(u_\theta\) peaks are of \(\mathcal{O}(5\times10^{-6}\) m/s), as shown later, and \(\rho\) is of \(\mathcal{O}(500\) kg/m\textsuperscript{3}), \(\Delta p_{peak}\sim\rho u_{\theta,peak}^2\) is of \(\mathcal{O}(10^{-13}\) bar). This follows the expected scaling to balance centrifugal force. Thus, in the micron scale where \(u^+\) provides an order of magnitude for the azimuthal velocity, \(\Delta p_{peak}\) would need to reach at most values of \(\mathcal{O}(0.5\) bar) -- still at least two orders of magnitude lower than the ambient pressures. \par 

\begin{figure}
\centering
\begin{subfigure}{0.33\textwidth}
  \centering
  \includegraphics[width=1.0\linewidth]{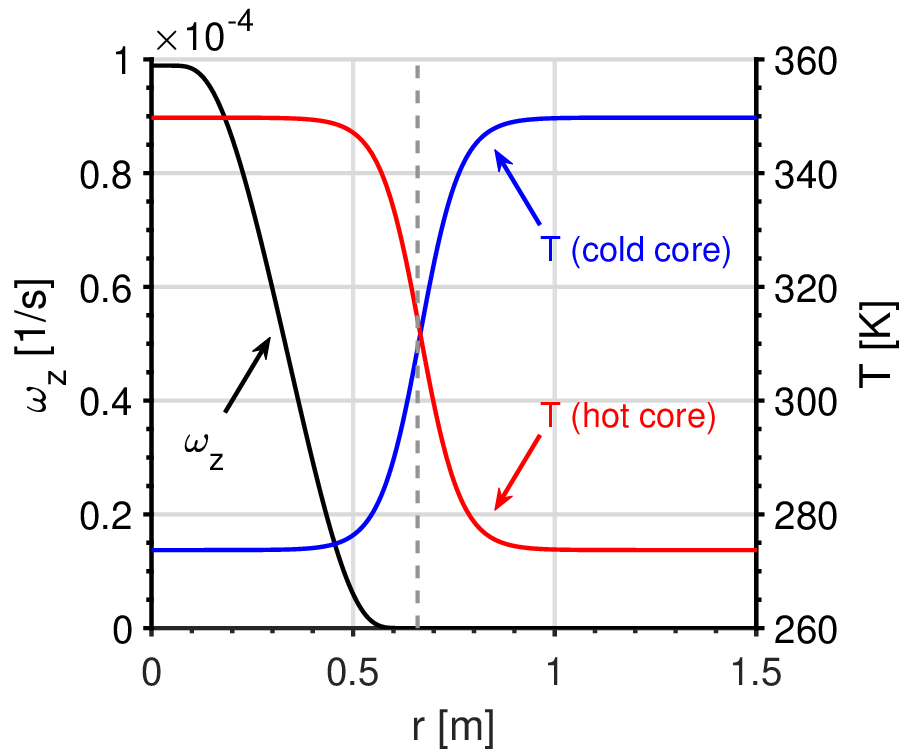}
  \caption{} 
  \label{subfig:initial_vor_temp_all}
\end{subfigure}%
\begin{subfigure}{0.33\textwidth}
  \centering
  \includegraphics[width=1.0\linewidth]{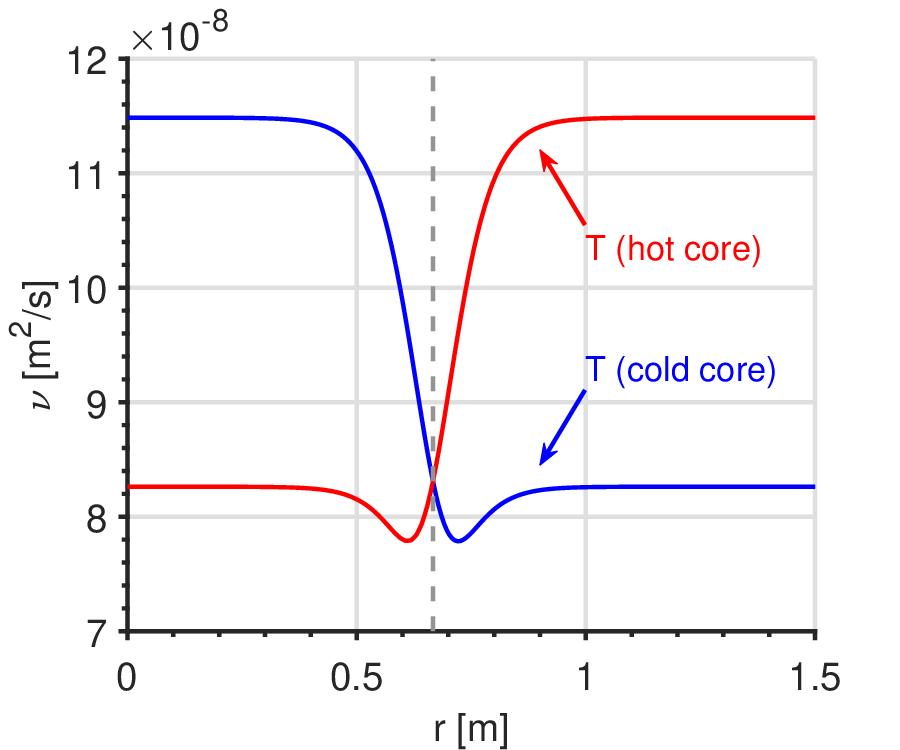}
  \caption{} 
  \label{subfig:initial_kinvis_all}
\end{subfigure}%
\begin{subfigure}{0.33\textwidth}
  \centering
  \includegraphics[width=1.0\linewidth]{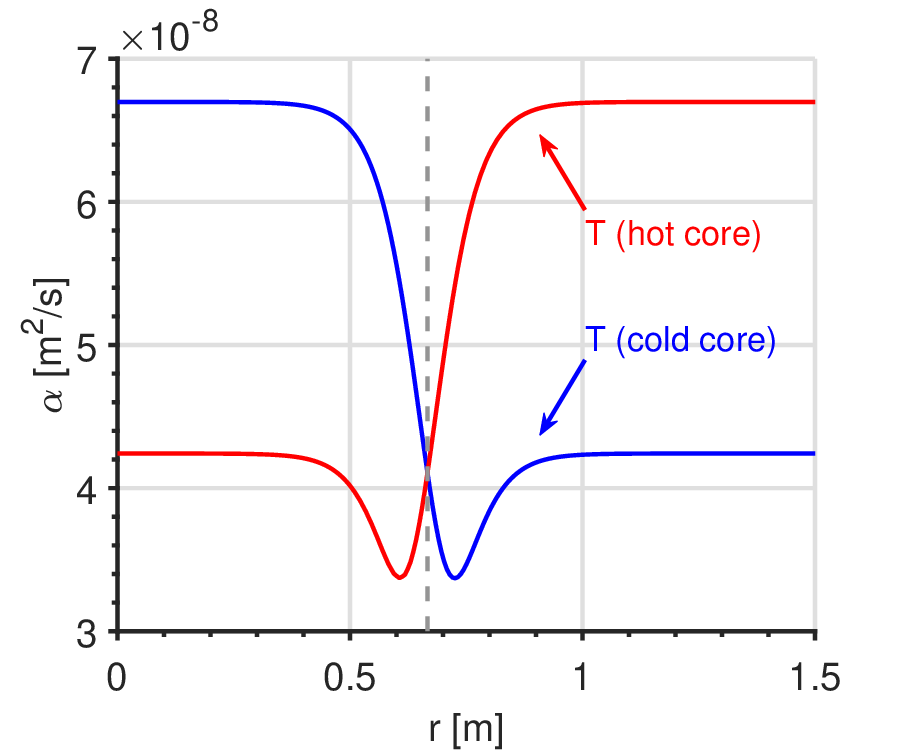}
  \caption{} 
  \label{subfig:initial_alpha_all}
\end{subfigure}%
\caption{Initial profiles of the sCO\(_2\) vortex column with \(Re_\text{max}=500\) and \(p_r=2\). The dashed vertical line represents \(r_c\). (a) \(\omega_z\) and \(T\); (b) \(\nu\); and (c) \(\alpha\).}
\label{fig:Fig4}
\end{figure}

As an example, figure \ref{fig:Fig4} presents the initial profiles of \(\omega_z\), \(T\), kinematic viscosity (or momentum diffusivity) \(\nu=\mu/\rho\), and thermal diffusivity \(\alpha=k/(\rho c_p)\) for both the cold and hot vortex core configurations with \(Re_\text{max}=500\) and \(p_r=2\). Note how the setup choice results in the strong non-linear relationship of the fluid properties with temperature across the pseudo-boiling line falling primarily either inside the vortex core (hot core) or outside (cold core), as visualised by the local minima in diffusivities. \par

\begin{table}
\begin{center}
\def~{\hphantom{0}}
\begin{tabular}{|r|r|r|} 
\multicolumn{1}{c}{\(Re_\text{max}\)} & 
\multicolumn{1}{c}{\(\Gamma_0\) [m\textsuperscript{2}/s]} & 
\multicolumn{1}{c}{\(\omega_0\) [1/s]}\\[3pt]
\multicolumn{1}{c}{100} & 
\multicolumn{1}{c}{\(7.5955\times 10^{-6}\)} & 
\multicolumn{1}{c}{\(1.9779\times 10^{-5}\)} \\
\multicolumn{1}{c}{200} & 
\multicolumn{1}{c}{\(1.5191\times 10^{-5}\)} & 
\multicolumn{1}{c}{\(3.9558\times 10^{-5}\)} \\
\multicolumn{1}{c}{400} & 
\multicolumn{1}{c}{\(3.0382\times 10^{-5}\)} & 
\multicolumn{1}{c}{\(7.9116\times 10^{-5}\)} \\
\multicolumn{1}{c}{500} & 
\multicolumn{1}{c}{\(3.7977\times 10^{-5}\)} & 
\multicolumn{1}{c}{\(9.8895\times 10^{-5}\)} \\
\multicolumn{1}{c}{1000} & 
\multicolumn{1}{c}{\(7.5955\times 10^{-5}\)} & 
\multicolumn{1}{c}{\(1.9779\times 10^{-4}\)} \\
\end{tabular}
\caption{Initialisation of the vortex parameters \(\Gamma_0\) and \(\omega_0\) to satisfy a \(Re_\text{max}\) for sCO\(_2\) at \(p_r=1.3\) between \(T_\text{min}=273.72\) K and \(T_\text{max}=349.75\) K.}
\label{tab:initial_vorticity}
\end{center}
\end{table}

\subsection{Axisymmetric Solution at Low Reynolds Numbers}
\label{subsec:solution_collapse}

This works aims at recovering the purely axisymmetric evolution of the unperturbed vortex column under supercritical fluid conditions to understand how thermal mixing across the pseudo-boiling line affects the vortex evolution. However, the multi-dimensional nature of the fluid solver prevents an axisymmetric solution to develop for any \(Re\) and \(p_r\). Only for \(Re_\text{max}<500\) we observe a stable axisymmetric solution. This issue is analysed in detail in \ref{apn:A} in terms of \(Re_\text{max}\), domain size \(L\) and uniform grid spacing \(\Delta x\), and relates to the interaction between non-axisymmetric boundaries, i.e., use of a computational square box, and flow initialisation with a baroclinic torque due to the large density gradients in the supercritical fluid. Further analysis in \ref{subapn:B1} corroborates this behaviour by validating the flow solver with an incompressible Oseen vortex at \(Re_\text{max}=1000\), for which no numerical destabilization and good agreement with the analytical solution was found. \par 

Regardless of the numerical origin of this destabilisation in our simulations, the observation highlights how easily vortices in SCF may destabilise due to baroclinicity, particularly close to the critical point, when subject to a physical perturbation, e.g., a pressure wave traversing the vortex. Additionally, centrifugal and Rayleigh-Taylor instabilities may also occur due to the centrifugal force in configurations with a denser vortex core, i.e., the cold core case \citep{2005_JFM_Sipp}. The use of a purely axisymmetric solver instead of a multi-dimensional solver would allow us to consider higher \(Re_\text{max}\) without suffering from numerically-induced perturbations. However, the existence of stable vortex columns in SCFs at higher \(Re_\text{max}\) might be questionable as they might quickly destabilise under small perturbations. Therefore, one may argue that the physical problem we aim to study emerges only at sufficiently low \(Re_\text{max}\) where pressure gradients are small. Like the Oseen vortex, this is a fundamental study of an axisymmetric vortex in a supercritical fluid. \par 

Henceforth, all reported radial profiles have been obtained by averaging the numerical solution at different radial locations on the Cartesian grid. Based on \ref{apn:A}, the domain size is \(L=16\pi\) m. This ensures that the solution remains fairly axisymmetric, as highlighted in following lines. \par

\begin{figure}
\centering
\begin{subfigure}{0.33\textwidth}
  \centering
  \includegraphics[width=1.05\linewidth]{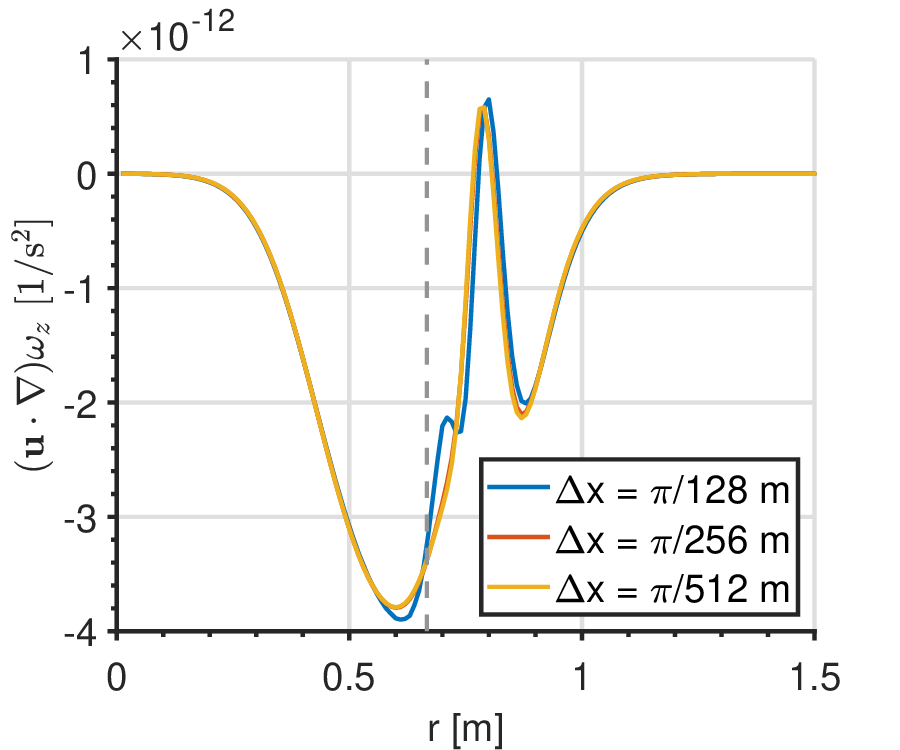}
  \caption{} 
  \label{subfig:convergence_convectiveterm}
\end{subfigure}%
\begin{subfigure}{0.33\textwidth}
  \centering
  \includegraphics[width=1.05\linewidth]{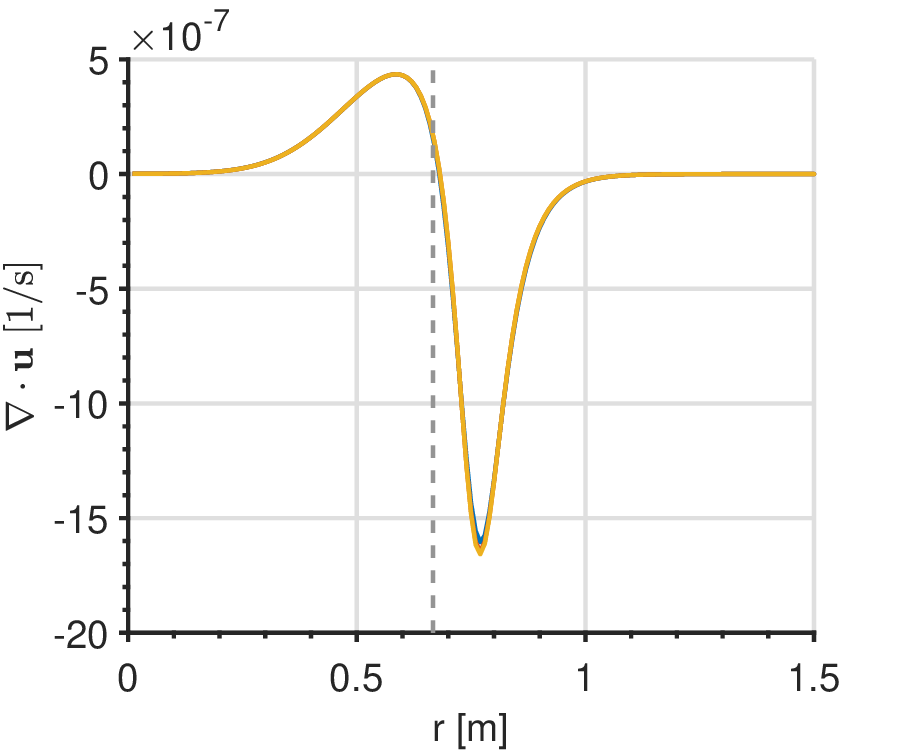}
  \caption{} 
  \label{subfig:convergence_dilatationterm}
\end{subfigure}%
\begin{subfigure}{0.33\textwidth}
  \centering
  \includegraphics[width=1.05\linewidth]{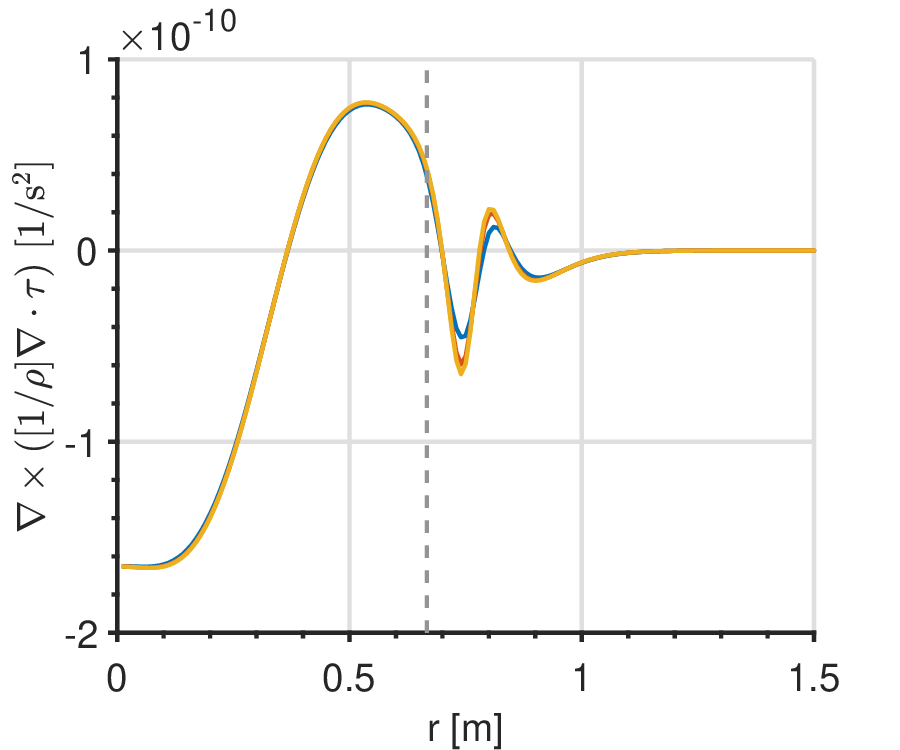}
  \caption{} 
  \label{subfig:convergence_viscousterm}
\end{subfigure}%
\caption{Grid convergence of relevant terms in the vorticity equation (\ref{eqn:vorticityZ_2}) for the vortex with a cold core, \(Re_\text{max}=400\) and \(L=16\pi\) m at \(t^*=4.32\). The dashed vertical line represents \(r_c\). (a) \((\boldsymbol{u}\boldsymbol{\cdot}\boldsymbol{\nabla})\omega_z\); (b) \(\boldsymbol{\nabla}\boldsymbol{\cdot}\boldsymbol{u}\); and (c) \(\boldsymbol{\nabla\times}\big([1/\rho]\boldsymbol{\nabla}\boldsymbol{\cdot}\boldsymbol{\tau}\big)\).}
\label{fig:Fig5}
\end{figure}

Grid convergence is analysed by looking at the terms in the vorticity equation

\begin{equation}
\label{eqn:vorticityZ_2}
\begin{split}
\frac{\partial\omega_z}{\partial t} = - (\boldsymbol{u}\boldsymbol{\cdot}\boldsymbol{\nabla}) \omega_z - \omega_z (\boldsymbol{\nabla}\boldsymbol{\cdot}\boldsymbol{u}) + \bigg[\boldsymbol{\nabla\times}\bigg(\frac{1}{\rho}\boldsymbol{\nabla}\boldsymbol{\cdot}\boldsymbol{\tau}\bigg)\bigg]\boldsymbol{\cdot}\hat{\textbf{e}}_z
\end{split}
\end{equation}

\noindent
where \(\hat{\textbf{e}}_z=(0,0,1)\) and the effects of baroclinicity are neglected under the assumption that density and pressure gradients are perfectly aligned in the axisymmetric flow. Figure \ref{fig:Fig5} shows that \(\Delta x = \pi/256\) m is enough for the cold vortex core with \(Re_\text{max}=400\) and \(p_r=2\) to achieve grid convergence of the large variations of the relevant terms in (\ref{eqn:vorticityZ_2}) early in the simulation (\(t^*=[\Gamma_0/r_c^2]t=4.32\)). These are the convective term \((\boldsymbol{u}\boldsymbol{\cdot}\boldsymbol{\nabla})\omega_z\), the volume dilatation \(\boldsymbol{\nabla}\boldsymbol{\cdot}\boldsymbol{u}\) from the compressible vorticity stretching term \(- \omega_z (\boldsymbol{\nabla}\boldsymbol{\cdot}\boldsymbol{u})\), and the compressible viscous term \(\boldsymbol{\nabla\times}\big([1/\rho]\boldsymbol{\nabla}\boldsymbol{\cdot}\boldsymbol{\tau}\big)\). Note that grid convergence focuses on (\ref{eqn:vorticityZ_2}) since part of our objective is to explain vorticity evolution in SCF. Since these terms are obtained from the solution of primitive variables, the convergence of, e.g., velocity, is also guaranteed. \par 

Next, we consider the following cases given by \(Re_\text{max}=[100, 200, 400]\) with \(p_r=2\) and the vortex core hotter or colder than the surrounding fluid to verify that the numerical solution collapses to an axisymmetric solution when properly scaled. Then, the analysis of how heat transfer affects the evolution of a radially stratified axisymmetric vortex in a SCF can be performed because of the limited evolution of non-axisymmetric modes. \par

\begin{figure}
\centering
\begin{subfigure}{0.33\textwidth}
  \centering
  \includegraphics[width=1.05\linewidth]{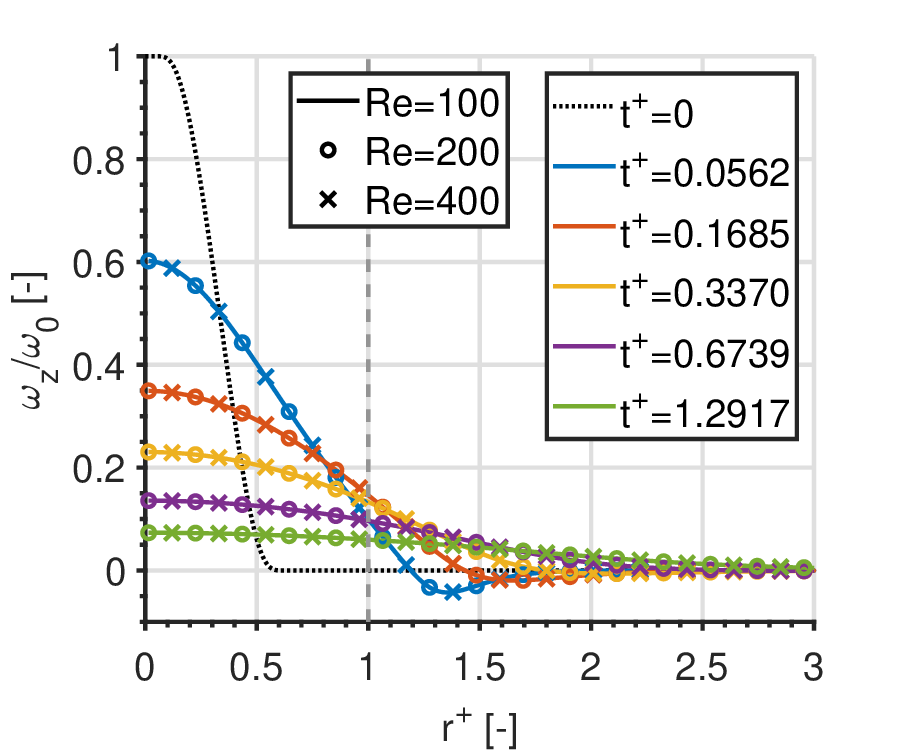}
  \caption{} 
  \label{subfig:collapse_vorticity_cold}
\end{subfigure}%
\begin{subfigure}{0.33\textwidth}
  \centering
  \includegraphics[width=1.05\linewidth]{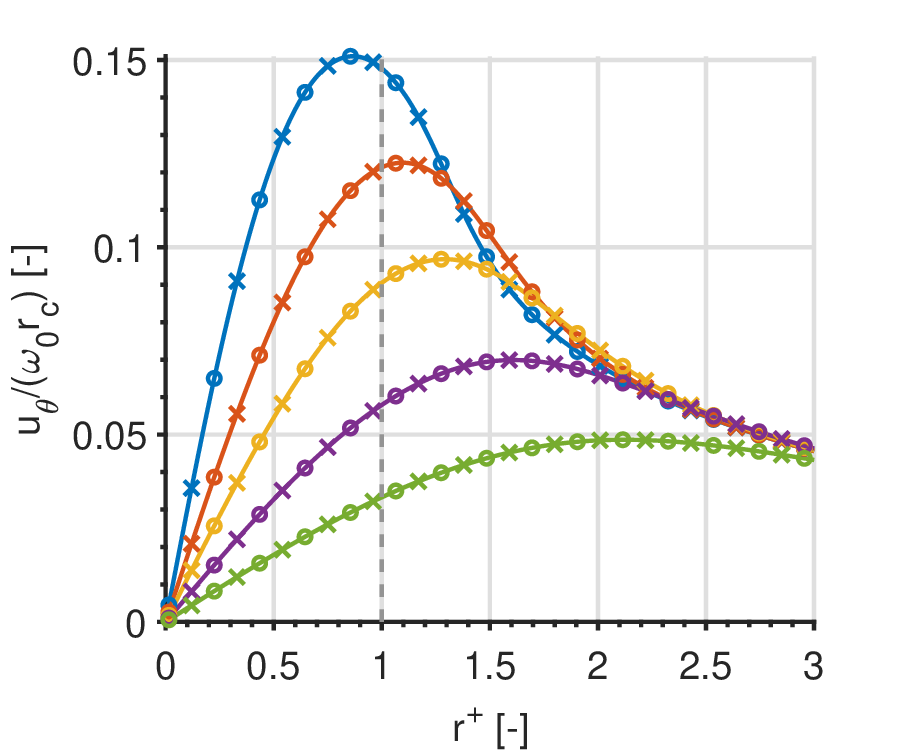}
  \caption{} 
  \label{subfig:collapse_azimuthalvelocity_cold}
\end{subfigure}%
\begin{subfigure}{0.33\textwidth}
  \centering
  \includegraphics[width=1.05\linewidth]{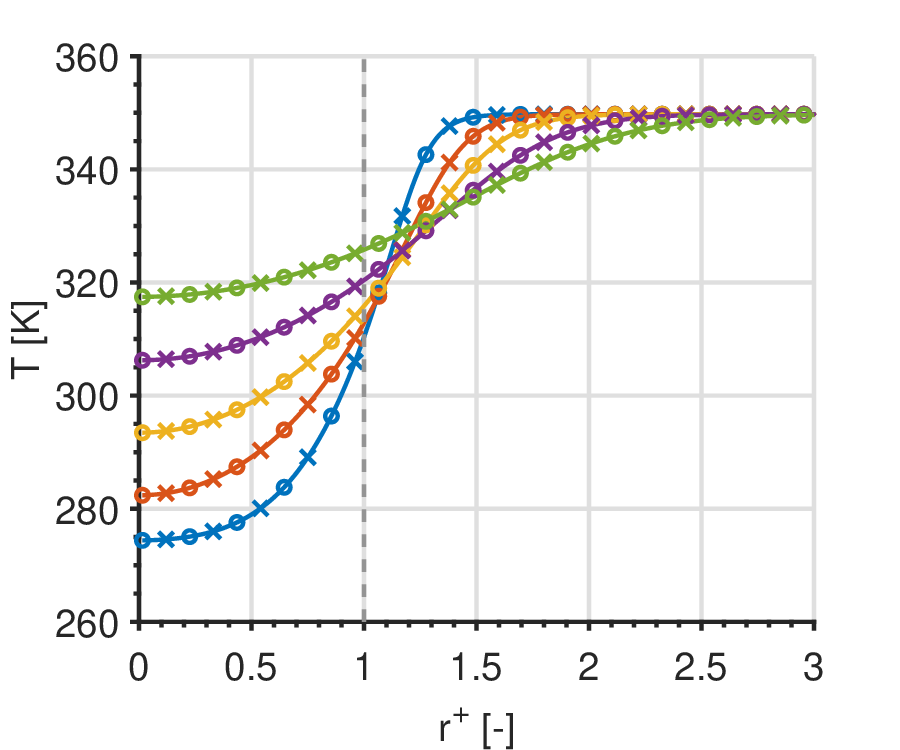}
  \caption{} 
  \label{subfig:collapse_temp_cold}
\end{subfigure}%
\\
\begin{subfigure}{0.33\textwidth}
  \centering
  \includegraphics[width=1.05\linewidth]{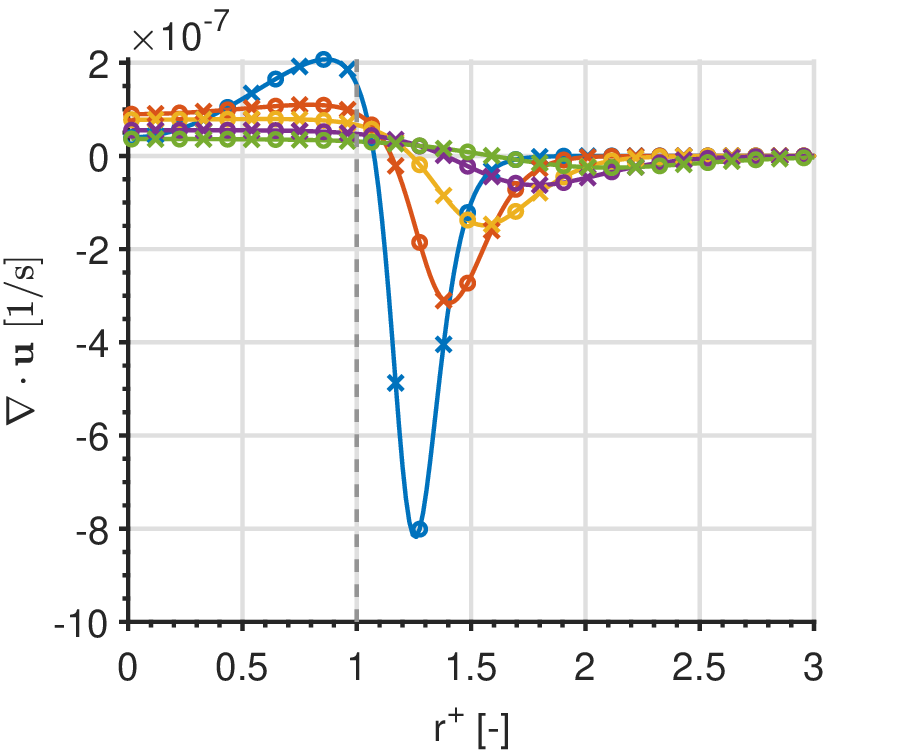}
  \caption{} 
  \label{subfig:collapse_divergence_cold}
\end{subfigure}%
\begin{subfigure}{0.33\textwidth}
  \centering
  \includegraphics[width=1.05\linewidth]{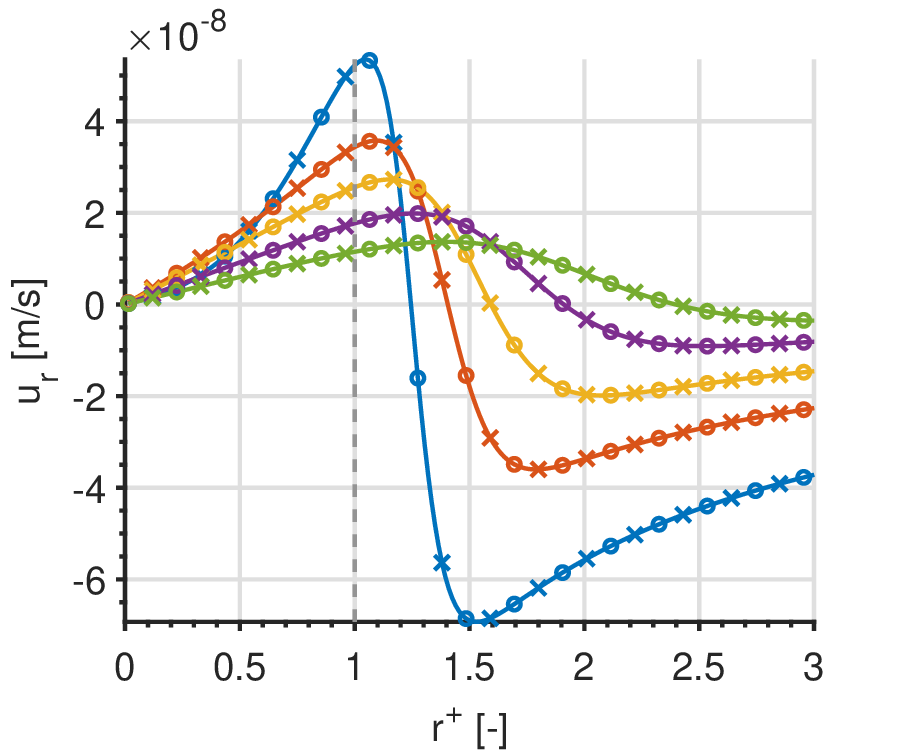}
  \caption{} 
  \label{subfig:collapse_radialvelocity_cold}
\end{subfigure}%
\begin{subfigure}{0.33\textwidth}
  \centering
  \includegraphics[width=1.05\linewidth]{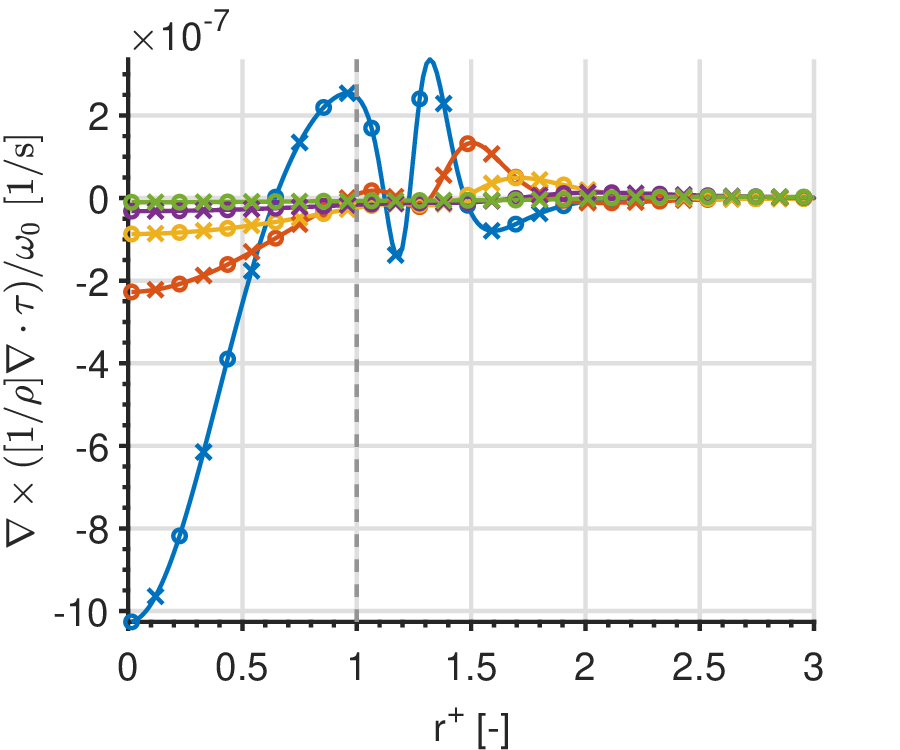}
  \caption{} 
  \label{subfig:collapse_viscousterm_cold}
\end{subfigure}%
\caption{Collapse of the numerical solution of the vortex with a cold core obtained with \(p_r=2\), \(L=16\pi\) m and \(\Delta x = \pi/256\) m for different \(Re_\text{max}\) at various \(t^+\). The dashed vertical line represents \(r_c\). (a) \(\omega_z/\omega_0\); (b) \(u_\theta/(\omega_0r_c)\); (c) \(T\); (d) \(\boldsymbol{\nabla}\boldsymbol{\cdot}\boldsymbol{u}\); (e) \(u_r\); and (f) \(\boldsymbol{\nabla\times}\big([1/\rho]\boldsymbol{\nabla}\boldsymbol{\cdot}\boldsymbol{\tau}\big)/\omega_0\).}
\label{fig:Fig6}
\end{figure}

Regarding the symmetry of the solution, table \ref{tab:RSD_analysed_configurations} presents the relative standard deviation (RSD) of the azimuthal velocity \(u_\theta\) extracted along \(r=r_c\) toward the end of the simulation at \(t^*=548.67\) (see (\ref{eqn:mean+RSD})). The RSD remains well below 0.1\% for all the cases studied. Moreover, the numerical solution collapses independent of \(Re_\text{max}\) when properly scaled, as shown in figure \ref{fig:Fig6} for the cold vortex core. This is analogous to Oseen's analytical solution \citep{1912_AMA_Oseen}.

\begin{equation}
\label{eqn:mean+RSD}\bar{u}_\theta=\frac{1}{N}\sum_{i=1}^{N} u_{\theta,i} \quad \quad \quad \quad \text{RSD} = \frac{1}{\bar{u}_\theta}\sqrt{\frac{1}{N}\sum_{i=1}^{N}(u_{\theta,i}-\bar{u}_\theta)^2}
\end{equation}

\begin{table}
\begin{center}
\def~{\hphantom{0}}
\begin{tabular}{|r|r|r|} 
\multicolumn{1}{c}{\(Re_\text{max}\)} & 
\multicolumn{1}{c}{Cold Vortex Core} & 
\multicolumn{1}{c}{Hot Vortex Core} \\[3pt]
\multicolumn{1}{c}{100} & 
\multicolumn{1}{c}{0.0212\%} & 
\multicolumn{1}{c}{0.0279\%} \\ 
\multicolumn{1}{c}{200} & 
\multicolumn{1}{c}{0.0142\%} & 
\multicolumn{1}{c}{0.0174\%} \\ 
\multicolumn{1}{c}{400} & 
\multicolumn{1}{c}{0.0259\%} & 
\multicolumn{1}{c}{0.0119\%} \\ 
\end{tabular}
\caption{RSD of \(u_\theta\) extracted along \(r_c\) for different \(Re_\text{max}\) at \(p_r=2\) toward the end of the simulation at \(t^*=548.67\) with \(L=16\pi\) m and \(\Delta x = \pi/256\) m.}
\label{tab:RSD_analysed_configurations}
\end{center}
\end{table}

The various snapshots in figure \ref{fig:Fig6} are presented in terms of another non-dimensional time \(t^+=(\bar{\nu}/r_c^2)t\) scaling with an average momentum diffusivity \(\bar{\nu}\) rather than \(\Gamma_0\). Since the vortex evolution is mainly diffusion-driven -- a small \(|u_r|\ll|u_\theta|\) occurs due to the fluid's compressibility -- this may offer a better comparison among different thermodynamic states, i.e., ambient pressures, when comparing vortices with the same circulation. For \(p_r=2\), \(\bar{\nu}=\frac{1}{2}(\nu_{T_{\text{min}}}+\nu_{T_{\text{max}}})=9.8737\times 10^{-8}\) m\textsuperscript{2}/s. Note that scaling with an average thermal diffusivity would result in similar time scales since \(\nu\) and \(\alpha\) are comparable in magnitude and distribution, as shown in figure \ref{fig:Fig4}. Additionally, the radial profiles are represented in terms of a non-dimensional distance \(r^+=r/r_c\). Heat transfer across the vortex is independent of the vortex strength. It causes the evolution of the temperature profiles (figure \ref{subfig:collapse_temp_cold}), the local volume dilatation (figure \ref{subfig:collapse_divergence_cold}) and the resulting small radial velocity (figure \ref{subfig:collapse_radialvelocity_cold}). In contrast, the evolution of the vorticity profiles (figure \ref{subfig:collapse_vorticity_cold}), azimuthal velocity (figure \ref{subfig:collapse_azimuthalvelocity_cold}), and the viscous term (figure \ref{subfig:collapse_viscousterm_cold}) in (\ref{eqn:vorticityZ_2}) need to be scaled by \(\omega_0\) to collapse. Although not reported here for brevity, similar observations hold for the hot vortex core and other ambient pressures. This confirms that the numerical solutions are axisymmetric at low \(Re_\text{max}\). \par

The next part of this paper considers six configurations defined by a vortex strength given by \(Re_\text{max}=200\) or \(\Gamma_0=1.5191\times10^{-5}\) m\textsuperscript{2}/s, different ambient pressures, i.e., \(p_r=[1.3, 1.5, 2]\), and with cold or hot vortex cores. Given that at lower \(p_r\) grid resolution is more important to capture the larger gradients in fluid properties and mitigate numerical destabilisation, the results are obtained with \(L=16\pi\) m and \(\Delta x = \pi/512\) m. The following average kinematic viscosities are used to evaluate \(t^+\): \(\bar{\nu}_{(p_r=1.3)}=1.0867\times 10^{-7}\) m\textsuperscript{2}/s, \(\bar{\nu}_{(p_r=1.5)}=1.0360\times 10^{-7}\) m\textsuperscript{2}/s and \(\bar{\nu}_{(p_r=2)}=9.8737\times 10^{-8}\) m\textsuperscript{2}/s. \par 

Table \ref{tab:RSD_analysed_pr_configurations} shows the RSD of \(u_\theta\) along \(r_c\) at \(t^+\approx1.29\) or toward the end of the simulations. \ref{apn:C} shows contour plots of \(\omega_z\) for all configurations at different \(t^+\) to visualise the axisymmetric behaviour of the numerical solution. For cases with a hot core, the solution remains axisymmetric with RSD values well below 0.01\% for all analysed pressures. With the cold core, the RSD increases as pressure decreases due to the larger density gradients enhancing the numerically-induced baroclinic torque and subsequent instability. Nonetheless, the RSD remains below 0.1\% for \(p_r=1.5\) and \(2\). For \(p_r=1.3\), the RSD is closer to 1\% and a small non-axisymmetric behaviour can also be seen in the contour plots (see figure \ref{subfig:wz_Re200_512_16PI_pr1p3_cold_042} in \ref{apn:C}), especially in the region \(r\lesssim  r_c\). In this study, we still accept the average of any variable extracted at each fixed radial location as the true axisymmetric solution of the cold vortex core case at \(p_r=1.3\). \par 

\begin{table}
\begin{center}
\def~{\hphantom{0}}
\begin{tabular}{|r|r|r|} 
\multicolumn{1}{c}{\(p_r\)} & 
\multicolumn{1}{c}{Cold Vortex Core} & 
\multicolumn{1}{c}{Hot Vortex Core} \\[3pt]
\multicolumn{1}{c}{1.3} & 
\multicolumn{1}{c}{0.7837\%} & 
\multicolumn{1}{c}{0.0030\%} \\ 
\multicolumn{1}{c}{1.5} & 
\multicolumn{1}{c}{0.0951\%} & 
\multicolumn{1}{c}{0.0021\%} \\ 
\multicolumn{1}{c}{2} & 
\multicolumn{1}{c}{0.0027\%} & 
\multicolumn{1}{c}{0.0028\%} \\ 
\end{tabular}
\caption{RSD of \(u_\theta\) extracted along \(r_c\) at various \(p_r\) values and \(t^+\approx1.29\) with \(L=16\pi\) m and \(\Delta x = \pi/512\) m.}
\label{tab:RSD_analysed_pr_configurations}
\end{center}
\end{table}

\subsection{Vortex Evolution Based on Core Temperature and Thermodynamic Pressure}
\label{subsec:cold_vs_hot}

This section compares the evolution of the vortex column between the cold and the hot vortex cores obtained with \(Re_\text{max}=200\), and analyses the effects of the highly varying properties across the thermal mixing layer at different thermodynamic pressures. \par 

\begin{figure}
\centering
\begin{subfigure}{0.33\textwidth}
  \centering
  \includegraphics[width=1.05\linewidth]{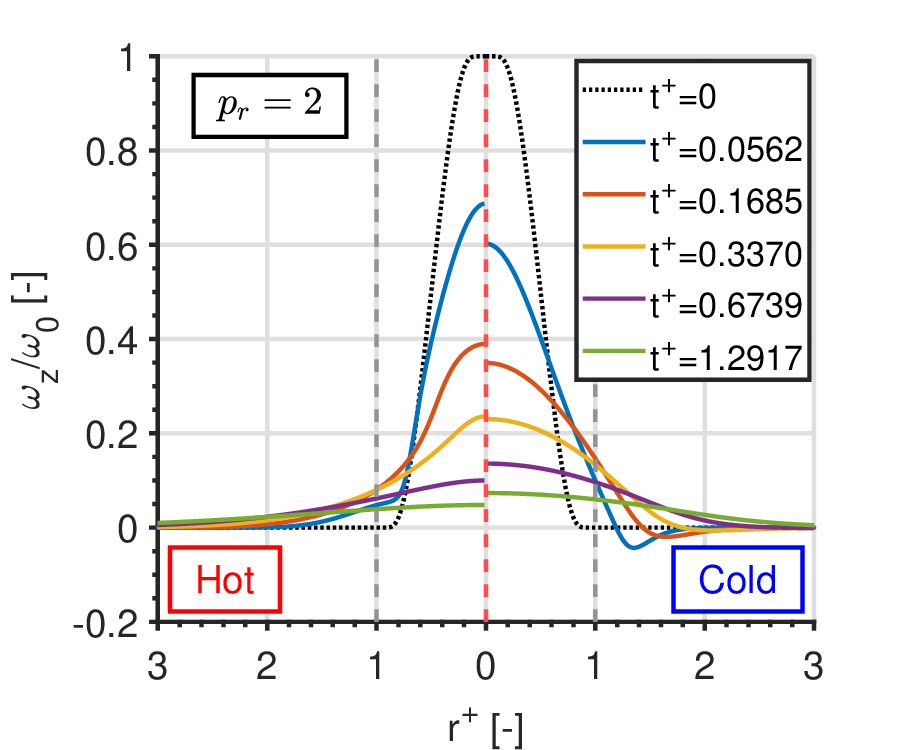}
  \caption{} 
  \label{subfig:vorticity_Re200_pr2_cold_vs_hot}
\end{subfigure}%
\begin{subfigure}{0.33\textwidth}
  \centering
  \includegraphics[width=1.05\linewidth]{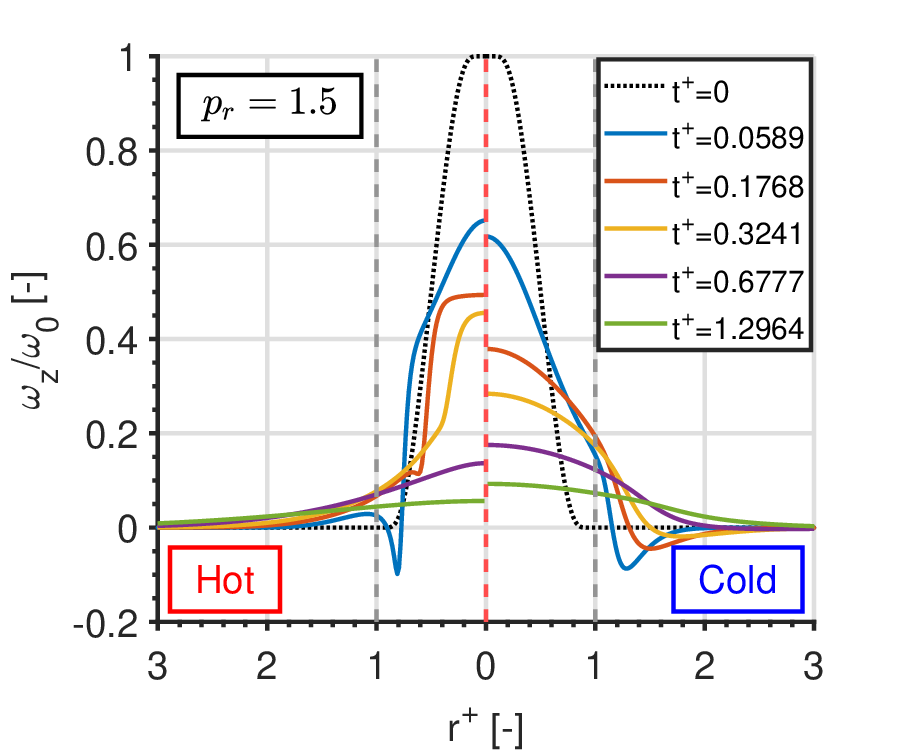}
  \caption{} 
  \label{subfig:vorticity_Re200_pr1p5_cold_vs_hot}
\end{subfigure}%
\begin{subfigure}{0.33\textwidth}
  \centering
  \includegraphics[width=1.05\linewidth]{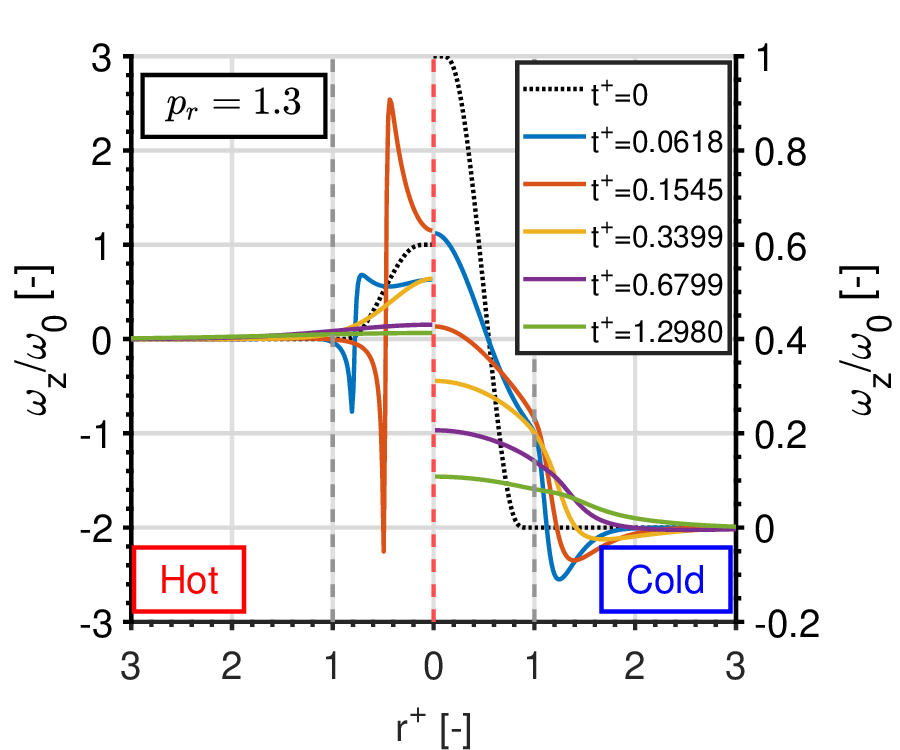}
  \caption{} 
  \label{subfig:vorticity_Re200_pr1p3_cold_vs_hot}
\end{subfigure}%
\caption{Evolution with \(t^+\) of the vorticity \(\omega_z/\omega_0\) of the hot core (left) and the cold core (right) at different pressures. The dashed vertical lines represent the vortex centre (red) and \(r_c\) (grey). (a) \(p_r=2\); (b) \(p_r=1.5\); and (c) \(p_r=1.3\).}
\label{fig:Fig7}
\end{figure}

Figure \ref{fig:Fig7} presents the distributions of the non-dimensional vorticity \(\omega_z/\omega_0\) at various \(t^+\) for each configuration. Compared to the classical Oseen vortex solution shown in \ref{subapn:B1}, the non-linear behaviour of the fluid properties causing large gradients across the pseudo-boiling line quickly emerges and vorticity profiles clearly differ from Gaussian-like distributions. However, this is a transient process only lasting the thermal mixing process involving pseudo-boiling phenomena. Note that \(\omega_z\) recovers a quasi-Gaussian distribution at later \(t^+\) in some configurations, e.g., the hot core at \(p_r=2\). That is, once the heating (or cooling) of the vortex has mostly occurred and the fluid thermodynamic state is no longer near pseudo-boiling conditions. Extremely sharp vorticity gradients are observed throughout. In the cold core cases, the main feature consists of a region of negative or reverse vorticity outside \(r_c\) where initially no vorticity existed. As the thermodynamic pressure decreases from \(p_r=2\) to \(1.3\), reverse vorticity in the cold core cases is accentuated. In contrast, the hot core configurations show richer phenomena. At \(p_r=2\), fluid rotation remains consistent during the evolution of the vortex flow. However, lower pressures result in a sharpening of the vorticity profile and the emergence of vorticity of opposite sign inside the vortex. More interestingly, a nearly discontinuous non-monotonic distribution of \(\omega_z\) is seen at \(p_r=1.3\) where a seemingly ``travelling" front is observed with vorticity decreasing sharply toward negative values on the outer side and increasing sharply on the inner side.  \par 

\begin{figure}
\centering
\begin{subfigure}{0.5\textwidth}
  \centering
  \includegraphics[width=1.05\linewidth]{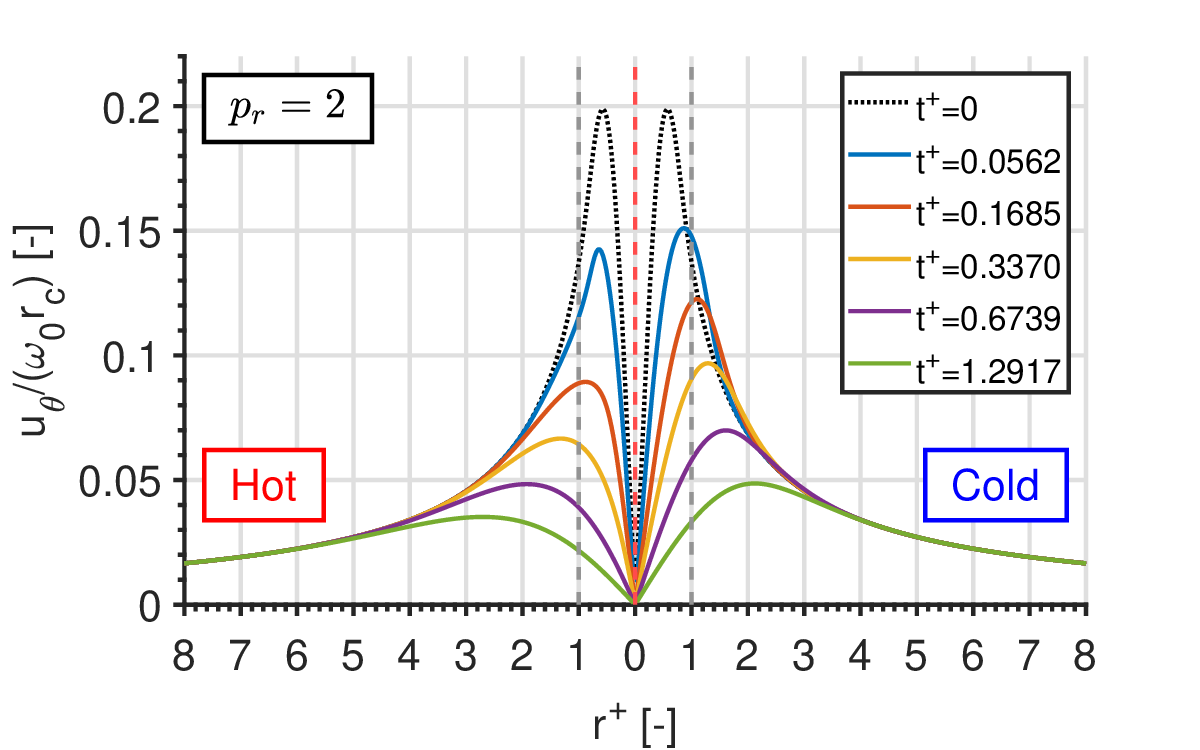}
  \caption{} 
  \label{subfig:utheta_Re200_pr2_cold_vs_hot}
\end{subfigure}%
\begin{subfigure}{0.5\textwidth}
  \centering
  \includegraphics[width=1.05\linewidth]{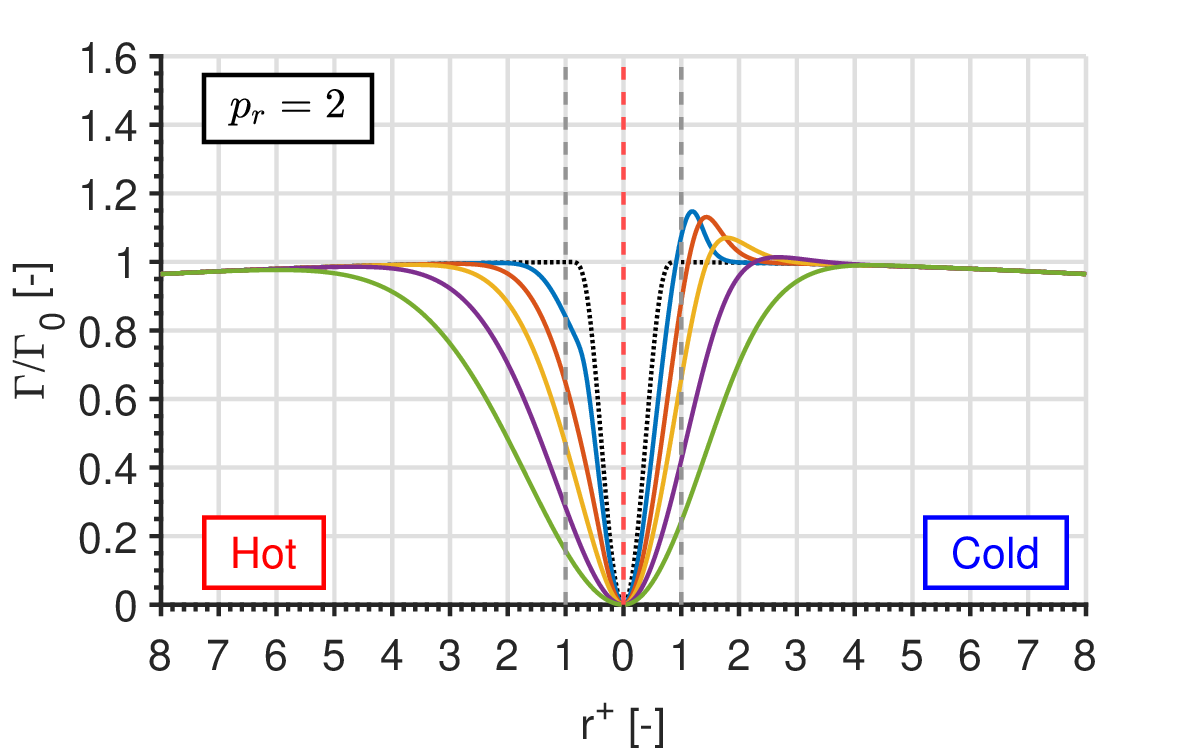}
  \caption{} 
  \label{subfig:circulation_Re200_pr2_cold_vs_hot}
\end{subfigure}%
\\
\begin{subfigure}{0.5\textwidth}
  \centering
  \includegraphics[width=1.05\linewidth]{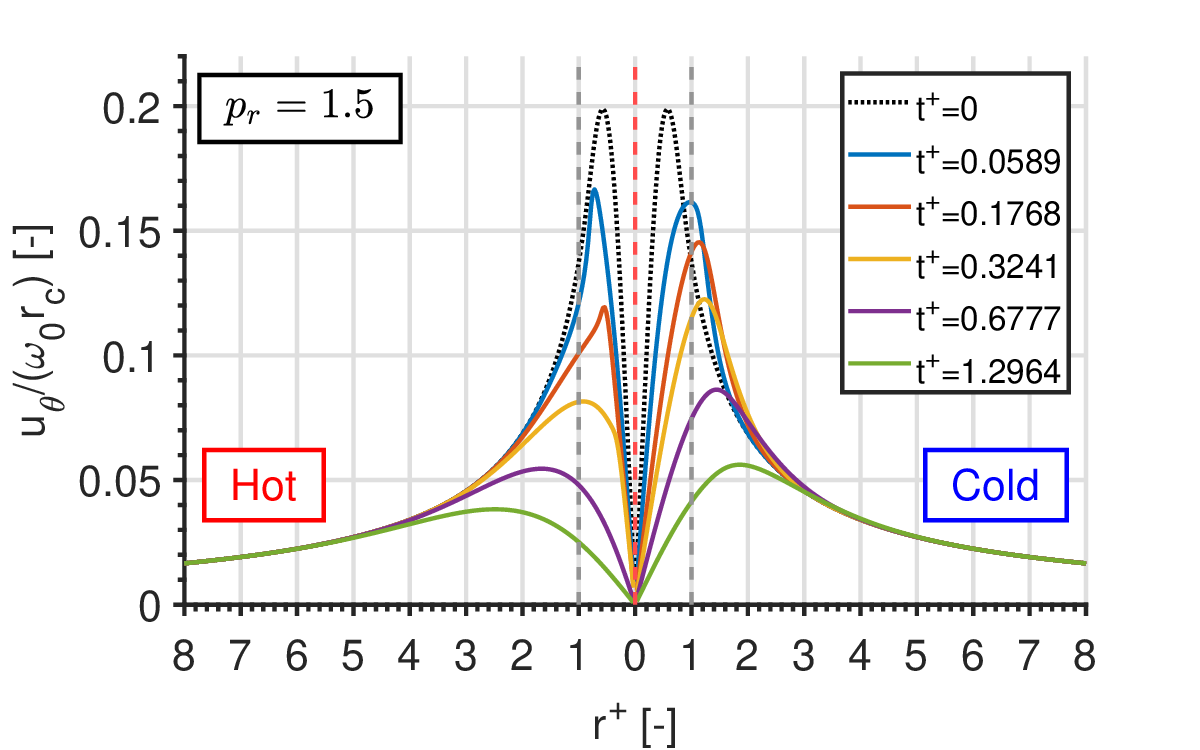}
  \caption{} 
  \label{subfig:utheta_Re200_pr1p5_cold_vs_hot}
\end{subfigure}%
\begin{subfigure}{0.5\textwidth}
  \centering
  \includegraphics[width=1.05\linewidth]{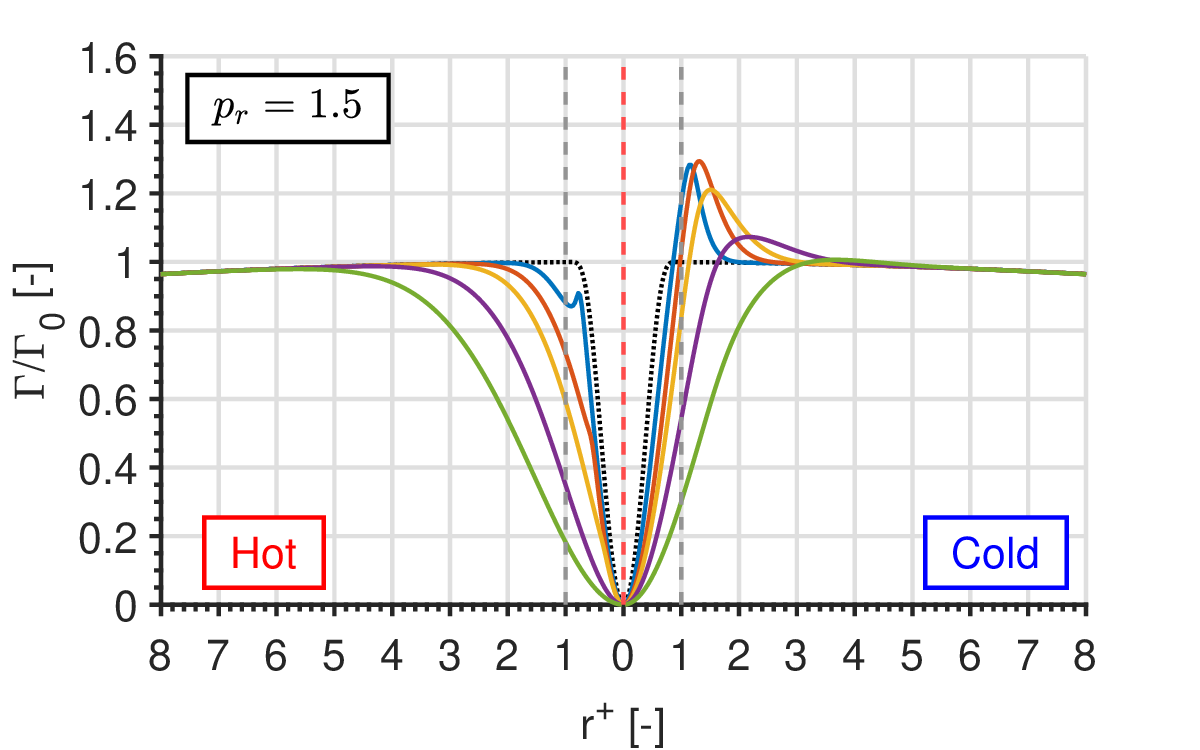}
  \caption{} 
  \label{subfig:circulation_Re200_pr1p5_cold_vs_hot}
\end{subfigure}%
\\
\begin{subfigure}{0.5\textwidth}
  \centering
  \includegraphics[width=1.05\linewidth]{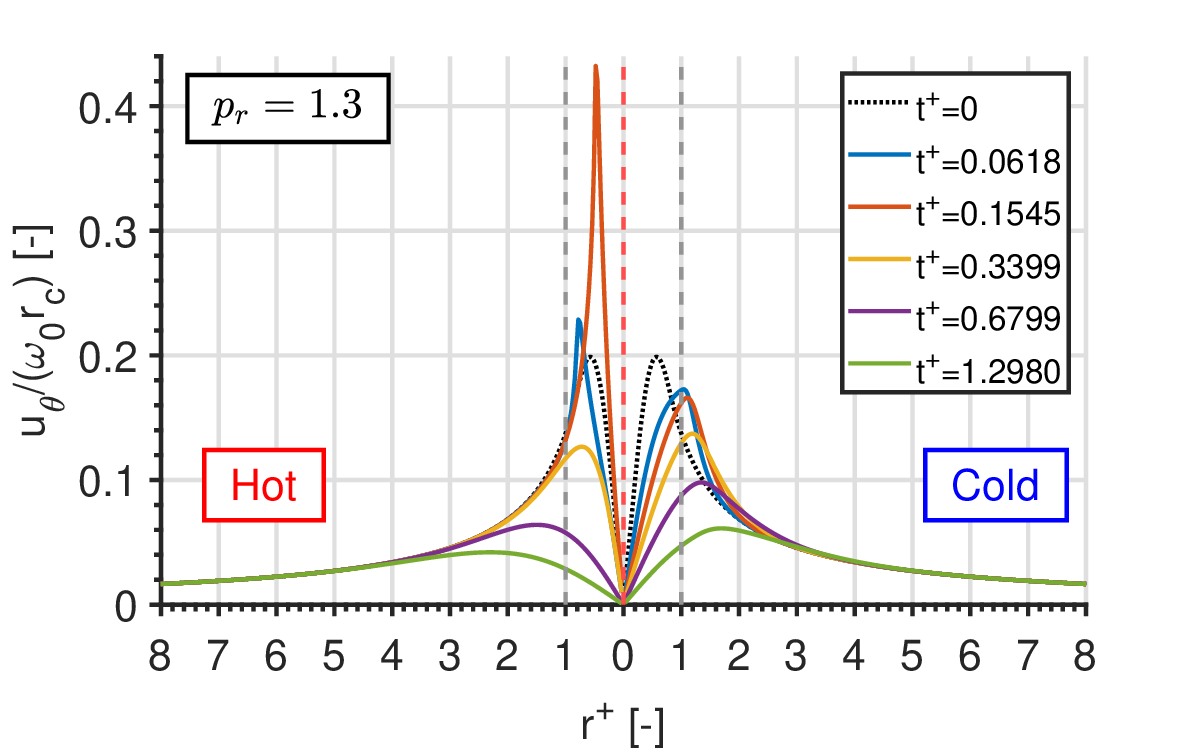}
  \caption{} 
  \label{subfig:utheta_Re200_pr1p3_cold_vs_hot}
\end{subfigure}%
\begin{subfigure}{0.5\textwidth}
  \centering
  \includegraphics[width=1.05\linewidth]{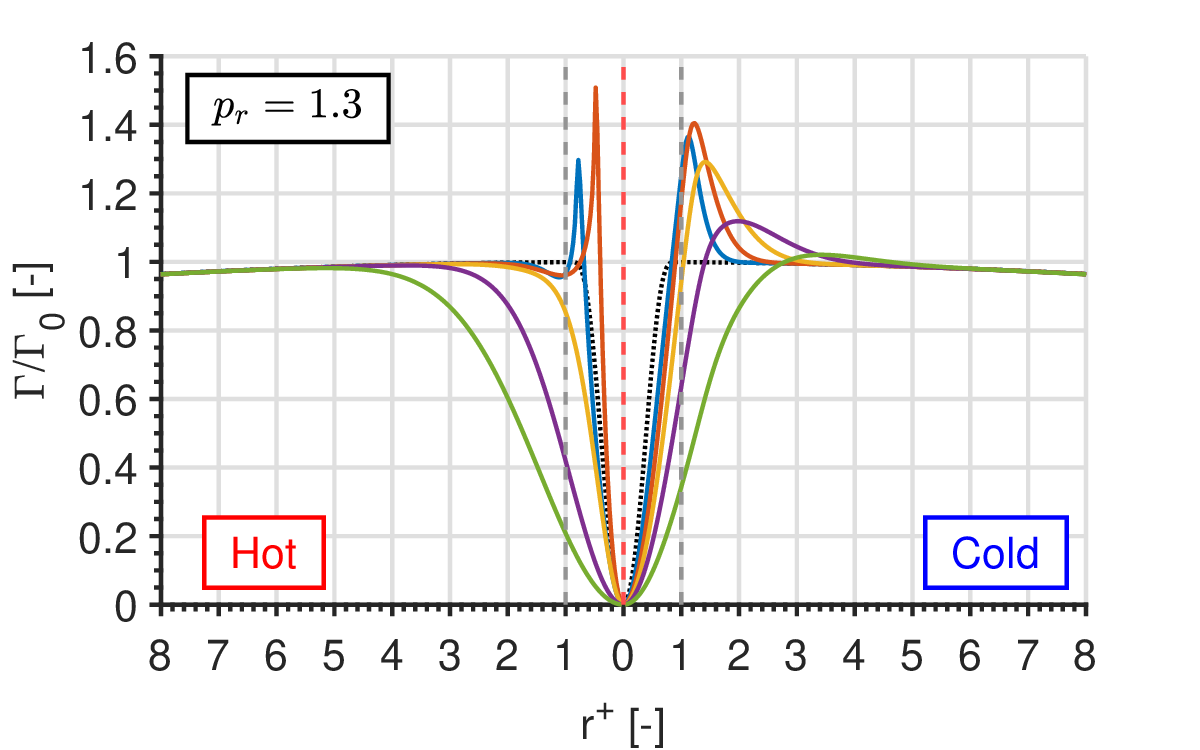}
  \caption{} 
  \label{subfig:circulation_Re200_pr1p3_cold_vs_hot}
\end{subfigure}%
\caption{Evolution with \(t^+\) of the azimuthal velocity and circulation of the hot core (left) and the cold core (right) at different pressures. The dashed vertical lines represent the vortex centre (red) and \(r_c\) (grey). (a)-(b) \(u_\theta/(\omega_0r_c)\) and \(\Gamma/\Gamma_0\) for \(p_r=2\); (c)-(d) \(u_\theta/(\omega_0r_c)\) and \(\Gamma/\Gamma_0\) for \(p_r=1.5\); and (e)-(f) \(u_\theta/(\omega_0r_c)\) and \(\Gamma/\Gamma_0\) for \(p_r=1.3\).}
\label{fig:Fig8}
\end{figure}

Further insights can be obtained from the evolution with \(t^+\) of the distributions of the azimuthal velocity \(u_\theta/(\omega_0r_c)\) (figures~\ref{subfig:utheta_Re200_pr2_cold_vs_hot}, \ref{subfig:utheta_Re200_pr1p5_cold_vs_hot} and \ref{subfig:utheta_Re200_pr1p3_cold_vs_hot}) and circulation \(\Gamma/\Gamma_0\), i.e., \(\Gamma=2\pi r u_\theta\) (figures \ref{subfig:circulation_Re200_pr2_cold_vs_hot}, \ref{subfig:circulation_Re200_pr1p5_cold_vs_hot} and \ref{subfig:circulation_Re200_pr1p3_cold_vs_hot}). Similarly to \(\omega_z\), \(u_\theta\) and \(\Gamma\) evolve differently for each vortex configuration. For the cold vortex, the \(p_r=2\) case is more representative. One observes higher peak azimuthal velocities in the cold vortex compared to the hot vortex located in the region where \(\omega_z\) goes from positive to negative. This behaviour reflects a local acceleration of the fluid in the azimuthal direction due to local vorticity production mechanisms compared to the more diffusion-dominated vorticity evolution in the hot vortex. These physical mechanisms are identified in section \ref{subsec:vorticity_analysis}. As a result, circulation also peaks above \(\Gamma_0\). While vorticity diffusion seems to dominate in the hot vortex at \(p_r=2\), additional vorticity mechanisms are greatly enhanced as \(p_r\) decreases (see section \ref{subsec:vorticity_analysis}). A particularly rapid azimuthal acceleration of the fluid occurs around \(t^+=0.1545\) at \(p_r=1.3\), also with \(\Gamma>\Gamma_0\) locally, as a result of the opposing vorticity trends seen in figure \ref{subfig:vorticity_Re200_pr1p3_cold_vs_hot}. \par 

Note that the issues related to the use of a multi-dimensional solver with a finite computational domain become evident in figures \ref{subfig:circulation_Re200_pr2_cold_vs_hot}, \ref{subfig:circulation_Re200_pr1p5_cold_vs_hot} and \ref{subfig:circulation_Re200_pr1p3_cold_vs_hot}. Despite aiming at approximating unbounded flow, circulation away from the vortex core is not constant (\(\Gamma \neq \Gamma_0\)) and decays with \(r\) \citep{2004_JFM_Pradeep}. However, the error is deemed small for the purpose of the analyses in this work focusing on the flow evolution near the vortex core (\(r^+<8\)). \par

Figures \ref{fig:Fig7} and \ref{fig:Fig8} suggest that the evolution of the vortex core radius differs significantly from incompressible vortex models such as Oseen's \citep{1912_AMA_Oseen}. The initial \(r_c\) defined in section \ref{sec:setup} follows a different definition of vortex core radius than classical models such as Gaussian vorticity distributions. Thus, in order to make a direct comparison with Oseen's model, a vortex core radius \(r_{core}\) is defined as the distance from the vortex centre where \(u_\theta\) peaks, i.e., \(\frac{\partial u_\theta}{\partial r}(r_{core},t)=0\). Some researchers define the vortex core radius as the distance from the vortex centre where \(\omega_z\) has decreased by a factor \(e\), i.e., \(\omega_z(r_{core},t)=\omega_{z,\text{max}}/e\). However, given the evolution of \(\omega_z\) shown in figure \ref{fig:Fig7}, where vorticity does not monotonically decrease from its maximum at the centre, this definition is ill-defined for our enhanced problem complexity. In Oseen's model given by (\ref{eqn:OseenVor}) and (\ref{eqn:OseenVor2}) in \ref{subapn:B1}, the former results in \(r_{core}=2.24182\sqrt{\nu t}\). Certainly, \(\nu\) varies significantly across the vortex during the early stages, where its evolution is affected by heat transfer. Therefore, the aim is to compare the evolution of \(r_{core}\) against Oseen's model in the asymptotic state as the temperature becomes uniform. That is, \(\nu_{T_{\text{min}}}\) is selected for the hot core and \(\nu_{T_{\text{max}}}\) for the cold core. \par 

\begin{figure}
\centering
\begin{subfigure}{0.5\textwidth}
  \centering
  \includegraphics[width=0.8\linewidth]{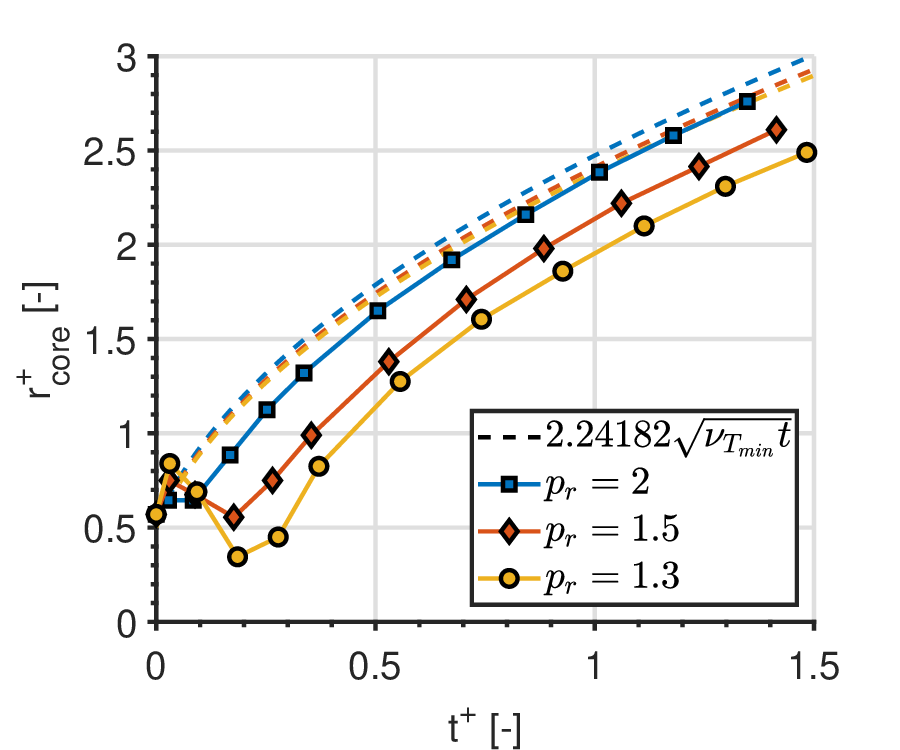}
  \caption{} 
  \label{subfig:coreradius_hot}
\end{subfigure}%
\begin{subfigure}{0.5\textwidth}
  \centering
  \includegraphics[width=0.8\linewidth]{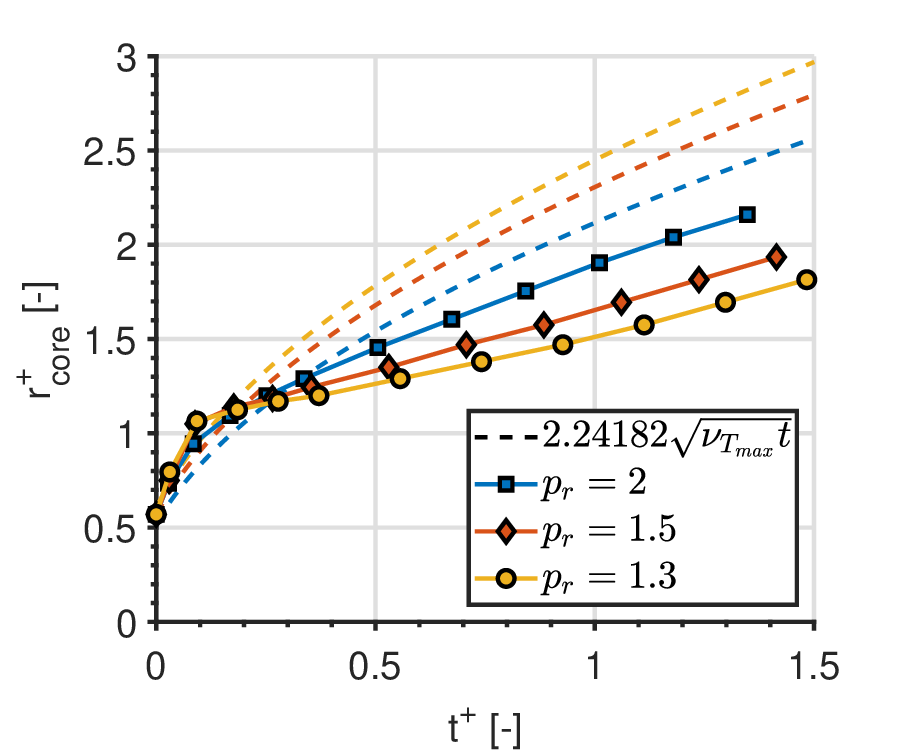}
  \caption{} 
  \label{subfig:coreradius_cold}
\end{subfigure}%
\caption{Evolution with \(t^+\) of the vortex core radius \(r_{core}\) at different pressures compared against the incompressible Oseen's model. (a) hot core; and (b) cold core.}
\label{fig:Fig9}
\end{figure}

Figure \ref{fig:Fig9} presents the evolution of \(r^+_{core}\). Despite the transient process in the SCF across the pseudo-boiling line, the effects on the vortex core radius are clear. Each core temperature configuration evolves differently, showing substantial differences from the incompressible solution -- where vorticity evolution is diffusion-driven only. The hot core at \(p_r=2\) initially expands very little. Although the lower momentum diffusivity due to the larger temperature does contribute to a lower growth rate compared to the asymptotic Oseen solution, the core also experiences a strong compression as shown in figure \ref{fig:Fig11}. Nevertheless, the compressible effects and cooling of the vortex occur rather quickly, and the evolution of the vortex core radius approaches the asymptotic solution. In contrast, lower pressures display a distinct transient process where \(r^+_{core}\) rapidly increases to then contract, even below the initial radius, e.g., for \(p_r=1.3\). After the transient, growth rates recover those of the asymptotic solution albeit an offset is created in the process. The cold core exhibits a completely different behaviour. For all \(p_r\), the vortex initially expands faster than the asymptotic solution due to the combined effects of higher momentum diffusivity and fluid expansion (see figure \ref{fig:Fig11}). In the longer term, the growth rate of \(r^+_{core}\) does not approach the asymptotic solution. Growth rates decay and a nearly linear relation with \(t^+\) seems to drive the growth of \(r^+_{core}\). Moreover, an opposite trend is observed where the vortex at \(p_r=1.3\) grows slower than at higher pressures compared to the asymptotic solution. These observations suggest that the effects of the large variations in fluid properties around the pseudo-boiling line remain in the flow for longer times in the cold vortex core compared to the hot core. \par 

\subsection{Evolution of Temperature, Fluid Properties and Radial Velocity}
\label{subsec:other_variables}

\begin{figure}
\centering
\begin{subfigure}{0.33\textwidth}
  \centering
  \includegraphics[width=1.05\linewidth]{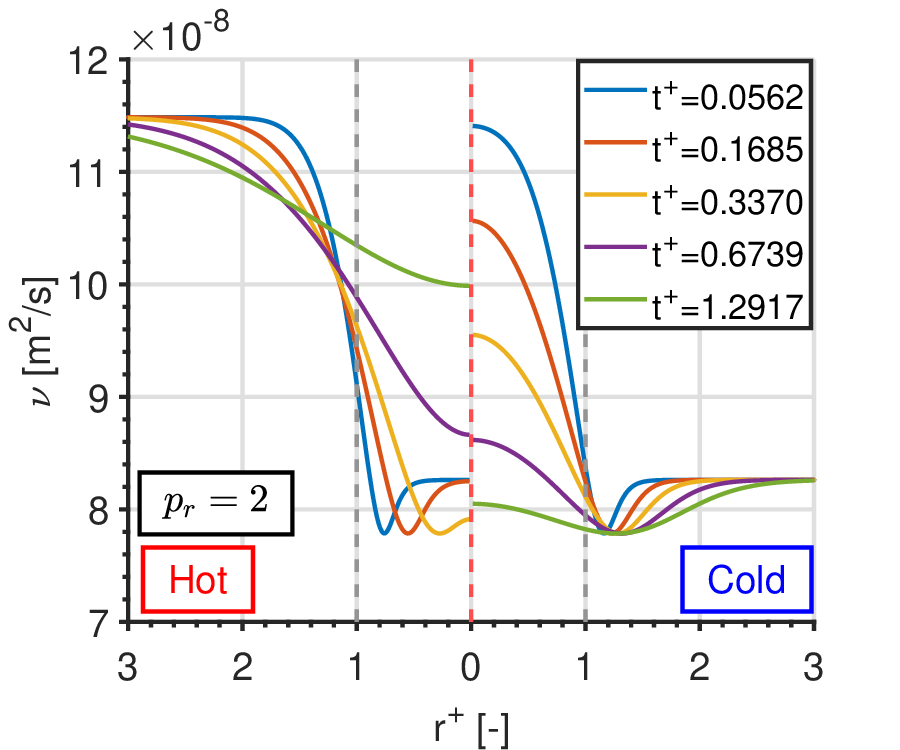}
  \caption{} 
  \label{subfig:kinvis_Re200_pr2_cold_vs_hot}
\end{subfigure}%
\begin{subfigure}{0.33\textwidth}
  \centering
  \includegraphics[width=1.05\linewidth]{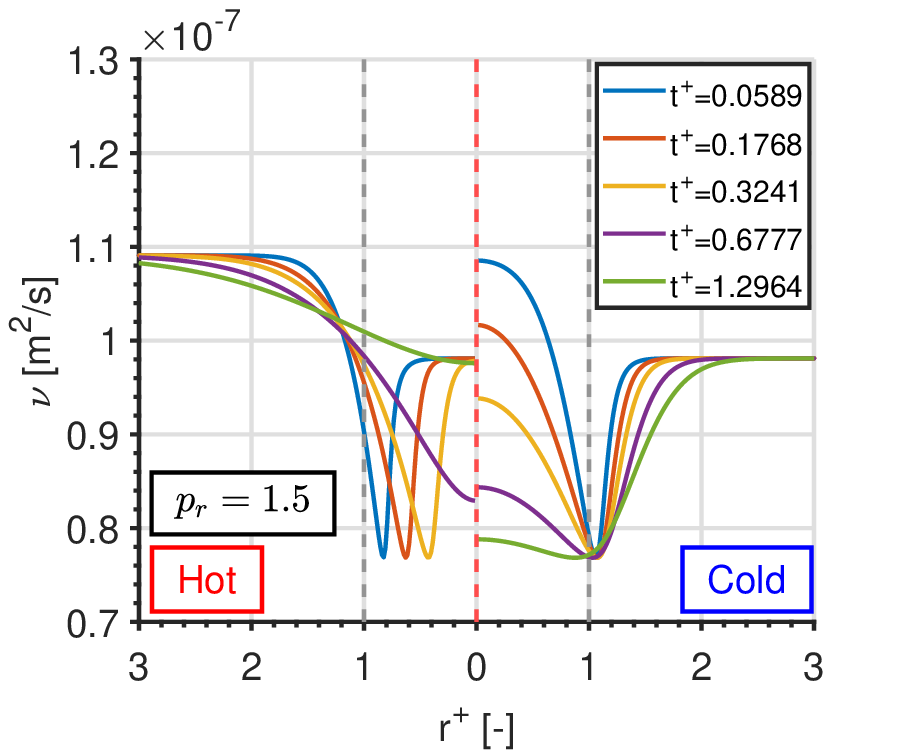}
  \caption{} 
  \label{subfig:kinvis_Re200_pr1p5_cold_vs_hot}
\end{subfigure}%
\begin{subfigure}{0.33\textwidth}
  \centering
  \includegraphics[width=1.05\linewidth]{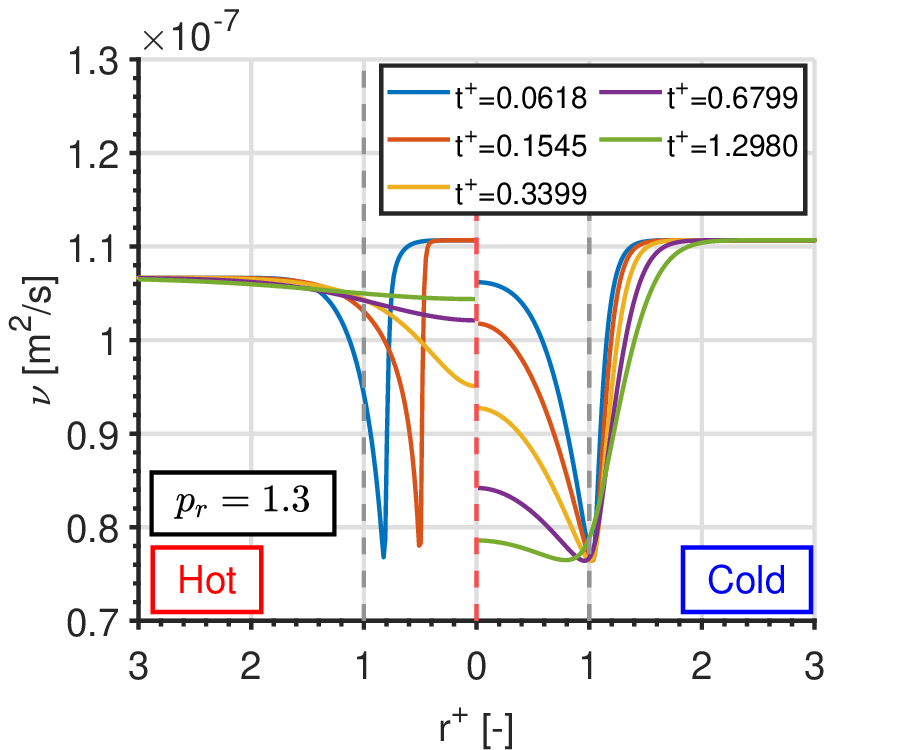}
  \caption{} 
  \label{subfig:kinvis_Re200_pr1p3_cold_vs_hot}
\end{subfigure}%
\\
\begin{subfigure}{0.33\textwidth}
  \centering
  \includegraphics[width=1.05\linewidth]{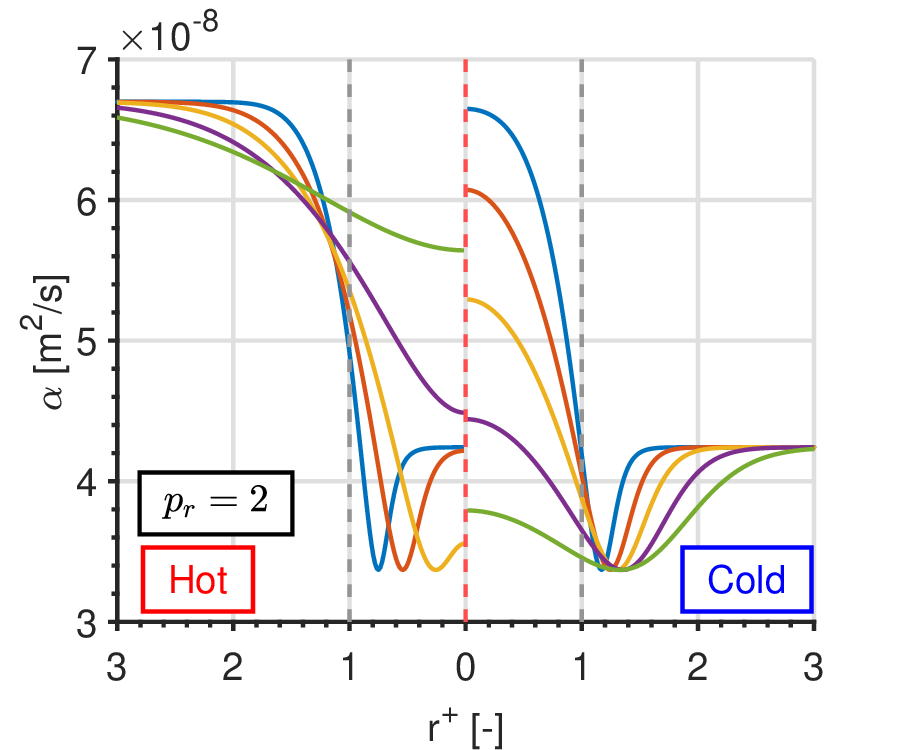}
  \caption{} 
  \label{subfig:alpha_Re200_pr2_cold_vs_hot}
\end{subfigure}%
\begin{subfigure}{0.33\textwidth}
  \centering
  \includegraphics[width=1.05\linewidth]{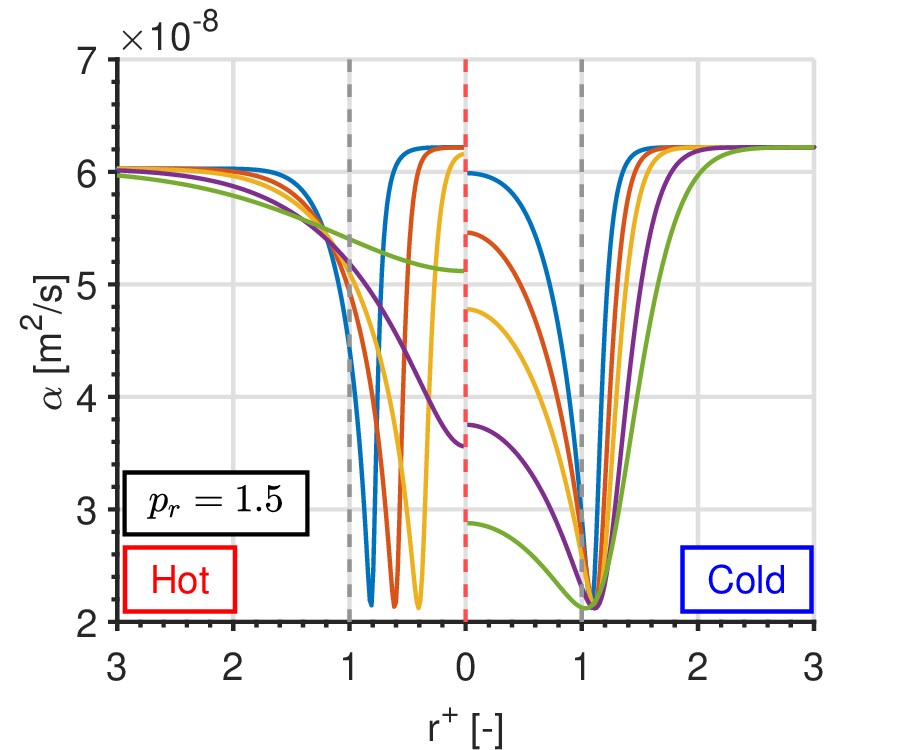}
  \caption{} 
  \label{subfig:alpha_Re200_pr1p5_cold_vs_hot}
\end{subfigure}%
\begin{subfigure}{0.33\textwidth}
  \centering
  \includegraphics[width=1.05\linewidth]{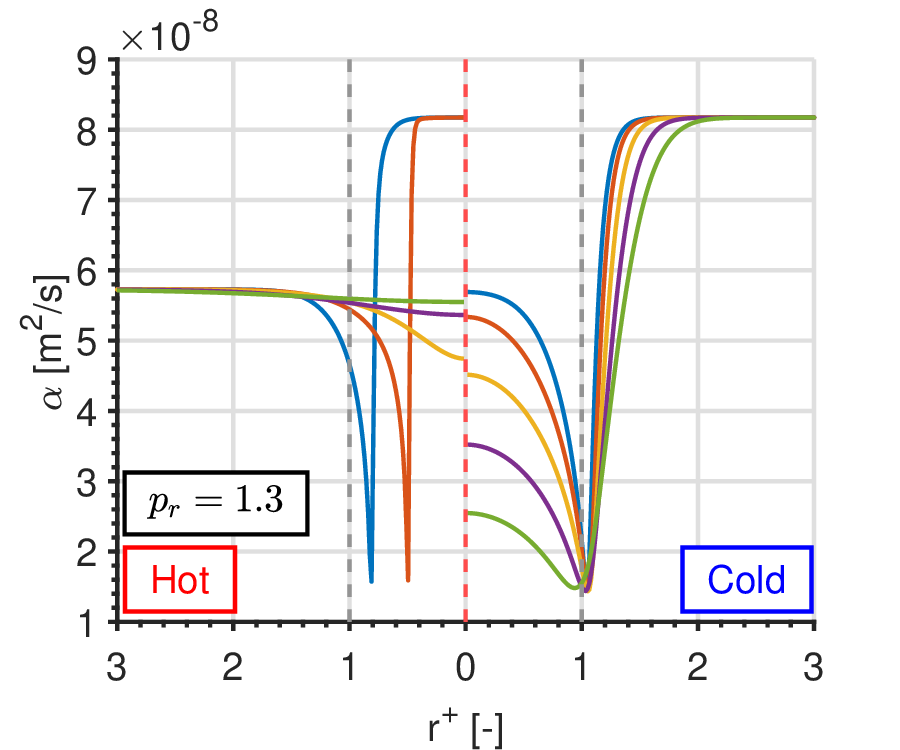}
  \caption{} 
  \label{subfig:alpha_Re200_pr1p3_cold_vs_hot}
\end{subfigure}%
\\
\begin{subfigure}{0.33\textwidth}
  \centering
  \includegraphics[width=1.05\linewidth]{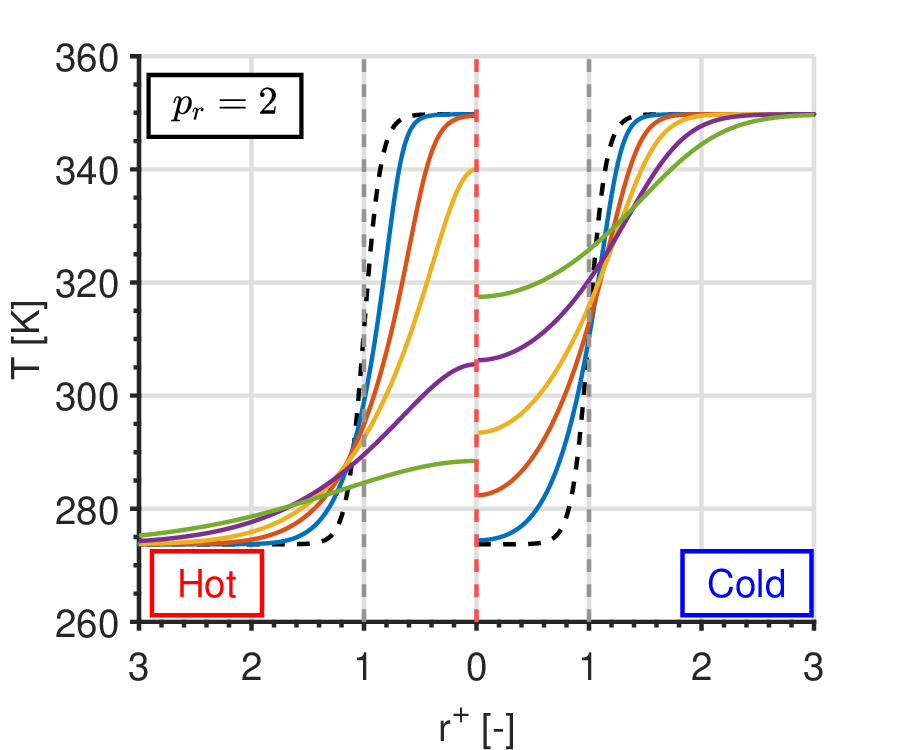}
  \caption{} 
  \label{subfig:temp_Re200_pr2_cold_vs_hot}
\end{subfigure}%
\begin{subfigure}{0.33\textwidth}
  \centering
  \includegraphics[width=1.05\linewidth]{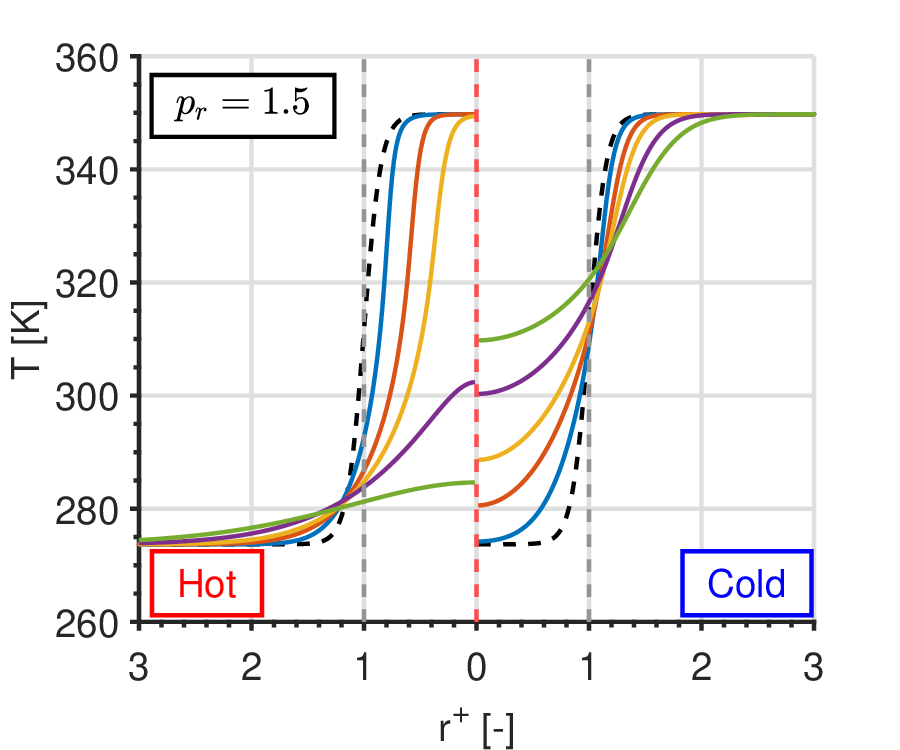}
  \caption{} 
  \label{subfig:temp_Re200_pr1p5_cold_vs_hot}
\end{subfigure}%
\begin{subfigure}{0.33\textwidth}
  \centering
  \includegraphics[width=1.05\linewidth]{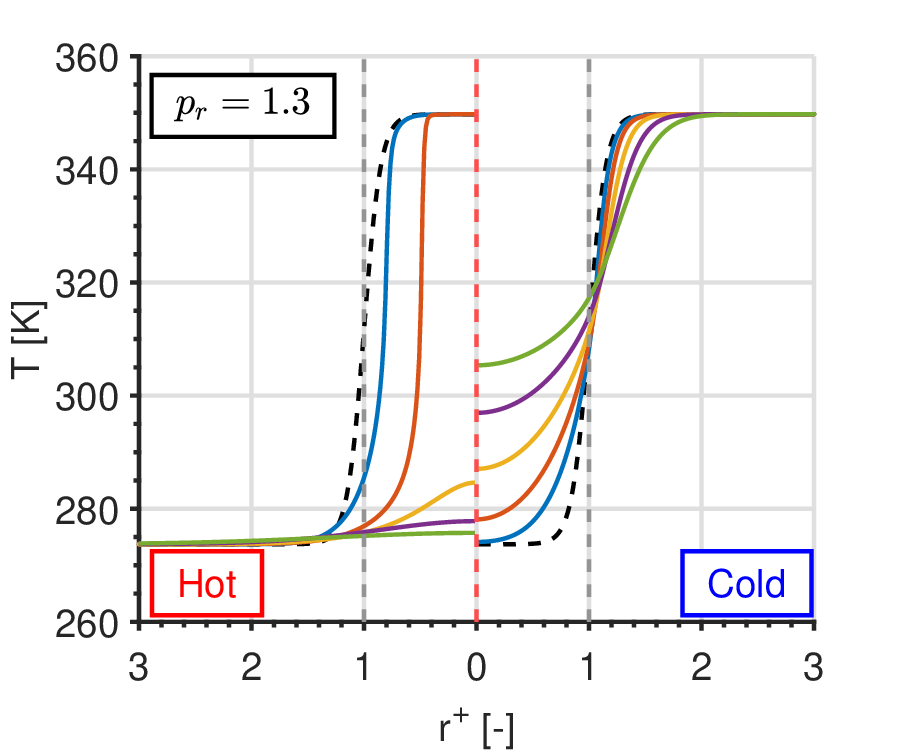}
  \caption{} 
  \label{subfig:temp_Re200_pr1p3_cold_vs_hot}
\end{subfigure}%
\caption{Evolution with \(t^+\) of relevant variables in the compressible vortex for the hot core (left) and the cold core (right) at different pressures. The dashed vertical lines represent the vortex centre (red) and \(r_c\) (grey). (a)-(c) \(\nu\); (d)-(f) \(\alpha\); and (g)-(i) \(T\).}
\label{fig:Fig10}
\end{figure}

\begin{figure}
\centering
\begin{subfigure}{0.33\textwidth}
  \centering
  \includegraphics[width=1.05\linewidth]{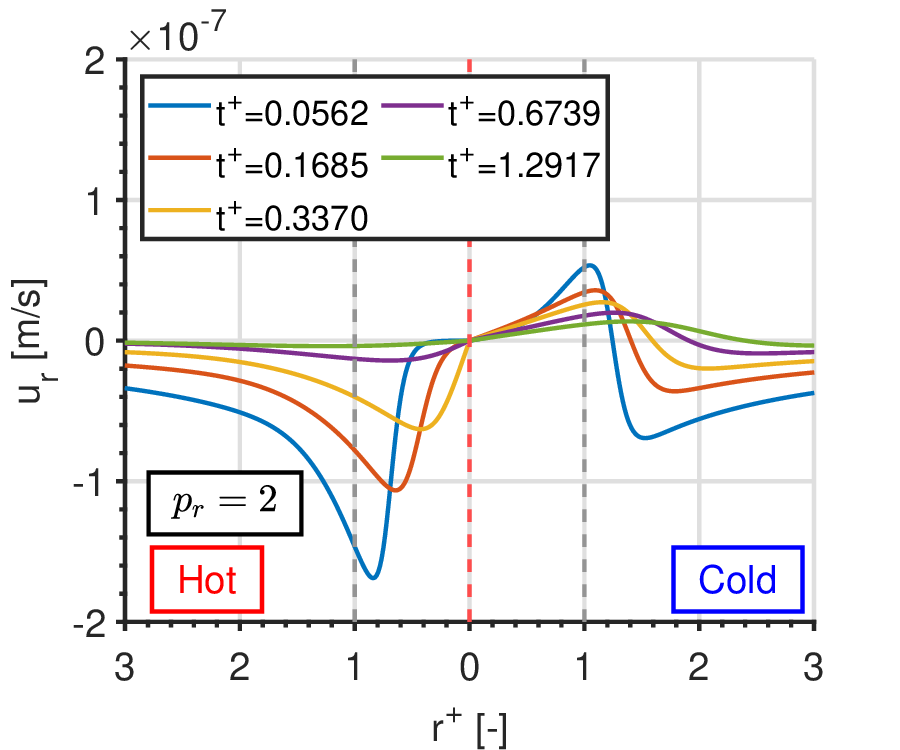}
  \caption{} 
  \label{subfig:radialvelocity_Re200_pr2_cold_vs_hot}
\end{subfigure}%
\begin{subfigure}{0.33\textwidth}
  \centering
  \includegraphics[width=1.05\linewidth]{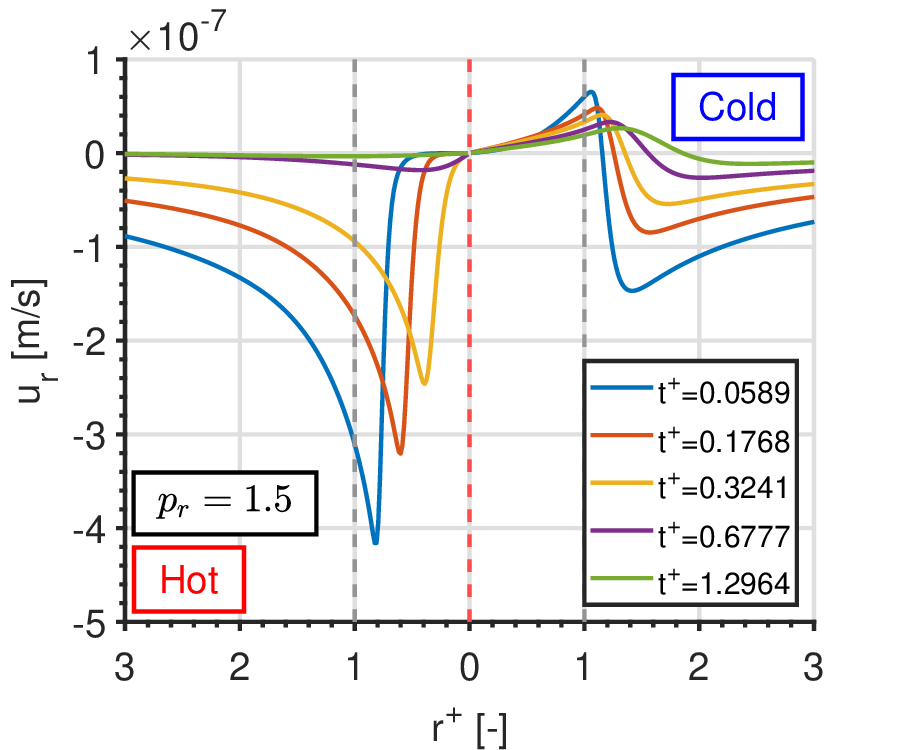}
  \caption{} 
  \label{subfig:radialvelocity_Re200_pr1p5_cold_vs_hot}
\end{subfigure}%
\begin{subfigure}{0.33\textwidth}
  \centering
  \includegraphics[width=1.05\linewidth]{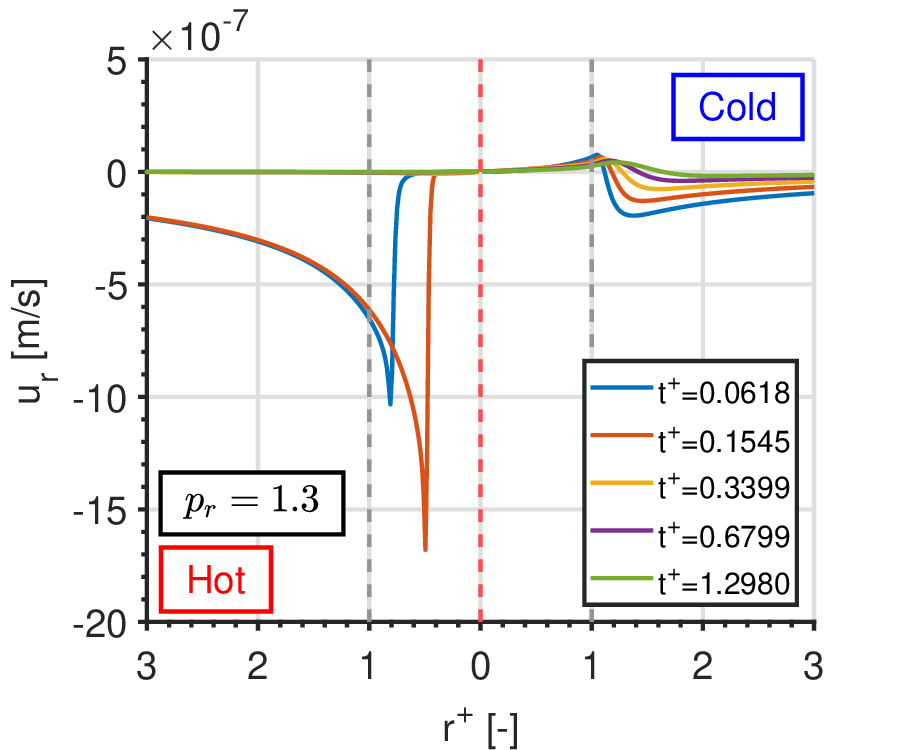}
  \caption{} 
  \label{subfig:radialvelocity_Re200_pr1p3_cold_vs_hot}
\end{subfigure}%
\\
\begin{subfigure}{0.33\textwidth}
  \centering
  \includegraphics[width=1.05\linewidth]{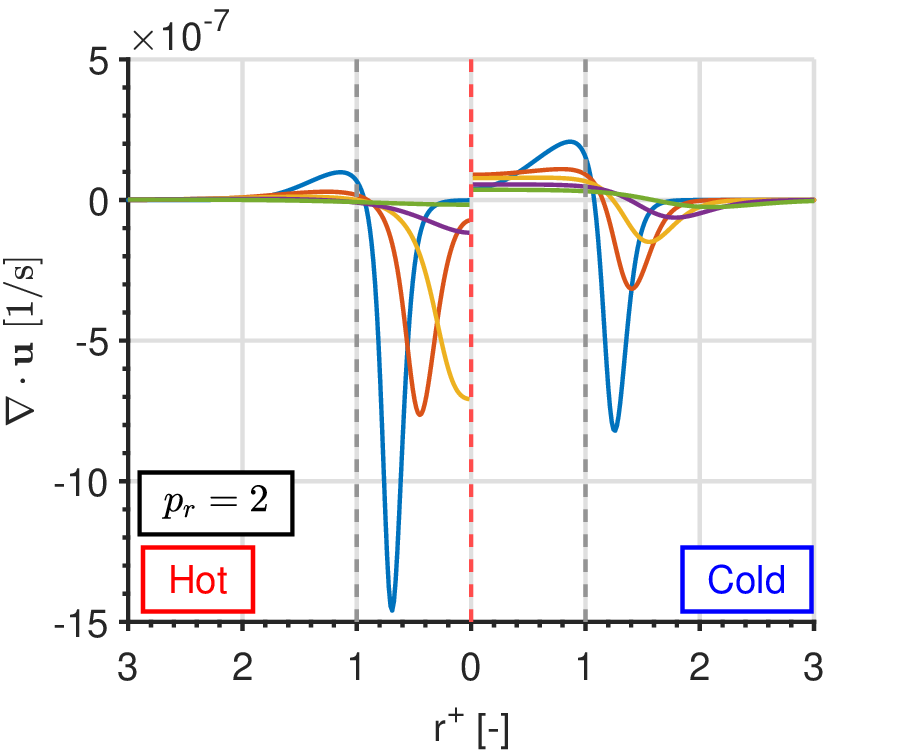}
  \caption{} 
  \label{subfig:divergence_Re200_pr2_cold_vs_hot}
\end{subfigure}%
\begin{subfigure}{0.33\textwidth}
  \centering
  \includegraphics[width=1.05\linewidth]{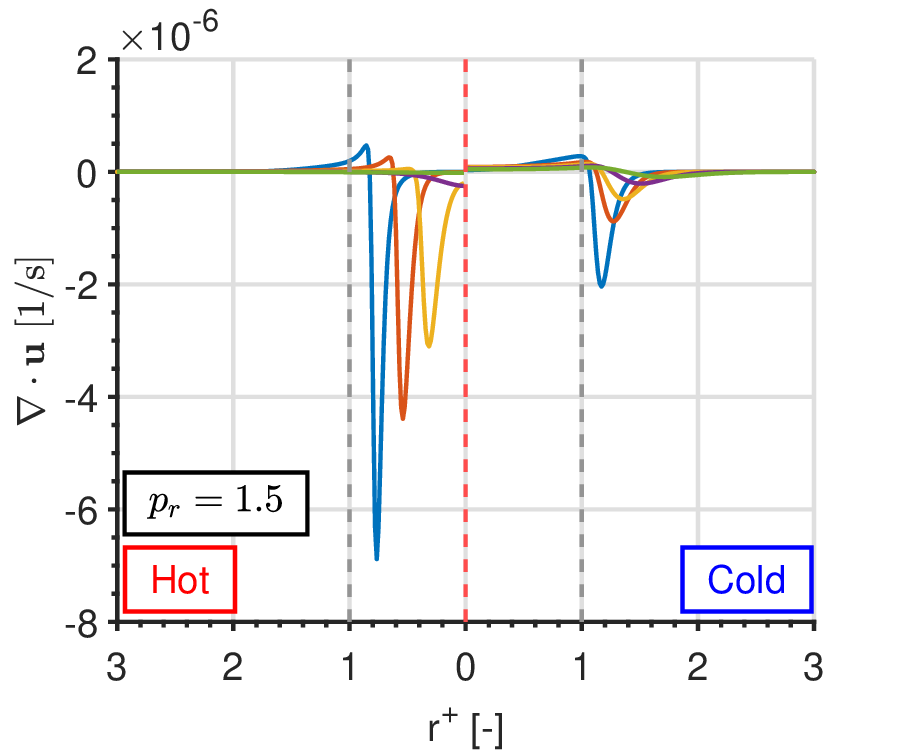}
  \caption{} 
  \label{subfig:divergence_Re200_pr1p5_cold_vs_hot}
\end{subfigure}%
\begin{subfigure}{0.33\textwidth}
  \centering
  \includegraphics[width=1.05\linewidth]{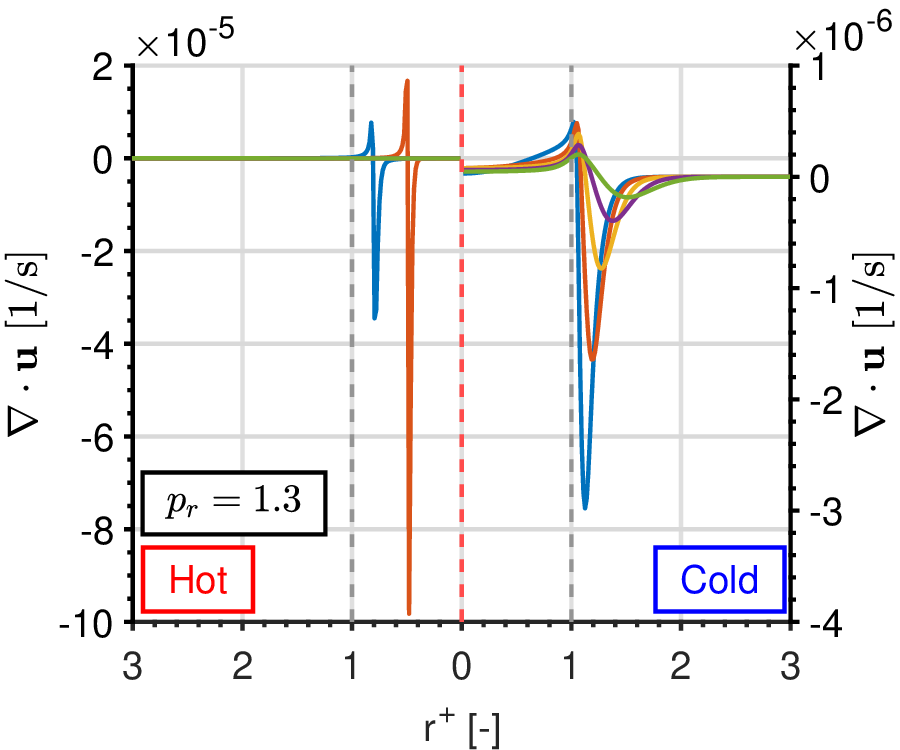}
  \caption{} 
  \label{subfig:divergence_Re200_pr1p3_cold_vs_hot}
\end{subfigure}%
\caption{Evolution with \(t^+\) of relevant variables in the compressible vortex for the hot core (left) and the cold core (right) at different pressures. The dashed vertical lines represent the vortex centre (red) and \(r_c\) (grey). (a)-(c) \(u_r\); and (d)-(f) \(\boldsymbol{\nabla}\boldsymbol{\cdot}\boldsymbol{u}\).}
\label{fig:Fig11}
\end{figure}

The temporal evolution of specific terms shown in section \ref{subsec:cold_vs_hot} and the related vorticity mechanisms discussed in section \ref{subsec:vorticity_analysis} are a consequence of the thermal mixing process in a SCF near pseudo-boiling conditions. Thus, we show here the evolution of temperature and other thermodynamic variables to describe the supercritical fluid behaviour. These terms are \(\nu\), \(\alpha\) and \(T\) (figure \ref{fig:Fig10}), and \(u_r\) and \(\boldsymbol{\nabla}\boldsymbol{\cdot}\boldsymbol{u}\) (figure \ref{fig:Fig11}). \par

Focusing first on the cold vortex, figure \ref{fig:Fig10} shows that the minima of momentum and thermal diffusivities are typically pinned at the edge of the initial vortex or \(r_c\), i.e., \(r^+=1\). In other words, the pseudo-boiling temperature where the fluid transitions from liquid-like to gas-like properties is nearly stationary in space while the vortex core heats up. The heat transfer behaviour resembles that of droplet heating and evaporation where the liquid-gas interface, in this case being the pseudo-boiling line in the continuous fluid, remains at an interfacial equilibrium temperature, with heat diffusing on both sides of the interface \citep{1993_IJHMT_Delplanque,2018_IJHMT_Poblador}. That is, the liquid-like fluid is inside the vortex, while the gas-like fluid surrounds it. When looking at figure \ref{fig:Fig11}, one observes that, much like the droplet evaporation phenomenon at high pressures and for multi-component fluids \citep{2003_CST_Zhang,2017_JCP_He,2024_AIAA_Poblador}, heating inside the vortex causes the fluid to expand, i.e., \(u_r>0\), while cooling outside the vortex causes the fluid to compress, i.e., \(u_r<0\). Despite these effects being usually negligible at lower pressures for single-component fluids where only the Stefan flow induces a radial velocity, here they play an important role where, in combination with local minima in diffusivities, the pseudo-boiling region remains nearly frozen in space. A similar self-stabilisation mechanism has been identified by \citet{2026_IJTS_Yin}. Since diffusivities are smaller on the colder side of the fluid, the diffusive nature of the problem is hindered by the fluid properties of the SCF, resulting in a slower growth rate for \(r^+_{core}\) as shown in figure \ref{fig:Fig9}. Moreover, a reversal in \(\nu\) and \(\alpha\) is observed whereby at larger \(p_r\), the diffusivities inside the vortex core are larger than outside; at lower \(p_r\) the opposite occurs. Additionally, although the minima in momentum diffusivities remains similar across all pressures, thermal diffusivity drops more as the ambient pressure gets closer the critical pressure, i.e., specific heat peaks as seen in figure \ref{subfig:validation_cp}. This could explain why \(r^+_{core}\) grows faster at \(p_r=2\) than at lower \(p_r\). Although our simulations do not run beyond \(t^+\approx1.3\), it is expected that an eventual equilibration of temperature would occur and that \(r^+_{core}\) would approach the asymptotic growth rates shown in figure \ref{fig:Fig9}. \par

The hot vortex core displays even more critical phenomena. At \(p_r=2\), diffusivities inside the core are much lower than outside. Therefore, \(r^+_{core}\) grows much slower than the asymptotic case at early times as seen in figure \ref{fig:Fig9}. This situation is reversed as pressure decreases with both \(\nu\) and \(\alpha\) being larger in the vortex core than outside; thus, \(r^+_{core}\) growing faster than the asymptotic solution. However, a more complex behaviour rapidly evolves. The cooling of the vortex core resembles now the condensation and rapid collapse, without surface tension effects, of a hot bubble surrounded by a cooler liquid, i.e., the gas-like fluid is inside the vortex while surrounded by the liquid-like fluid \citep{1955_JMP_Zwick,1963_JASA_Hickling}. Thus, the SCF expands outside the vortex core but strongly compresses inside the core (see figure \ref{fig:Fig11}). This results in a strong radial velocity into the vortex (\(u_r<0\)). At \(p_r=2\), this advection of cooler liquid into the vortex, balanced by thermal diffusion, cool the vortex reasonably fast compared to the heating of the corresponding cold core case. Then, pseudo-boiling effects vanish and the evolution of \(\omega_z\) and \(r^+_{core}\) resemble more the corresponding incompressible case at larger \(t^+\). As pressure decreases closer to \(p_c\), fluid compression in the hot vortex increases while the thermal diffusivity reaches lower minima across the pseudo-boiling line. This acts as a thermal barrier impeding heat conduction, leading to a steepening of the thermal front (see figure \ref{subfig:temp_Re200_pr1p3_cold_vs_hot}) -- this intensifies compressibility effects and the pseudo-condensation of the SCF. Naturally, this raises the question as to whether this compression of the thermal front, which tends to a discontinuity as \(p\rightarrow p_c\), and the subsequent effects on the SCF are well captured by the mesh resolution. To the best of our knowledge, this is well achieved in our simulations. As observed in figure \ref{fig:Fig7}, this compression of the thermal front around pseudo-boiling conditions has critical consequences for the evolution of vorticity as other vorticity production mechanisms become significant. The nature of vorticity evolution is discussed in section \ref{subsec:vorticity_analysis}. \par

\subsection{Vorticity Generation and Decay via Viscous Mechanisms}
\label{subsec:vorticity_analysis}

The mechanisms of vorticity generation and decay in the SCF are analysed first for the cases with \(p_r=2\) due to their less extreme behaviour. Section \ref{subsubsec:vorticityeq} shows the contribution to \(\partial\omega_z/\partial t\) of each term in the vorticity equation (\ref{eqn:vorticityZ_2}), highlighting viscous mechanisms, and section \ref{subsubsec:viscouseffects} explains vorticity diffusion and generation/decay mechanisms arising due to the strong variation of fluid properties across the pseudo-boiling line. Lastly, section \ref{subsubsec:pseudoboiling} discusses the enhancement of those mechanisms as \(p\rightarrow p_c\), coupled with the evolution of the SCF discussed in section \ref{subsec:other_variables}. \par

\subsubsection{Analysis of the Vorticity Terms}
\label{subsubsec:vorticityeq}

\begin{figure}
\centering
\begin{subfigure}{0.33\textwidth}
  \centering
  \includegraphics[width=1.05\linewidth]{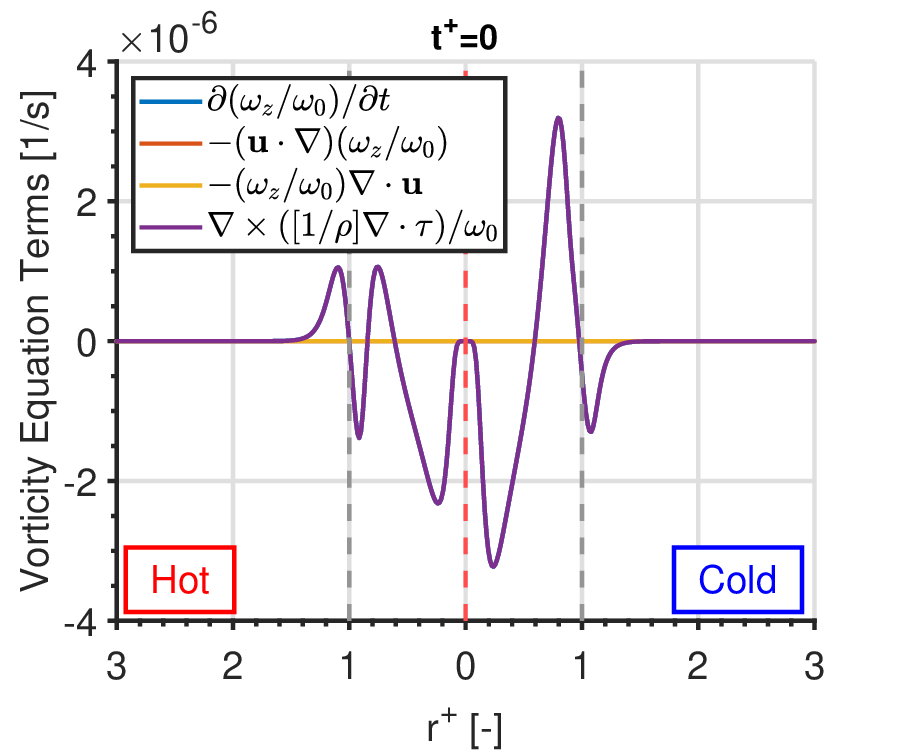}
  \caption{} 
  \label{subfig:vorticityeq_Re200_pr2_File000_cold_vs_hot}
\end{subfigure}%
\begin{subfigure}{0.33\textwidth}
  \centering
  \includegraphics[width=1.05\linewidth]{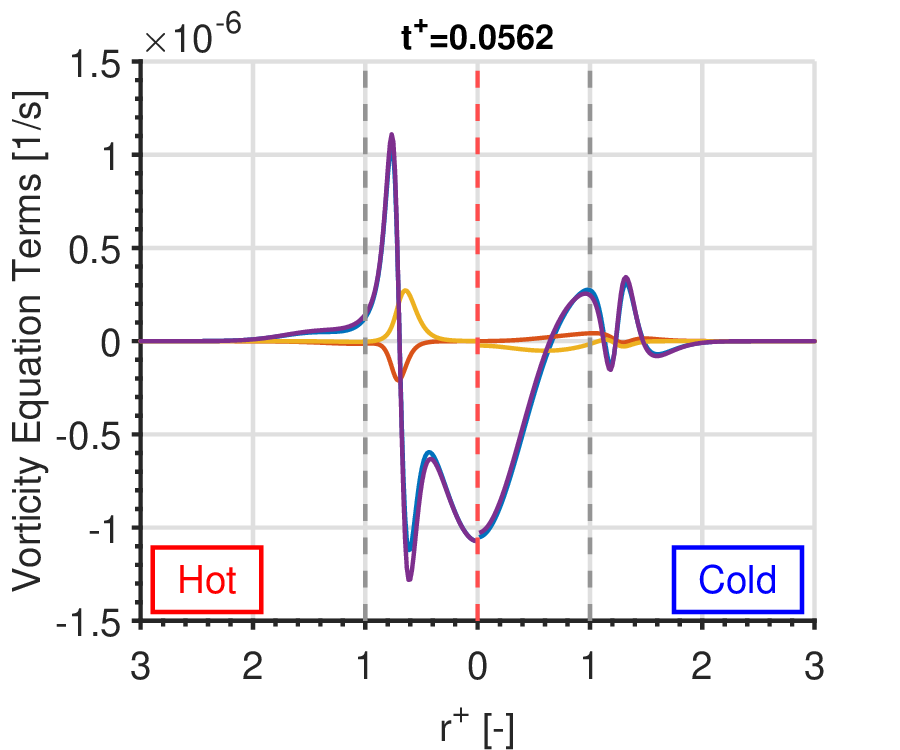}
  \caption{} 
  \label{subfig:vorticityeq_Re200_pr2_File002_cold_vs_hot}
\end{subfigure}%
\begin{subfigure}{0.33\textwidth}
  \centering
  \includegraphics[width=1.05\linewidth]{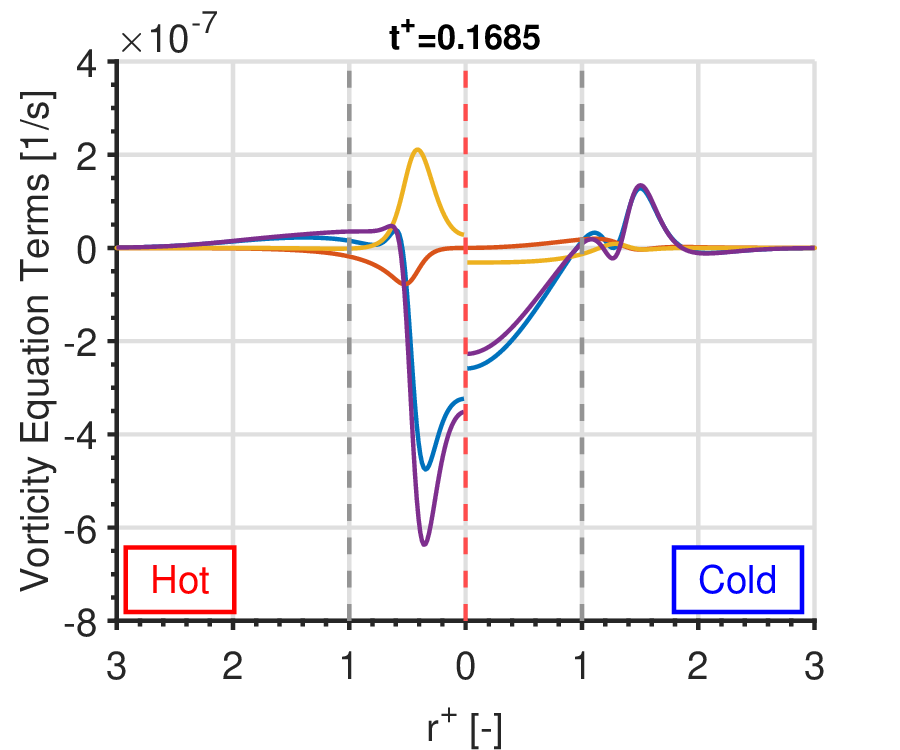}
  \caption{} 
  \label{subfig:vorticityeq_Re200_pr2_File006_cold_vs_hot}
\end{subfigure}%
\\
\begin{subfigure}{0.33\textwidth}
  \centering
  \includegraphics[width=1.05\linewidth]{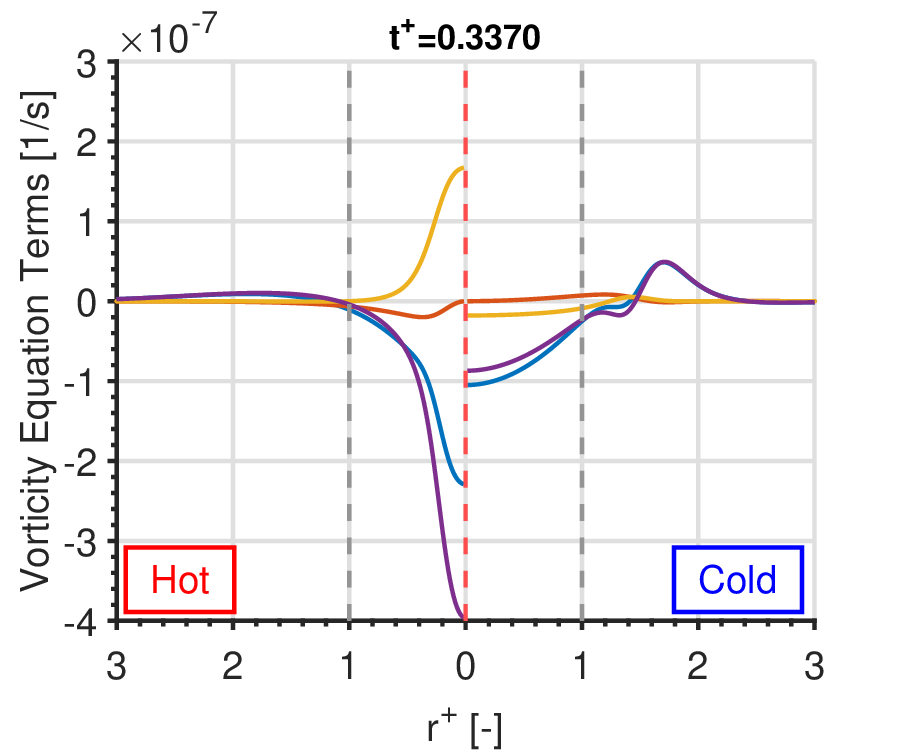}
  \caption{} 
  \label{subfig:vorticityeq_Re200_pr2_File012_cold_vs_hot}
\end{subfigure}%
\begin{subfigure}{0.33\textwidth}
  \centering
  \includegraphics[width=1.05\linewidth]{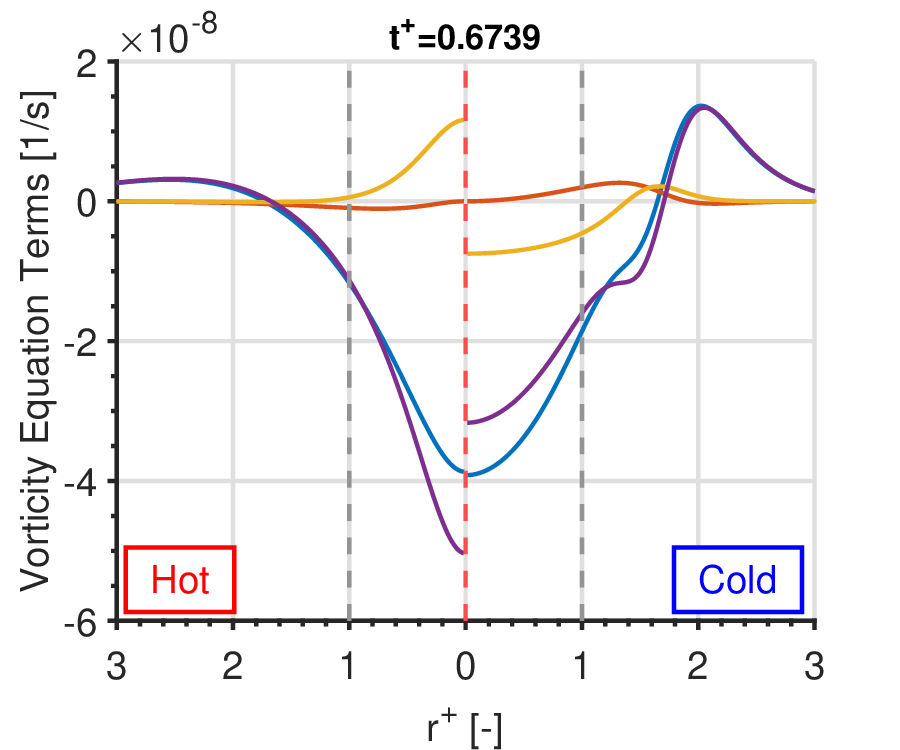}
  \caption{} 
  \label{subfig:vorticityeq_Re200_pr2_File024_cold_vs_hot}
\end{subfigure}%
\begin{subfigure}{0.33\textwidth}
  \centering
  \includegraphics[width=1.05\linewidth]{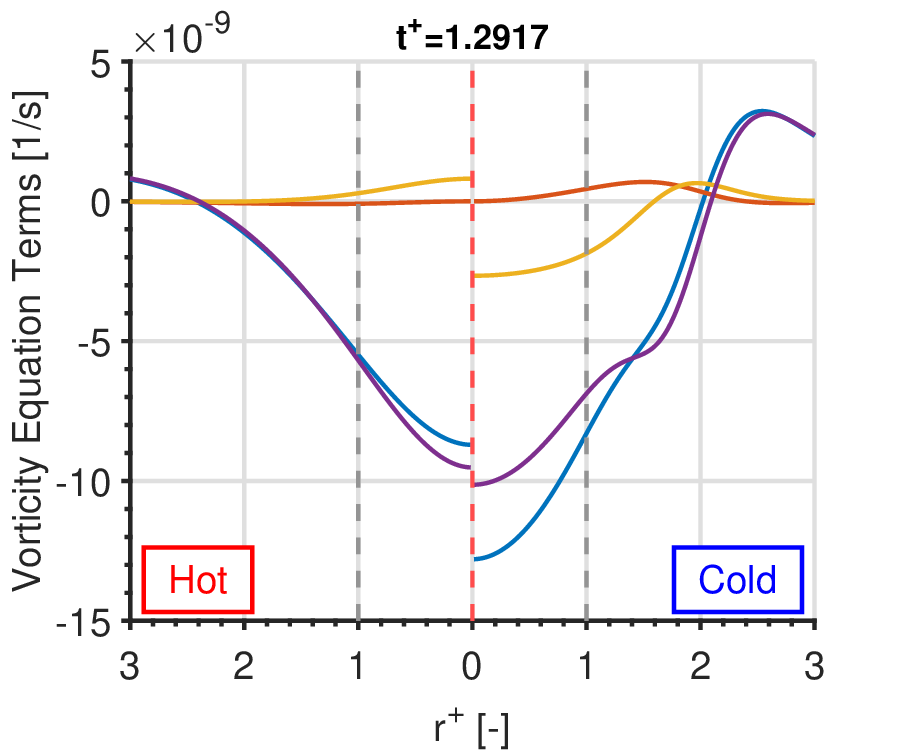}
  \caption{} 
  \label{subfig:vorticityeq_Re200_pr2_File046_cold_vs_hot}
\end{subfigure}%
\caption{Evolution with \(t^+\) of the terms in the vorticity equation non-dimensionalized with \(\omega_0\) for the hot core (left) and the cold core (right) at \(p_r=2\). The dashed vertical lines represent the vortex centre (red) and \(r_c\) (grey). (a) \(t^+=0\); (b) \(t^+=0.0562\); (c) \(t^+=0.1685\); (d) \(t^+=0.3370\); (e) \(t^+=0.6739\); and (f) \(t^+=1.2917\).}
\label{fig:Fig12}
\end{figure}

Figure \ref{fig:Fig12} shows the various terms of (\ref{eqn:vorticityZ_2}), scaled by \(\omega_0\), at the selected \(t^+\) for each configuration at \(p_r=2\) that determine the local evolution of \(\omega_z\). In addition, the initial profiles at \(t^+=0\) are shown. The change in \(\omega_z\) is mainly driven by the viscous term, i.e., vorticity decays and the vortex spreads due to viscous diffusion in the incompressible limit. In the SCF, the viscous term presents large oscillations about the pseudo-boiling line -- where fluid properties vary strongly -- which occurs inside (hot core) or outside (cold core) the initial vortex depending on the configuration. In fact, the viscous term is responsible for the production of negative vorticity along the edge of the vortex in the cold core configuration, as depicted in figure \ref{subfig:vorticityeq_Re200_pr2_File000_cold_vs_hot}. Initially, \(\omega_z=0\) outside the vortex core (\(r>r_c\)) and viscous effects other than diffusion generate negative vorticity in the region, whereas the opposite is observed for the hot core. The physical mechanisms causing this vorticity production outside the vortex core are explained in section \ref{subsubsec:viscouseffects}. \par

Vorticity transport due to radial advection and vorticity stretching due to the fluid's compressibility are not negligible. Particularly, vorticity stretching is dominant in the hot vortex during the early times of the core cooling. Stretching counteracts viscous effects near the vortex centre, increasing the time scales for the decay of vorticity. In contrast, fluid expansion in the cold vortex enhances the decay of \(\omega_z\), which is more noticeable at later \(t^+\). Overall, the compression of the hot vortex is faster than the expansion of the cold vortex, as discussed in sections \ref{subsec:cold_vs_hot} and \ref{subsec:other_variables}. At pressures closer to the critical point, compressibility effects may be accentuated and thus compete more strongly with the viscous term. However, this is not observed as discussed in section \ref{subsubsec:pseudoboiling}. \par 

\subsubsection{Viscous Effects across the Pseudo-boiling Line}
\label{subsubsec:viscouseffects}

Given the importance of the viscous term, further insight into the mechanisms for vorticity generation and decay can be obtained by rewriting (\ref{eqn:vorticityZ_2}) in cylindrical coordinates and imposing axisymmetry. Thus, identifying the relevant components in \(\boldsymbol{\nabla\times}\big([1/\rho]\boldsymbol{\nabla}\boldsymbol{\cdot}\boldsymbol{\tau}\big)\). The terms on the RHS of (\ref{eqn:vorticityZ_2}) are

\begin{equation}
\label{eqn:vorticityZ_3}
\begin{split}
(\boldsymbol{u}\boldsymbol{\cdot}\boldsymbol{\nabla}) \omega_z = u_r\frac{\partial\omega_z}{\partial r}
\end{split}
\end{equation}
\begin{equation}
\label{eqn:vorticityZ_4}
\begin{split}
\omega_z (\boldsymbol{\nabla}\boldsymbol{\cdot}\boldsymbol{u}) = \omega_z\bigg(\frac{\partial u_r}{\partial r}+\frac{u_r}{r}\bigg)
\end{split}
\end{equation}
\begin{equation}
\label{eqn:vorticityZ_5}
\begin{split}
\bigg[\boldsymbol{\nabla\times}\bigg(\frac{1}{\rho}\boldsymbol{\nabla}\boldsymbol{\cdot}\boldsymbol{\tau}\bigg)\bigg]\boldsymbol{\cdot}\hat{\textbf{e}}_z & = \frac{\mu}{\rho} \bigg(\frac{\partial^3u_\theta}{\partial r^3}+\frac{2}{r}\frac{\partial^2u_\theta}{\partial r^2}-\frac{1}{r^2}\frac{\partial u_\theta}{\partial r}+\frac{u_\theta}{r^3}\bigg) + \frac{1}{\rho}\frac{\partial^2\mu}{\partial r^2}\bigg(\frac{\partial u_\theta}{\partial r}-\frac{u_\theta}{r}\bigg) \\
& + \frac{1}{\rho}\frac{\partial \mu}{\partial r}\bigg(2\frac{\partial^2u_\theta}{\partial r^2}+\frac{1}{r}\frac{\partial u_\theta}{\partial r}-\frac{u_\theta}{r^2}\bigg) + \frac{\mu}{\rho^2}\frac{\partial\rho}{\partial r}\bigg(-\frac{\partial^2 u_\theta}{\partial r^2}-\frac{1}{r}\frac{\partial u_\theta}{\partial r}+\frac{u_\theta}{r^2}\bigg) \\
& + \frac{1}{\rho^2}\frac{\partial\rho}{\partial r}\frac{\partial\mu}{\partial r}\bigg(-\frac{\partial u_\theta}{\partial r}+\frac{u_\theta}{r}\bigg)
\end{split}
\end{equation}

It is evident that, for an incompressible fluid (\(u_r=0\)), convective transport and compressible vortex stretching do not occur. Despite the fact that radial and azimuthal viscous stresses are not zero, i.e., \(\tau_{rr}=\mu\big(\frac{4}{3}\frac{\partial u_r}{\partial r}-\frac{2}{3}\frac{u_r}{r}\big)\neq0\) and \(\tau_{\theta\theta}=\mu\big(\frac{4}{3}\frac{u_r}{r}-\frac{2}{3}\frac{\partial u_r}{\partial r}\big)\neq0\), (\ref{eqn:vorticityZ_5}) confirms that the viscous effects on the evolution of the vorticity result only from the torque generated by shear between fluid elements moving at different \(u_\theta\) given by \(\tau_{r\theta}=\tau_{\theta r}=\mu\big(\frac{\partial u_\theta}{\partial r}-\frac{u_\theta}{r}\big)\) and the radial gradients in the fluid properties. That is, additional terms emerge in the compressible vortex due to the varying fluid properties and include the first- and second-order radial derivatives of \(\rho\) and \(\mu\). \par 

\begin{figure}
\centering
\includegraphics[width=0.6\linewidth]{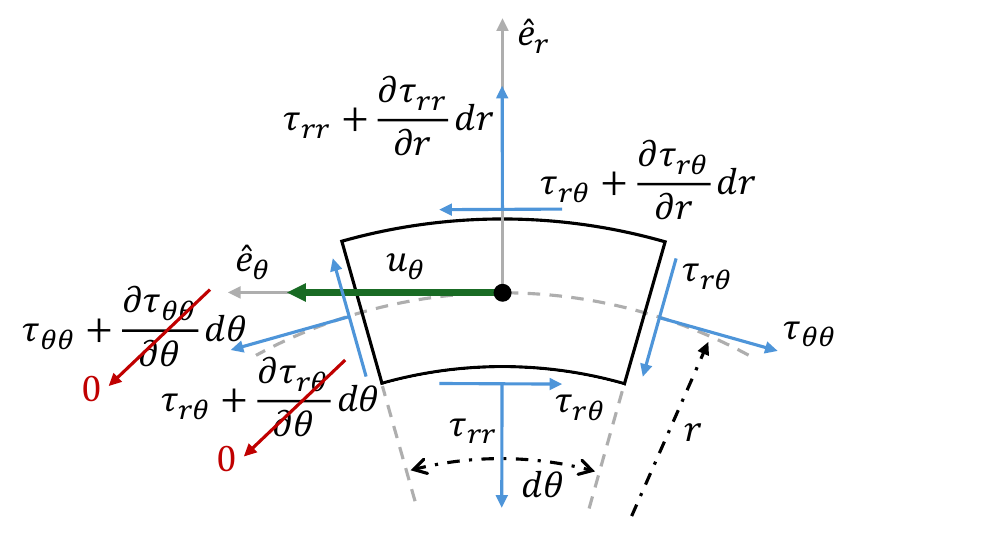}
\caption{Viscous stresses acting on a two-dimensional fluid element in cylindrical coordinates under axisymmetric conditions.}
\label{fig:Fig13}
\end{figure}

Figure \ref{fig:Fig13} sketches the viscous stresses that act on a two-dimensional infinitesimal fluid element in cylindrical coordinates. Axisymmetry yields the viscous force per unit volume 

\begin{equation}
\label{eqn:stress}
\boldsymbol{\nabla}\boldsymbol{\cdot}\boldsymbol{\tau}=\bigg(\frac{\partial\tau_{rr}}{\partial r}+\frac{1}{r}(\tau_{rr}-\tau_{\theta\theta})\bigg)\hat{\textbf{e}}_r+\bigg(\frac{\partial\tau_{r\theta}}{\partial r}+\frac{2\tau_{r\theta}}{r}\bigg)\hat{\textbf{e}}_\theta
\end{equation}

\noindent
with the radial stress causing a zero net torque. The symmetry brought about by \(\frac{\partial}{\partial\theta}=0\) ensures that the resulting forces from \(\frac{\partial\tau_{rr}}{\partial r}\) and \(\frac{\tau_{rr}}{r}\) are perfectly aligned with the centre of mass, while the torques caused by the two resultant lateral stresses from \(\tau_{\theta\theta}\) perfectly balance each other. Thus, the net torque changing the vorticity can come only from the azimuthal viscous stresses. Expanding on \(\boldsymbol{\nabla\times}\big([1/\rho]\boldsymbol{\nabla}\boldsymbol{\cdot}\boldsymbol{\tau}\big)\) as shown in (\ref{eqn:stress2}), two contributions to the variation of \(\omega_z\) are identified: a torque caused by the radially-varying azimuthal viscous stresses and a torque resulting from the inertial response of the fluid to these stresses, which is similar to the baroclinic torque but is linked to viscous forces instead of the pressure gradient. Unlike the incompressible fluid behaviour, the radial variations of \(\mu\) and \(\rho\) affect the distribution of mass (inertia) and shear resistance of the viscous fluid, generating the additional terms.

\begin{equation}
\label{eqn:stress2}
\bigg[\boldsymbol{\nabla\times}\bigg(\frac{1}{\rho}\boldsymbol{\nabla}\boldsymbol{\cdot}\boldsymbol{\tau}\bigg)\bigg]\boldsymbol{\cdot}\hat{\textbf{e}}_z = \bigg[\frac{1}{\rho}\boldsymbol{\nabla\times\nabla\cdot\tau}\bigg]\boldsymbol{\cdot}\hat{\textbf{e}}_z-\bigg[\frac{1}{\rho^2}\boldsymbol{\nabla}\rho\boldsymbol{\times\nabla\cdot\tau}\bigg]\boldsymbol{\cdot}\hat{\textbf{e}}_z = \underbrace{\frac{1}{\rho}\frac{1}{r}\frac{\partial}{\partial r}\bigg(r\bigg(\frac{\partial\tau_{r\theta}}{\partial r}+\frac{2\tau_{r\theta}}{r}\bigg)\bigg)}_{\text{Varying viscous stress torque}}-\underbrace{\frac{1}{\rho^2}\frac{\partial\rho}{\partial r}\bigg(\frac{\partial\tau_{r\theta}}{\partial r}+\frac{2\tau_{r\theta}}{r}\bigg)}_{\text{Inertial response torque}}
\end{equation}

To identify the major mechanisms represented by the viscous term, (\ref{eqn:vorticityZ_5}) is rewritten in terms of \(\omega_z\) and its derivatives. For the axisymmetric problem, the following relations hold 

\begin{equation}
\label{eqn:vorticityZ_6}
\begin{split}
\omega_z=\frac{1}{r}\frac{\partial}{\partial r}(ru_\theta)=\frac{\partial u_\theta}{\partial r}+\frac{u_\theta}{r} \rightarrow \frac{\partial u_\theta}{\partial r}=\omega_z-\frac{u_\theta}{r}
\end{split}
\end{equation}
\begin{equation}
\label{eqn:vorticityZ_7}
\begin{split}
\frac{\partial^2u_\theta}{\partial r^2}=\frac{\partial\omega_z}{\partial r}-\frac{\omega_z}{r}+\frac{2u_\theta}{r^2}
\end{split}
\end{equation}
\begin{equation}
\label{eqn:vorticityZ_8}
\begin{split}
\frac{\partial^3u_\theta}{\partial r^3}=\frac{\partial^2\omega_z}{\partial r^2}-\frac{1}{r}\frac{\partial\omega_z}{\partial r}+\frac{3\omega_z}{r^2}-\frac{6u_\theta}{r^3}
\end{split}
\end{equation}

\noindent
Then, combining (\ref{eqn:vorticityZ_6})-(\ref{eqn:vorticityZ_8}) with (\ref{eqn:vorticityZ_5}) results in

\begin{equation}
\label{eqn:vorticityZ_9}
\begin{split}
\bigg[\boldsymbol{\nabla\times}\bigg(\frac{1}{\rho}\boldsymbol{\nabla}\boldsymbol{\cdot}\boldsymbol{\tau}\bigg)\bigg]\boldsymbol{\cdot}\textbf{e}_z & = \underbrace{\frac{\mu}{\rho} \Bigg[\frac{1}{r}\frac{\partial}{\partial r}\bigg(r\frac{\partial\omega_z}{\partial r}\bigg)\Bigg]}_\text{(A) Vorticity viscous diffusion} + \underbrace{\omega_z\Bigg[\frac{1}{\rho}\bigg(\frac{1}{r}\frac{\partial}{\partial r}\bigg(r\frac{\partial\mu}{\partial r}\bigg)\bigg)-\frac{2}{\rho}\bigg(\frac{1}{r}\frac{\partial\mu}{\partial r}\bigg)-\frac{1}{\rho^2}\frac{\partial\rho}{\partial r}\frac{\partial\mu}{\partial r}\Bigg]}_\text{(B) Vorticity stretching from viscosity gradients} \\
& + \underbrace{\Bigg[\frac{2}{\rho}\frac{\partial\mu}{\partial r}-\frac{\mu}{\rho^2}\frac{\partial\rho}{\partial r}\Bigg]\frac{\partial\omega_z}{\partial r}}_\text{(C) Vorticity-viscosity gradient alignment} - \underbrace{\frac{2u_\theta}{r}\Bigg[\frac{1}{\rho}\bigg(\frac{1}{r}\frac{\partial}{\partial r}\bigg(r\frac{\partial\mu}{\partial r}\bigg)\bigg)-\frac{2}{\rho}\bigg(\frac{1}{r}\frac{\partial\mu}{\partial r}\bigg)-\frac{1}{\rho^2}\frac{\partial\rho}{\partial r}\frac{\partial\mu}{\partial r}\Bigg]}_\text{(D) Vorticity production from azimuthal velocity}
\end{split}
\end{equation}

Four physical mechanisms are identified in (\ref{eqn:vorticityZ_9}) emerging from the viscous term in the compressible vorticity equation. The term (A) represents the viscous diffusion of vorticity -- analogous to \(\nu\boldsymbol{\nabla^2}\omega_z\) in the incompressible limit. Sharing similarities with the compressible vorticity stretching term in (\ref{eqn:vorticityZ_2}), term (B) captures vorticity stretching due to viscosity and density gradients. Term (C) denotes the generation or decay of vorticity by the alignment of the vorticity gradient with the gradients of viscosity and density. Lastly, term (D) is a vorticity production term (source or sink) resulting from the fluid moving across a region with varying fluid properties. In this case, moving around the vortex centre with \(u_\theta\). Given the highly varying properties across the pseudo-boiling line, the other terms (B)-(D) become important. Note that while terms (A)-(C) are effectively redistributing existing vorticity, term (D) generates vorticity. \par

\begin{figure}
\centering
\begin{subfigure}{0.33\textwidth}
  \centering
  \includegraphics[width=1.0\linewidth]{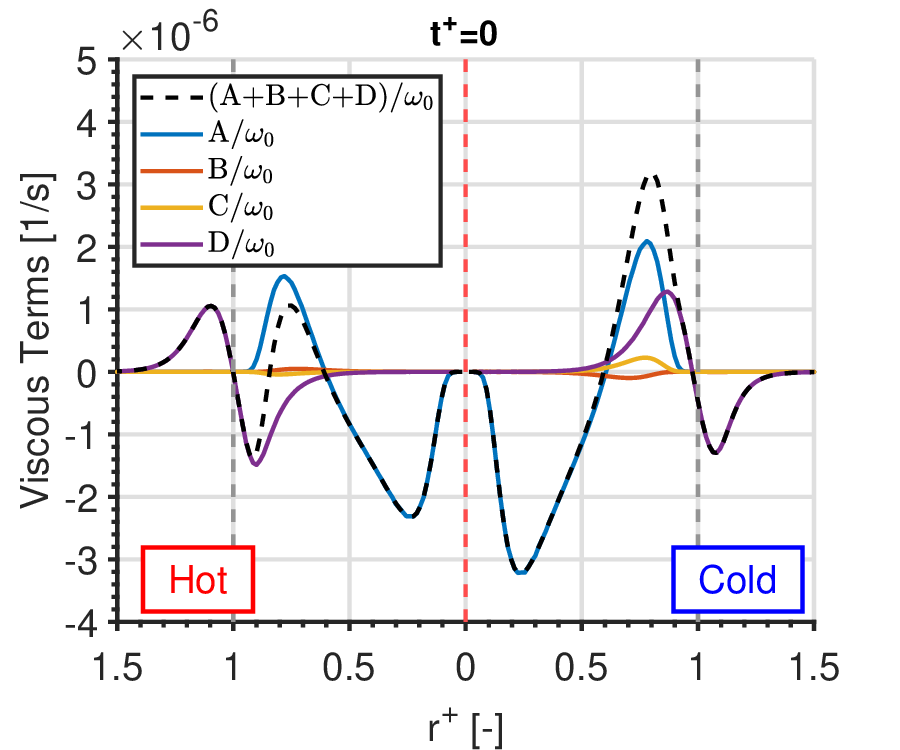}
  \caption{} 
  \label{subfig:vorticitydiffusion_Re200_pr2_File000_cold_vs_hot}
\end{subfigure}%
\begin{subfigure}{0.33\textwidth}
  \centering
  \includegraphics[width=1.0\linewidth]{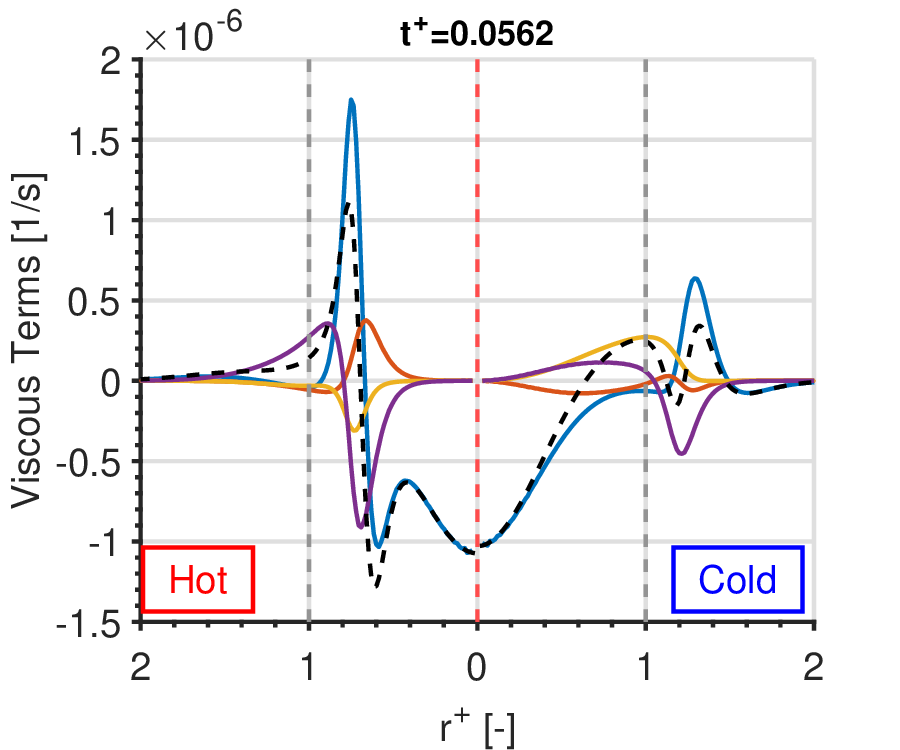}
  \caption{} 
  \label{subfig:vorticitydiffusion_Re200_pr2_File002_cold_vs_hot}
\end{subfigure}%
\begin{subfigure}{0.33\textwidth}
  \centering
  \includegraphics[width=1.0\linewidth]{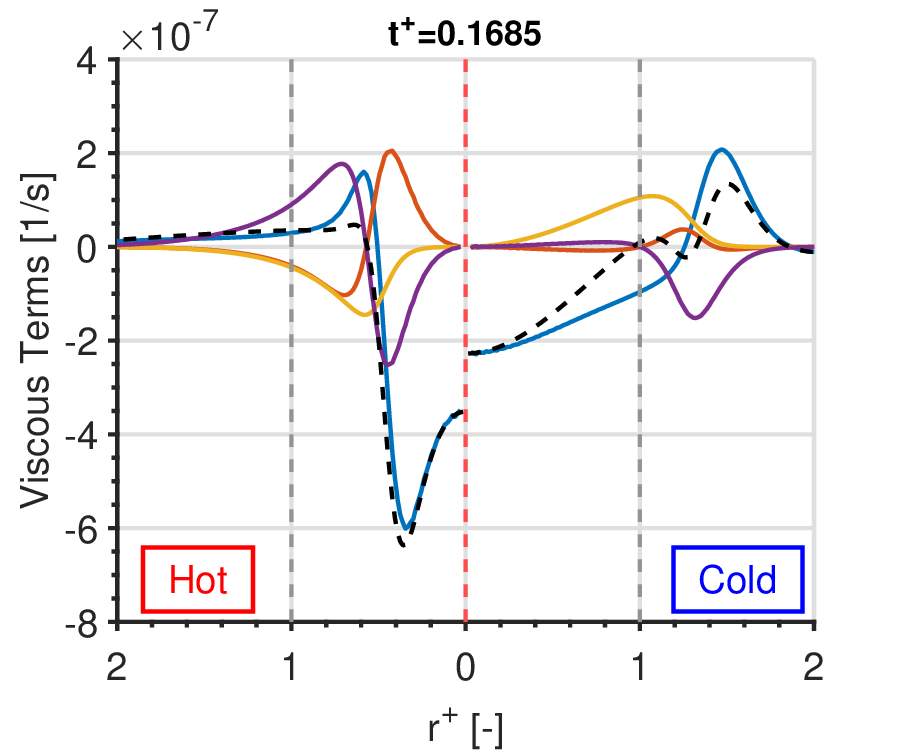}
  \caption{} 
  \label{subfig:vorticitydiffusion_Re200_pr2_File006_cold_vs_hot}
\end{subfigure}%
\\
\begin{subfigure}{0.33\textwidth}
  \centering
  \includegraphics[width=1.0\linewidth]{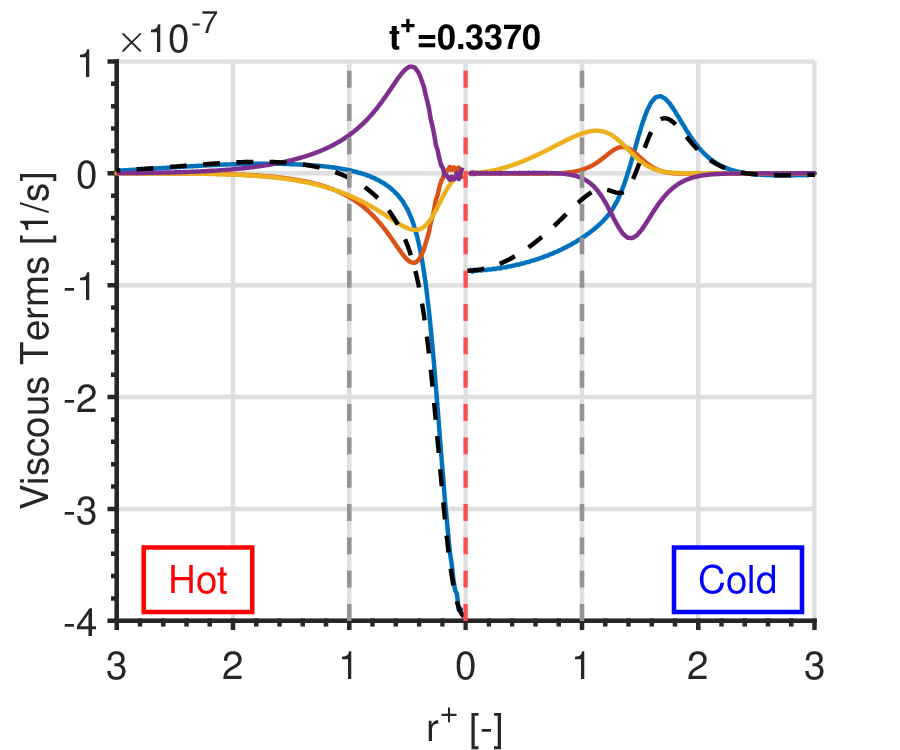}
  \caption{} 
  \label{subfig:vorticitydiffusion_Re200_pr2_File012_cold_vs_hot}
\end{subfigure}%
\begin{subfigure}{0.33\textwidth}
  \centering
  \includegraphics[width=1.0\linewidth]{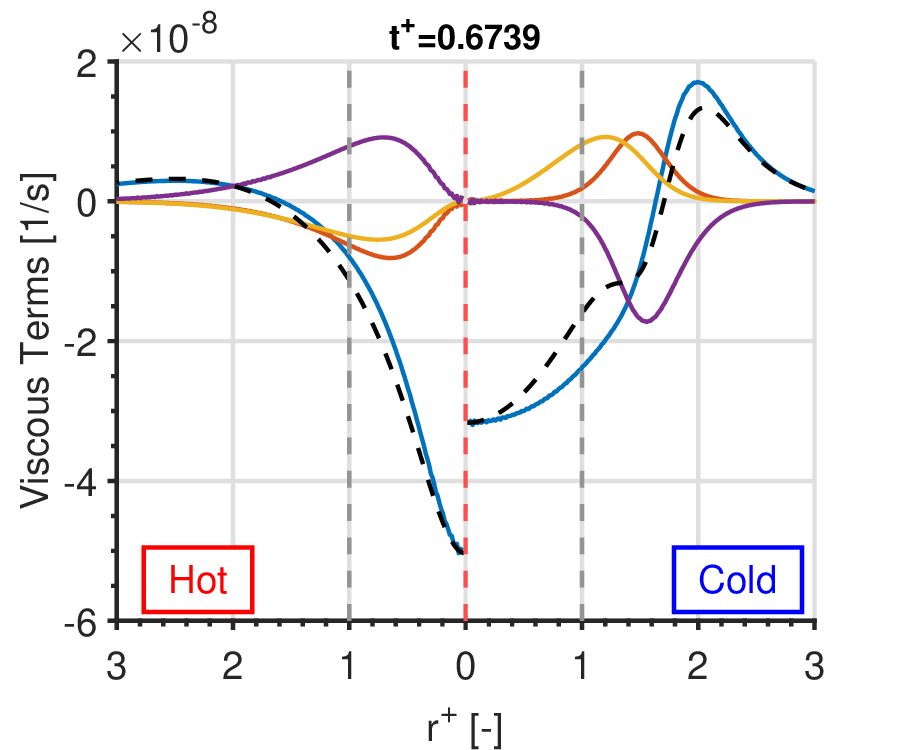}
  \caption{} 
  \label{subfig:vorticitydiffusion_Re200_pr2_File024_cold_vs_hot}
\end{subfigure}%
\begin{subfigure}{0.33\textwidth}
  \centering
  \includegraphics[width=1.0\linewidth]{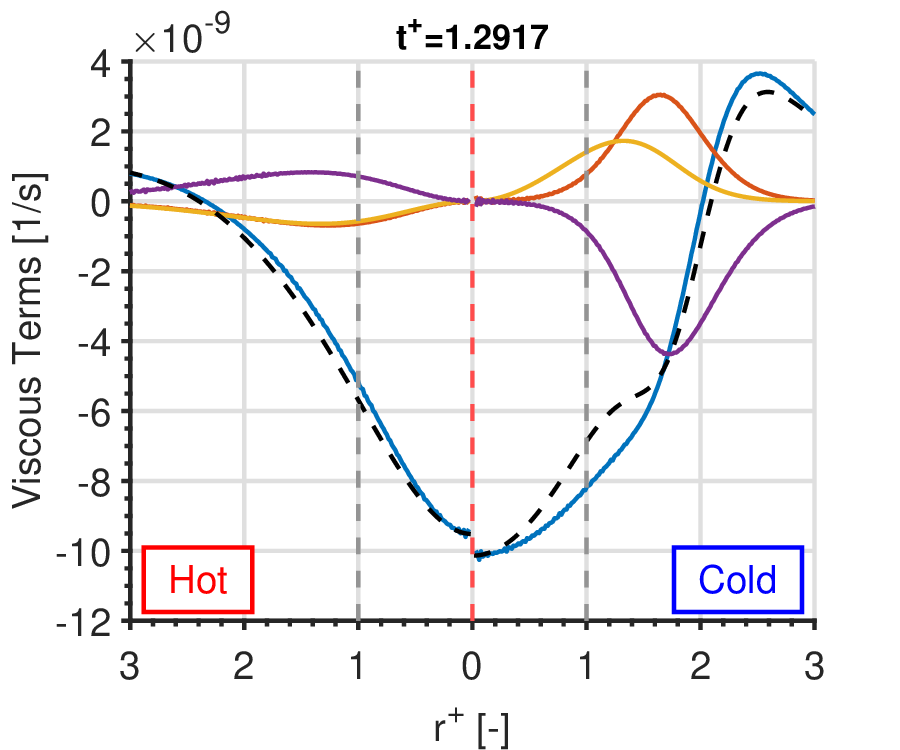}
  \caption{} 
  \label{subfig:vorticitydiffusion_Re200_pr2_File046_cold_vs_hot}
\end{subfigure}%
\caption{Evolution with \(t^+\) of the various viscous terms in (\ref{eqn:vorticityZ_9}) non-dimensionalised with \(\omega_0\) for the hot core (left) and the cold core (right) at \(p_r=2\). The dashed vertical lines represent the vortex centre (red) and \(r_c\) (grey). (a) \(t^+=0\); (b) \(t^+=0.0562\); (c) \(t^+=0.1685\); (d) \(t^+=0.3370\); (e) \(t^+=0.6739\); and (f) \(t^+=1.2917\).}
\label{fig:Fig14}
\end{figure}

Figure \ref{fig:Fig14} presents the contributions of terms (A)-(D) to the viscous term \(\boldsymbol{\nabla\times}\big([1/\rho]\boldsymbol{\nabla}\boldsymbol{\cdot}\boldsymbol{\tau}\big)\) at \(p_r=2\). Some minor oscillations, particularly around the vortex centre, appear because of the numerical differentiation of the discrete solution to obtain the relevant derivatives. Although viscous diffusion, or term (A), dominates in certain regions of the vortex, i.e., near the vortex centre, terms (B)-(D) are comparable in magnitude to (A) in other regions. Viscous diffusion behaves very similarly to the incompressible Oseen vortex (see \ref{subapn:B2}). Some initial differences in its evolution occur due to the initialisation of the vortex with a compact support instead of a Gaussian distribution, e.g., \(\nu\boldsymbol{\nabla^2}\omega_z=0\) at \(r=0\), and the effects of terms (B)-(D) on the vorticity distribution over time. Terms (B)-(D) tend to cancel each other at later times once the thermal mixing region across the vortex has diffused enough and the sharp variations of fluid properties are mitigated. As a result, figures \ref{subfig:vorticitydiffusion_Re200_pr2_File012_cold_vs_hot} to \ref{subfig:vorticitydiffusion_Re200_pr2_File046_cold_vs_hot} show that the distribution of \(\boldsymbol{\nabla\times}\big([1/\rho]\boldsymbol{\nabla}\boldsymbol{\cdot}\boldsymbol{\tau}\big)\) is mainly a slightly perturbed viscous diffusion term. This suggests that the vorticity evolves to a certain solution where its production mechanisms linked to the varying fluid properties -- terms (B)-(D) -- cancel each other.  Even at \(t^+\) where pseudo-boiling effects are still strong, the viscous effect is mainly diffusion, especially for the hot core. However, the contributions of terms (B)-(D) to the viscous term may become important in specific flow regions and also become larger for thermodynamic conditions closer to the critical point, e.g., \(p_r=1.3\), as the gradients of density and viscosity across the pseudo-boiling line are accentuated (see figure \ref{fig:Fig2}). This can explain the strong differences in the evolution of vortex-related quantities shown in section \ref{subsec:cold_vs_hot}. \par 

The following describes in more detail the behaviour of each sub-term within terms (B), (C) and (D), by decomposing \(\boldsymbol{\nabla\times}\big([1/\rho]\boldsymbol{\nabla}\boldsymbol{\cdot}\boldsymbol{\tau}\big)\) into various physical mechanisms responsible for generating vorticity. Each sub-term is identified as follows

\begin{equation}
\label{eqn:vorticityZ_10}
\text{Term (B)} =\underbrace{\omega_z\frac{1}{\rho}\bigg(\frac{1}{r}\frac{\partial}{\partial r}\bigg(r\frac{\partial\mu}{\partial r}\bigg)\bigg)}_\text{(B1)}-\underbrace{\omega_z\frac{2}{\rho}\bigg(\frac{1}{r}\frac{\partial\mu}{\partial r}\bigg)}_\text{(B2)} -\underbrace{\omega_z\frac{1}{\rho^2}\frac{\partial\rho}{\partial r}\frac{\partial\mu}{\partial r}}_\text{(B3)}
\end{equation}
\begin{equation}
\label{eqn:vorticityZ_11}
\text{Term (C)}=\underbrace{\frac{2}{\rho}\frac{\partial\mu}{\partial r}\frac{\partial\omega_z}{\partial r}}_\text{(C1)}-\underbrace{\frac{\mu}{\rho^2}\frac{\partial\rho}{\partial r}\frac{\partial\omega_z}{\partial r}}_\text{(C2)}
\end{equation}
\begin{equation}
\label{eqn:vorticityZ_12}
\text{Term (D)}=- \underbrace{\frac{2u_\theta}{r}\frac{1}{\rho}\bigg(\frac{1}{r}\frac{\partial}{\partial r}\bigg(r\frac{\partial\mu}{\partial r}\bigg)\bigg)}_\text{(D1)}+\underbrace{\frac{4u_\theta}{r}\frac{1}{\rho}\bigg(\frac{1}{r}\frac{\partial\mu}{\partial r}\bigg)}_\text{(D2)} +\underbrace{\frac{2u_\theta}{r}\frac{1}{\rho^2}\frac{\partial\rho}{\partial r}\frac{\partial\mu}{\partial r}}_\text{(D3)}
\end{equation}

\noindent
with terms (B) and (D) kept separate although both are related to the shear stress \(\tau_{r\theta}=\mu\big(\omega_z-\frac{2u_\theta}{r}\big)\) as \(\text{(B)}+\text{(D)}=\frac{\tau_{r\theta}}{\mu}\Big[\frac{1}{\rho}\big(\frac{1}{r}\frac{\partial}{\partial r}\big(r\frac{\partial\mu}{\partial r}\big)\big)-\frac{2}{\rho}\big(\frac{1}{r}\frac{\partial\mu}{\partial r}\big)-\frac{1}{\rho^2}\frac{\partial\rho}{\partial r}\frac{\partial\mu}{\partial r}\Big]\). \par

\begin{figure}
\centering
\begin{subfigure}{0.33\textwidth}
  \centering
  \includegraphics[width=1.0\linewidth]{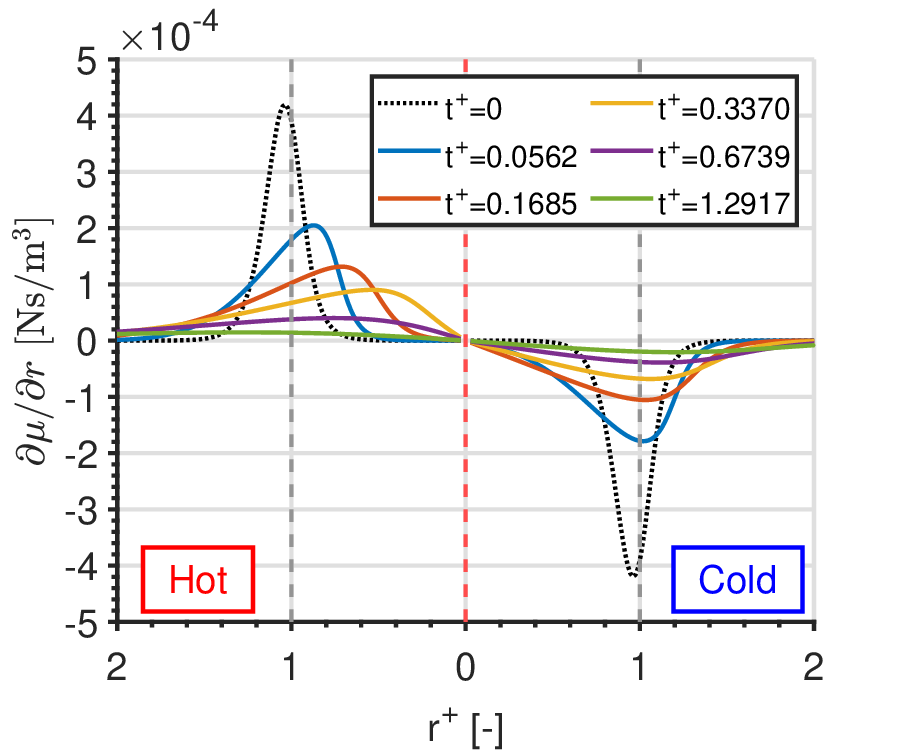}
  \caption{} 
  \label{subfig:fluidpropertiesgradient1_Re200_pr2_cold_vs_hot}
\end{subfigure}%
\begin{subfigure}{0.33\textwidth}
  \centering
  \includegraphics[width=1.0\linewidth]{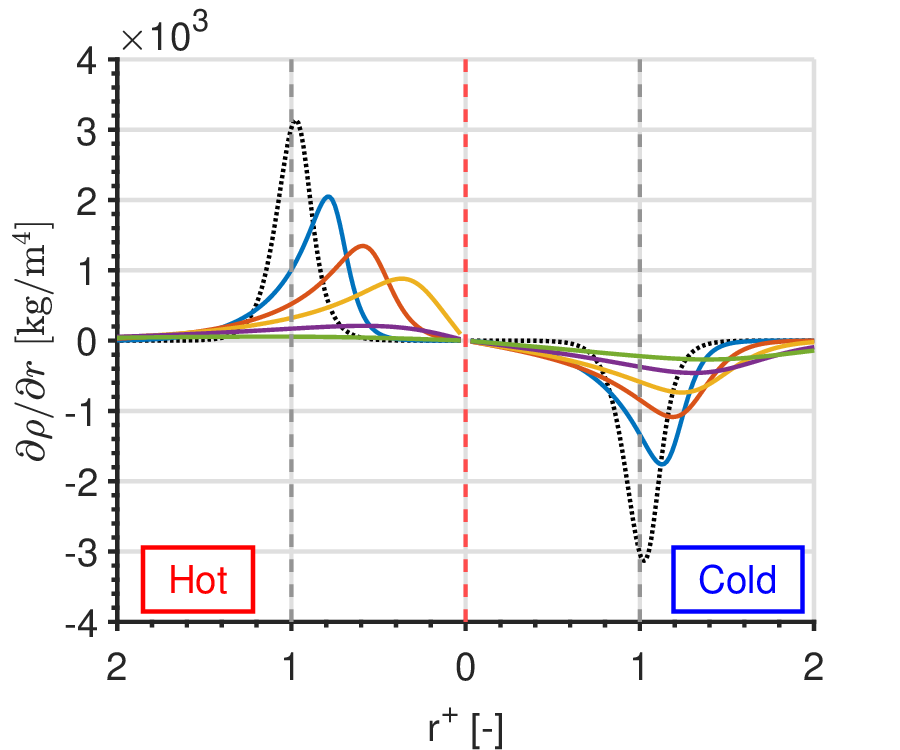}
  \caption{} 
  \label{subfig:fluidpropertiesgradient2_Re200_pr2_cold_vs_hot}
\end{subfigure}%
\begin{subfigure}{0.33\textwidth}
  \centering
  \includegraphics[width=1.0\linewidth]{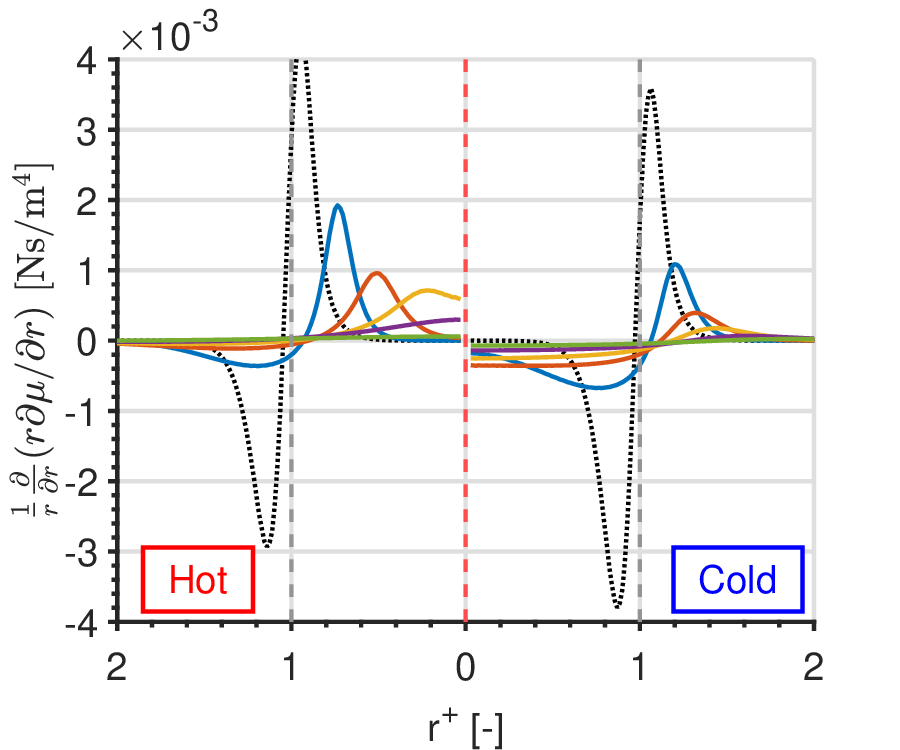}
  \caption{} 
  \label{subfig:fluidpropertiesgradient3_Re200_pr2_cold_vs_hot}
\end{subfigure}%
\caption{Evolution with \(t^+\) of the partial derivatives of viscosity and density found in (\ref{eqn:vorticityZ_9}) for the hot core (left) and the cold core (right) at \(p_r=2\). The dashed vertical lines represent the vortex centre (red) and \(r_c\) (grey). (a) \(\frac{\partial\mu}{\partial r}\); (b) \(\frac{\partial\rho}{\partial r}\); and (c) \(\frac{1}{r}\frac{\partial}{\partial r}\big(r\frac{\partial\mu}{\partial r}\big)\).}
\label{fig:Fig15}
\end{figure}

Define \(r_+\) as the region of the fluid immediately above \(r\) and \(r_-\) as the region of the fluid immediately below \(r\). Due to the decomposition of the viscous term into various isolated effects, note that the actual distribution of vorticity \(\omega_z(r)\) is decomposed into a ``constant" part \(\omega_z\) concerning term (B), and a ``varying" part \(\frac{\partial\omega_z}{\partial r}\) concerning term (C). Moreover, figure \ref{fig:Fig15} shows the evolution of the derivatives of \(\mu\) and \(\rho\) to explain the role of the sub-terms in (B), (C) and (D). Although the second-order derivative of \(\mu\) oscillates sharply across the pseudo-boiling line from negative to positive values, \(\frac{\partial\mu}{\partial r}\) and \(\frac{\partial\rho}{\partial r}\) are always negative in the cold core as the vortex heats up and are always positive in the hot core as the vortex cools down. This has consequences for vorticity generation and decay, as discussed below. \par

Term (B) is a vorticity stretching term resulting from the contribution of \(\omega_z\) to the viscous shear stress \(\tau_{r\theta}\), i.e., \(\mu\omega_z\), and its pointwise alteration by the varying fluid properties. Term (B2) only concerns the first-order effects of radial variations in viscosity under a ``constant" vorticity across the fluid. It represents the increase in the rotation of the fluid, i.e., vorticity magnitude \(|\omega_z|\), due to the varying viscosity causing a shear stress imbalance across a fluid element. If \(\frac{\partial\mu}{\partial r}>0\), the stresses exerted on the fluid element are higher at \(r_+\) than at \(r_-\). This results in a net torque that opposes the existing vorticity of the fluid, i.e., decreases \(|\omega_z|\). In contrast, if \(\frac{\partial\mu}{\partial r}<0\), the stresses are higher at \(r_-\) than at \(r_+\) and cause a reverse viscous torque that increases \(|\omega_z|\). In other words, there is more shear resistance in \(r_+\) or \(r_-\) opposing or favouring the rotation of a fluid element depending on how viscosity varies. The behaviour of term (B2) is the same regardless of the sign of \(\omega_z\). That is, the impact on \(|\omega_z|\) only depends on the sign of \(\frac{\partial\mu}{\partial r}\) since the direction of the viscous stresses changes with the sign of \(\omega_z\). \par 

\begin{figure}
\centering
\includegraphics[width=0.8\linewidth]{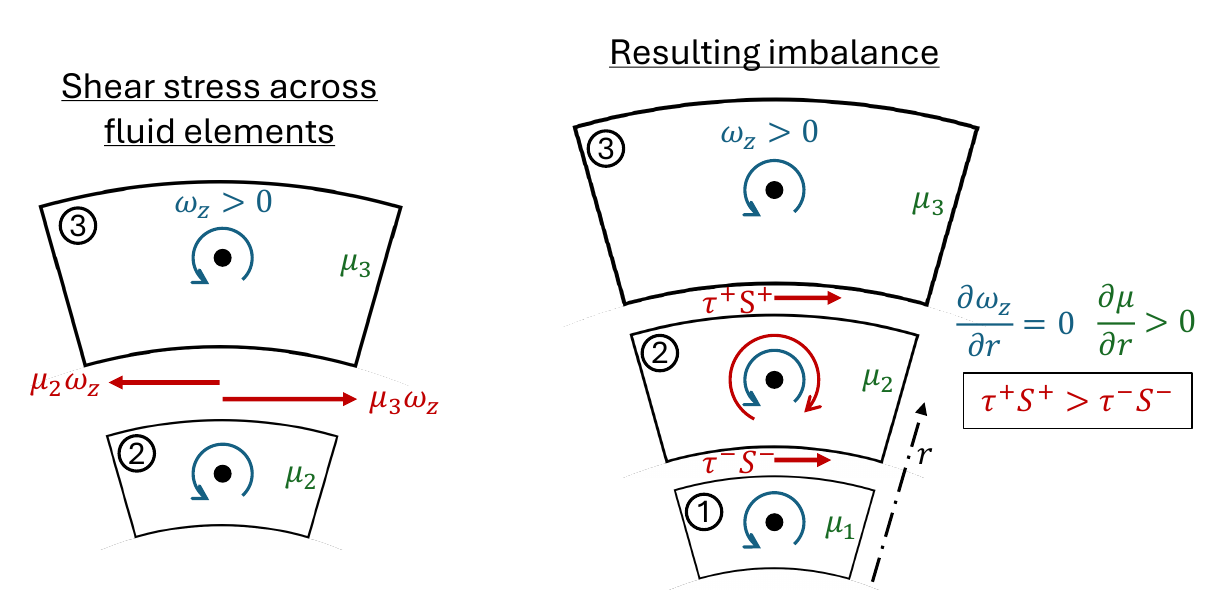}
\caption{Sketch of viscous mechanism represented by term (B2) for stretching or squeezing vorticity. The resultant viscous forces on the fluid element are represented.}
\label{fig:Fig16}
\end{figure}

A schematic of this mechanism is presented in figure \ref{fig:Fig16} where \(\omega_z>0\) and \(\frac{\partial\mu}{\partial r}>0\). Three consecutive fluid elements, each with a different viscosity but the same \(\omega_z\), are represented with the resultant viscous forces between them given by \(\tau^+S^+\) and \(\tau^-S^-\). For example, the rotation of element 2 causes the shear stress \(\mu_2\omega_z\) on element 3, and the rotation of element 3 causes stress \(\mu_3\omega_z\) on element 2. The net effect is \(\tau^+=(\mu_3-\mu_2)\omega_z\). Similarly, \(\tau^-=(\mu_2-\mu_1)\omega_z\). Without loss of generality, \(\frac{\partial\mu}{\partial r}\) is approximately constant across the fluid elements in the infinitesimal limit; thus, \((\mu_3-\mu_2)\approx(\mu_2-\mu_1)\). Given that \(\tau^+\) and \(\tau^-\) act on an area defined, respectively, by \(S^+=(r+dr)d\theta\) and \(S^-=rd\theta\), the net viscous torque around element 2 is given by \(-(\mu_3-\mu_2)\omega_zdrd\theta\), decreasing vorticity when \(\omega_z>0\) and \(\mu_3>\mu_2\). Similar, the behaviour of term (B2) for other signs of \(\omega_z\) and \(\frac{\partial\mu}{\partial r}\) follows. \par 

Next, term (B1) captures the second-order effects of radial variations in viscosity on the vorticity evolution, i.e., whether the fluid is becoming increasingly more or less viscous. For example, vorticity stretching occurs if viscosity varies more sharply in \(r_+\) than in \(r_-\) with \(\frac{\partial \mu}{\partial r}>0\) and \(\omega_z>0\) -- this results in \(\frac{1}{r}\big(\frac{\partial}{\partial r}\big(r\frac{\partial\mu}{\partial r}\big)\big)>0\) and term (B1) increasing \(\omega_z\). Physically, this behaviour is closely related to the effects of term (B2). This is because different \(\frac{\partial \mu}{\partial r}\) decrease vorticity faster in \(r_+\) than in \(r_-\). This translates into a lower viscous stress at \(r_+\) than at \(r_-\), which causes a net torque to increase vorticity. Similarly, vorticity stretching also occurs if \(\frac{\partial \mu}{\partial r}<0\) and viscosity varies more sharply in \(r_-\) than in \(r_+\). However, the viscous torque in this case is generated from \(\omega_z\) increasing faster in \(r_-\) than in \(r_+\) via the mechanism (B2). A similar reasoning may describe scenarios where vorticity decreases when \(\frac{1}{r}\big(\frac{\partial}{\partial r}\big(r\frac{\partial\mu}{\partial r}\big)\big)<0\) or the behaviour of term (B1) when \(\omega_z<0\). \par

Lastly, term (B3) arises from the alignment of viscosity and density gradients. The interplay between the differential stresses caused by the viscosity gradient and the varying fluid inertia caused by the density gradient results in a net torque on a fluid element that always opposes the existing rotation, since \(\frac{\partial\mu}{\partial r}\) and \(\frac{\partial\rho}{\partial r}\) are always aligned. In other words, the increase in viscous shear stress across a fluid element in tandem with an increase in fluid inertia generates an imbalance in which the region of the fluid (\(r_+\) or \(r_-\)) experiencing higher stresses also responds less favourably to maintaining fluid rotation due to the increased density. Thus, reducing \(|\omega_z|\). \par 

\begin{figure}
\centering
\begin{subfigure}{0.33\textwidth}
  \centering
  \includegraphics[width=1.0\linewidth]{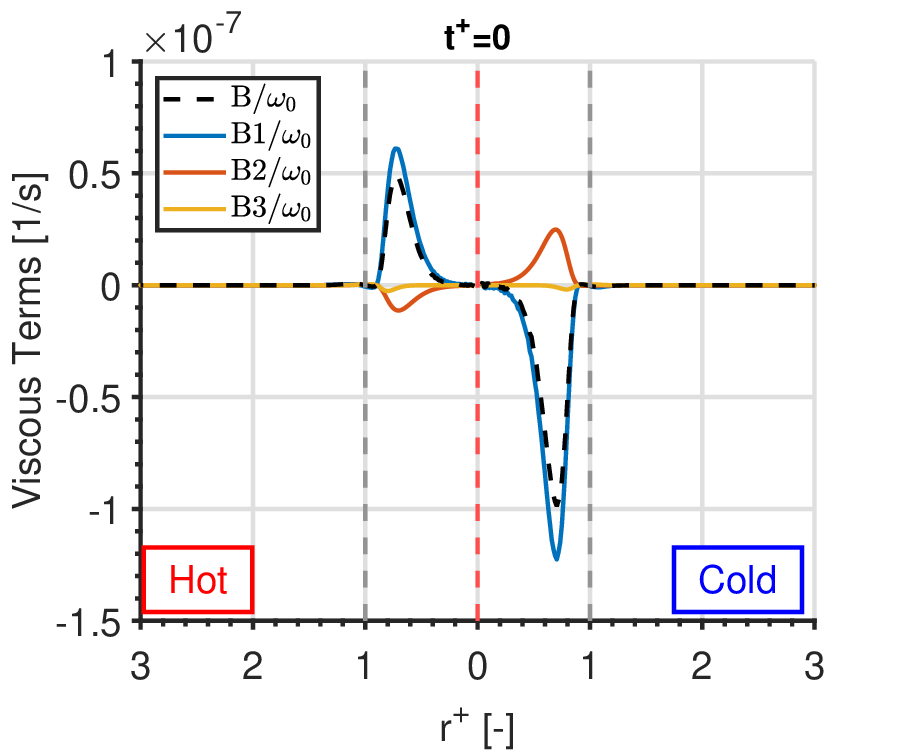}
  \caption{} 
  \label{subfig:vorticity_termB_Re200_pr2_File000_cold_vs_hot}
\end{subfigure}%
\begin{subfigure}{0.33\textwidth}
  \centering
  \includegraphics[width=1.0\linewidth]{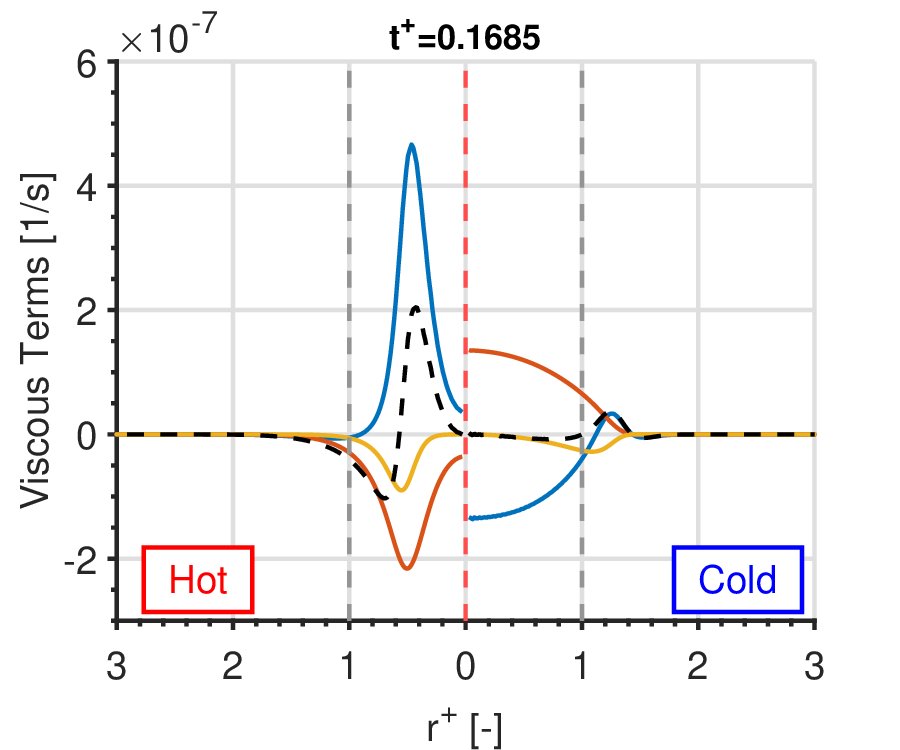}
  \caption{} 
  \label{subfig:vorticity_termB_Re200_pr2_File006_cold_vs_hot}
\end{subfigure}%
\begin{subfigure}{0.33\textwidth}
  \centering
  \includegraphics[width=1.0\linewidth]{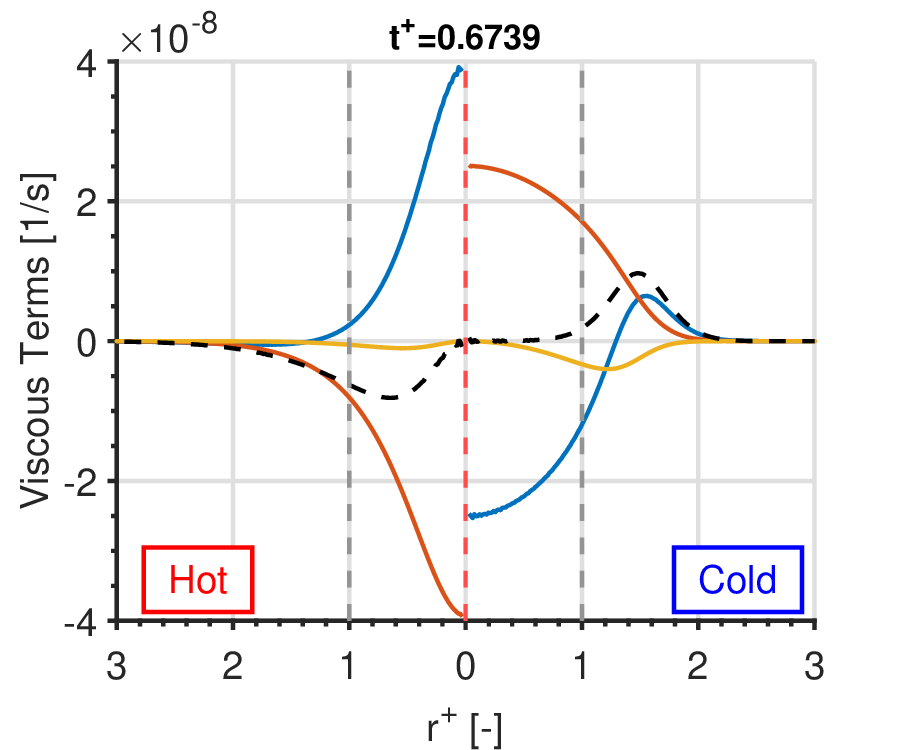}
  \caption{} 
  \label{subfig:vorticity_termB_Re200_pr2_File024_cold_vs_hot}
\end{subfigure}%
\caption{Evolution with \(t^+\) of the various sub-terms from term (B) given by (\ref{eqn:vorticityZ_10}) and non-dimensionalised with \(\omega_0\) for the hot core (left) and the cold core (right) at \(p_r=2\). The dashed vertical lines represent the vortex centre (red) and \(r_c\) (grey). (a) \(t^+=0\); (b) \(t^+=0.1685\); and (c) \(t^+=0.6739\).}
\label{fig:Fig17}
\end{figure}

Figure \ref{fig:Fig17} shows the radial profiles of term (B) and its sub-terms at selected \(t^+\). The contribution of term (B3) or the inertial response of the fluid with varying density is minor compared to terms (B1) and (B2). Vorticity stretching is strongly linked to the viscosity gradients. Primarily, the second-order effects of viscosity variations cause vorticity stretching in the hot core configuration and the opposite in the cold core configuration. In contrast, first-order effects show an opposite behaviour causing vorticity stretching in the cold core but compressing vorticity in the hot core. Given that \(\omega_z\) is always positive (except for a narrow region in the cold core case as shown in figure \ref{subfig:vorticity_Re200_pr2_cold_vs_hot}), the behaviour of terms (B1) and (B2) is tightly coupled to the derivatives of fluid properties shown in figure \ref{fig:Fig15}. (B1) and (B2) tend to cancel each other out, particularly near the vortex centre. Nonetheless, second-order effects in the viscosity variations, i.e., term (B1), seem to dominate, especially during the early evolution of the vortex when a pseudo-boiling line is well defined. \par

\begin{figure}
\centering
\includegraphics[width=0.9\linewidth]{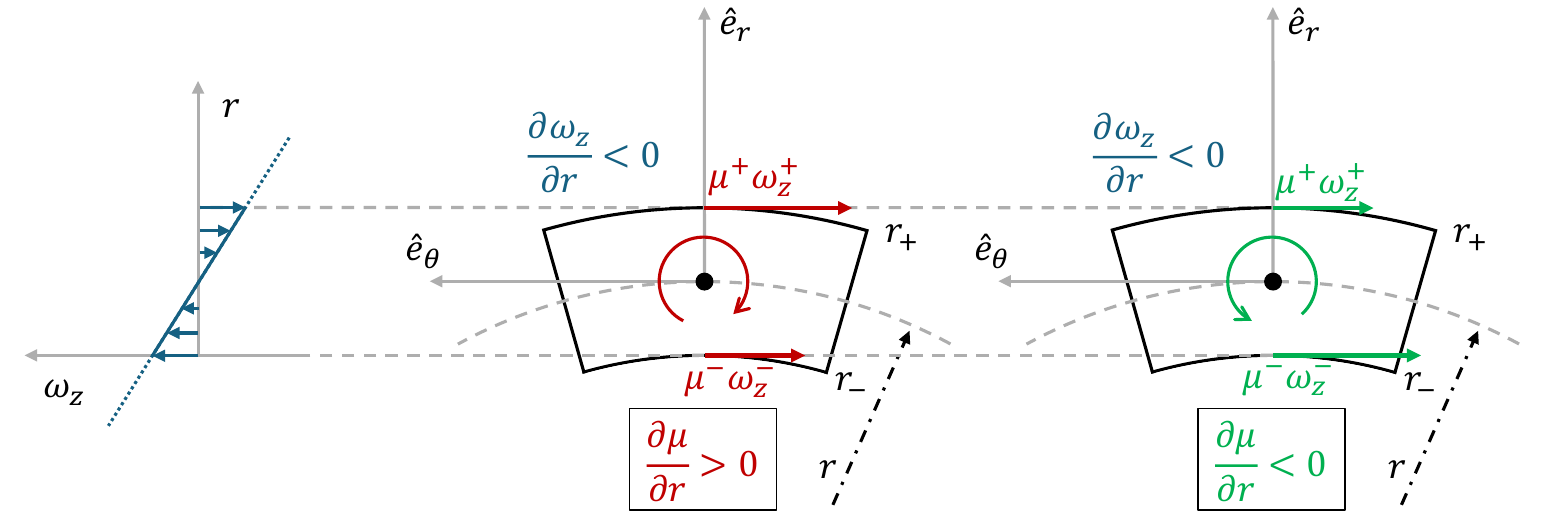}
\caption{Sketch of viscous mechanism represented by term (C1) for redistributing vorticity based on the alignment between the viscosity and vorticity gradients. The resultant viscous forces on the fluid element are represented.}
\label{fig:Fig18}
\end{figure}

Term (C) refers to the changes in vorticity caused by the alignment of the vorticity gradient with the viscosity/density gradient. All are only in the radial direction, and term (C) concerns the balance between viscosity gradient effect, or term (C1), and density gradient effect, or term (C2). Here, the effects of the distribution of \(\omega_z\) or \(\frac{\partial\omega_z}{\partial r}\) are isolated from \(\omega_z\). Term (C1) describes how positive vorticity is generated when viscosity and vorticity gradients are aligned. That is, when both the higher \(\mu\) and \(|\omega_z|\) coincide in the same region (\(r_+\) or \(r_-\)) and the resulting stresses on a fluid element cause a positive torque. If the viscosity and vorticity gradients point in opposite directions, reverse vorticity is generated. Figure \ref{fig:Fig18} sketches this mechanism in the common case where \(\frac{\partial\omega_z}{\partial r}<0\). Since \(\frac{\partial\mu}{\partial r}\) and \(\frac{\partial\rho}{\partial r}\) are always aligned, term (C2) opposes this torque due to the increased inertia of the fluid in regions of higher density, similar to term (B3). \par

\begin{figure}
\centering
\begin{subfigure}{0.33\textwidth}
  \centering
  \includegraphics[width=1.0\linewidth]{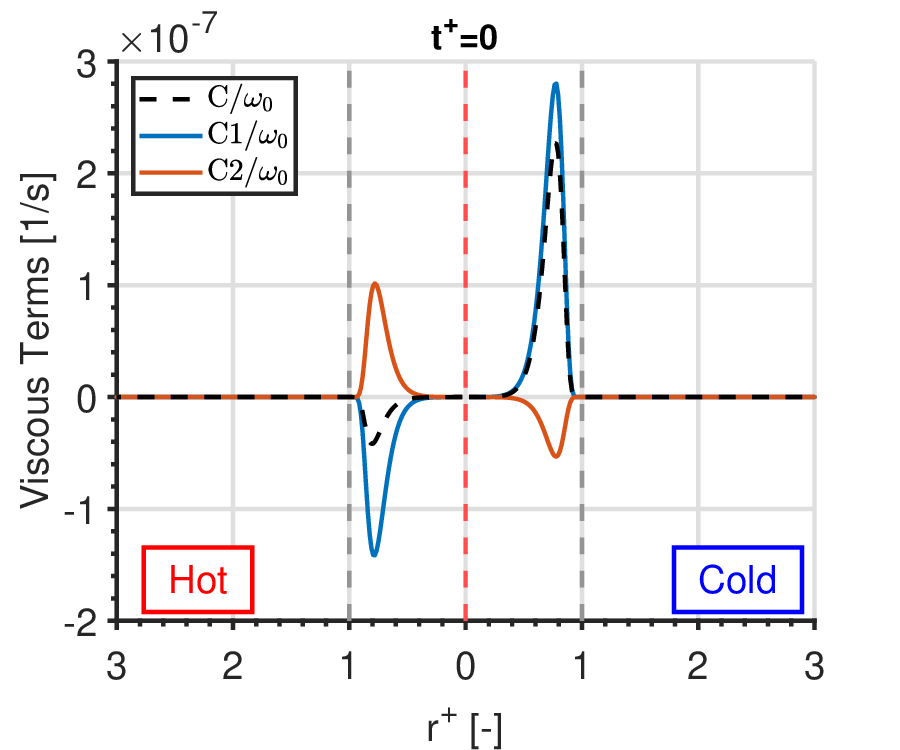}
  \caption{} 
  \label{subfig:vorticity_termC_Re200_pr2_File000_cold_vs_hot}
\end{subfigure}%
\begin{subfigure}{0.33\textwidth}
  \centering
  \includegraphics[width=1.0\linewidth]{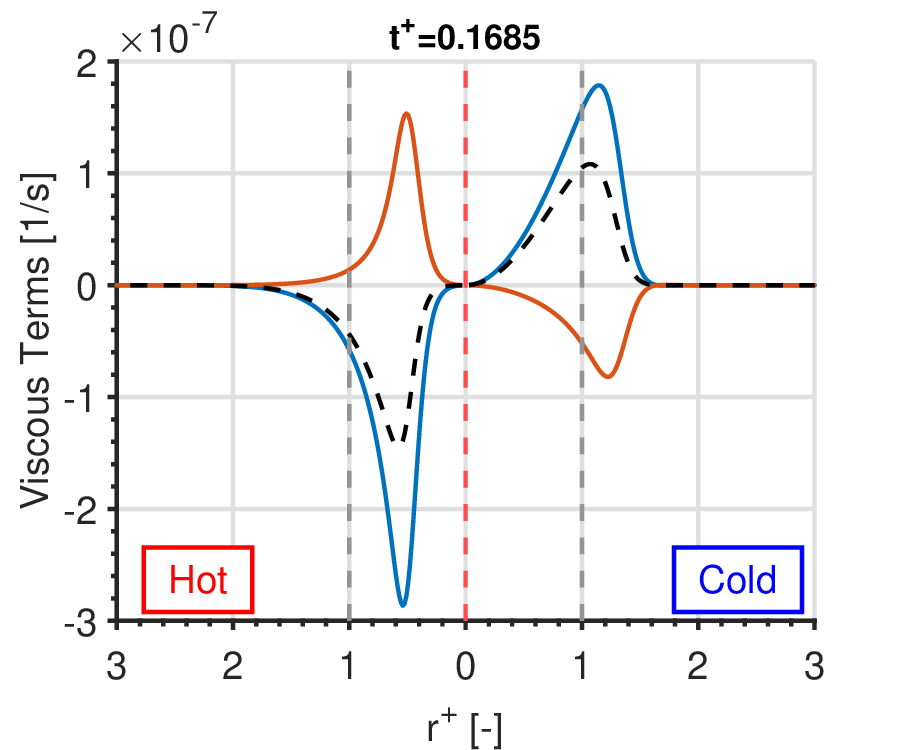}
  \caption{} 
  \label{subfig:vorticity_termC_Re200_pr2_File006_cold_vs_hot}
\end{subfigure}%
\begin{subfigure}{0.33\textwidth}
  \centering
  \includegraphics[width=1.0\linewidth]{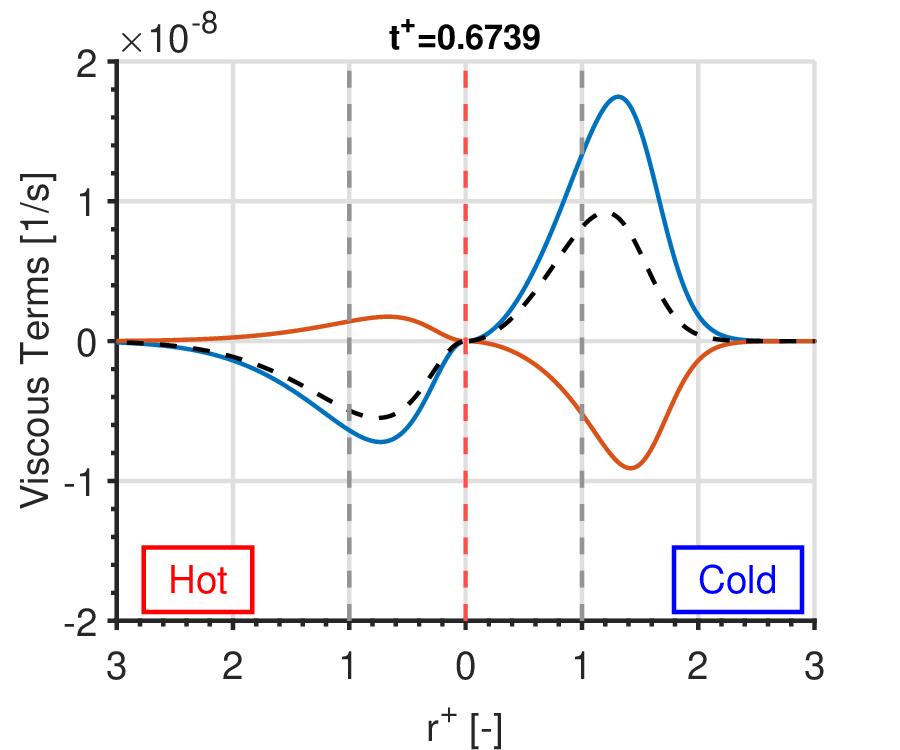}
  \caption{} 
  \label{subfig:vorticity_termC_Re200_pr2_File024_cold_vs_hot}
\end{subfigure}%
\caption{Evolution with \(t^+\) of the various sub-terms from term (C) given by (\ref{eqn:vorticityZ_11}) and non-dimensionalised with \(\omega_0\) for the hot core (left) and the cold core (right) at \(p_r=2\). The dashed vertical lines represent the vortex centre (red) and \(r_c\) (grey). (a) \(t^+=0\); (b) \(t^+=0.1685\); and (c) \(t^+=0.6739\).}
\label{fig:Fig19}
\end{figure}

The radial profiles of the sub-terms (C1) and (C2), as well as term (C), are presented in figure \ref{fig:Fig19} for various \(t^+\). Since \(\frac{\partial\omega_z}{\partial r}\) is negative (again except for a narrow region in the cold core case as shown in figure \ref{subfig:vorticity_Re200_pr2_cold_vs_hot}), term (C1) refers to the viscosity gradient and increases vorticity in the cold core while it decreases vorticity in the hot core. The inertial response due to the density gradient or term (C2) opposes these vorticity changes. Although similar in magnitude, the viscous effect dominates term (C) in both configurations at \(p_r=2\). \par  

\begin{figure}
\centering
\begin{subfigure}{0.33\textwidth}
  \centering
  \includegraphics[width=1.0\linewidth]{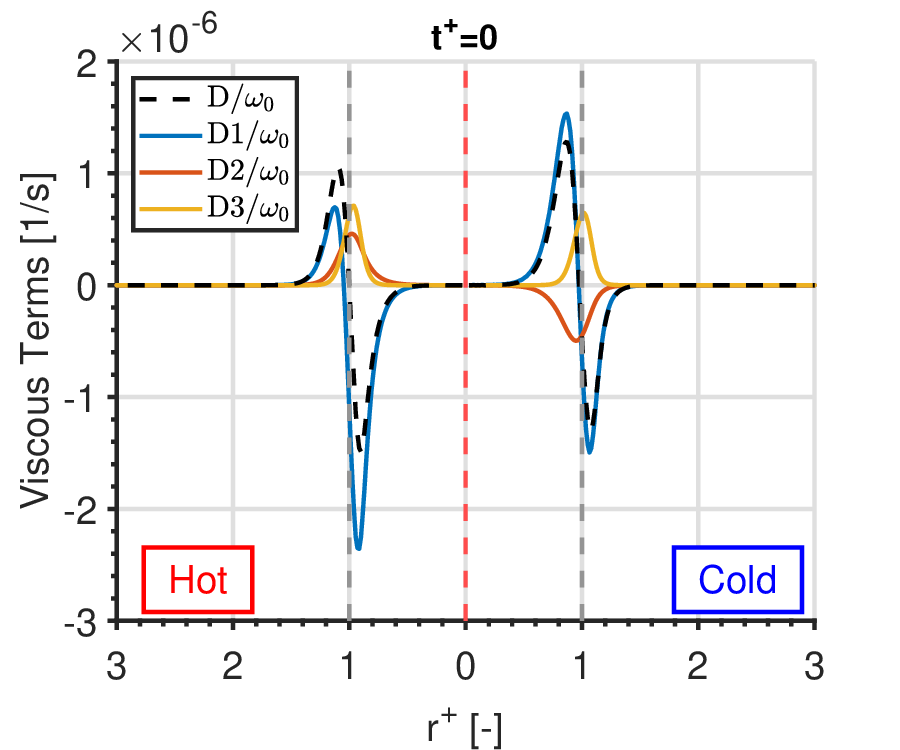}
  \caption{} 
  \label{subfig:vorticity_termD_Re200_pr2_File000_cold_vs_hot}
\end{subfigure}%
\begin{subfigure}{0.33\textwidth}
  \centering
  \includegraphics[width=1.0\linewidth]{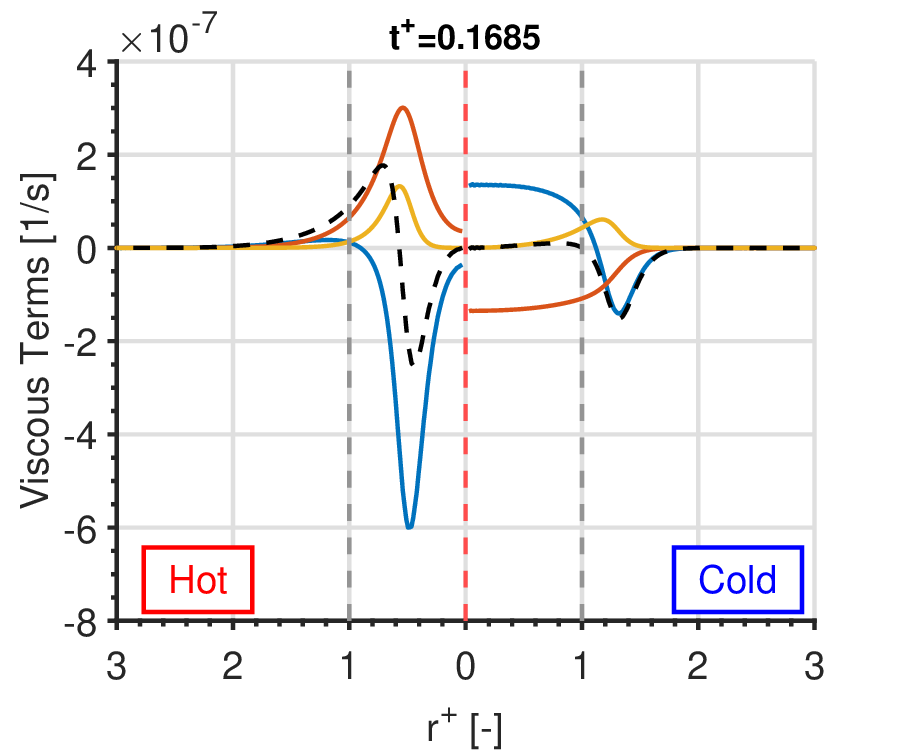}
  \caption{} 
  \label{subfig:vorticity_termD_Re200_pr2_File006_cold_vs_hot}
\end{subfigure}%
\begin{subfigure}{0.33\textwidth}
  \centering
  \includegraphics[width=1.0\linewidth]{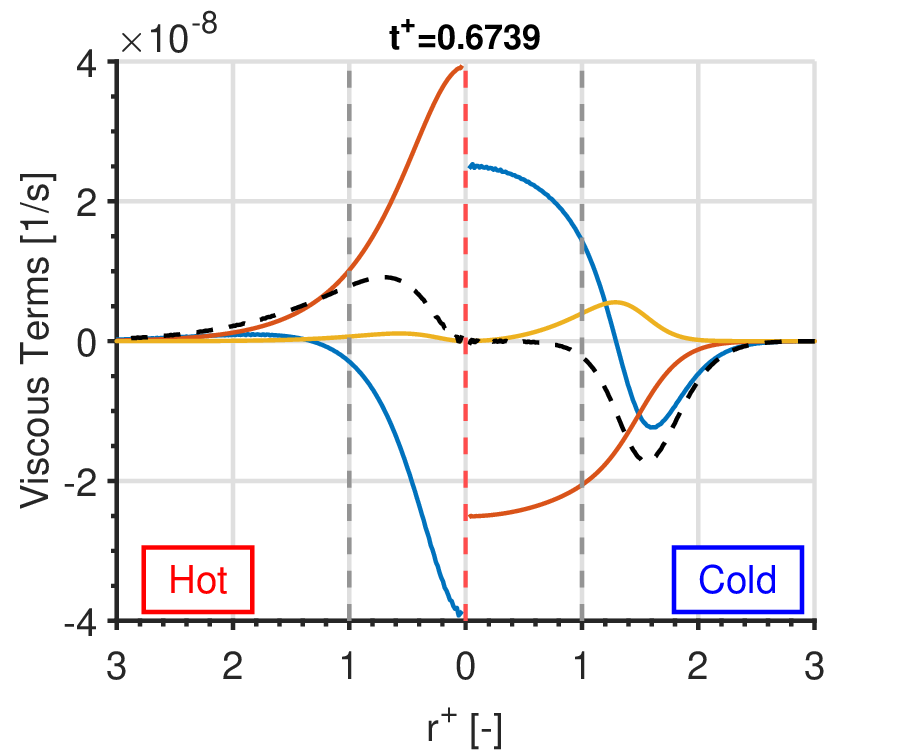}
  \caption{} 
  \label{subfig:vorticity_termD_Re200_pr2_File024_cold_vs_hot}
\end{subfigure}%
\caption{Evolution with \(t^+\) of the various sub-terms from term (D) given by (\ref{eqn:vorticityZ_12}) and non-dimensionalized with \(\omega_0\) for the hot core (left) and the cold core (right) at \(p_r=2\). The dashed vertical lines represent the vortex centre (red) and \(r_c\) (grey). (a) \(t^+=0\); (b) \(t^+=0.1685\); and (c) \(t^+=0.6739\).}
\label{fig:Fig20}
\end{figure}

Lastly, the physical mechanisms for vorticity generation or decay described by terms (D1) and (D2) based on differential stresses are the same as in terms (B1) and (B2). However, term (D) captures the contribution of the azimuthal velocity to the viscous shear stress \(\tau_{r\theta}\), given by \(-\mu\frac{2u_\theta}{r}\), resulting in a reversed effect on the vorticity evolution due to the stress acting in the opposite direction. Regarding term (D3), it always generates positive vorticity (\(\omega_z>0\)) because of the inertia of the denser and more viscous fluid layers. As shown in figure \ref{fig:Fig20}, term (D3) is comparable in magnitude to terms (D1) and (D2), as opposed to the contribution of term (B3). Resulting from the motion of a fluid element across a region with varying properties, term (D) produces vorticity. A clear consequence is observed in the negative vorticity emerging outside the vortex core in the cold case, which is caused primarily by second-order effects in the viscosity gradient. Similarly to term (B), terms (D1) and (D2) tend to cancel each other close to the vortex centre. \par 

\subsubsection{Pseudo-boiling Enhancement near the Critical Point}
\label{subsubsec:pseudoboiling}

\begin{figure}
\centering
\begin{subfigure}{0.33\textwidth}
  \centering
  \includegraphics[width=1.0\linewidth]{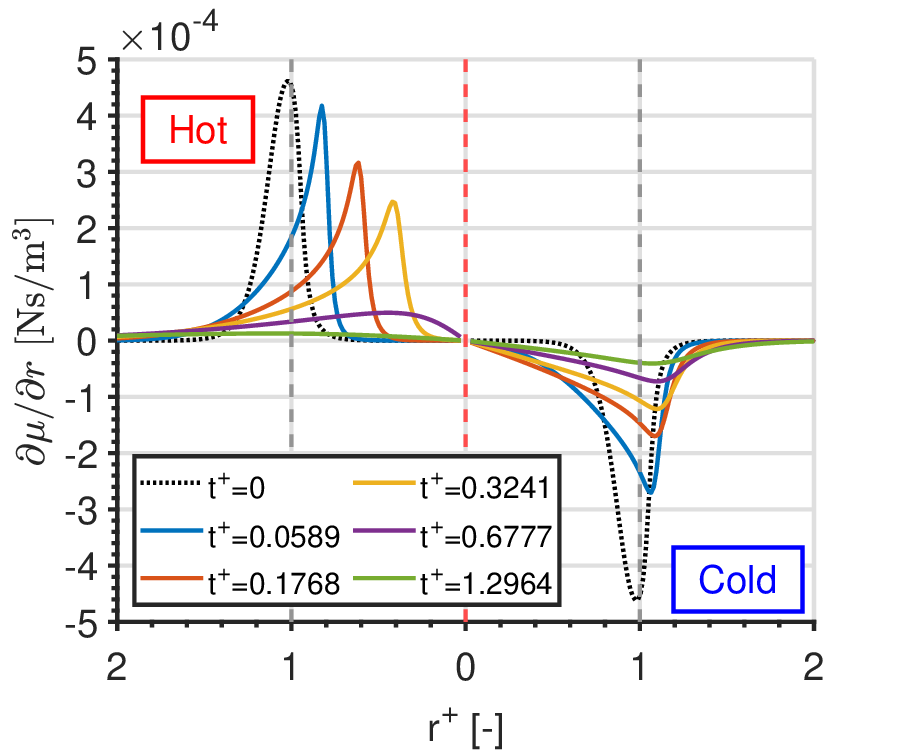}
  \caption{} 
  \label{subfig:fluidpropertiesgradient1_Re200_pr1p5_cold_vs_hot}
\end{subfigure}%
\begin{subfigure}{0.33\textwidth}
  \centering
  \includegraphics[width=1.0\linewidth]{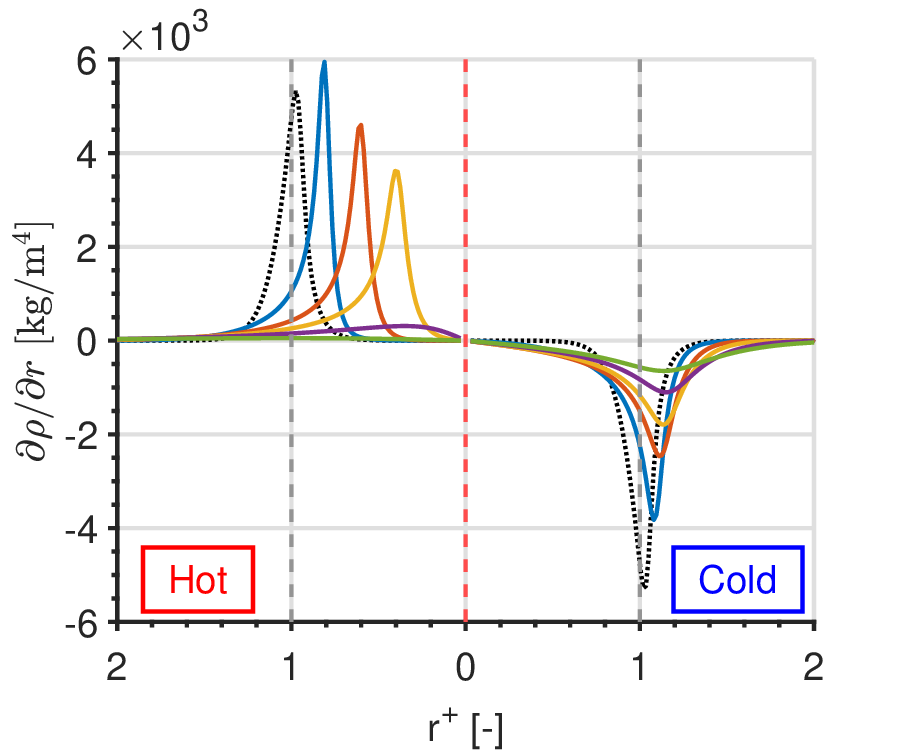}
  \caption{} 
  \label{subfig:fluidpropertiesgradient2_Re200_pr1p5_cold_vs_hot}
\end{subfigure}%
\begin{subfigure}{0.33\textwidth}
  \centering
  \includegraphics[width=1.0\linewidth]{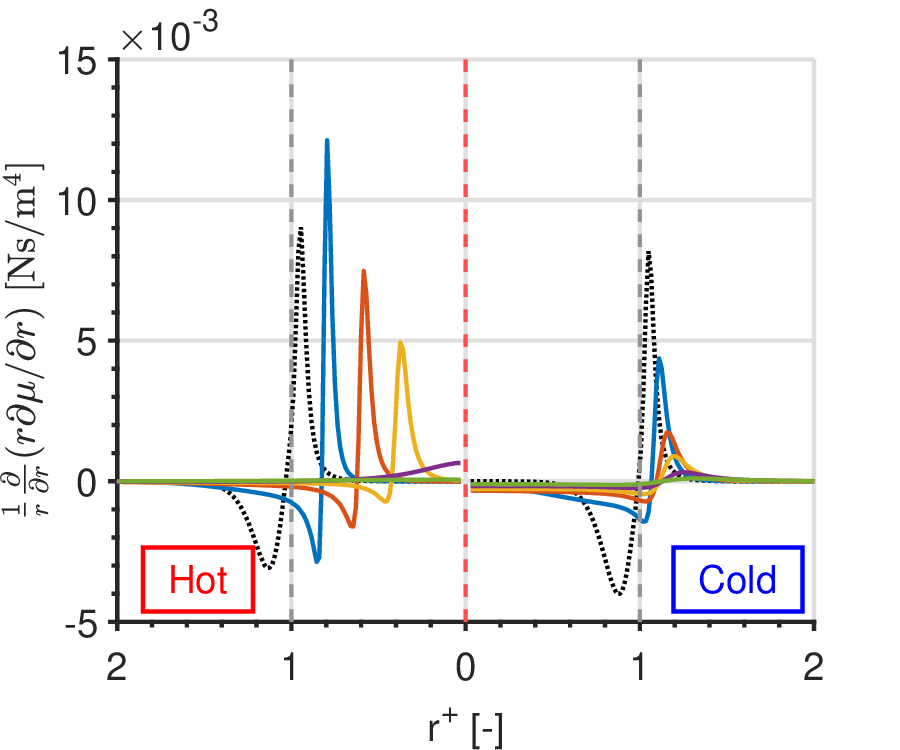}
  \caption{} 
  \label{subfig:fluidpropertiesgradient3_Re200_pr1p5_cold_vs_hot}
\end{subfigure}%
\\
\begin{subfigure}{0.33\textwidth}
  \centering
  \includegraphics[width=1.0\linewidth]{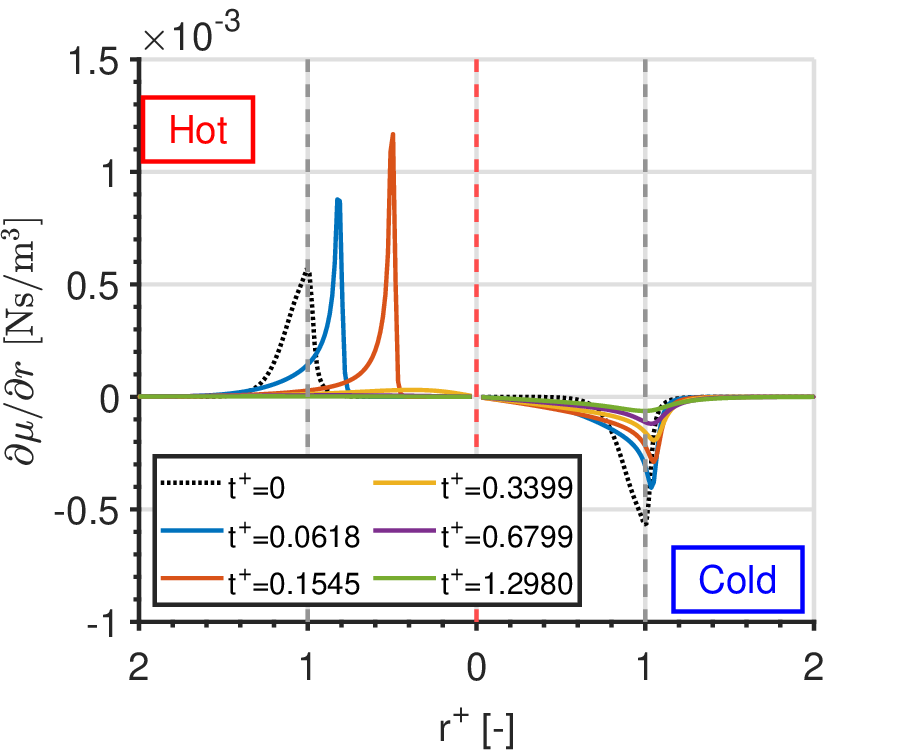}
  \caption{} 
  \label{subfig:fluidpropertiesgradient1_Re200_pr1p3_cold_vs_hot}
\end{subfigure}%
\begin{subfigure}{0.33\textwidth}
  \centering
  \includegraphics[width=1.0\linewidth]{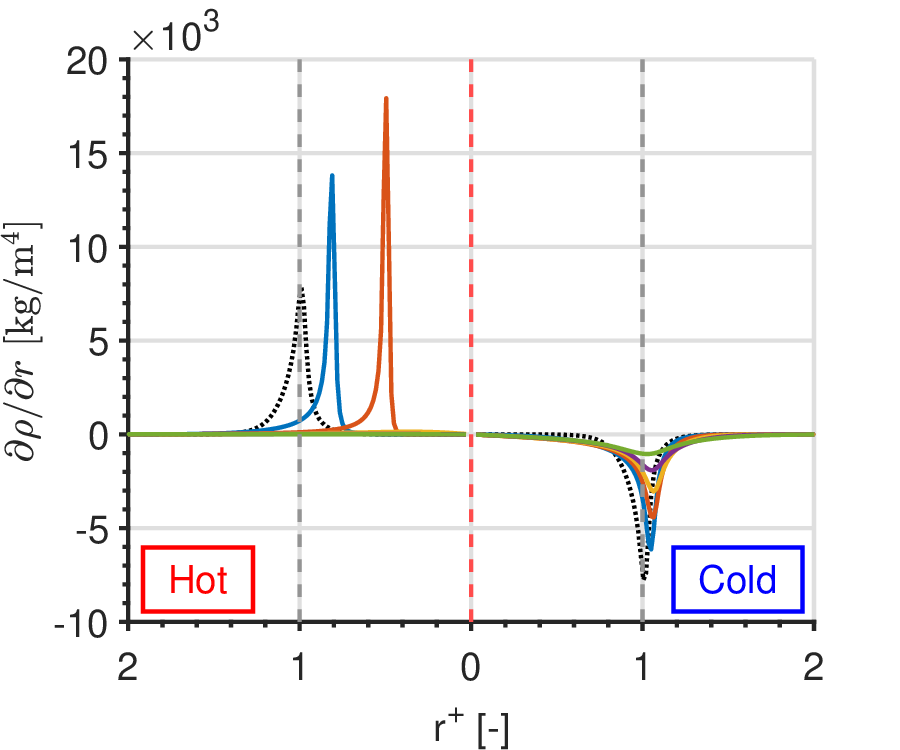}
  \caption{} 
  \label{subfig:fluidpropertiesgradient2_Re200_pr1p3_cold_vs_hot}
\end{subfigure}%
\begin{subfigure}{0.33\textwidth}
  \centering
  \includegraphics[width=1.0\linewidth]{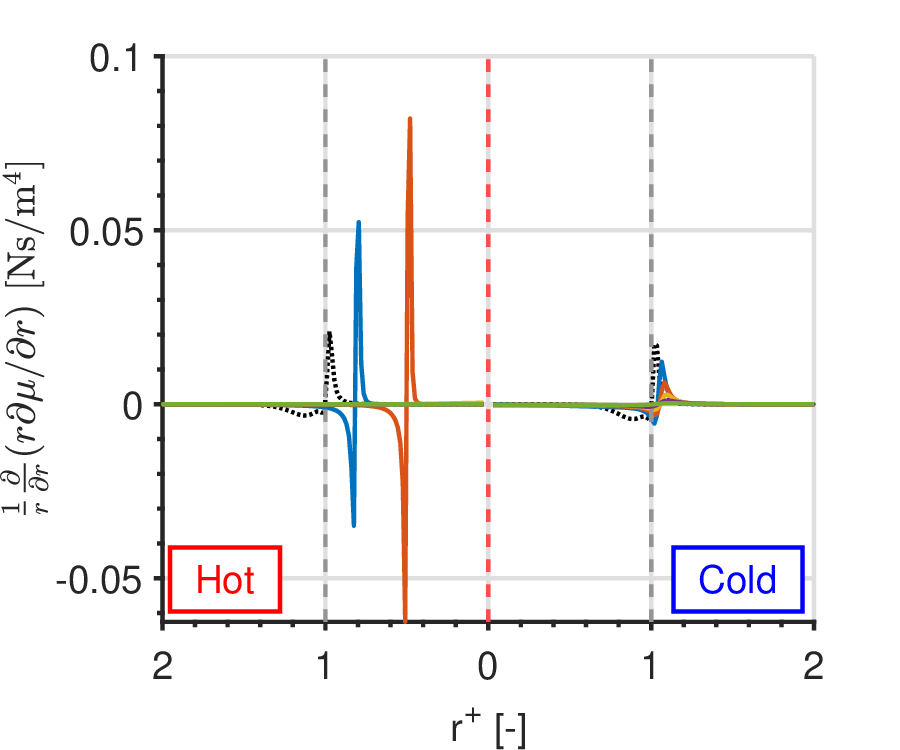}
  \caption{} 
  \label{subfig:fluidpropertiesgradient3_Re200_pr1p3_cold_vs_hot}
\end{subfigure}%
\caption{Evolution with \(t^+\) of the various partial derivatives of viscosity and density found in (\ref{eqn:vorticityZ_9}) for the hot core (left) and the cold core (right). The dashed vertical lines represent the vortex centre (red) and \(r_c\) (grey). At \(p_r=1.5\): (a) \(\frac{\partial\mu}{\partial r}\); (b) \(\frac{\partial\rho}{\partial r}\); (c) \(\frac{1}{r}\frac{\partial}{\partial r}\big(r\frac{\partial\mu}{\partial r}\big)\); at \(p_r=1.3\): (d) \(\frac{\partial\mu}{\partial r}\); (e) \(\frac{\partial\rho}{\partial r}\); and (f) \(\frac{1}{r}\frac{\partial}{\partial r}\big(r\frac{\partial\mu}{\partial r}\big)\).}
\label{fig:Fig21}
\end{figure}

As pressure drops closer to \(p_c\), the effects of varying fluid properties on the evolution of the vortex are increasingly enhanced. Figure \ref{fig:Fig21} shows the derivatives of \(\mu\) and \(\rho\) affecting vorticity production, i.e., first- and second-order derivatives. Compared to \(p_r=2\), a five-fold increase is generally observed for \(p_r=1.5\), and about a ten-fold increase for \(p_r=1.3\). However, the steepening of the thermal front in the hot core case as \(p\rightarrow p_c\) increases the magnitude of the derivatives compared to the corresponding cold core case. At \(p_r=1.3\), this steepening results in derivatives five times larger that in the cold vortex case. Note that, especially for figures \ref{subfig:fluidpropertiesgradient2_Re200_pr1p3_cold_vs_hot} and \ref{subfig:fluidpropertiesgradient3_Re200_pr1p3_cold_vs_hot}, the profiles for the cold vortex core are left unmagnified to visualise the stronger derivatives in the hot vortex core. In the cold case, the profiles evolve very similar to those at lower pressures, albeit with larger magnitudes as discussed above, since a similar stationary state of the pseudo-boiling line occurs at all pressures, i.e., the temperature distributions being pinned about the pseudo-boiling temperature around \(r^+=1\). \par 

\begin{figure}
\centering
\begin{subfigure}{0.33\textwidth}
  \centering
  \includegraphics[width=1.0\linewidth]{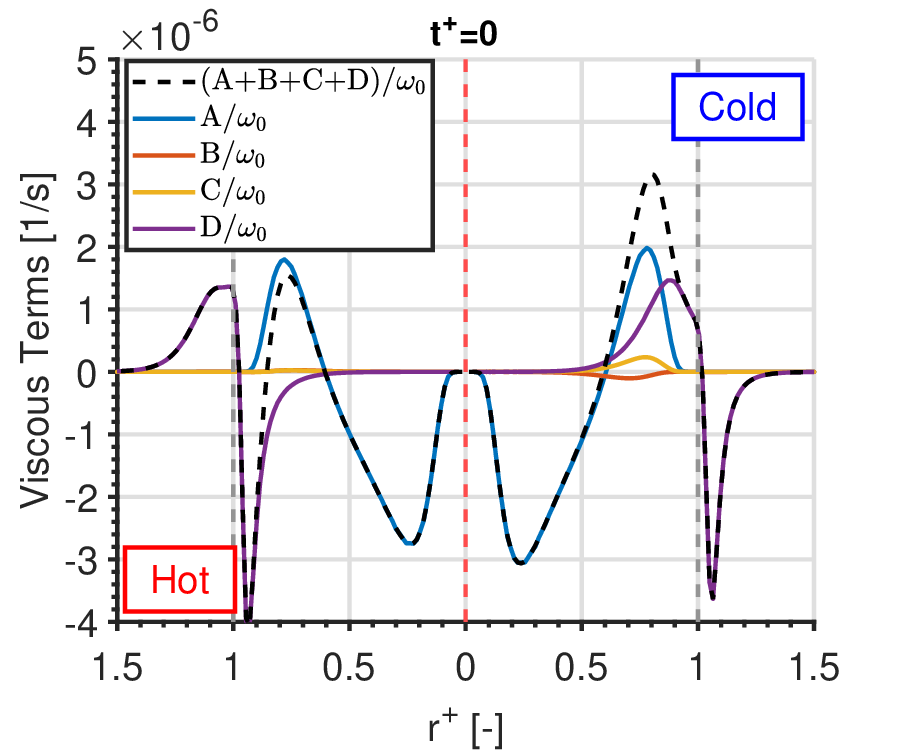}
  \caption{} 
  \label{subfig:vorticitydiffusion_Re200_pr1p5_File000_cold_vs_hot}
\end{subfigure}%
\begin{subfigure}{0.33\textwidth}
  \centering
  \includegraphics[width=1.0\linewidth]{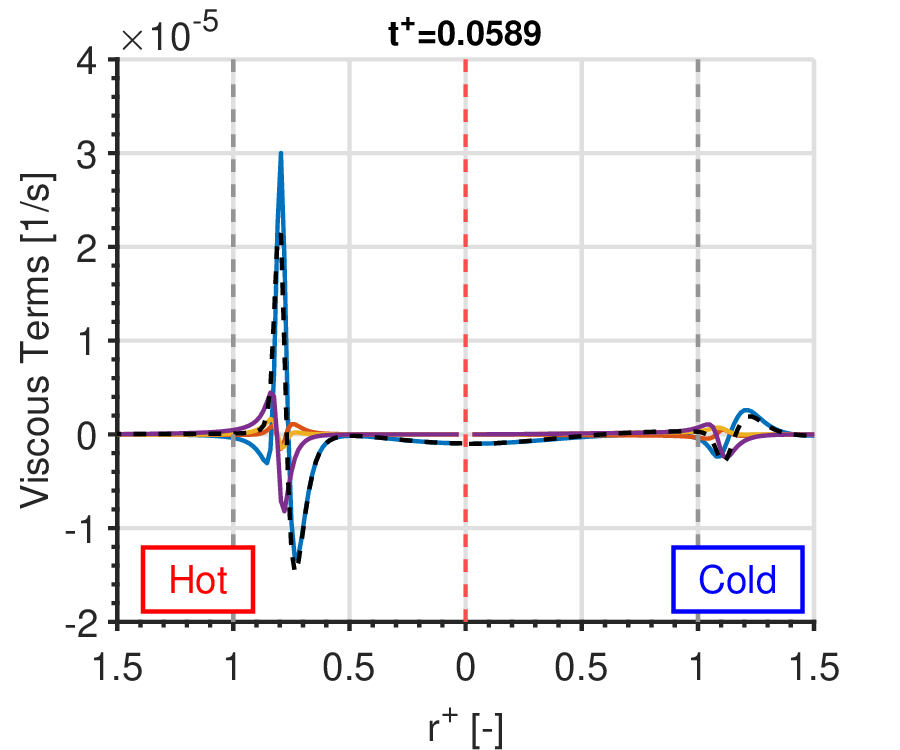}
  \caption{} 
  \label{subfig:vorticitydiffusion_Re200_pr1p5_File002_cold_vs_hot}
\end{subfigure}%
\begin{subfigure}{0.33\textwidth}
  \centering
  \includegraphics[width=1.0\linewidth]{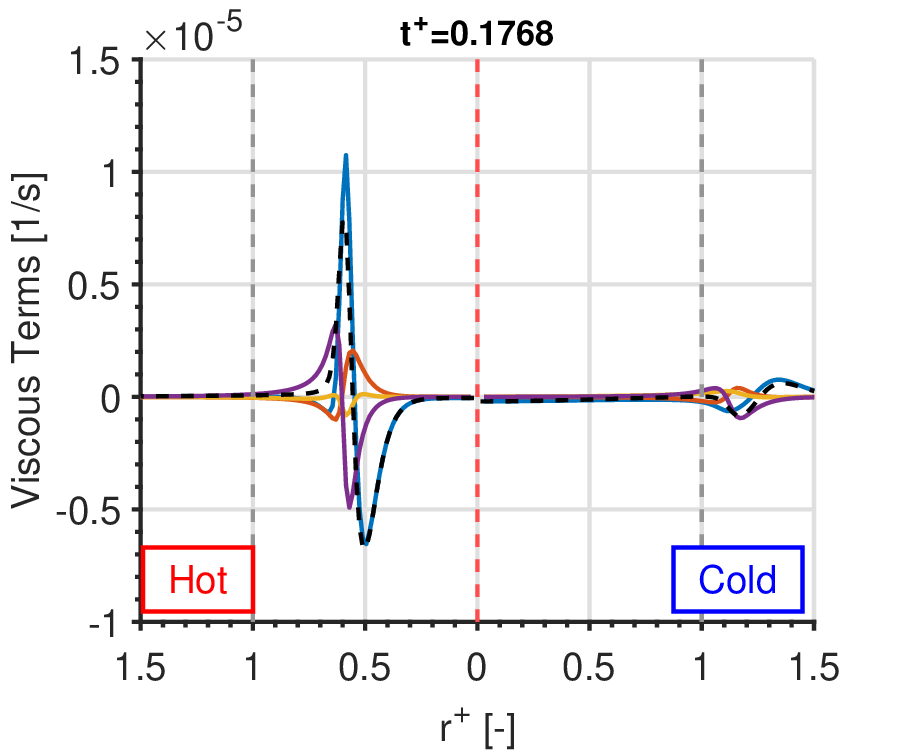}
  \caption{} 
  \label{subfig:vorticitydiffusion_Re200_pr1p5_File006_cold_vs_hot}
\end{subfigure}%
\\
\begin{subfigure}{0.33\textwidth}
  \centering
  \includegraphics[width=1.0\linewidth]{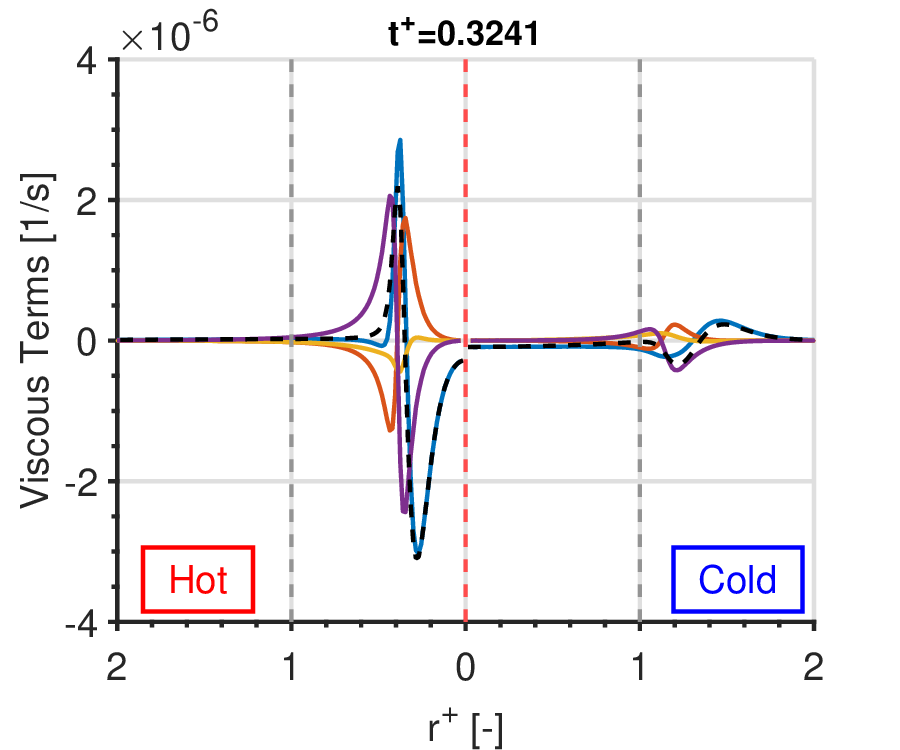}
  \caption{} 
  \label{subfig:vorticitydiffusion_Re200_pr1p5_File011_cold_vs_hot}
\end{subfigure}%
\begin{subfigure}{0.33\textwidth}
  \centering
  \includegraphics[width=1.0\linewidth]{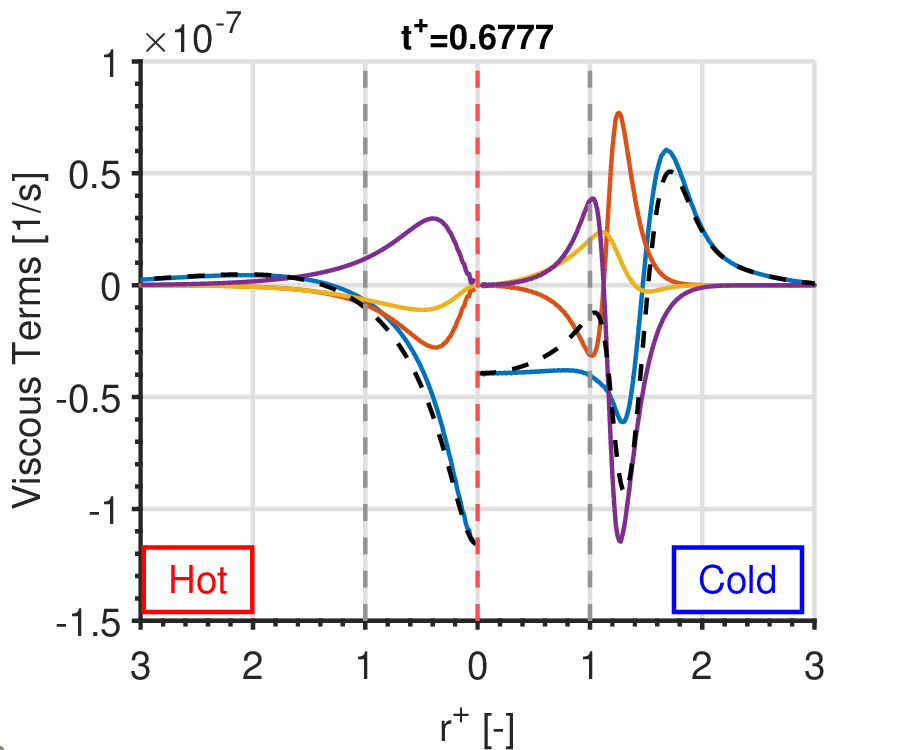}
  \caption{} 
  \label{subfig:vorticitydiffusion_Re200_pr1p5_File023_cold_vs_hot}
\end{subfigure}%
\begin{subfigure}{0.33\textwidth}
  \centering
  \includegraphics[width=1.0\linewidth]{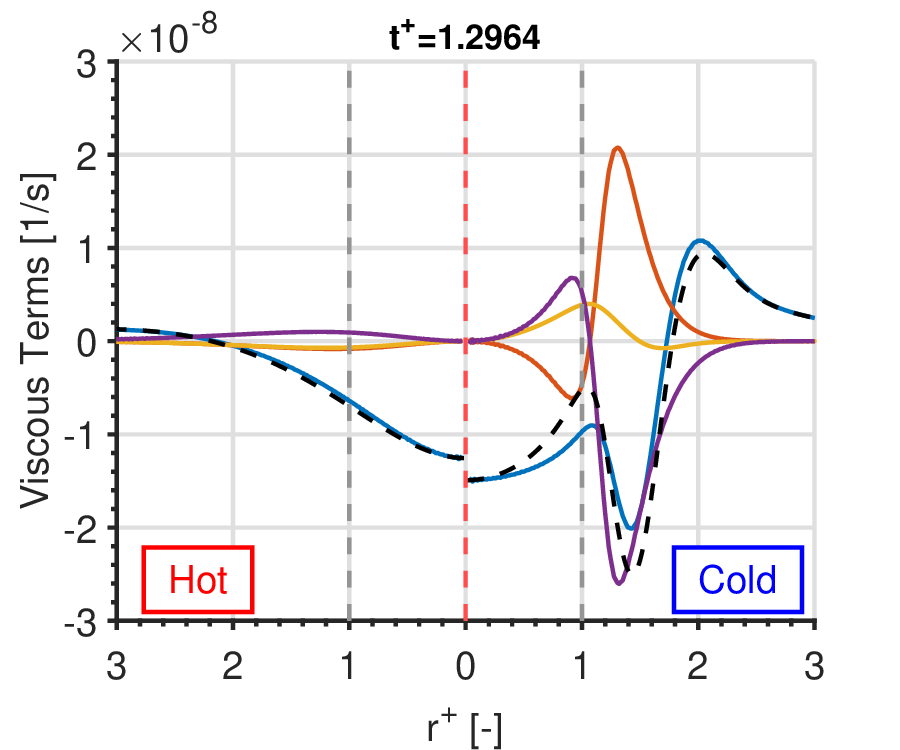}
  \caption{} 
  \label{subfig:vorticitydiffusion_Re200_pr1p5_File044_cold_vs_hot}
\end{subfigure}%
\caption{Evolution with \(t^+\) of the various viscous terms in (\ref{eqn:vorticityZ_9}) non-dimensionalised with \(\omega_0\) for the hot core (left) and the cold core (right) at \(p_r=1.5\). The dashed vertical lines represent the vortex centre (red) and \(r_c\) (grey). (a) \(t^+=0\); (b) \(t^+=0.0589\); (c) \(t^+=0.1768\); (d) \(t^+=0.3241\); (e) \(t^+=0.6777\); and (f) \(t^+=1.2964\).}
\label{fig:Fig22}
\end{figure}

\begin{figure}
\centering
\begin{subfigure}{0.33\textwidth}
  \centering
  \includegraphics[width=1.0\linewidth]{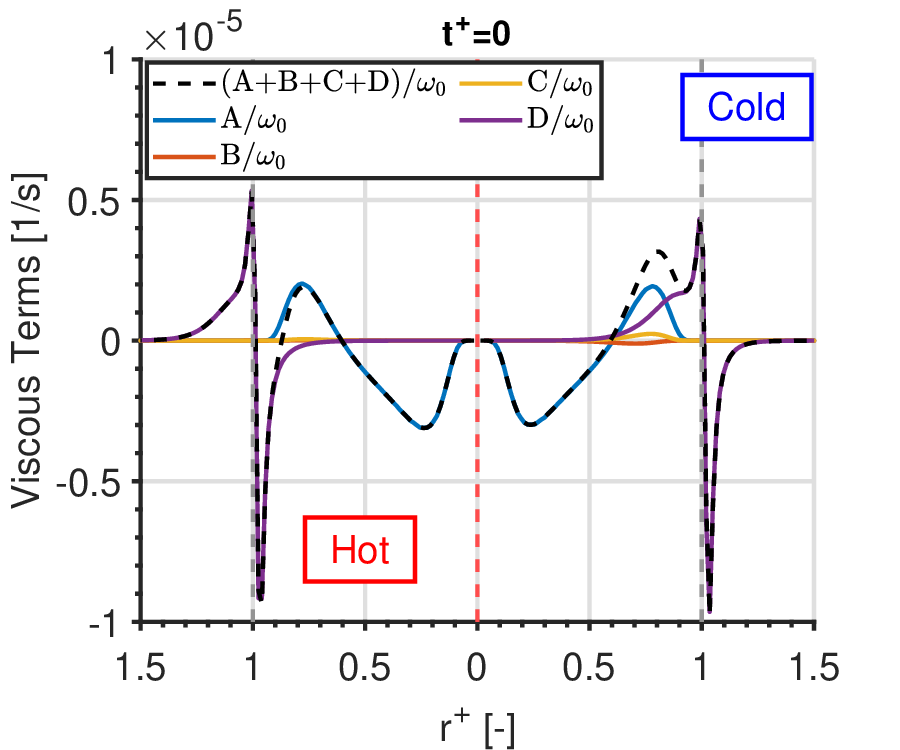}
  \caption{} 
  \label{subfig:vorticitydiffusion_Re200_pr1p3_File000_cold_vs_hot}
\end{subfigure}%
\begin{subfigure}{0.33\textwidth}
  \centering
  \includegraphics[width=1.0\linewidth]{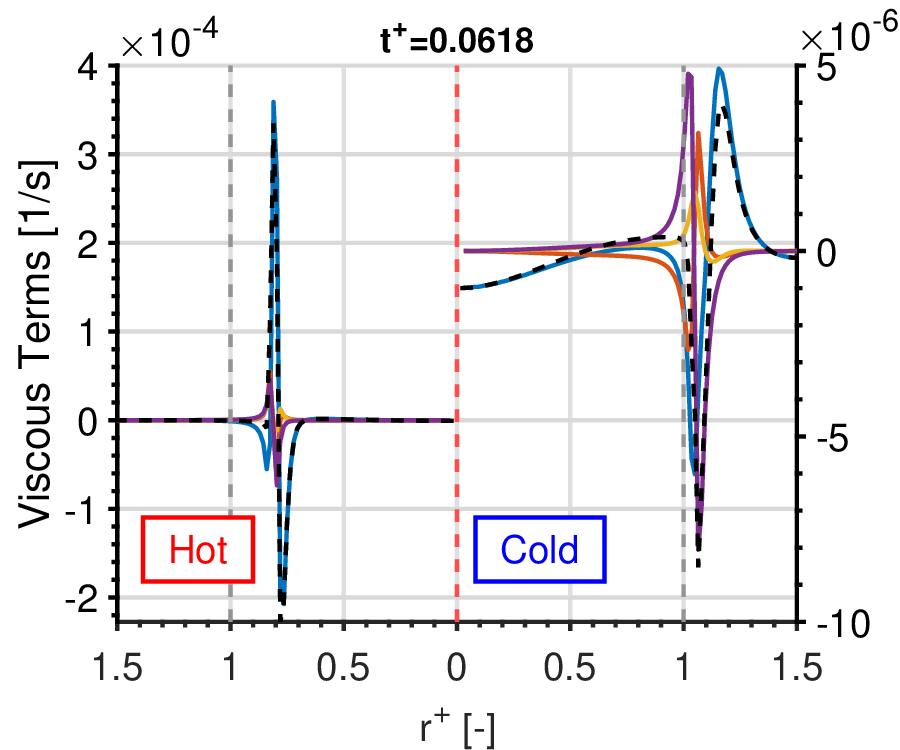}
  \caption{} 
  \label{subfig:vorticitydiffusion_Re200_pr1p3_File002_cold_vs_hot}
\end{subfigure}%
\begin{subfigure}{0.33\textwidth}
  \centering
  \includegraphics[width=1.0\linewidth]{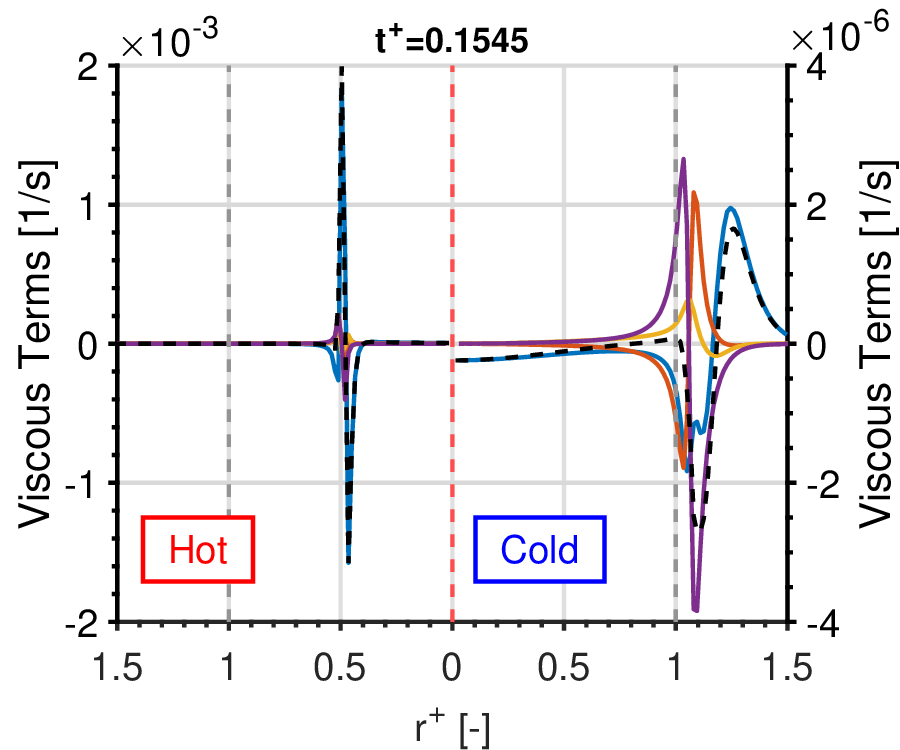}
  \caption{} 
  \label{subfig:vorticitydiffusion_Re200_pr1p3_File005_cold_vs_hot}
\end{subfigure}%
\\
\begin{subfigure}{0.33\textwidth}
  \centering
  \includegraphics[width=1.0\linewidth]{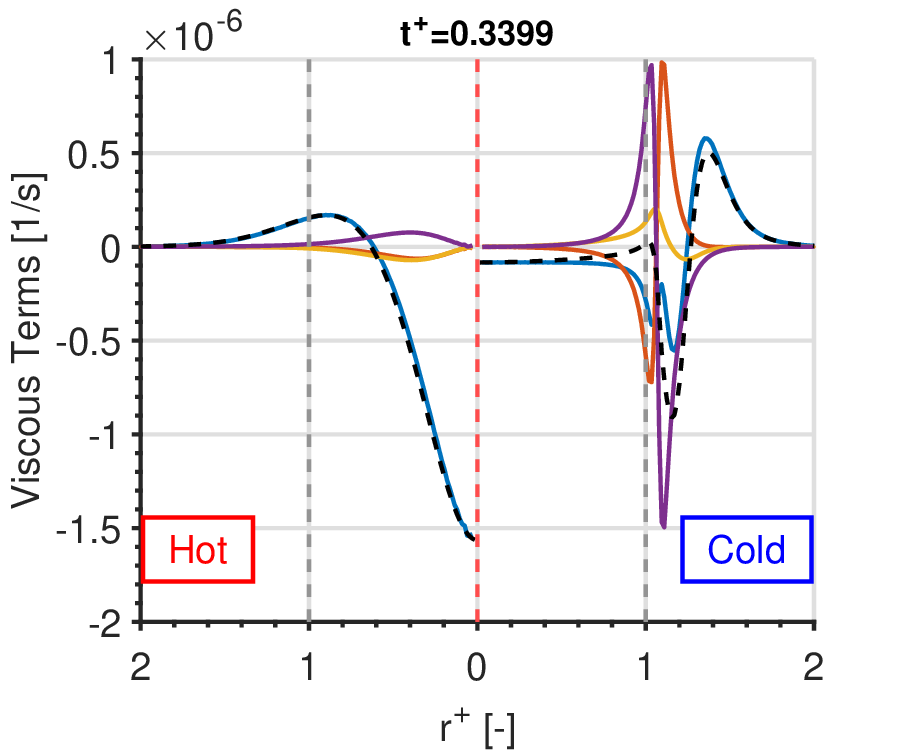}
  \caption{} 
  \label{subfig:vorticitydiffusion_Re200_pr1p3_File011_cold_vs_hot}
\end{subfigure}%
\begin{subfigure}{0.33\textwidth}
  \centering
  \includegraphics[width=1.0\linewidth]{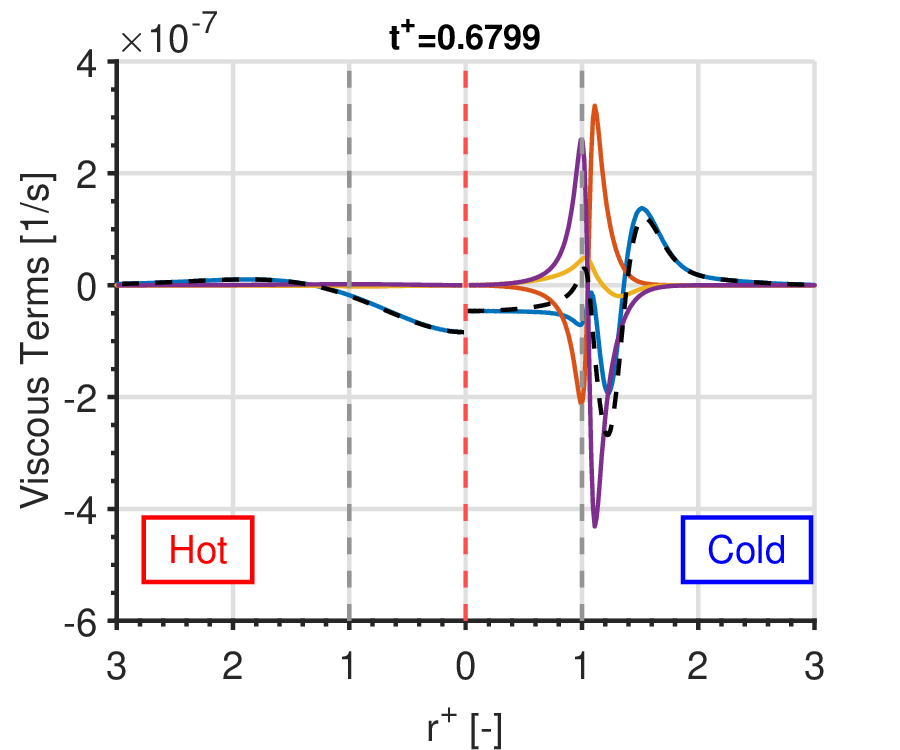}
  \caption{} 
  \label{subfig:vorticitydiffusion_Re200_pr1p3_File022_cold_vs_hot}
\end{subfigure}%
\begin{subfigure}{0.33\textwidth}
  \centering
  \includegraphics[width=1.0\linewidth]{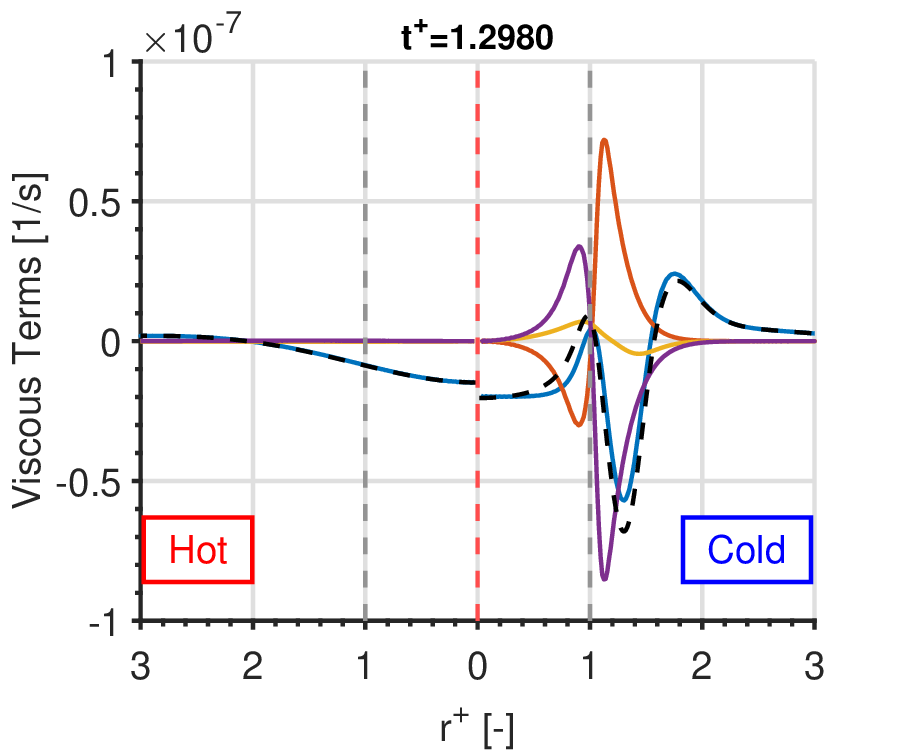}
  \caption{} 
  \label{subfig:vorticitydiffusion_Re200_pr1p3_File042_cold_vs_hot}
\end{subfigure}%
\caption{Evolution with \(t^+\) of the various viscous terms in (\ref{eqn:vorticityZ_9}) non-dimensionalised with \(\omega_0\) for the hot core (left) and the cold core (right) at \(p_r=1.3\). The dashed vertical lines represent the vortex centre (red) and \(r_c\) (grey). (a) \(t^+=0\); (b) \(t^+=0.0618\); (c) \(t^+=0.1545\); (d) \(t^+=0.3399\); (e) \(t^+=0.6799\); and (f) \(t^+=1.2980\).}
\label{fig:Fig23}
\end{figure}

Despite the lower pressures resulting in stronger compressibility and radial velocities as shown in figure \ref{fig:Fig11}, viscous mechanisms still dominate the evolution of \(\omega_z\), similar to figure \ref{fig:Fig12}. Therefore, the focus on the mechanisms generating the vorticity profiles and related vortex quantities discussed in section \ref{subsec:cold_vs_hot} is centred around the terms in equation (\ref{eqn:vorticityZ_9}). Figures \ref{fig:Fig22} and \ref{fig:Fig23} show the evolution of terms (A) to (D) in both hot and cold cores for \(p_r=1.5\) and \(1.3\), respectively. The initial distributions at \(t^+=0\) show that while the magnitude of viscous diffusion, vorticity stretching and vorticity-viscosity gradient alignment remain similar, the vorticity production term linked to the fluid translation with \(u_\theta\) increases drastically. For term (A), this is expected since the initial \(\omega_z\) distribution is identical across the various \(p_r\) values and the initial distributions of \(\nu=\mu/\rho\) are similar and of the same order of magnitude. In the case of terms (B) and (C), the larger derivatives of fluid properties are important. However, their enhancement occurs near the pseudo-boiling line around \(r_c\) where both \(\omega_z\) and its derivative approach zero; thus, no major pressure effect is observed in the initial distributions of terms (B) and (C). This is not the case for term (D). Since a localised vortex induces \(u_\theta\) outside the core, vorticity production due to the fluid motion as it traverses the pseudo-boiling line is magnified as \(p\rightarrow p_c\) with the increased gradients of fluid properties. This immediately translates into the following effects. For the cold vortex core, a stronger negative vorticity is generated outside the vortex as \(p\rightarrow p_c\), while \(\omega_z\) rises sharply in the vicinity of \(r_c\). In contrast, \(\omega_z>0\) emerges strongly outside the initial vortex in the hot core as \(p\rightarrow p_c\), while negative vorticity can arise inside the vortex near its edge as the fluid, with weak \(\omega_z>0\), experiences viscous mechanisms that oppose its vorticity and reverses it. Of course, these dynamics are partially influenced by the choice of initial temperature distribution in our study. Nonetheless, they highlight the importance of different vorticity production via viscous mechanisms directly linked to the strong variations of fluid properties across the pseudo-boiling line in SCF. \par 

As \(t^+\) increases, the evolution of the viscous term in the cold core highlights the stronger fluid properties gradients across the pseudo-boiling line. Viscous diffusion dominates the overall trend, especially at later \(t^+\). Yet, the evolution of \(\omega_z\) differs from classical Oseen theory, which is reflected in the diffusion mechanism across the vortex, i.e., non-monotonic behaviour. Note that the case at \(p_r=2\) has weaker pseudo-boiling effects and recovers a more monotonic behaviour (see figure \ref{fig:Fig14}). Similarly, the distributions of terms (B), (C) and (D) show a richer behaviour as \(p\rightarrow p_c\). Since pseudo-boiling effects vanish quicker at higher pressures, the distributions of these viscous terms observed toward \(t^+\approx1\) at \(p_r=1.5\) or \(1.3\) resemble those at \(t^+\approx0.05\) at \(p_r=2\). In other words, pseudo-boiling effects remain in the flow for longer at pressures closer to the critical pressure for two reasons: the enhanced gradients in fluid properties and the thermal barrier emerging due to the sharp decrease in thermal diffusivity. \par

The hot core configurations display features of an emerging discontinuity at the thermal front with the viscous term in the vorticity equation dominated, in general, by viscous diffusion, which quickly spikes several orders of magnitude compared to the cold core cases; i.e., about one order of magnitude greater at \(p_r=1.5\) and about two to three orders of magnitude greater at \(p_r=1.3\). This has direct implications in the vorticity distributions at \(p_r=1.3\), showing a nearly discontinuous distribution across the vortex (see figure \ref{subfig:vorticity_Re200_pr1p3_cold_vs_hot}). The corresponding process is explained in the following paragraph. The faster collapse of the hot vortex as \(p\rightarrow p_c\) translates into viscous diffusion being the main mechanism for vorticity redistribution, which is particularly evident at \(p_r=1.3\), e.g., see figures \ref{subfig:vorticitydiffusion_Re200_pr1p3_File022_cold_vs_hot} and \ref{subfig:vorticitydiffusion_Re200_pr1p3_File042_cold_vs_hot}. The evolution of each sub-term inside terms (B)-(D) for \(p_r=1.5\) and \(1.3\) is not shown here for brevity. These sub-terms are linked to the physical mechanisms described in section \ref{subsubsec:viscouseffects} and behave very similar to those shown for \(p_r=2\), albeit responding to the stronger derivatives in fluid properties seen in figure \ref{fig:Fig21}. \par 

\begin{figure}
\centering
\begin{subfigure}{0.33\textwidth}
  \centering
  \includegraphics[width=1.0\linewidth]{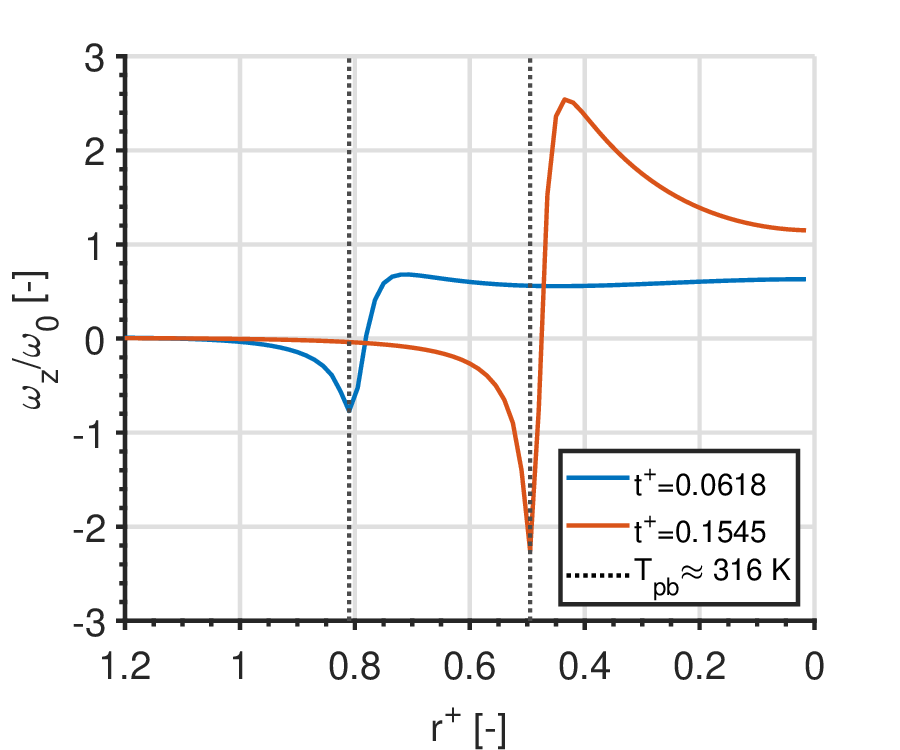}
  \caption{} 
  \label{subfig:vorticity_Re200_pr1p3_File002_File005_hot}
\end{subfigure}%
\begin{subfigure}{0.33\textwidth}
  \centering
  \includegraphics[width=1.0\linewidth]{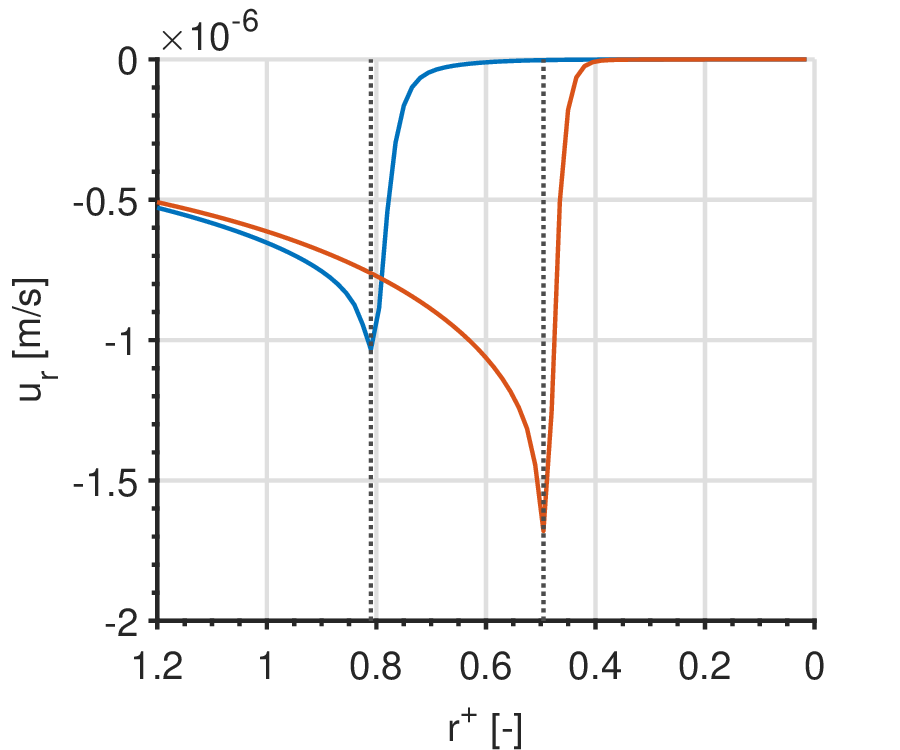}
  \caption{} 
  \label{subfig:radialvelocity_Re200_pr1p3_File002_File005_hot}
\end{subfigure}%
\caption{Distributions at \(t^+=0.0618\) and \(t^+=0.1545\) of \(\omega_z/\omega_0\) and \(u_r\) for the hot core at \(p_r=1.3\). The dotted vertical lines represent the location of the pseudo-boiling line given by \(T_{pb}\approx 316\) K. (a) \(\omega_z/\omega_0\); and (b) \(u_r\).}
\label{fig:Fig24}
\end{figure}

We now focus on the hot vortex core at \(p_r=1.3\) to explain the emerging quasi-discontinuity in the vorticity distribution. Given the trends observed with decreasing \(p_r\), this behaviour results from the combined effects of the pseudo-boiling phenomena. Figure \ref{fig:Fig24} shows the distributions of vorticity and radial velocity at \(t^+=0.0618\) and \(0.1545\). Additionally, the location where temperature matches the pseudo-boiling temperature, i.e., \(T(r)=T_{pb}\) is shown. \(T_{pb}\approx 316\) K is defined by the temperature where the maximum specific heat \(c_p\) occurs at \(p_r=1.3\). As discussed in \citet{2015_JSF_Banuti}, \(c_p\) is an appropriate marker to define the pseudo-boiling line since the peak in specific heat resembles the diverging behaviour of \(c_p\) across a subcritical interface (see figure \ref{fig:Fig1}). Remarkably, the location where the sharp decrease in vorticity and radial velocity occurs coincides with the pseudo-boiling line, i.e., where the pseudo-condensation of the gas-like fluid interface occurs. \par 

\begin{figure}
\centering
\begin{subfigure}{0.33\textwidth}
  \centering
  \includegraphics[width=1.0\linewidth]{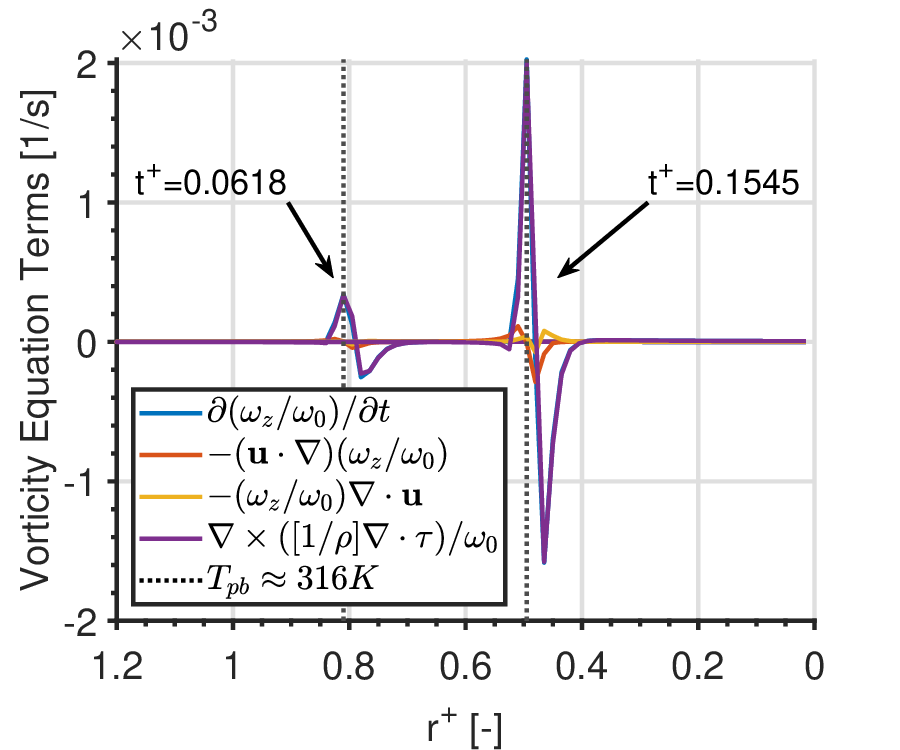}
  \caption{} 
  \label{subfig:vorticityeq_Re200_pr1p3_File002_File005_hot}
\end{subfigure}%
\begin{subfigure}{0.33\textwidth}
  \centering
  \includegraphics[width=1.0\linewidth]{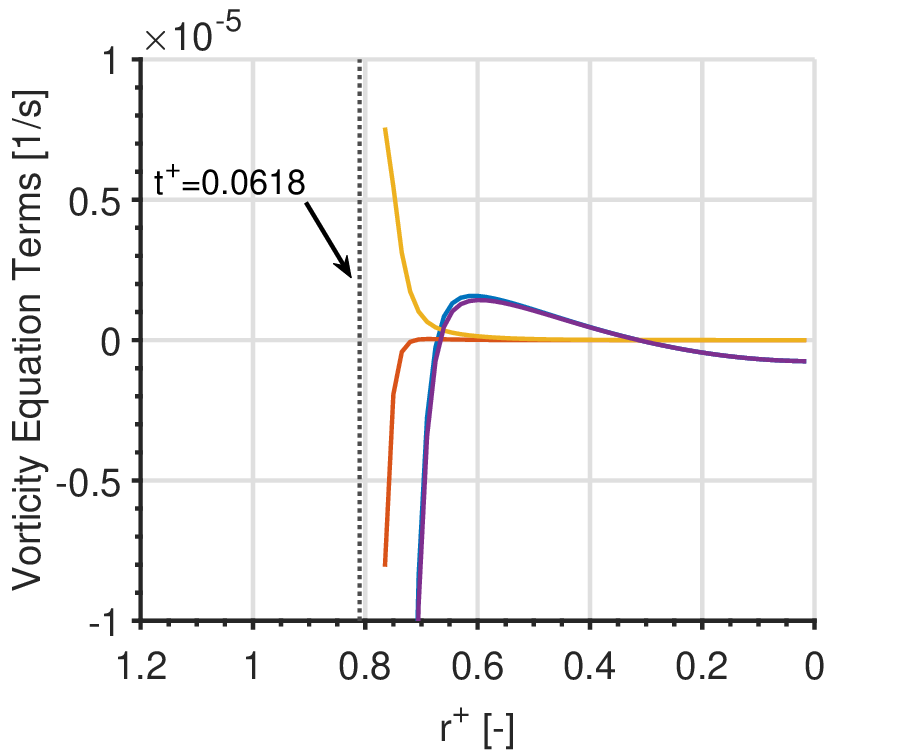}
  \caption{} 
  \label{subfig:vorticityeq_Re200_pr1p3_File002_hot_closeup}
\end{subfigure}%
\begin{subfigure}{0.33\textwidth}
  \centering
  \includegraphics[width=1.0\linewidth]{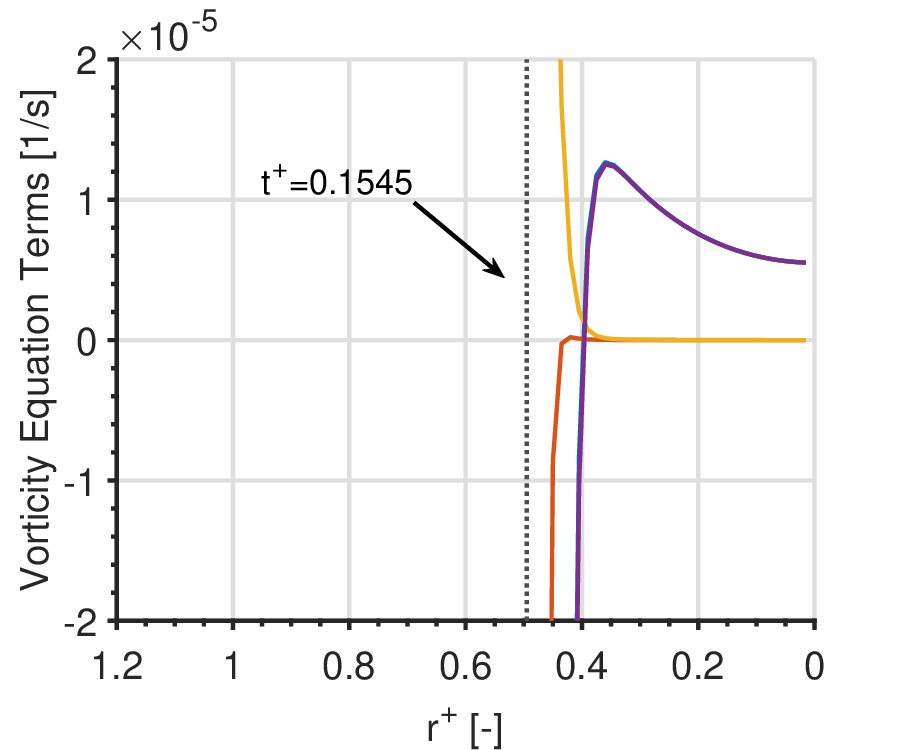}
  \caption{} 
  \label{subfig:vorticityeq_Re200_pr1p3_File005_hot_closeup}
\end{subfigure}%
\caption{Distribution at \(t^+=0.0618\) and \(t^+=0.1545\) of the terms in the vorticity equation given by (\ref{eqn:vorticityZ_9}) non-dimensionalised with \(\omega_0\) for the hot core at \(p_r=1.3\). The dotted vertical lines represent the location of the pseudo-boiling line given by \(T_{pb}\approx 316\) K. (a) global distributions; (b) distribution in the gas-like side of the vortex at \(t^+=0.0618\); and (c) distribution in the gas-like side of the vortex at \(t^+=0.1545\).}
\label{fig:Fig25}
\end{figure}

\begin{figure}
\centering
\begin{subfigure}{0.33\textwidth}
  \centering
  \includegraphics[width=1.0\linewidth]{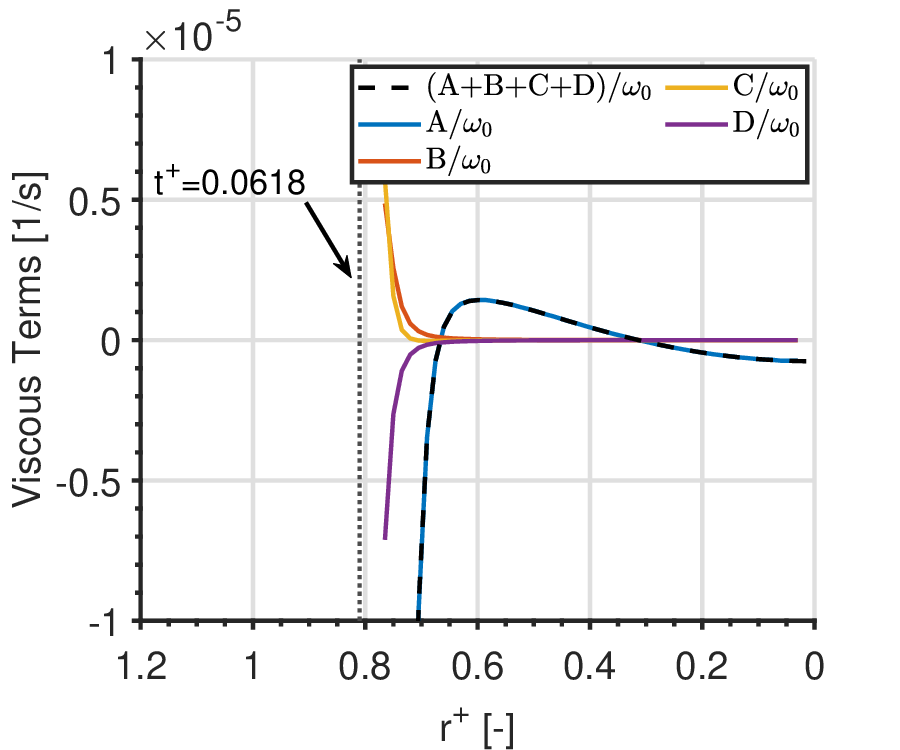}
  \caption{} 
  \label{subfig:vorticitydiffusion_Re200_pr1p3_File002_hot_closeup}
\end{subfigure}%
\begin{subfigure}{0.33\textwidth}
  \centering
  \includegraphics[width=1.0\linewidth]{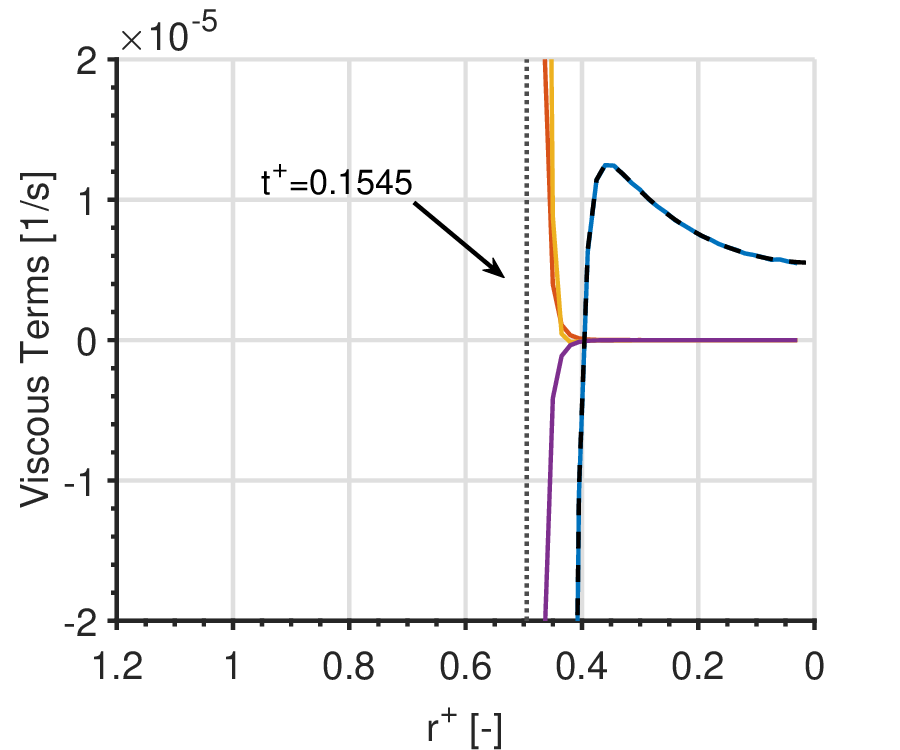}
  \caption{} 
  \label{subfig:vorticitydiffusion_Re200_pr1p3_File005_hot_closeup}
\end{subfigure}%
\caption{Distributions in the gas-like side of the vortex of the various viscous terms in (\ref{eqn:vorticityZ_9}) non-dimensionalised with \(\omega_0\) for the hot core at \(p_r=1.3\). The dotted vertical lines represent the location of the pseudo-boiling line given by \(T_{pb}\approx 316\) K. (a) \(t^+=0.0618\); and (b) \(t^+=0.1545\).}
\label{fig:Fig26}
\end{figure}

Figure \ref{fig:Fig25} depicts the various terms responsible for modifying vorticity at \(t^+=0.0618\) and \(0.1545\), as well as the location where \(T(r)=T_{pb}\). As previously discussed, the main mechanism describing the evolution of \(\omega_z\) is the viscous term \(\boldsymbol{\nabla\times}\big([1/\rho]\boldsymbol{\nabla}\boldsymbol{\cdot}\boldsymbol{\tau}\big)\). Which peaks around the pseudo-boiling temperature and sharply decreases on the gas-like side of the SCF. Moreover, the close-up profiles in the gas-like side of the vortex reveal that the increase or decrease of vorticity inside the vortex away from \(T_{pb}\) seen in figure \ref{subfig:vorticity_Re200_pr1p3_cold_vs_hot} is also driven by viscous mechanisms. Radial advection and compressible vortex stretching are only important in the vicinity of \(T_{pb}\), but tend to cancel each other. Looking at the contributions of each viscous mechanism, i.e., terms (A)-(D) (see figure \ref{fig:Fig26}), viscous diffusion dominates while the other mechanisms linked to the variation of fluid properties are only active near \(T_{pb}\). \par 

These results highlight that, for the hot vortex core, the steepening of the thermal front indeed occurs at the pseudo-boiling temperature as \(p\rightarrow p_c\). Thus, all vorticity generation mechanisms directly linked to the compressibility of the fluid and non-linear variation of fluid properties concentrate around \(T_{pb}\), with viscous diffusion dominating the evolution of \(\omega_z\) away from the thermal front. The evolution of \(\omega_z\) shown in figure \ref{subfig:vorticity_Re200_pr1p3_cold_vs_hot} follows from the initial reverse vorticity generated by term (D) near \(r_c\). Inherently, the viscous diffusion term (A) captures the curvature in the vorticity distribution, which can become extremely large due to the contributions of term (D). In these scenarios, viscous diffusion is not strong enough to redistribute vorticity smoothly and instead sharpens the distribution of \(\omega_z\) and increases its magnitude inside the gas-like region of the vortex. Compared to the corresponding cold core configuration, this is a direct result of the localised sharp minimum in momentum diffusivity around \(T_{pb}\) with nearly an order of magnitude larger diffusivity in the core (see figure \ref{subfig:kinvis_Re200_pr1p3_cold_vs_hot}). \par 

This diverging behaviour suggests that a centrifugal instability develops as \(p\rightarrow p_c\) and that stable hot vortex columns cannot exist near \(p_c\), independently of baroclinicity effects. Yet, there is a range of thermodynamic states where the increase in circulation of the vortex column resulting from a locally negative vorticity remains balanced by viscous diffusion. In the cold vortex column, the self-stabilisation of the pseudo-boiling line allows the ``spreading" of pseudo-boiling effects. Thus, viscous diffusion may prevent a centrifugal instability from occurring despite the emergence of negative vorticity, even at pressures much closer to \(p_c\).  \par

\section{Concluding Remarks}
\label{sec:conclusions}

This work explores supercritical fluid vortex dynamics in a radially stratified axisymmetric vortex column where the effects on the vorticity evolution of compressible stretching and varying viscous stresses are isolated from the baroclinic torque. Radial stratification is achieved via a cylindrical thermal mixing layer about the pseudo-boiling line for each thermodynamic pressure. Vortices described by a sufficiently low \(Re\) (\(Re_\text{max}\equiv\Gamma_0\big(\frac{\rho}{\mu}\big)_\text{max}<400\) for sCO\(_2\)) collapse to an axisymmetric solution independent of \(Re_\text{max}\). Two thermal configurations, with the vortex core colder or hotter than the surrounding fluid, are considered for sCO\(_2\) over three thermodynamic pressures close but above its critical pressure. In all cases, the large variations of fluid properties across the pseudo-boiling line occur in the vicinity of the vortex core. \par

Vortex-related quantities, i.e., vorticity and circulation, evolve differently in each case and show substantial differences from the incompressible solution of \citet{1912_AMA_Oseen}. Sharp vorticity gradients are observed as \(p\rightarrow p_c\), especially for the hot vortex, and regions of negative (or counter-rotating) vorticity are generated, which translate into an azimuthal acceleration of the fluid and a local overshoot of circulation. As pressure drops from \(p_r=2\) to \(1.3\), the gradients of the fluid properties are magnified and the radial derivatives of density and viscosity increase by up to an order of magnitude. Moreover, pseudo-boiling effects become clearer. The evolution of the cold vortex shows similarities with droplet evaporation phenomena in transcritical fluids, i.e., a cooler liquid-like fluid surrounded by a hotter gas-like fluid. The pseudo-boiling line remains nearly fixed in space as thermal diffusion occurs on both sides, showing a self-stabilisation mechanism, also shown in \citet{2026_IJTS_Yin}. In contrast, the hot vortex evolves resembling the rapid condensation and collapse of a vapour bubble, i.e., a hotter gas-like fluid core surrounded by a cooler liquid-like fluid. As pressure drops, strong radial velocities into the vortex are generated at the same time that thermal diffusivity drops at the pseudo-boiling line. This results in an effective blocking of thermal conduction (or the formation of a thermal barrier) and a steepening of the temperature profiles, i.e., a formation of a thermal front around the pseudo-boiling temperature occurs that is advected inwards. Altogether, substantially different vorticity evolutions are observed between the cold and hot vortex configurations. Specifically, the diverging nature of the vorticity distribution and sharp rise in circulation in the hot vortex as pressure drops suggests that a centrifugal instability can easily emerge; i.e., hot vortex columns cannot exist near \(p_c\). This is not the case for cold vortex columns as the self-stabilisation of the pseudo-boiling line allows a balance between the increased circulation and viscous diffusion. \par

The analysis of the vorticity equation reveals that such strong differences in vorticity evolution are a direct result of the varying properties. Due to the weak compressibility, viscous mechanisms dominate. By decomposing the viscous term in the vorticity equation, three additional mechanisms beyond viscous diffusion of vorticity are identified and examined in detail: (1) a vorticity stretching term resulting from the viscosity variations, (2) a term linked to the alignment between vorticity and viscosity/density gradients, and (3) a vorticity production (source or sink) related to the interaction of the fluid elements moving around the vortex axis with the varying properties in the fluid. Although the first two mechanisms effectively redistribute existing vorticity, the third describes vorticity generation directly linked to the varying properties across the fluid. The identified mechanisms become important across the pseudo-boiling line of supercritical fluids and are comparable in magnitude to viscous diffusion. In particular, the third mechanism is responsible for the generation of the negative or counter-rotating vorticity and the subsequent local increase in circulation. \par 

However, the newly described viscous mechanisms are only important in the vicinity of the pseudo-boiling line and viscous diffusion dominates most of the vortex evolution. Still, these mechanisms are strong enough to affect vorticity distribution. In the cold vortex, their strength decays as thermal mixing occurs, but at the same time they do not vanish completely as they spread around the pseudo-boiling line. In the hot vortex, these mechanisms concentrate around the pseudo-boiling line while a thermal front develops with decreasing pressure. Since the thermal front is rapidly advected inwards, pseudo-boiling effects vanish rather rapidly and the hot vortex evolves similar to an incompressible vortex at later times. As a result, the growth of the hot core radius approaches asymptotically the growth rates of the corresponding incompressible vortex. This is not the case for the cold vortex, whose core radius evolves at much slower rates as self-stabilisation of the pseudo-boiling line occurs. \par 

The constraints imposed on the vortex column problem in this work allow us to highlight the complexities brought about by SCF and the additional mechanisms, beyond baroclinicity, that must be considered to understand vorticity dynamics under supercritical fluid conditions. Future work is aimed at extending the analysis to more intricate problems such as the interaction of vortex rings and the process of vortex reconnection. \par

\section*{Declaration of Interests}

The authors report no conflict of interest.

\section*{Author Contributions}

J. Poblador-Ibanez conceived the study, conducted the numerical simulations, analysed the raw data, produced the results and wrote the original draft of the manuscript. F. Hussain revised the results and refined the manuscript.

\section*{Acknowledgements}

The authors acknowledge the use of computational resources of DelftBlue supercomputer, provided by Delft High Performance Computing Centre (https://www.tudelft.nl/dhpc). \par

\appendix

\section{Numerical Symmetry Breaking of the Vortex}
\label{apn:A}

\setcounter{figure}{0}

Following the initialisation described in section \ref{subsec:initialisation_sCO2}, the sensitivity of the solution for different \(Re_\text{max}\), \(L\in[2\pi,4\pi,8\pi,16\pi]\) m, and uniform grid sizes \(\Delta x \in [\pi/128, \pi/256,\pi/512]\) m is investigated. This is necessary given the use of a multi-dimensional solver and a Cartesian mesh to define the computational domain and discretise the problem. That is, to ensure that the study correctly represents the axisymmetric solution and is free from numerical destabilisation. \par 

\begin{figure}
\centering
\begin{subfigure}{0.33\textwidth}
  \centering
  \includegraphics[width=1.0\linewidth]{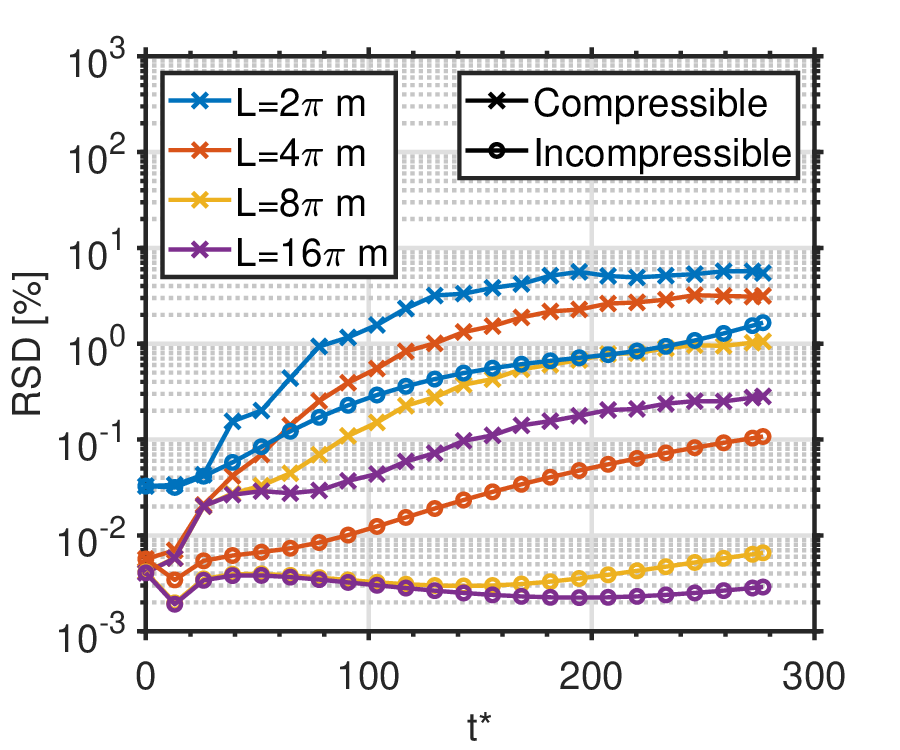}
  \caption{} 
  \label{subfig:rsd_Re500_utheta_rc_incomp}
\end{subfigure}%
\begin{subfigure}{0.33\textwidth}
  \centering
  \includegraphics[width=1.0\linewidth]{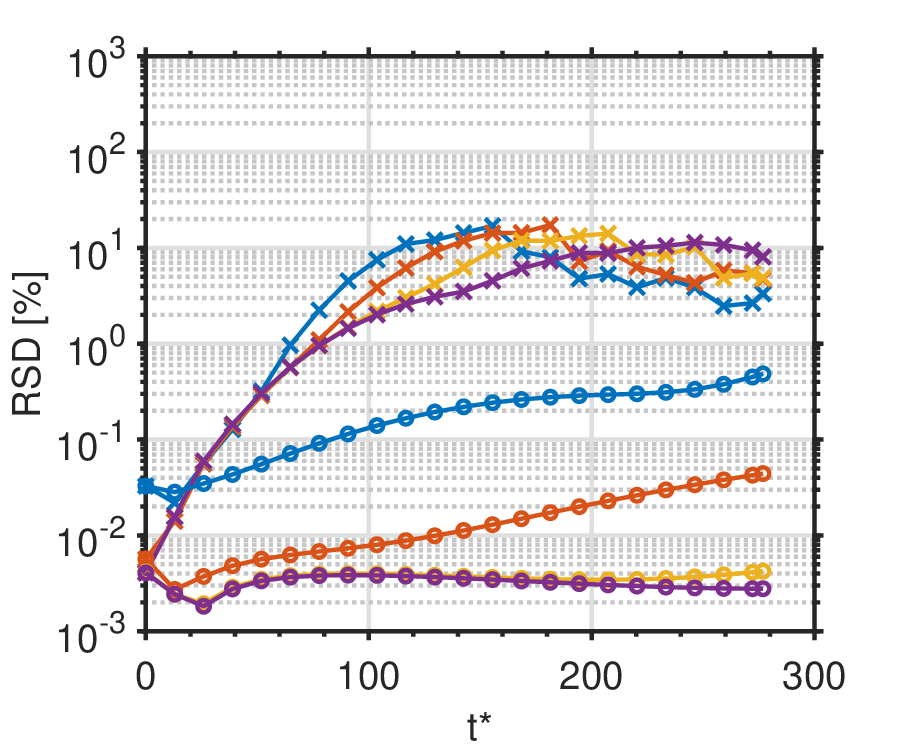}
  \caption{} 
  \label{subfig:rsd_Re1000_utheta_rc_incomp}
\end{subfigure}%
\caption{Comparison of the RSD of the azimuthal velocity \(u_\theta\) extracted along \(r_c\) with \(\Delta x = \pi/256\) m between the compressible and incompressible vortices with the same \(Re_\text{max}\). (a) \(Re_\text{max}=500\); and (b) \(Re_\text{max}=1000\).}
\label{fig:Fig27}
\end{figure}

The axisymmetry breaking is evident from the RSD of the azimuthal velocity \(u_\theta\) extracted along \(r=r_c\), given by (\ref{eqn:mean+RSD}). Figure \ref{fig:Fig27} shows the evolution of the RSD of \(u_\theta\) for both the compressible and incompressible vortices with \(\Delta x=\pi/256\) m and initialised with the same \(Re_\text{max}\). A uniform temperature is used in the incompressible case. Further, a non-dimensional time defined as \(t^*=(\Gamma_0/r_c^2)t\) is used to properly measure the effects of increasing \(Re_\text{max}\) on the numerical solution destabilisation time scales. The destabilisation of the compressible vortex is augmented compared to the incompressible vortex. Depending on \(L\) and \(Re_\text{max}\), the RSD of the incompressible case can be one to three orders of magnitude lower than when compressible. In particular, the RSD remains reasonably constant for domain sizes with \(L\geq 8\pi\) m, ensuring a stable evolution of the vortex despite the appearance of non-axisymmetric modes. Therefore, multi-dimensional flow solvers may be used to analyse a wider range of \(Re_\text{max}\) in axisymmetric incompressible flows if the domain size is large enough. A validation of the incompressible vortex against Oseen's analytical solution \citep{1912_AMA_Oseen} is presented in \ref{subapn:B1} for \(Re_\text{max}=1000\). \par

\begin{figure}
\centering
\begin{subfigure}{0.25\textwidth}
  \centering
  \includegraphics[width=1.0\linewidth]{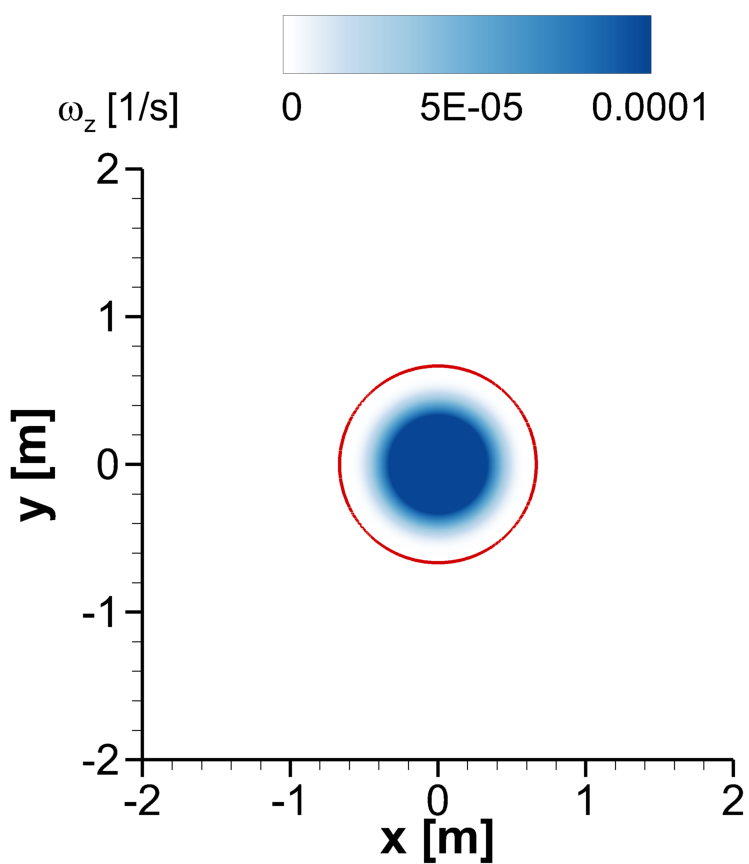}
  \caption{} 
  \label{subfig:wz_Re1000_512_8PI_000}
\end{subfigure}%
\begin{subfigure}{0.25\textwidth}
  \centering
  \includegraphics[width=1.0\linewidth]{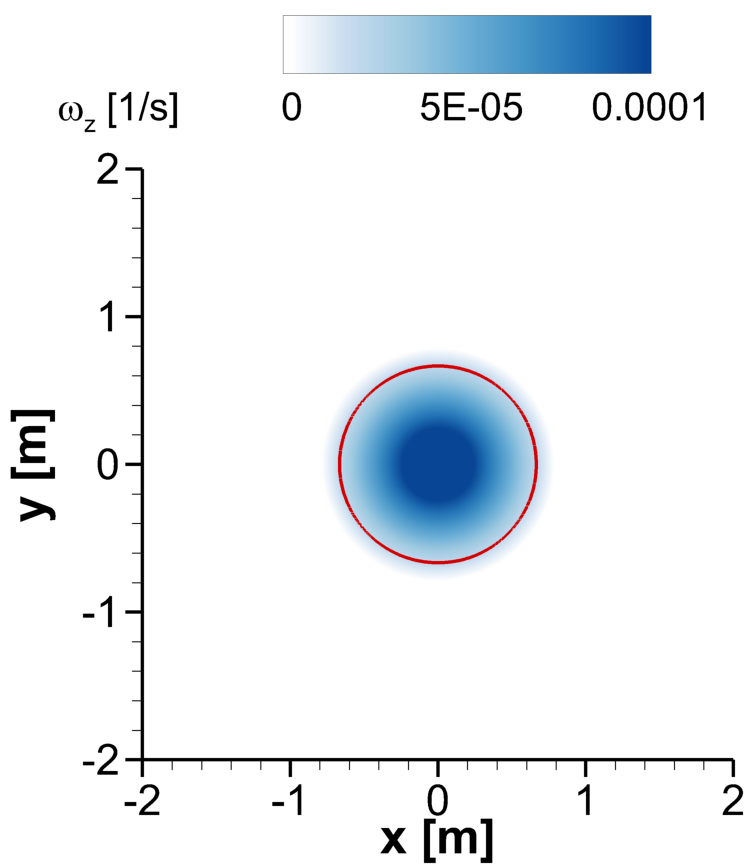}
  \caption{} 
  \label{subfig:wz_Re1000_512_8PI_009}
\end{subfigure}%
\begin{subfigure}{0.25\textwidth}
  \centering
  \includegraphics[width=1.0\linewidth]{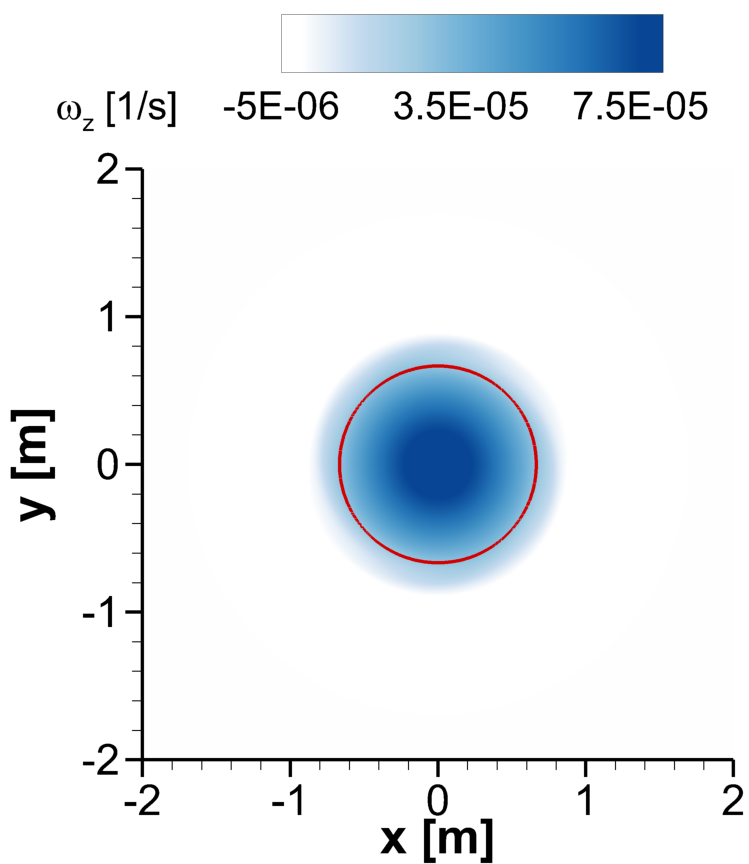}
  \caption{} 
  \label{subfig:wz_Re1000_512_8PI_021}
\end{subfigure}%
\begin{subfigure}{0.25\textwidth}
  \centering
  \includegraphics[width=1.0\linewidth]{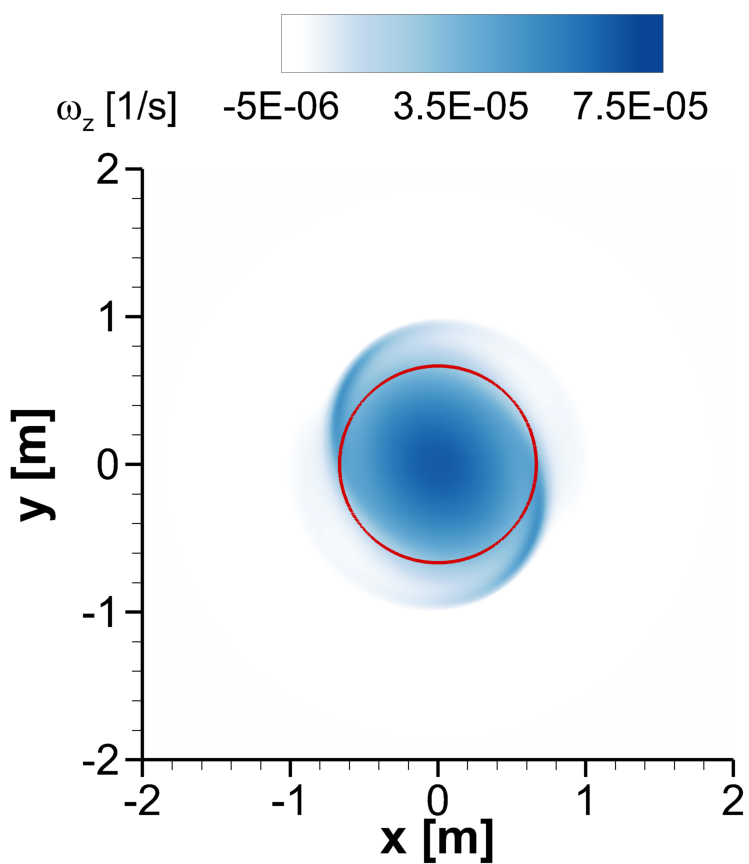}
  \caption{} 
  \label{subfig:wz_Re1000_512_8PI_030}
\end{subfigure}%
\\
\begin{subfigure}{0.25\textwidth}
  \centering
  \includegraphics[width=1.0\linewidth]{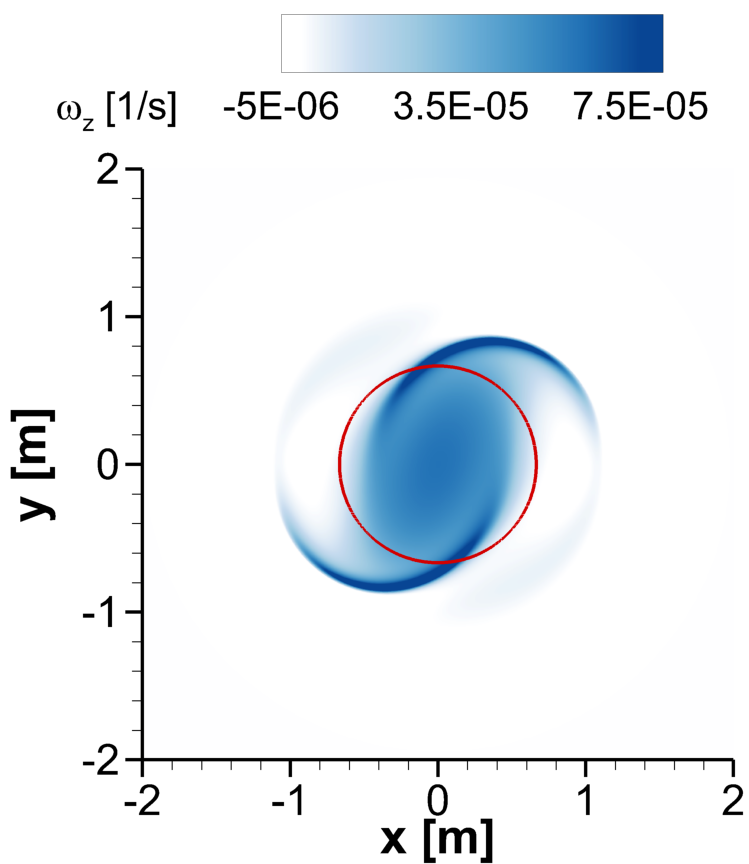}
  \caption{} 
  \label{subfig:wz_Re1000_512_8PI_036}
\end{subfigure}%
\begin{subfigure}{0.25\textwidth}
  \centering
  \includegraphics[width=1.0\linewidth]{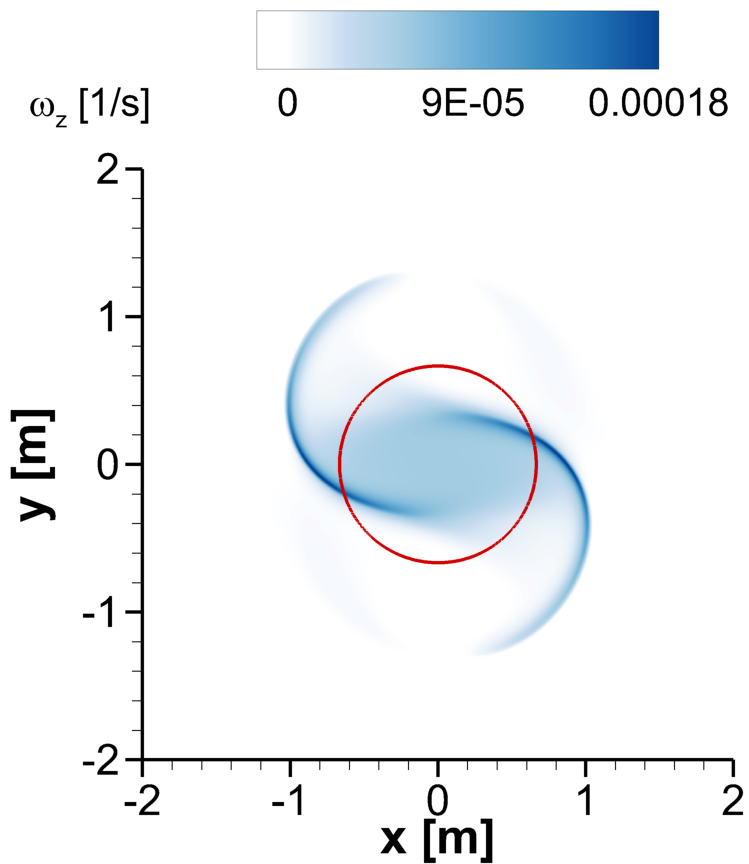}
  \caption{} 
  \label{subfig:wz_Re1000_512_8PI_042}
\end{subfigure}%
\begin{subfigure}{0.25\textwidth}
  \centering
  \includegraphics[width=1.0\linewidth]{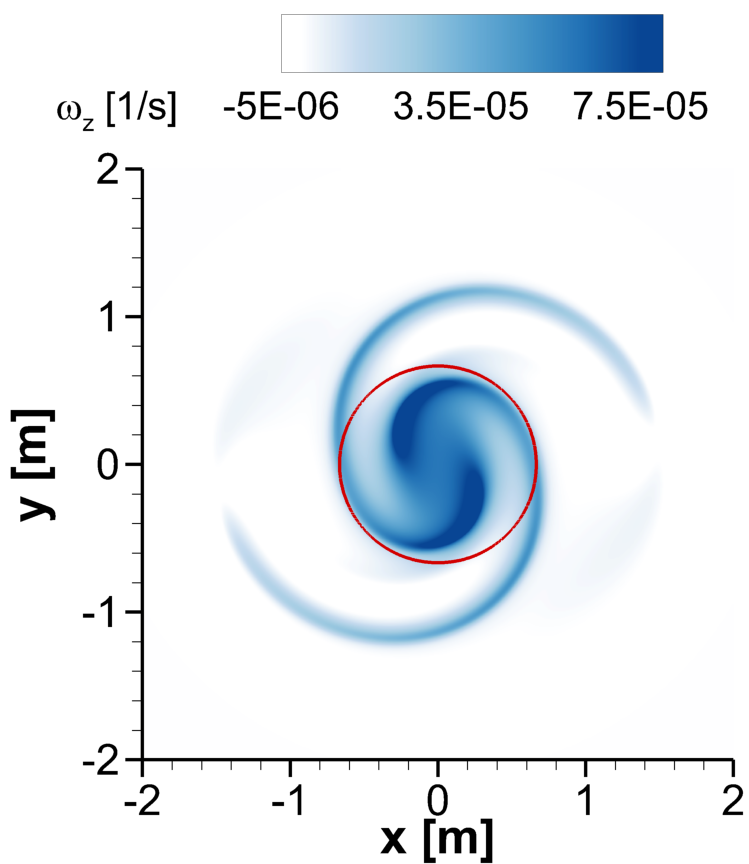}
  \caption{} 
  \label{subfig:wz_Re1000_512_8PI_051}
\end{subfigure}%
\begin{subfigure}{0.25\textwidth}
  \centering
  \includegraphics[width=1.0\linewidth]{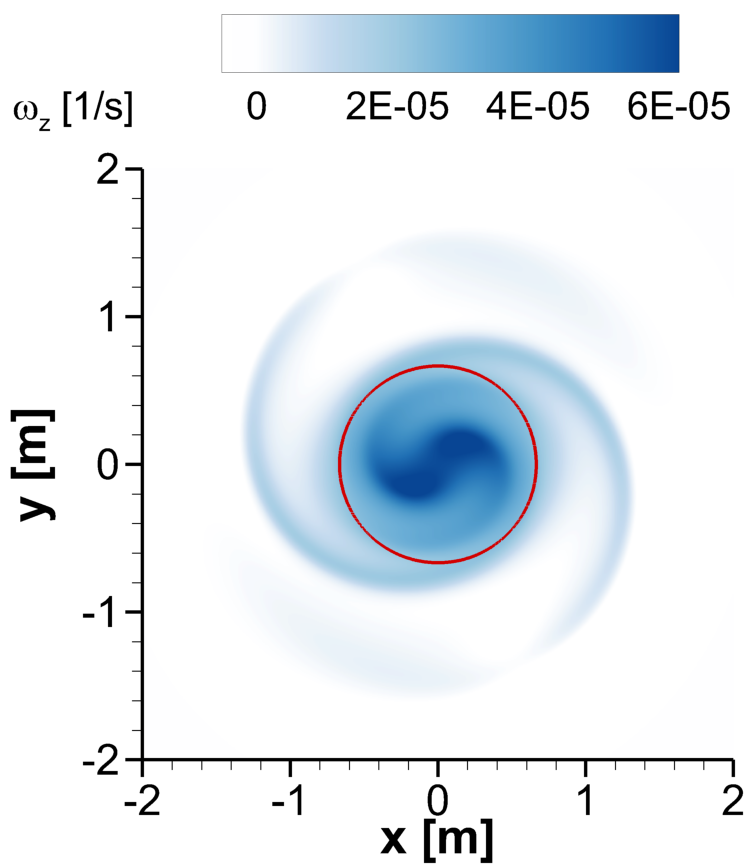}
  \caption{} 
  \label{subfig:wz_Re1000_512_8PI_063}
\end{subfigure}%
\caption{Contours of \(\omega_z\) for \(Re_\text{max}=1000\), \(L=8\pi\) m and \(\Delta x = \pi/512\) m. The red iso-contour represents \(r_c\). (a) \(t^*=0\); (b) \(t^*=38.88\); (c) \(t^*=90.72\); (d) \(t^*=129.61\); (e) \(t^*=155.53\); (f) \(t^*=181.45\); (g) \(t^*=220.33\); and (h) \(t^*=272.17\).}
\label{fig:Fig28}
\end{figure}

For \(Re_\text{max}=500\), a stable, yet non-axisymmetric, evolution of the compressible vortex results if the domain is large enough. However, \(Re_\text{max}=1000\) exhibits a complete destabilisation of the vortex with large oscillations in the RSD values regardless of \(L\). This is visualized in figure \ref{fig:Fig28}, which shows the distribution of \(\omega_z\) in different snapshots for \(Re_\text{max}=1000\), \(L=8\pi\) m and \(\Delta x = \pi/512\) m. While at \(t^*=90.72\) the vorticity has decayed and remains fairly axisymmetric (see figure \ref{subfig:wz_Re1000_512_8PI_021}), the vortex quickly destabilises. \par 

Two perturbation sources are responsible for breaking the symmetry of the flow: the choice of initialisation of \(\boldsymbol{u}\) and the domain size (via the chosen BC). While the latter vanishes as \(L\) increases, the initialisation of \(\boldsymbol{u}\) with (\ref{eqn:initialvelocity}) may not satisfy the compatibility conditions between the initial value and the boundary conditions \citep{1973_CaF_Wu,2022_FDR_Nagy}, inducing a small non-axisymmetric perturbation to the velocity field. Particularly around \(r_c\) where \(\omega_z\) vanishes, \(\boldsymbol{\nabla\cdot u}\neq 0\) after initialisation and a small radial velocity is imposed (\(|u_r|\lll|u_\theta|\)). Similar initialisation errors in the velocity field are observed if the Biot-Savart equation is used. Therefore, the errors are caused by the discretization of the initial condition itself and vanish as the mesh is refined. The use of Biot-Savart's equation improves the velocity field near the boundaries of the computational domain, which is more consistent with the vortex column in infinite space compared to the initialisation with (\ref{eqn:initialvelocity}). Of course, the domain can be set big enough to isolate the vortex from boundary effects in favour of the computational efficiency of solving (\ref{eqn:initialvelocity}). \par

The mechanisms breaking the inherent axisymmetry of the Oseen-like vortex are analysed in terms of vorticity generation. The vorticity equation becomes

\begin{equation}
\label{eqn:vorticityZ}
\begin{split}
\frac{D\omega_z}{Dt} = &- \omega_z (\boldsymbol{\nabla \cdot u}) + \frac{1}{\rho^2}\big(\boldsymbol{\nabla}\rho\boldsymbol{\times}\boldsymbol{\nabla}p\big)\boldsymbol{\cdot}\hat{\textbf{e}}_z + \bigg[\boldsymbol{\nabla\times}\bigg(\frac{1}{\rho}\boldsymbol{\nabla}\boldsymbol{\cdot}\boldsymbol{\tau}\bigg)\bigg]\boldsymbol{\cdot}\hat{\textbf{e}}_z
\end{split}
\end{equation}

\noindent
where \(\hat{\textbf{e}}_z=(0,0,1)\) is the unit vector normal to the 2D plane. Note that the vortex stretching and tilting term \((\boldsymbol{\omega\cdot\nabla})\boldsymbol{u}\) is identically zero in a 2D flow. In the incompressible limit, only the viscous term remains on the right-hand side (RHS) of (\ref{eqn:vorticityZ}), which simplifies to \(\nu\boldsymbol{\nabla^2}\omega_z\). Thus, the incompressible vortex is mostly unaffected by the perturbation in the initial velocity due to viscous diffusion. \par

\begin{figure}
\centering
\begin{subfigure}{0.33\textwidth}
  \centering
  \includegraphics[width=0.85\linewidth]{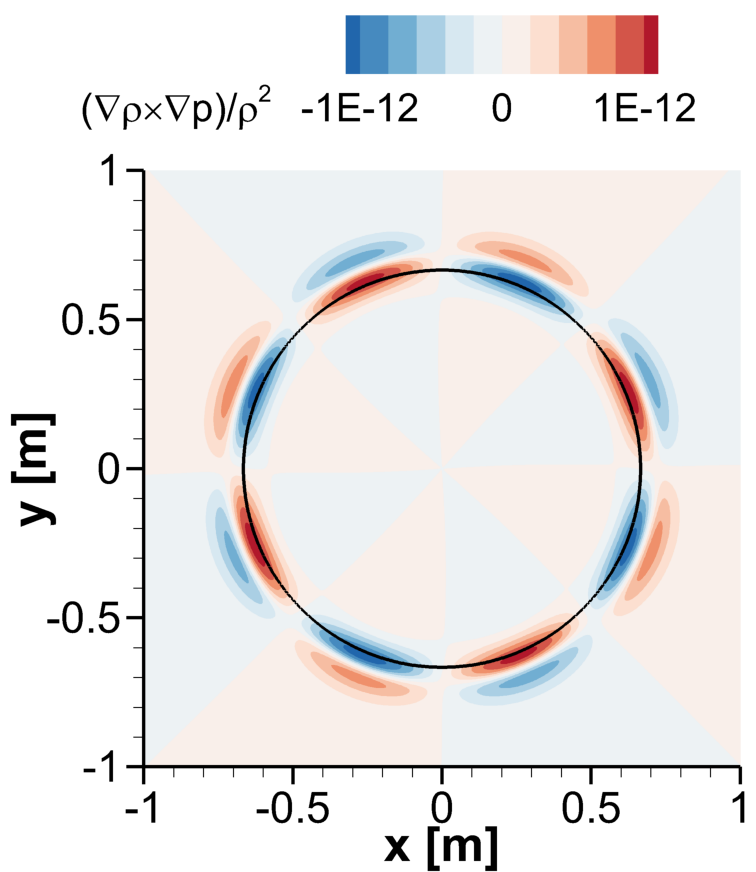}
  \caption{} 
  \label{subfig:baro_Re1000_512_16PI_initialoutput}
\end{subfigure}%
\begin{subfigure}{0.33\textwidth}
  \centering
  \includegraphics[width=1.0\linewidth]{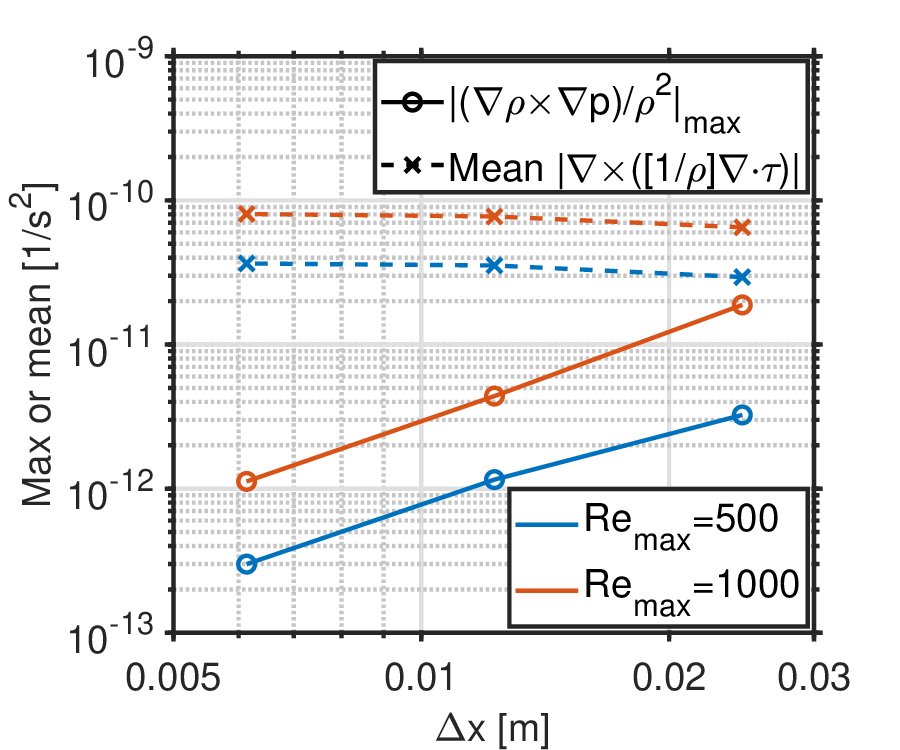}
  \caption{} 
  \label{subfig:MaxorMean_baro_visc_initialoutput}
\end{subfigure}%
\begin{subfigure}{0.33\textwidth}
  \centering
  \includegraphics[width=1.0\linewidth]{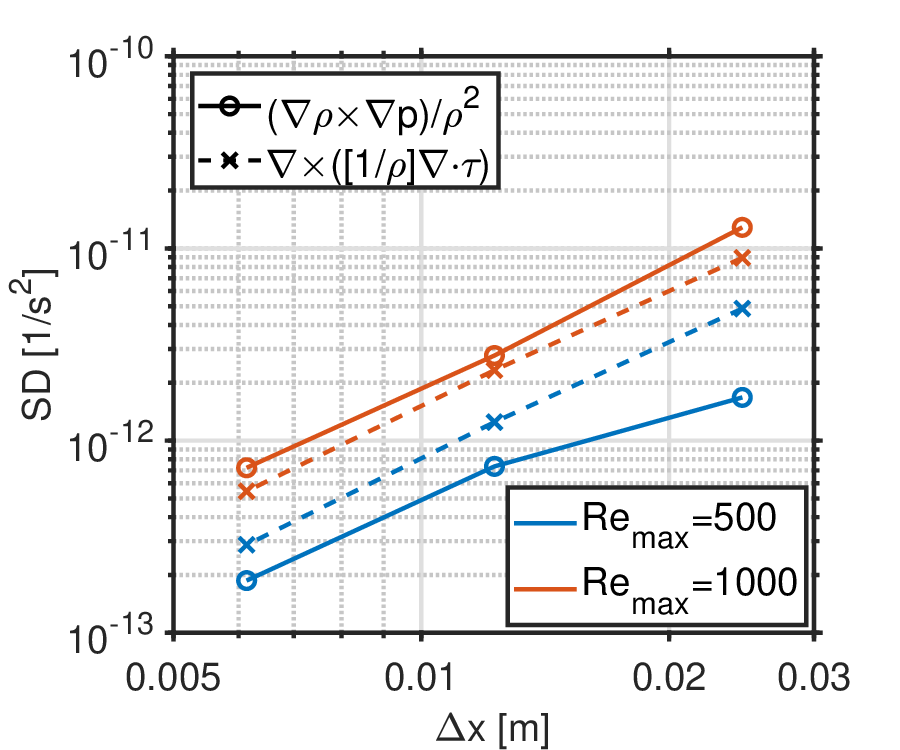}
  \caption{} 
  \label{subfig:SD_baro_visc_initialoutput}
\end{subfigure}%
\caption{RHS terms in the vorticity equation (\ref{eqn:vorticityZ}) at \(t^*\approx 0.085\) with a domain size of \(L=16\pi\) m. (a) \((\boldsymbol{\nabla}\rho\boldsymbol{\times}\boldsymbol{\nabla}p)/\rho^2\) for \(Re_\text{max}=1000\) and \(\Delta x=\pi/512\) m; (b) maximum absolute value of baroclinic torque in the vicinity of \(r_c\) and mean absolute value of viscous term along \(r_c\); and (c) SD of baroclinic torque and viscous term at \(r_c\).}
\label{fig:Fig29}
\end{figure}

Figure \ref{subfig:baro_Re1000_512_16PI_initialoutput} shows the baroclinic torque in the vicinity of the compressible vortex for \(Re_\text{max}=1000\), \(L=16\pi\) m and \(\Delta x=\pi/512\) m shortly after initialisation at \(t^*\approx 0.085\). Despite the perturbation in \(\boldsymbol{u}\), the compressible and viscous terms remain fairly axisymmetric due to the dominant effect of the radial stratification. Near the edge of the vortex with \(r\geq r_c\), the compressible term is negligible, but a competition between the destabilising effect of the baroclinic torque and the stabilising effect of viscous diffusion occurs, driving the destabilisation of the vortex -- separately from boundary effects. This destabilisation originates at \(r_c\) where the perturbations in the initial \(\boldsymbol{u}\) are stronger and the density gradient is the largest. Here, non-axisymmetric baroclinicity results from the misalignment between \(\boldsymbol{\nabla}\rho\) (which is approximately radial during the early times) and \(\boldsymbol{\nabla}p\) where \(p\) has non-axisymmetric modes due to the initial perturbation in \(\boldsymbol{u}\). \par

Figure \ref{subfig:MaxorMean_baro_visc_initialoutput} shows the evolution with \(\Delta x\) of the maximum absolute value of \((\boldsymbol{\nabla}\rho\boldsymbol{\times}\boldsymbol{\nabla}p)/\rho^2\) in the domain, which occurs near \(r_c\), as well as the mean absolute value of \(\boldsymbol{\nabla\times}([1/\rho]\boldsymbol{\nabla\cdot\tau})\) for \(Re_\text{max}=500\) and \(Re_\text{max}=1000\) at \(t^*\approx 0.085\). The standard deviations (SD) of both the baroclinic and viscous terms at \(r_c\) are shown in figure \ref{subfig:SD_baro_visc_initialoutput}. The increased vortex destabilisation is observed when the baroclinic torque is of similar magnitude to the viscous term, i.e., viscosity cannot diffuse \(\omega_z\) perturbations fast enough. As \(\Delta x\) is refined and the initial \(\boldsymbol{u}\) perturbation is smaller, both the maximum value of the baroclinic torque and the corresponding SD decrease. However, destabilisation occurs faster as \(Re\) increases. The viscous term doubles in magnitude if \(Re\) doubles, i.e., twice the vortex circulation results in twice the velocity field and its perturbations. In contrast, the baroclinic term becomes fourfold (the pressure field balances the centrifugal force scaling with \(u^2_\theta\)). Thus, the destabilising effect of \((\boldsymbol{\nabla}\rho\boldsymbol{\times}\boldsymbol{\nabla}p)/\rho^2\) can rapidly overcome the stabilising effect (via diffusion) of \(\boldsymbol{\nabla\times}([1/\rho]\boldsymbol{\nabla\cdot\tau})\). In other words, the baroclinic torque coupled to an initially perturbed velocity field is responsible for accelerating the vortex destabilisation. Note that with an operating pressure closer to the critical point, e.g., \(p_r=1.3\) instead of \(p_r=2\) in figure \ref{subfig:validation_den}, \(\boldsymbol{\nabla}\rho\) across the thermal mixing layer increases in magnitude and also enhances the destabilising effect of the baroclinic torque. Thus, lower \(Re_\text{max}\) vortices in pressure conditions closer to \(p_c\) may also become unstable. Such instability mechanism promotes the emergence of centrifugal or Rayleigh-Taylor instabilities in dense vortex cores, as shown in figure \ref{fig:Fig28}. Note that if the density is a function of pressure only, i.e., barotropic flow, the baroclinic torque is exactly zero, improving the axisymmetric behaviour of the numerical solution. \par 

This brief analysis proves that robust numerical approaches are required to solve the governing equations and impose accurate boundary conditions, which are usually overlooked in the incompressible limit, and avoid the generation of non-physical turbulence resulting from the interaction between physical mechanisms and ill-defined numerical setups. This is particularly problematic if one aims to study, e.g., anti-parallel vortex reconnection \citep{2022_ARFM_Yao}, in SCF to understand the role of vorticity dynamics in turbulence cascade. \par

\section{Incompressible Oseen Vortex}
\label{apn:B}

\setcounter{figure}{0}

\subsection{Numerical Setup Validation}
\label{subapn:B1}

\begin{figure}
\centering
\begin{subfigure}{0.33\textwidth}
  \centering
  \includegraphics[width=1.0\linewidth]{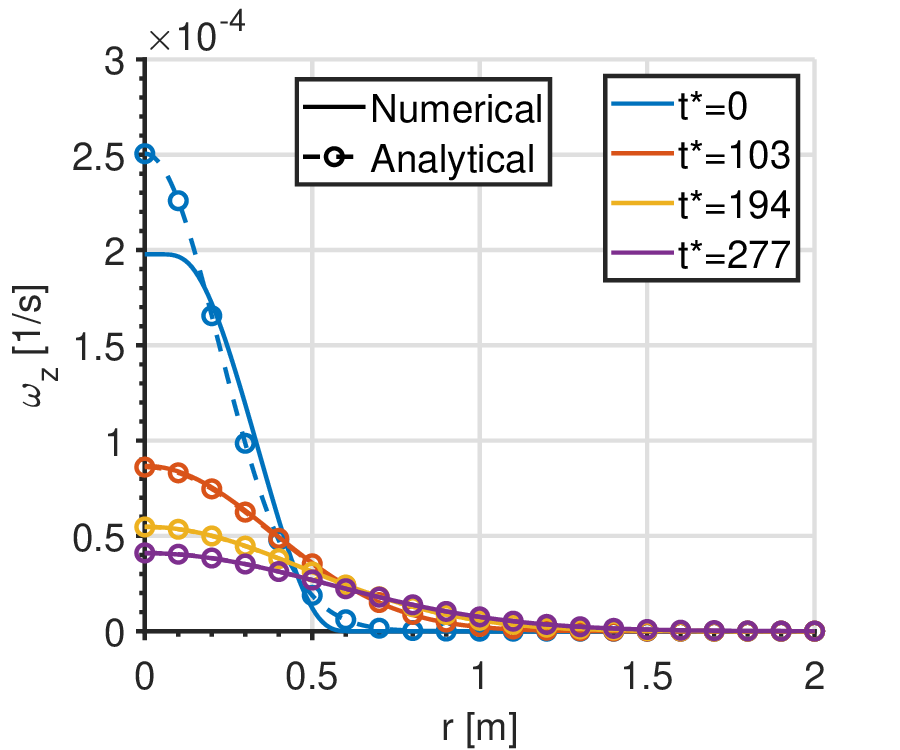}
  \caption{} 
  \label{subfig:vor_profile_incomp}
\end{subfigure}%
\begin{subfigure}{0.33\textwidth}
  \centering
  \includegraphics[width=1.0\linewidth]{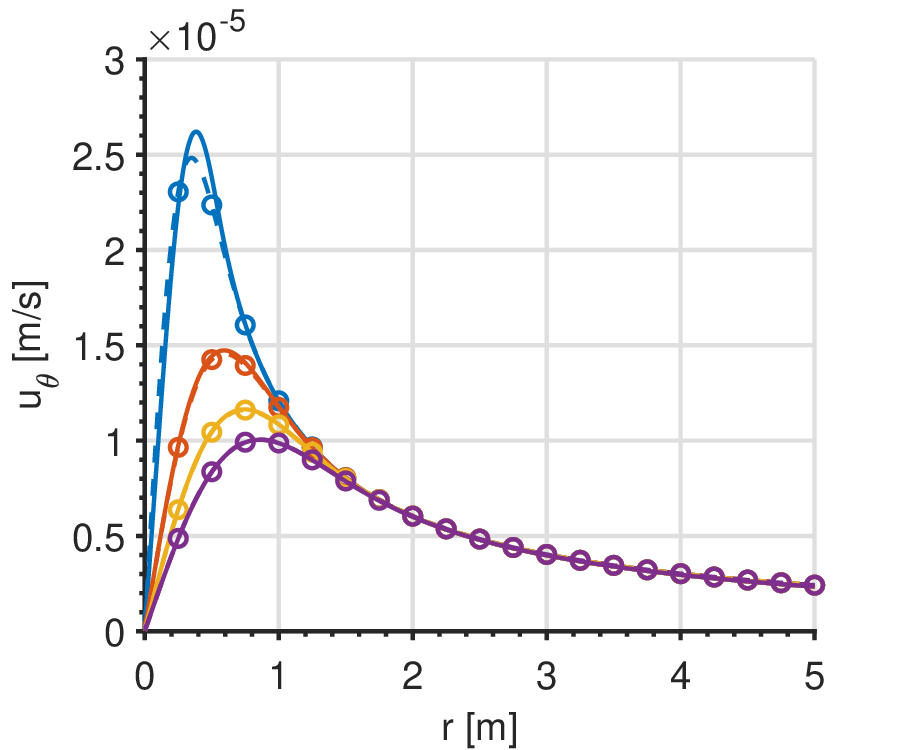}
  \caption{} 
  \label{subfig:utheta_profile_incomp}
\end{subfigure}%
\begin{subfigure}{0.33\textwidth}
  \centering
  \includegraphics[width=1.0\linewidth]{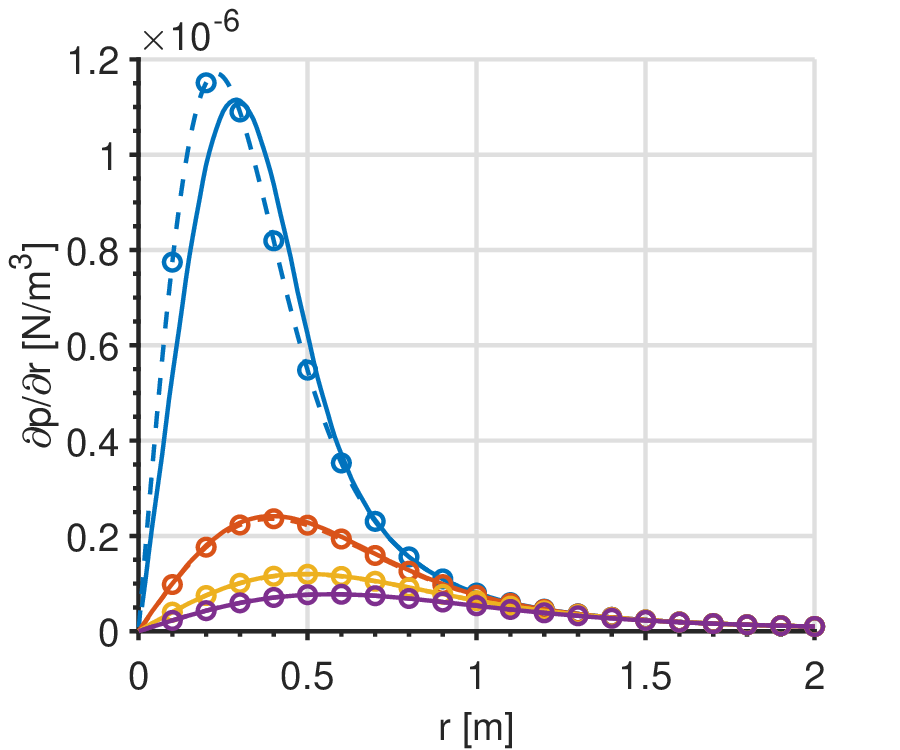}
  \caption{} 
  \label{subfig:dpdr_profile_incomp}
\end{subfigure}%
\caption{Validation of the incompressible vortex with \(Re_\text{max}=1000\), \(L=16\pi\) m and \(\Delta x = \pi/256\) m against Oseen's analytical solution. (a) \(\omega_z\); (b) \(u_\theta\); and (c) \(\partial p/\partial r\).}
\label{fig:Fig30}
\end{figure}

The incompressible vortex is validated against Oseen's analytical solution \citep{1912_AMA_Oseen} in figure \ref{fig:Fig30} for \(Re_\text{max}=1000\) with a domain size of \(L=16\pi\) m and \(\Delta x = \pi/256\) m. The initial conditions at \(t^*=0\) differ due to the compact support vorticity profile used in this work, that is, Oseen's formulation assumes a Gaussian vorticity distribution given by (\ref{eqn:OseenVor}) with the resulting azimuthal velocity given by (\ref{eqn:OseenVor2}). An initial time \(t_0\) for the analytical solution is defined by imposing \(\omega_z(r_c,t_0)=0.01\omega_z(0,t_0)\), which closely matches the initial distribution of \(u_\theta\) from the compact vorticity form. In addition, the pressure gradient provides the centripetal force to keep a fluid particle moving in a circular path, i.e., \(\frac{\partial p}{\partial r}(r,t)=\rho(u_\theta^2/r)\). Despite the initial difference between the Gaussian and compact forms, the numerical solution relaxes to Oseen's solution and good agreement is observed.

\begin{equation}
\label{eqn:OseenVor}
\omega_z(r,t)=\frac{\Gamma_0}{4\pi\nu t}\exp{[-r^2/(4\nu t)]}
\end{equation}
\begin{equation}
\label{eqn:OseenVor2}
u_\theta(r,t)=\frac{\Gamma_0}{2\pi r}(1-\exp{[-r^2/(4\nu t)]}) 
\end{equation}

\subsection{Vorticity Viscous Diffusion}
\label{subapn:B2}

The vorticity diffusion term due to viscous effects in the vorticity equation (\ref{eqn:vorticityZ}) is \(\nu\boldsymbol{\nabla^2}\omega_z\) in the incompressible limit. In cylindrical coordinates and under axisymmetry conditions, it can be written as

\begin{equation}
\label{eqn:OseenVor3}
\nu\boldsymbol{\nabla^2}\omega_z = \frac{1}{r}\frac{\partial}{\partial r}\bigg(r\frac{\partial\omega_z}{\partial r}\bigg)=\frac{\partial^2\omega_z}{\partial r^2}+\frac{1}{r}\frac{\partial \omega_z}{\partial r}
\end{equation}

Applied to the Oseen vortex \citep{1912_AMA_Oseen} with \(\omega_z\) given by (\ref{eqn:OseenVor}), the partial derivatives of \(\omega_z\) in (\ref{eqn:OseenVor3}) are simplified to

\begin{equation}
\label{eqn:OseenVor4}
\frac{\partial\omega_z}{\partial r}(r,t)=-\omega_z(r,t)\bigg(\frac{r}{2\nu t}\bigg)
\end{equation}
\begin{equation}
\label{eqn:OseenVor5}
\frac{\partial^2\omega_z}{\partial r^2}(r,t)=\frac{\omega_z(r,t)}{2\nu t}\bigg(\frac{r^2}{2\nu t}-1\bigg)
\end{equation}

\begin{figure}
\centering
\includegraphics[width=0.4\linewidth]{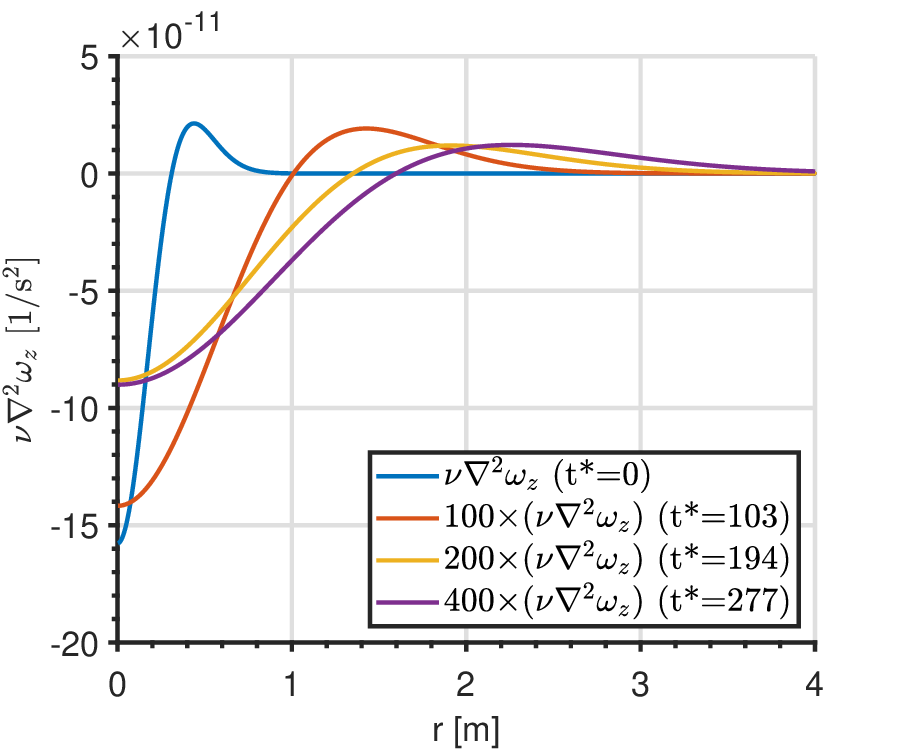}
\caption{Evolution of \(\nu\boldsymbol{\nabla^2}\omega_z\) with \(t^*\) in the Oseen vortex for \(Re_\text{max}=200\). A scaling factor is included to magnify the profiles at later \(t^*\).}
\label{fig:Fig31}
\end{figure}

The vorticity viscous diffusion term is plotted in figure \ref{fig:Fig31} at various \(t^*\) for the vortex with \(Re_\text{max}=200\). The vortex is initialised at time \(t_0\) as described in \ref{apn:A}. Then, \(t^*\) refers to the time from \(t^*_0\), i.e., \(t^*=0\) at \(t^*_0\). Since the magnitude of \(\nu\boldsymbol{\nabla^2}\omega_z\) decays rapidly as vorticity diffuses, the profiles in figure \ref{fig:Fig31} are scaled by a given amount for visualisation purposes. \par 

\section{Vorticity Contour Plots}
\label{apn:C}

\setcounter{figure}{0}

\begin{figure}
\centering
\begin{subfigure}{0.25\textwidth}
  \centering
  \includegraphics[width=1.0\linewidth]{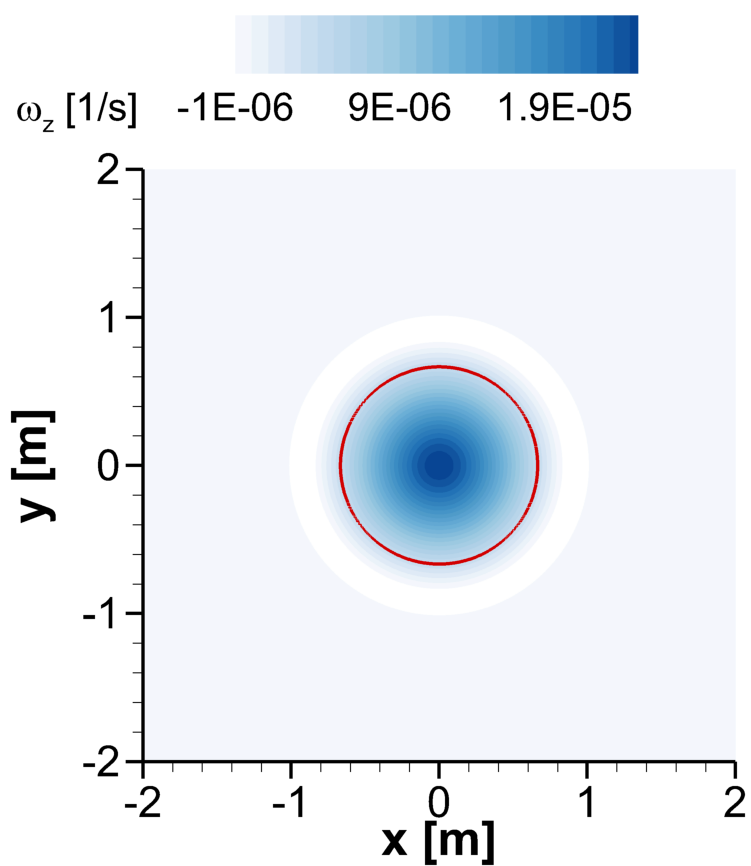}
  \caption{} 
  \label{subfig:wz_Re200_512_16PI_pr2_cold_002}
\end{subfigure}%
\begin{subfigure}{0.25\textwidth}
  \centering
  \includegraphics[width=1.0\linewidth]{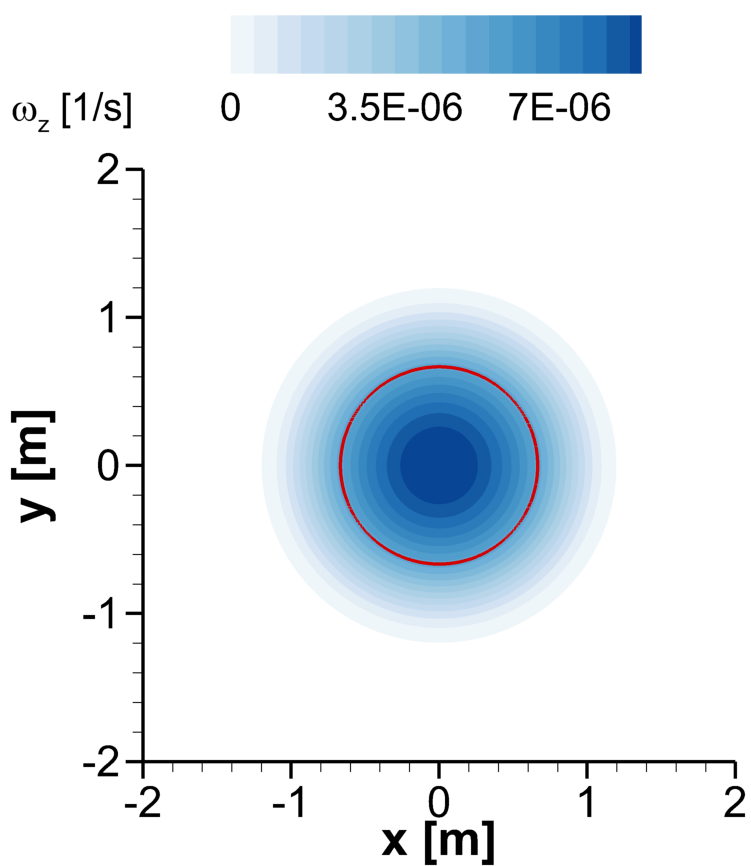}
  \caption{} 
  \label{subfig:wz_Re200_512_16PI_pr2_cold_012}
\end{subfigure}%
\begin{subfigure}{0.25\textwidth}
  \centering
  \includegraphics[width=1.0\linewidth]{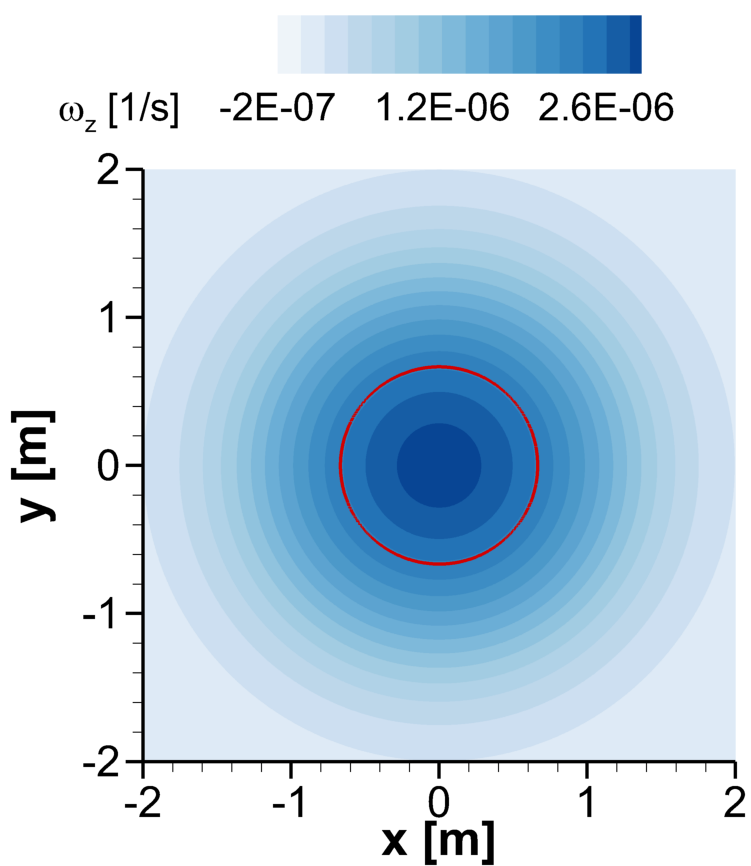}
  \caption{} 
  \label{subfig:wz_Re200_512_16PI_pr2_cold_046}
\end{subfigure}%
\\
\begin{subfigure}{0.25\textwidth}
  \centering
  \includegraphics[width=1.0\linewidth]{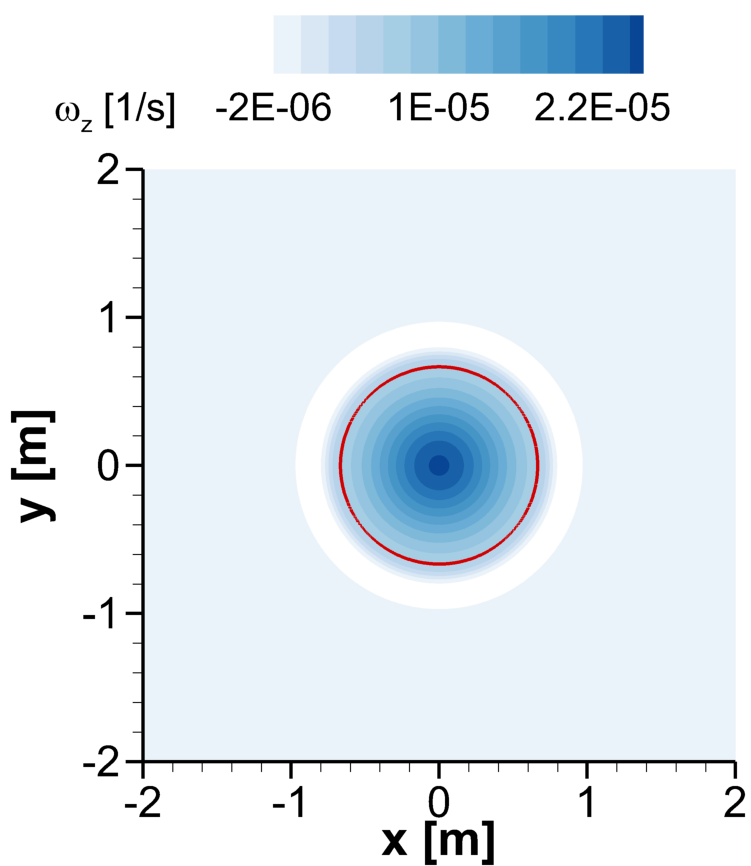}
  \caption{} 
  \label{subfig:wz_Re200_512_16PI_pr1p5_cold_002}
\end{subfigure}%
\begin{subfigure}{0.25\textwidth}
  \centering
  \includegraphics[width=1.0\linewidth]{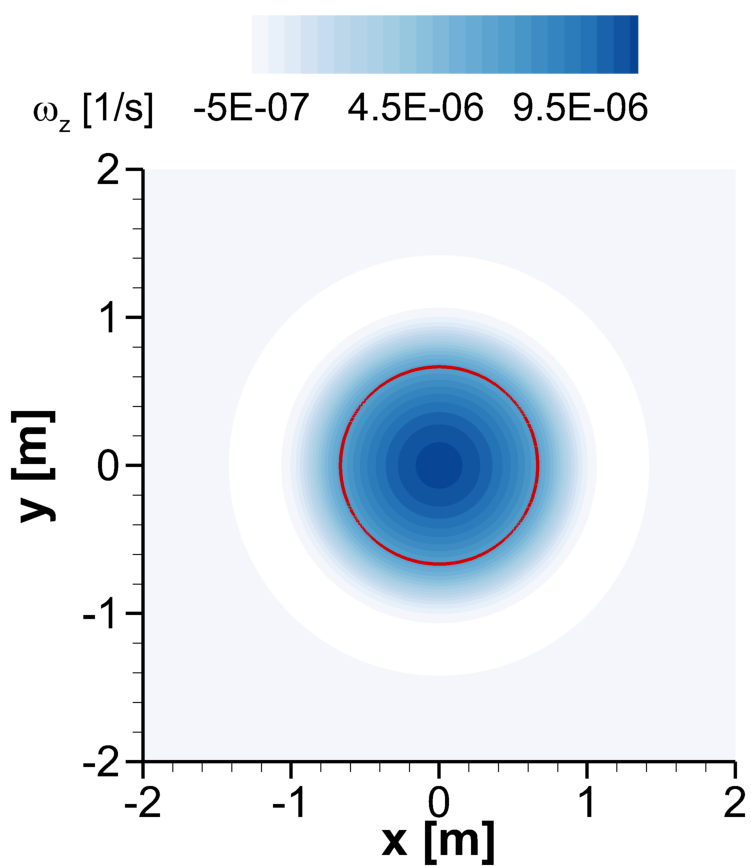}
  \caption{} 
  \label{subfig:wz_Re200_512_16PI_pr1p5_cold_011}
\end{subfigure}%
\begin{subfigure}{0.25\textwidth}
  \centering
  \includegraphics[width=1.0\linewidth]{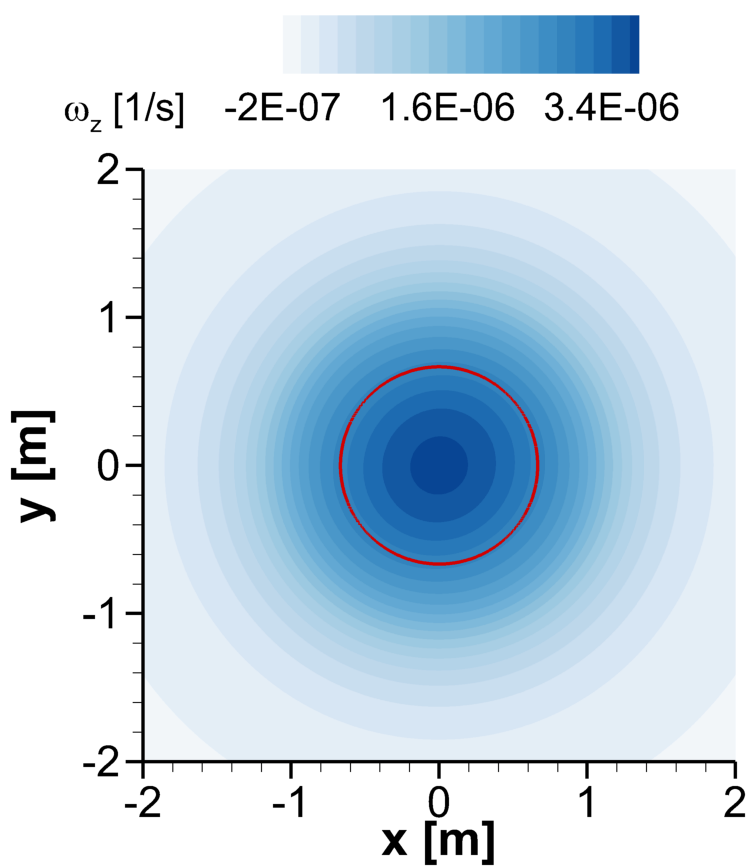}
  \caption{} 
  \label{subfig:wz_Re200_512_16PI_pr1p5_cold_044}
\end{subfigure}%
\\
\begin{subfigure}{0.25\textwidth}
  \centering
  \includegraphics[width=1.0\linewidth]{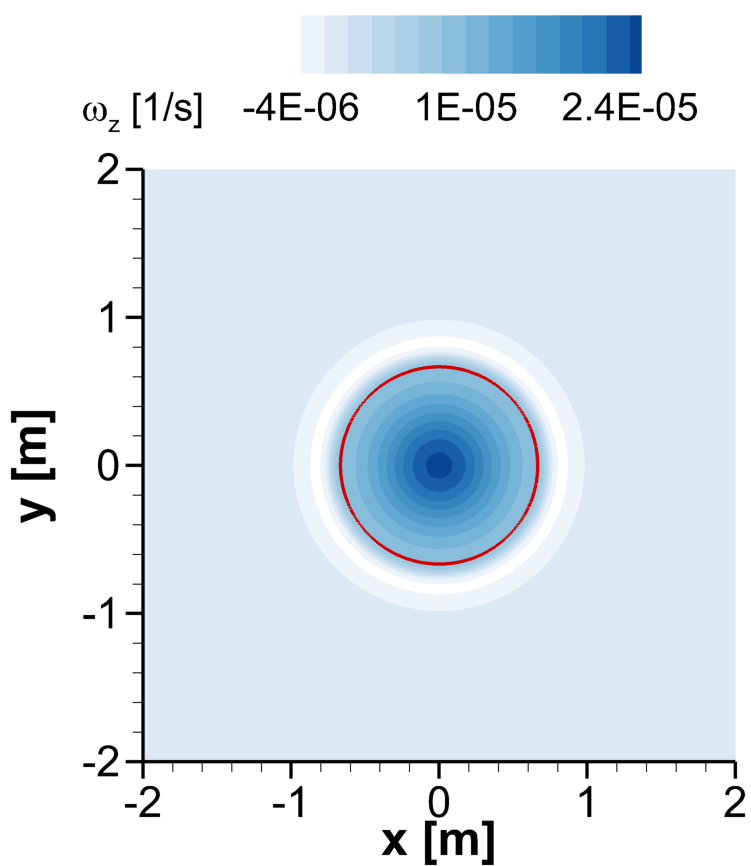}
  \caption{} 
  \label{subfig:wz_Re200_512_16PI_pr1p3_cold_002}
\end{subfigure}%
\begin{subfigure}{0.25\textwidth}
  \centering
  \includegraphics[width=1.0\linewidth]{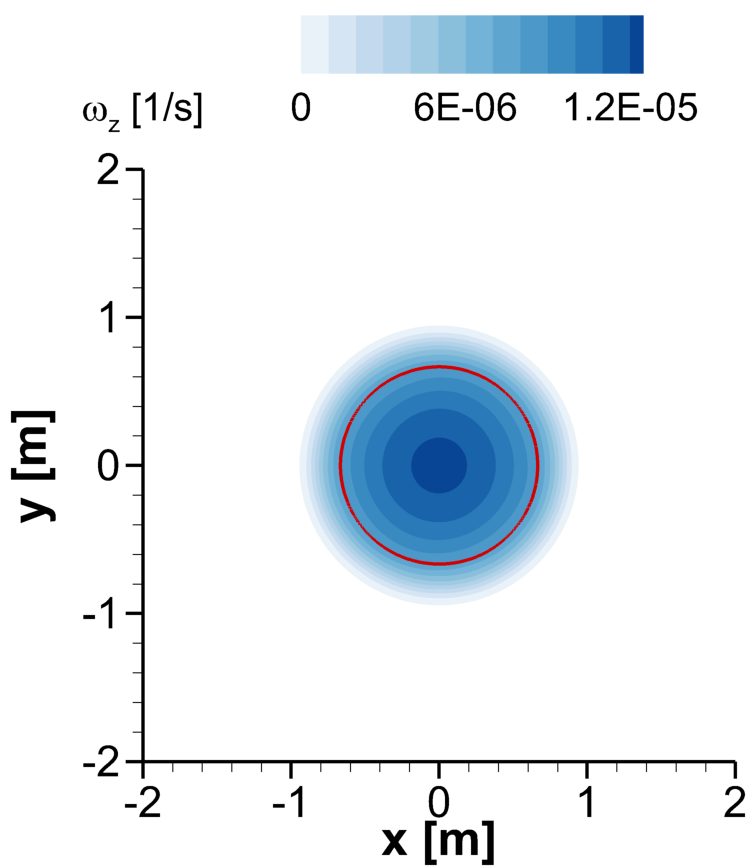}
  \caption{} 
  \label{subfig:wz_Re200_512_16PI_pr1p3_cold_011}
\end{subfigure}%
\begin{subfigure}{0.25\textwidth}
  \centering
  \includegraphics[width=1.0\linewidth]{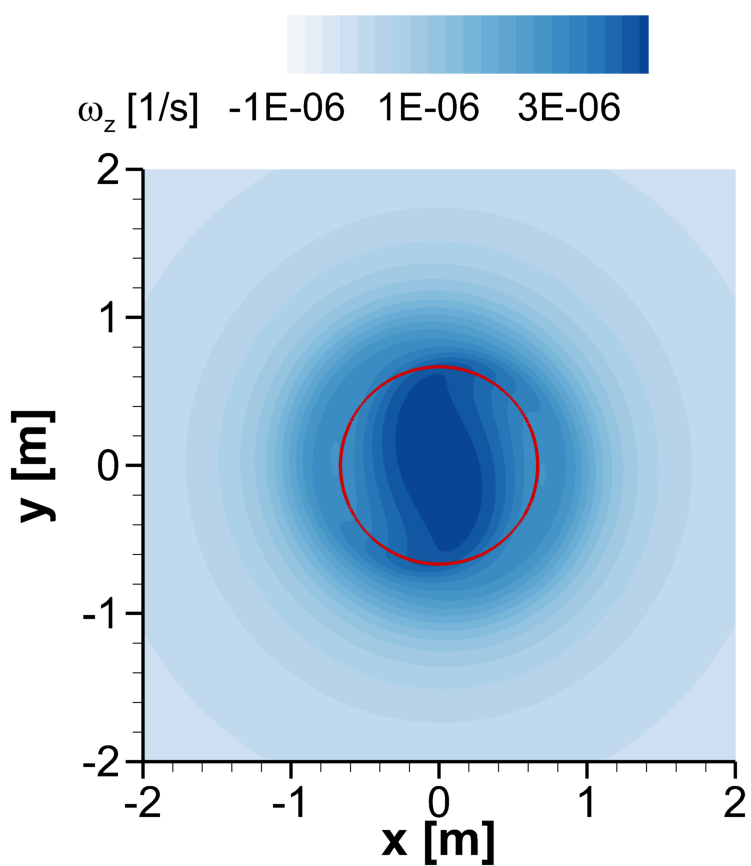}
  \caption{} 
  \label{subfig:wz_Re200_512_16PI_pr1p3_cold_042}
\end{subfigure}%
\caption{Contours of \(\omega_z\) for the cold core. The red iso-contour represents \(r_c\). (a) \(t^+=0.0562\) at \(p_r=2\); (b) \(t^+=0.3370\) at \(p_r=2\); (c) \(t^+=1.2917\) at \(p_r=2\); (d) \(t^+=0.0589\) at \(p_r=1.5\); (e) \(t^+=0.3241\) at \(p_r=1.5\); (f) \(t^+=1.2964\) at \(p_r=1.5\); (g) \(t^+=0.0618\) at \(p_r=1.3\); (h) \(t^+=0.3399\) at \(p_r=1.3\); and (i) \(t^+=1.2980\) at \(p_r=1.3\).}
\label{fig:Fig32}
\end{figure}

The multi-dimensional nature of the fluid solver used for this study is highlighted in the following figures displaying the contour plots of vorticity for all analysed cases. As discussed in previous text, the solution remains nearly axisymmetric for the analysed times and, therefore, radial profiles have been extracted for the study. The only exception is the \(p_r=1.3\) case with a cold vortex core, which develops clear non-axisymmetric modes as time advances (see figure \ref{subfig:wz_Re200_512_16PI_pr1p3_cold_042} and RSD values of \(u_\theta\) reported in table \ref{tab:RSD_analysed_pr_configurations}). As stated in section \ref{subsec:solution_collapse}, this deviation from a purely axisymmetric solution is accepted. \par 

\begin{figure}
\centering
\begin{subfigure}{0.25\textwidth}
  \centering
  \includegraphics[width=1.0\linewidth]{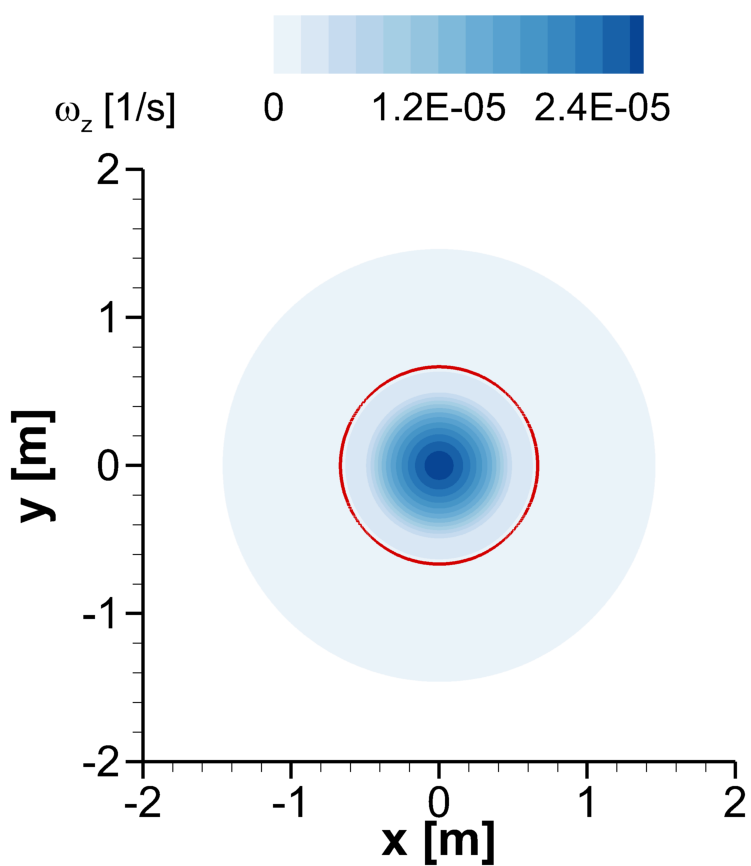}
  \caption{} 
  \label{subfig:wz_Re200_512_16PI_pr2_hot_002}
\end{subfigure}%
\begin{subfigure}{0.25\textwidth}
  \centering
  \includegraphics[width=1.0\linewidth]{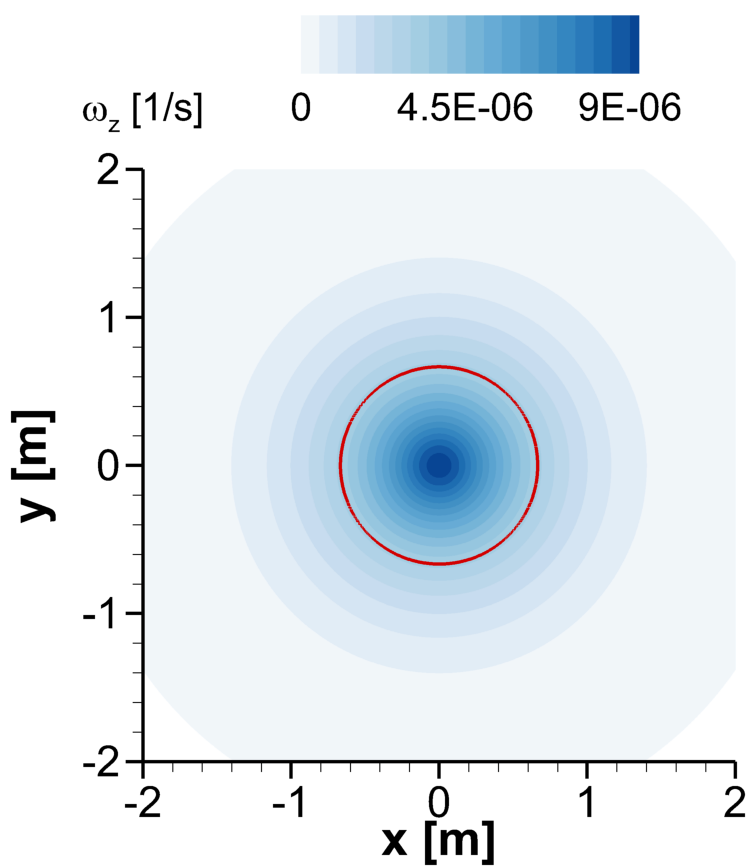}
  \caption{} 
  \label{subfig:wz_Re200_512_16PI_pr2_hot_012}
\end{subfigure}%
\begin{subfigure}{0.25\textwidth}
  \centering
  \includegraphics[width=1.0\linewidth]{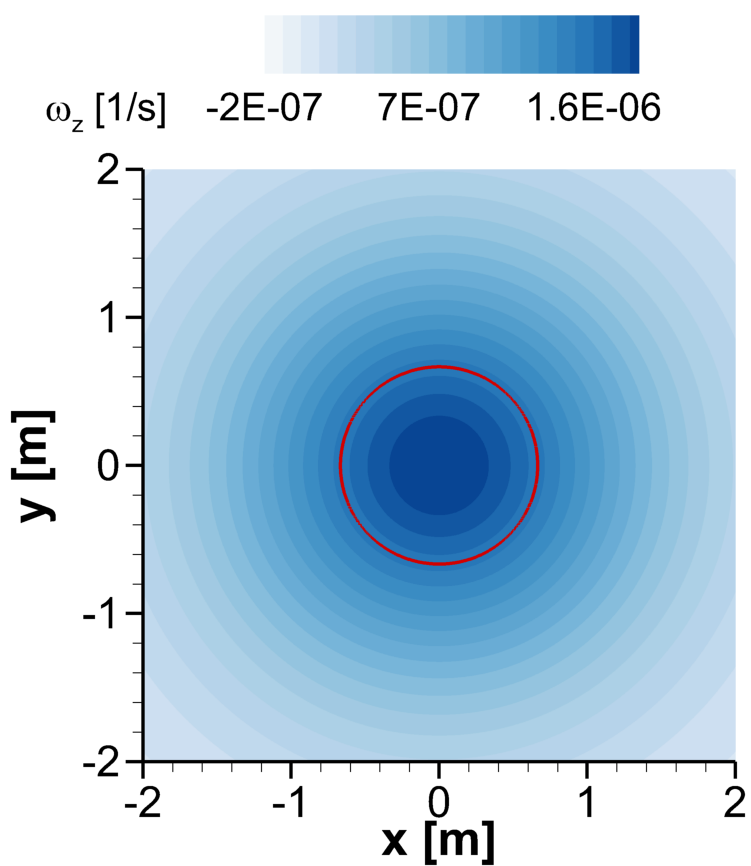}
  \caption{} 
  \label{subfig:wz_Re200_512_16PI_pr2_hot_046}
\end{subfigure}%
\\
\begin{subfigure}{0.25\textwidth}
  \centering
  \includegraphics[width=1.0\linewidth]{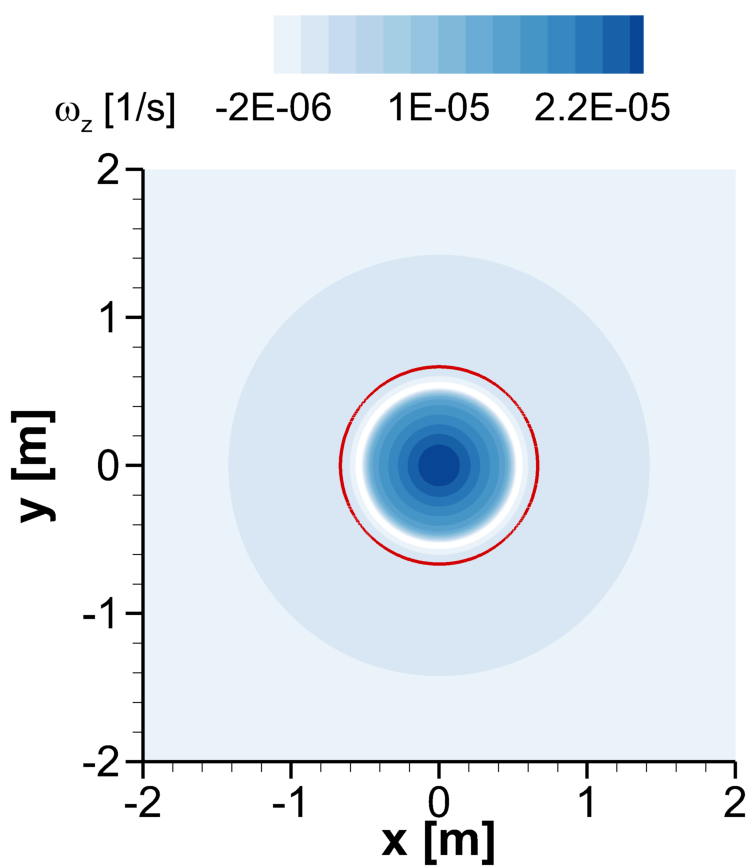}
  \caption{} 
  \label{subfig:wz_Re200_512_16PI_pr1p5_hot_002}
\end{subfigure}%
\begin{subfigure}{0.25\textwidth}
  \centering
  \includegraphics[width=1.0\linewidth]{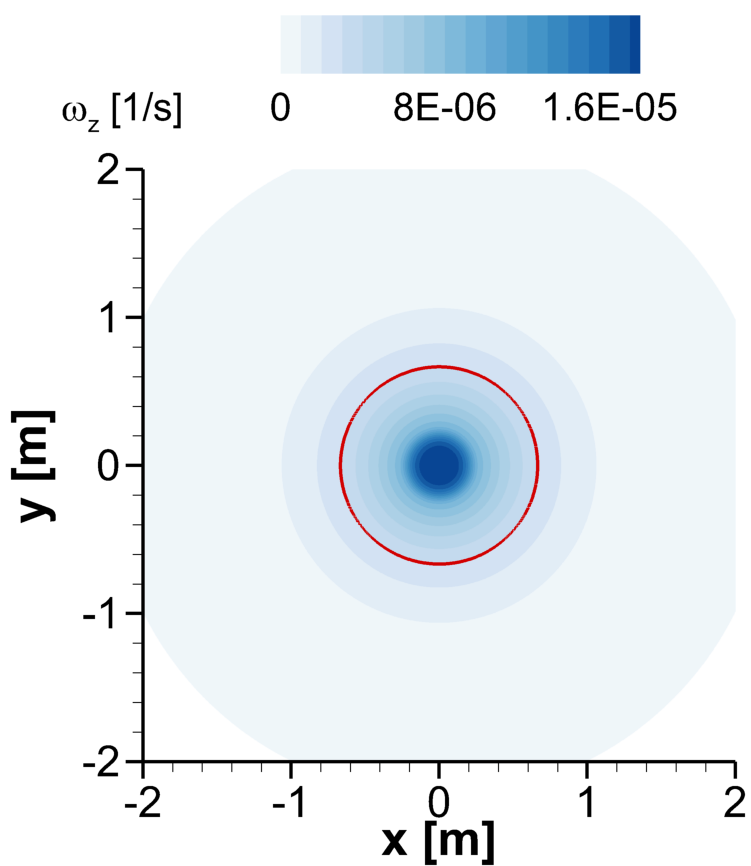}
  \caption{} 
  \label{subfig:wz_Re200_512_16PI_pr1p5_hot_011}
\end{subfigure}%
\begin{subfigure}{0.25\textwidth}
  \centering
  \includegraphics[width=1.0\linewidth]{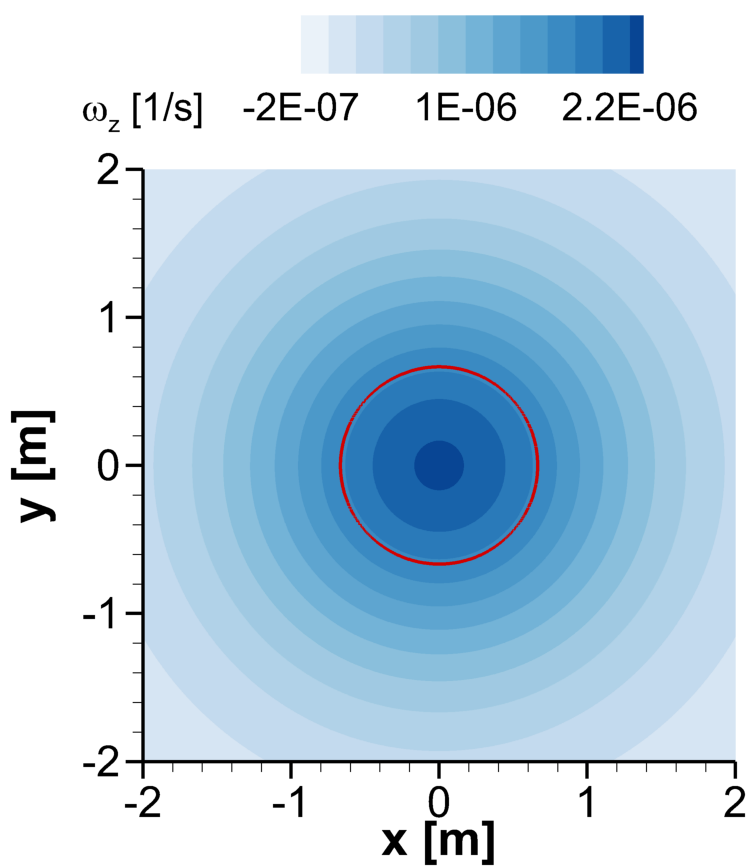}
  \caption{} 
  \label{subfig:wz_Re200_512_16PI_pr1p5_hot_044}
\end{subfigure}%
\\
\begin{subfigure}{0.25\textwidth}
  \centering
  \includegraphics[width=1.0\linewidth]{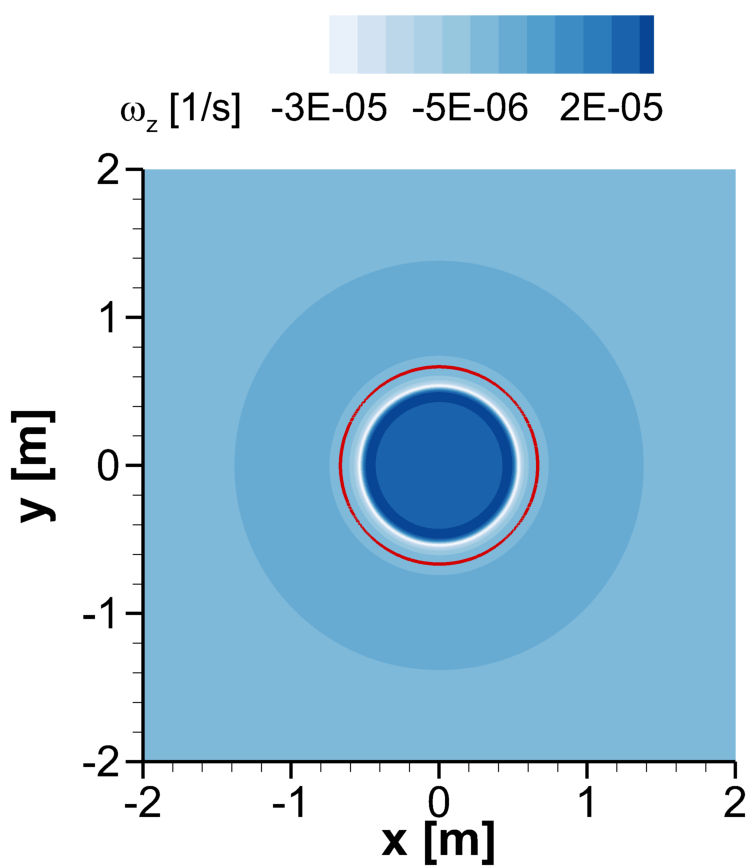}
  \caption{} 
  \label{subfig:wz_Re200_512_16PI_pr1p3_hot_002}
\end{subfigure}%
\begin{subfigure}{0.25\textwidth}
  \centering
  \includegraphics[width=1.0\linewidth]{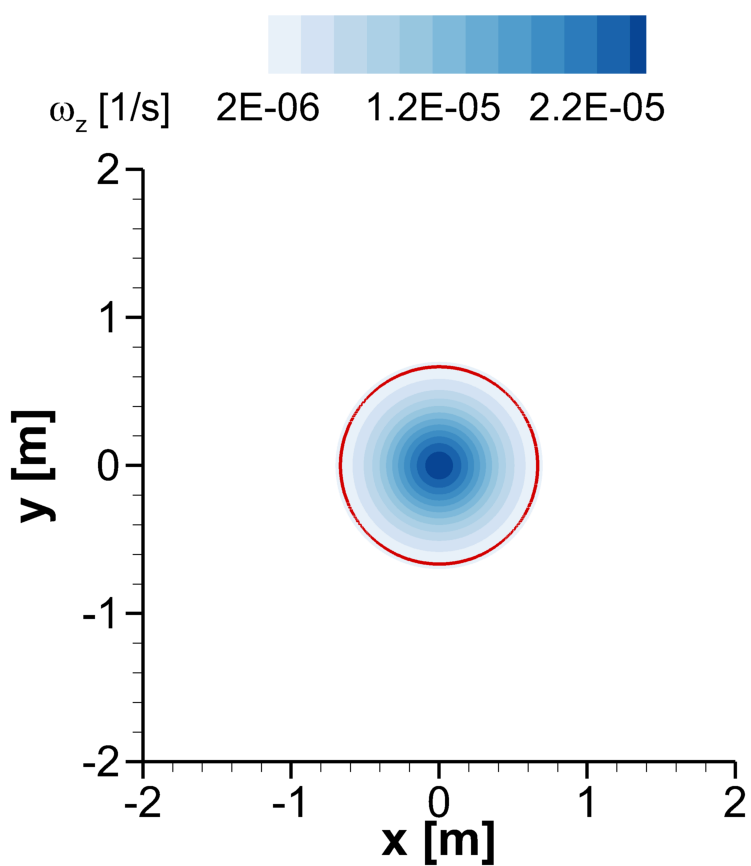}
  \caption{} 
  \label{subfig:wz_Re200_512_16PI_pr1p3_hot_011}
\end{subfigure}%
\begin{subfigure}{0.25\textwidth}
  \centering
  \includegraphics[width=1.0\linewidth]{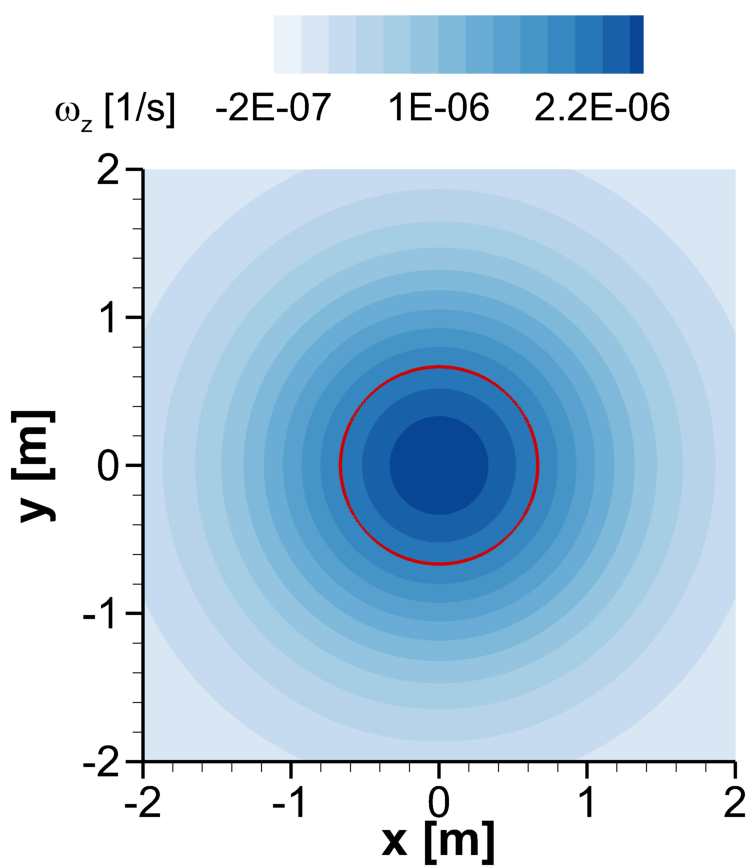}
  \caption{} 
  \label{subfig:wz_Re200_512_16PI_pr1p3_hot_042}
\end{subfigure}%
\caption{Contours of \(\omega_z\) for the hot core. The red iso-contour represents \(r_c\). (a) \(t^+=0.0562\) at \(p_r=2\); (b) \(t^+=0.3370\) at \(p_r=2\); (c) \(t^+=1.2917\) at \(p_r=2\); (d) \(t^+=0.0589\) at \(p_r=1.5\); (e) \(t^+=0.3241\) at \(p_r=1.5\); (f) \(t^+=1.2964\) at \(p_r=1.5\); (g) \(t^+=0.0618\) at \(p_r=1.3\); (h) \(t^+=0.3399\) at \(p_r=1.3\); and (i) \(t^+=1.2980\) at \(p_r=1.3\).}
\label{fig:Fig33}
\end{figure}

\newpage
\typeout{}
\bibliography{journal_bib}


\end{document}